\documentclass[11pt,a4paper]{article}

\usepackage{jcappub}
\usepackage[margin=0.95in]{geometry}

\usepackage[utf8]{inputenc}
\usepackage{graphicx}
\usepackage{tensor}
\usepackage{verbatim}
\usepackage{epsfig}
\usepackage{subfigure}
\usepackage{color}
\usepackage{multirow}
\usepackage{comment}
\usepackage{slashed}
\usepackage{amsmath}
\usepackage{ulem}
\usepackage{framed}

\usepackage{amssymb,amsmath,amsfonts,mathrsfs}
\usepackage[dvipsnames]{xcolor}
\usepackage{graphicx}
\usepackage{longtable}
\usepackage{verbatim}
\usepackage{color}
\usepackage[mathscr]{euscript}
\usepackage{framed}
\usepackage{mdframed} 

\usepackage{cancel}
\usepackage{color}
\usepackage{hyperref}
\hypersetup{colorlinks, citecolor=bluscuro, linkcolor=black, urlcolor=bluscuro}
\definecolor{rossos}{cmyk}{0,1,1,0.55}
\definecolor{bluscuro}{rgb}{0.15, 0.2, .85}
\definecolor{bluchiaro}{cmyk}{1,.3,0.,0.1}

\numberwithin{equation}{section}

\usepackage{tikz}

\usepackage{tcolorbox}

\newcommand{\llp}{\left [}
\newcommand{\rrp}{\right ]}

\newcommand{\lp}{\left (}
\newcommand{\rp}{\right )}
\newcommand{\GW}{\text{\tiny GW}}
\newcommand{\co}{\text{\tiny cut-off}}
\def\ii{{\text{\tiny i}}}

\def\PBH{\text{\tiny  PBH}}

\newcommand{\be}{\begin{equation}\begin{aligned}}
\newcommand{\ee}{\end{aligned}\end{equation}}

\newcommand{\bbe}{\begin{align}}
\newcommand{\eee}{\end{align}}

\newcommand{\bea}{\begin{eqnarray}}
\newcommand{\eea}{\end{eqnarray}}

\def\beq{\begin{equation}}
\def\eeq{\end{equation}}

\def\d{{\rm d}}

\def\beqa{\begin{eqnarray}}

	\def\eeqa{\end{eqnarray}}
	
\def\lsim{\mathrel{\rlap{\lower4pt\hbox{\hskip0.5pt$\sim$}}
		\raise1pt\hbox{$<$}}}     
\def\gsim{\mathrel{\rlap{\lower4pt\hbox{\hskip0.5pt$\sim$}}
		\raise1pt\hbox{$>$}}}     

\def\d{{\rm d}}

\def\d{{\rm d}}

\def\PBH{\text{\tiny \rm PBH}}

\def\eeqa{\end{eqnarray}}

\def\bq{\begin{quote}}
\def\eq{\end{quote}}

 at 10truept
\newcommand{\arXiv}[2]{\href{http://arxiv.org/pdf/#1}{{\tt [#2/#1]}}}
\newcommand{\arXivold}[1]{\href{http://arxiv.org/pdf/#1}{{\tt [#1]}}}

\def\eeqa{\end{eqnarray}}
\def\lsim{\mathrel{\rlap{\lower4pt\hbox{\hskip0.5pt$\sim$}}
  \raise1pt\hbox{$<$}}}     
\def\gsim{\mathrel{\rlap{\lower4pt\hbox{\hskip0.5pt$\sim$}}
  \raise1pt\hbox{$>$}}}     

\title{Primordial Black Holes Confront LIGO/Virgo data: Current situation}
\author[a]{V. De Luca,}
\author[a,b]{G. Franciolini,}
\author[c,d]{P. Pani,}
\author[a,d]{A. Riotto}

\affiliation[a]{
	Department of Theoretical Physics and Center for Astroparticle Physics (CAP) \\
			24 quai E. Ansermet, CH-1211 Geneva 4, Switzerland}
\affiliation[b]{Instituto de F\'isica Te\'orica UAM-CSIC, Universidad Aut\'onoma de Madrid, Cantoblanco, Madrid, 28049 Spain}
\affiliation[c]{Dipartimento di Fisica, “Sapienza” Università di Roma, Piazzale Aldo Moro 5, 00185, Roma, Italy}
\affiliation[d]{INFN, Sezione di Roma, Piazzale Aldo Moro 2, 00185, Roma, Italy}

\abstract{
The LIGO and Virgo Interferometers have so far provided 11 gravitational-wave~(GW) observations of black-hole 
binaries. Similar detections are bound to become very frequent in the near future. With the current and upcoming wealth 
of data, it is possible to confront specific formation models with observations. We investigate here whether current 
data are compatible with the hypothesis that LIGO/Virgo black holes are of primordial origin.
We compute in detail the mass and spin distributions of primordial black holes (PBHs), their merger rates, the 
stochastic background of unresolved coalescences, and confront them with current data from the first two observational 
runs, also including the recently discovered GW190412.
We compute the best-fit values for the parameters of the PBH mass distribution at formation that are compatible with 
current GW data. In all cases, the maximum fraction of PBHs in dark matter is constrained by these 
observations to be $f_\PBH\approx {\rm few}\times 10^{-3}$.
We discuss the predictions of the PBH scenario that can be directly tested as new data become available.
 In the most likely formation scenarios where  PBHs are born with negligible spin, the fact 
that at least one of the components of GW190412 is moderately spinning is incompatible with a primordial origin for 
this event, unless accretion or hierarchical mergers are significant.
In the absence of accretion, current non-GW constraints already exclude that LIGO/Virgo events are all of primordial 
origin, whereas in the presence of accretion the GW bounds on the PBH abundance are the most stringent ones in the 
relevant mass range.
A strong phase of accretion during the cosmic history would favour mass ratios close to unity, and a redshift-dependent 
correlation between high masses, high spins and nearly-equal mass binaries, with the secondary component spinning 
faster than the primary. Finally, we highlight that accretion can play 
an important role to relax current constraints on the PBH abundance, which calls for a better modelling of the mass and 
angular momentum accretion rates at redshift $z\lesssim30$. 
}
\emailAdd{valerio.deluca@unige.ch}
\emailAdd{gabriele.franciolini@unige.ch}
\emailAdd{paolo.pani@uniroma1.it}
\emailAdd{antonio.riotto@unige.ch}
\begin{document}

\maketitle
\flushbottom

\section{Introduction}
Thanks to the current available measurements of the Gravitational Waves~(GWs) from Black Holes~(BHs) mergers in the 
observational runs O1  and O2 \cite{LIGOScientific:2018mvr} and the more recent ongoing third phase 
\cite{LIGOScientific:2020stg} by the LIGO/Virgo collaborations, we have entered the era of GW astronomy. One of the most 
fundamental questions such observations raise is the nature of the BHs~\cite{Barack:2018yly}. One fascinating 
hypothesis is that the merging  BHs are of primordial origin, that 
is, they have formed early in the evolution of the universe (see Ref.~\cite{Sasaki:2018dmp} for a recent review). This 
possibility is also interesting as Primordial Black Holes~(PBHs) may comprise the totality or a fraction of the Dark 
Matter~(DM) in the universe~\cite{Bird:2016dcv,Sasaki:2016jop}.
	
In order to assess if the merger events may be ascribed to PBHs one needs to go through various steps:

\begin{enumerate} 
 \item First, one starts from a given mass function distribution (which might be theoretically justified
by a given PBH formation mechanism). Such mass function is determined by a set of parameters (typically two of them) 
parametrising the characteristic PBH mass scale and the width of the distribution. Their central values may be 
estimated from observations by requiring that all GW events detected so far (or a fraction thereof) are explained by 
PBHs. This requires fitting the key observables, namely the merger rate, the BH masses, and the redshift at which the 
event is produced, resulting in the best-fit value for the PBH abundance $f_\PBH$ in units of the DM one. 

\item Secondly, one has to check if the resulting model parameters and $f_\PBH$ are compatible with the current 
constraints from other observations (e.g. lensing and CMB distortion) \cite{Carr:2020gox}. This second step is necessary 
to check which fraction of the PBHs may form the DM and, above all, to see if the observed events are compatible with 
the PBH scenario, or if their primordial origin is already excluded by other observations. 

\item Thirdly, one can confront the theoretical predictions obtained through the PBH hypothesis of some key quantities, 
e.g. the binary chirp masses, mass ratios, spins, with the observed values, thus assessing whether the predictions are 
compatible or in tension with observations. 

\end{enumerate} 
The goal of this paper is to explore if the first two observational LIGO/Virgo runs --~also including the event 
GW190412 recently discovered in O3~--  are compatible with the hypothesis that the BHs are of primordial origin. In 
particular, we calculate the mass and spin distributions of PBHs, their merger rates, the stochastic background of 
unresolved coalescences, and confront them with current data. 

One particularly relevant phenomenon to take into account when performing such analysis is PBH mass accretion. Indeed,   
PBHs may  accrete efficiently during the cosmic history, see for instance~\cite{Ricotti:2007jk,Ricotti:2007au,zhang}.  
First of all accretion changes the PBH masses, their mass functions,  and their abundances. This is relevant when 
analysing the  constraints  at the present epoch for PBHs  with masses larger than a few solar masses \cite{paper2}. 
Furthermore, accretion may strongly influence the PBH merger rates, their final spins \cite{paper1},  as well as their 
mass ratios in the binaries. This is particularly important not only because the binary masses and effective spin 
parameter $\chi_{\text{\tiny eff}}$ can be measured, but also because the recent GW190412~\cite{LIGOScientific:2020stg} 
and future events will provide fundamental information on the individual BH spin and on mass ratio distributions with 
a large hierarchy between the BH masses.

The paper is organised as follows. In Section~\ref{sec:massespsins} we discuss the masses and the spins of the PBHs 
including the accretion phenomenon. In Section~\ref{sec:binaryevolution} we provide details about the binary evolution, 
while Section~\ref{sec:rates} is devoted to PBH merger rates. Section~\ref{sec:confrontation} represents the main bulk 
of our paper as it contains the comparison of the recent LIGO/Virgo data with the theoretical predictions. Finally, we 
conclude in Section~\ref{sec:keypredictions} by providing a list of our main findings that can be directly tested with 
current and future GW observations. 

A final note about notation. We are going to  use  the label ``i" for the quantities at formation time, to distinguish 
from those at the time of coalescence. Final values, e.g. PBH masses,  will carry no label. This distinction is relevant 
when accretion takes place. We use $G=c=1$ units throughout. 

\section{The masses and spins of PBHs}\label{sec:massespsins}
In this section we discuss the theoretical predictions for the masses and spins of binary BH components 
in the case the latter are of primordial origin.
We consider two situations: a)~accretion is negligible, hence the masses and spins of the binary components 
are those at formation; b)~baryonic mass accretion 
is modelled through the cosmic history, hence the masses and spins of isolated and binary PBHs are different from those 
at formation. In the latter case also the mass distribution and the PBH abundance are affected by accretion.

The reader interested in the final results can skip the details of the modelling and jump directly to 
Sec.~\ref{sec:summary}, where a summary of the theoretical predictions is presented. The limitations of the accretion 
model are discussed in Sec.~\ref{sec:limitations}.

\subsection{Mass distribution at formation} \label{sec:massfunctions}
One can consider several initial shapes for the mass function at high redshift, depending on the details of the PBH
formation mechanism. In Ref.~\cite{paper1} we considered a critical, spiky, lognormal, and power-law mass function.
Since the first two cases are unrealistic, we restrict here to the latter two distributions.
A distribution, motivated by the collapse of scale invariant perturbations, is given by a power-law mass function~\cite{Carr:1975qj,DeLuca:2020ioi,Byrnes:2018txb,qcd}
\begin{equation}
	\psi(M, z_\ii) = \frac{1}{2} \left( M_\text{\tiny min} ^{-\frac{1}{2}} - M_\text{\tiny max} ^{-\frac{1}{2}}\right) ^{-1}
M^{-\frac{3}{2}}. 
\end{equation}
 This is  described by two free parameters, $M_\text{\tiny min}$ and $M_\text{\tiny max}$.
An alternative, and maybe more popular, mass function is the lognormal one 
\be
\label{psi}
\psi (M,z_\ii) = \frac{1}{\sqrt{2 \pi} \sigma M} {\rm exp} \left(-\frac{{\rm log}^2(M/M_c)}{2 \sigma^2} 
\right)
\ee
expressed again by two parameters, the width $\sigma$ and the  peak reference mass $M_c$, first introduced in \cite{ds93}. It represents a frequent 
parametrisation for the cases of a PBH population arising from a symmetric peak in the primordial power spectrum, see 
for example Ref.~\cite{Carr:2017jsz}.

\subsection{Spin distribution at formation}\label{sec:spindistribution}
The requirement that the cosmological abundance of PBHs is less than the DM 
abundance sets a bound on the PBHs mass fraction, which in turn requires the collapse of density perturbations generating a PBH to be a rare event.  Applying the formalism  of peak theory \cite{bbks} in standard formation scenarios \cite{Blinnikov:2016bxu,juan,s1,s2,s3}, one finds that high (and rare) peaks in the density contrast, which eventually 
collapse to form PBHs, are primarily spherical. However, at first order in perturbation theory, the presence 
of small departures from spherical symmetry introduces torques induced by the surrounding matter 
perturbations. This leads to the generation of a small angular momentum before collapse. Due to the small time scales
which characterise the overdensity collapse, the action of the torque moments is indeed limited in time.

The estimated PBH spin at formation is~\cite{DeLuca:2019buf} 
\be
\chi_{\text{\tiny i}} = \frac{\Omega_\text{\tiny m}}{\pi} \sigma_\delta \sqrt{1-\gamma^2} \sim 10^{-2} 
\sqrt{1-\gamma^2},
\ee
where $\Omega_\text{\tiny m}\sim 0.3$ is the current DM abundance, $\sigma_\delta$ is the variance of the 
density perturbations at the horizon crossing time, and $\gamma$ parametrises the shape of the power spectrum of the 
density perturbations in terms of its variances (being $\gamma\sim1$ for very narrow power spectra). 
The suppression factor due to $\gamma$ arises because, as $\gamma$ approaches unity, the velocity shear tends to be 
more strongly aligned with the inertia tensor.
Thus, the initial spin of PBHs is expected to be below the percent level (see also Ref.~\cite{Mirbabayi:2019uph}). Non-standard 
scenarios for the PBH formation, for instance during an early matter-dominated epoch~\cite{Harada:2017fjm} 
following inflation or from the collapse of Q-balls~\cite{Cotner:2017tir}, may lead to higher values of the initial 
spin. 

\subsection{The role of accretion}
In this section we discuss the manifold roles of accretion onto PBHs during the cosmic history, reviewing and 
extending the analysis presented in previous work. Accretion can affect 
both the mass and spin of isolated and binary PBHs~\cite{paper1}. \footnote{In this paper we neglect  second-generation mergers of PBHs, which are 
those in which at least one of the two components of the binary results from the merger of a 
previous binary system~\cite{Gerosa:2017kvu,paper1}. This is justified by the fact that, in the case of PBH mergers, 
the fraction of second-generation events can be neglected in the LIGO/Virgo band~\cite{Liu:2019rnx,Wu:2020drm,paper1}.}
For the latter, it can also affect the binary 
evolution before GW emission becomes dominant (see Sec.~\ref{sec:accretion-driven} below). Furthermore, accretion 
modifies the mass distribution of PBHs and the fraction of PBHs in DM in a redshift-dependent fashion~\cite{paper2}.

\subsubsection{Accretion onto isolated PBHs}\label{sec:accretionisolated}
%
We model gas accretion onto an isolated PBH with mass $M$, moving with a relative velocity $v_\text{\tiny rel}$ with 
respect to the surrounding gas, through the Bondi-Hoyle mass accretion 
rate~\cite{ShapiroTeukolsky,Ricotti:2007jk,Ricotti:2007au}
\be
\label{R1}
\dot{M}_\text{\tiny B} = 4 \pi \lambda m_H n_{\rm gas} v_\text{\tiny eff} r_\text{\tiny B}^2
\ee
where $v_\text{\tiny eff} = \sqrt{v_\text{\tiny rel}^2 
	+ c_s^2}$ is the effective velocity, $c_s$ is the speed of sound, and the gas number density is $n_{\rm gas} \simeq 200  
(1+z/1000)^3\, {\rm cm}^{-3} $.
The Bondi-Hoyle radius reads
\be
r_\text{\tiny B} \equiv \frac{M}{v_\text{\tiny eff}^2} \simeq 1.3 \times 10^{-4}\lp \frac{M}{M_\odot} \rp 
\lp \frac{v_\text{\tiny eff}}{5.7 {\rm\, km \, s^{-1}}} \rp^{-2}\,  {\rm pc}. \label{rB}
\ee
For a gas in equilibrium at the temperature of the intergalactic medium, 
\be
c_s \simeq 5.7 \,  \left ( \frac{1+z}{1000}\right)^{1/2}
\llp \left (\frac{1+ z_\text{\tiny dec}}{1+z} \right)^\beta +1 \rrp^{-1/2 \beta}\, {\rm km \, s^{-1}},
\ee
with $\beta = 1.72$, and $z_\text{\tiny dec} \simeq 130$ being the redshift at which the baryonic matter decouples from the 
radiation fluid.
The accretion parameter $\lambda$ appearing in Eq.~\eqref{R1} keeps into account the effects of the Hubble expansion, 
the coupling of the CMB radiation to the gas through Compton scattering, and the gas 
viscosity~\cite{Ricotti:2007jk}. The main formulas to compute $\lambda$ are summarized in Appendix~B of 
Ref.~\cite{paper1}.

Current observational constraints imply that PBHs with masses larger than ${\cal O}(M_\odot)$ can 
comprise only a fraction of the DM~\cite{Carr:2020gox}. Thus, accretion onto PBHs should include the presence of an 
additional DM halo. While direct DM accretion onto the PBH is negligible~\cite{Ricotti:2007jk,zhang}, the halo acts 
as a catalyst enhancing the gas accretion rate. The DM halo has a typical spherical density profile $\rho 
\propto r^{-\alpha}$ (with approximately $\alpha \simeq 2.25$~\cite{Mack:2006gz, Adamek:2019gns}), truncated at a 
radius 
$ r_h \simeq 0.019 \, {\rm pc} (M/M_\odot)^{1/3} (1+z/1000)^{-1}$ and with total mass
\be
\label{halo mass}
M_h(z) = 3 M \lp \frac{1+z}{1000} \rp^{-1}\,.
\ee
The mass $M_h$ grows with time as long as the PBHs are isolated and eventually stops when all the available DM
has been accreted, i.e. approximately when $3 f_\PBH (1+z/1000)^{-1} = 1$. In the presence of a DM halo, the mass 
entering in the Bondi-Hoyle formula~\eqref{R1} is $M_h$, and this enhances the accretion rate. The halo extends much 
more than the Bondi radius ($r_h\gg r_\text{\tiny B}$) and this effect is taken into account by the accretion parameter 
$\lambda$~\cite{Ricotti:2007jk}.

It is customary to define the dimensionless accretion rate normalised to the Eddington one
\be
\dot m = \frac{\dot{M}_\text{\tiny B}}{\dot{M}_\text{\tiny Edd}}
\quad\text{with}\quad 
\dot{M}_\text{\tiny Edd}= 1.44 \times 10^{17} \lp \frac{M}{M_\odot}\rp \rm{g \, s^{-1}}, \label{mdot}
\ee
whose behaviour as a function of the redshift and PBH mass can be found in Refs.~\cite{Ricotti:2007jk,paper1}. 
It is noteworthy that $\dot m$ can be larger than unity (i.e., accretion can be super-Eddington) for $z\sim {\cal 
	O}(30)$ and for the masses of interest for this work.

The typical accretion time scale is given in terms of the Salpeter time by
\begin{equation}
	\tau_\text{\tiny acc} \equiv \frac{\tau_\text{\tiny Salp}}{\dot m}=\frac{\sigma_\text{\tiny T}}{4 \pi 
		m_\text{\tiny p}} \frac{1}{\dot m} =\frac{4.5 \times 10^8 \,{\rm yr}}{\dot m}\,, \label{tauACC}
\end{equation}
where $\sigma_\text{\tiny T}$ is the Thompson cross section and 
$m_\text{\tiny p}$ is the proton mass. For $z\lesssim 100$, $\tau_\text{\tiny acc}$ is smaller than the typical age of 
the universe. Accretion can therefore play an important role in the mass evolution of PBHs~\cite{Ricotti:2007au,paper1}.

The accretion rate~\eqref{R1} depends significantly on the relative velocity between the PBHs and the baryonic matter, 
and it is therefore sensitive to several physical processes that might increase the PBH characteristic velocities or the 
speed of sound in the gas, in turn reducing the accretion rate.
We discuss this point in Sec.~\ref{sec:limitations} below.
Given the large uncertainties in the modelling of the accretion rate at relatively small redshift, here we shall adopt 
an agnostic view and consider several cut-off values $z_\text{\tiny cut-off}=(15,10,7)$, below 
which accretion is negligible.

%

\subsubsection{Accretion onto binary PBHs}\label{sec:accretionbinary}
\begin{figure}[t!]
	\centering
	\includegraphics[width=0.6 \linewidth]{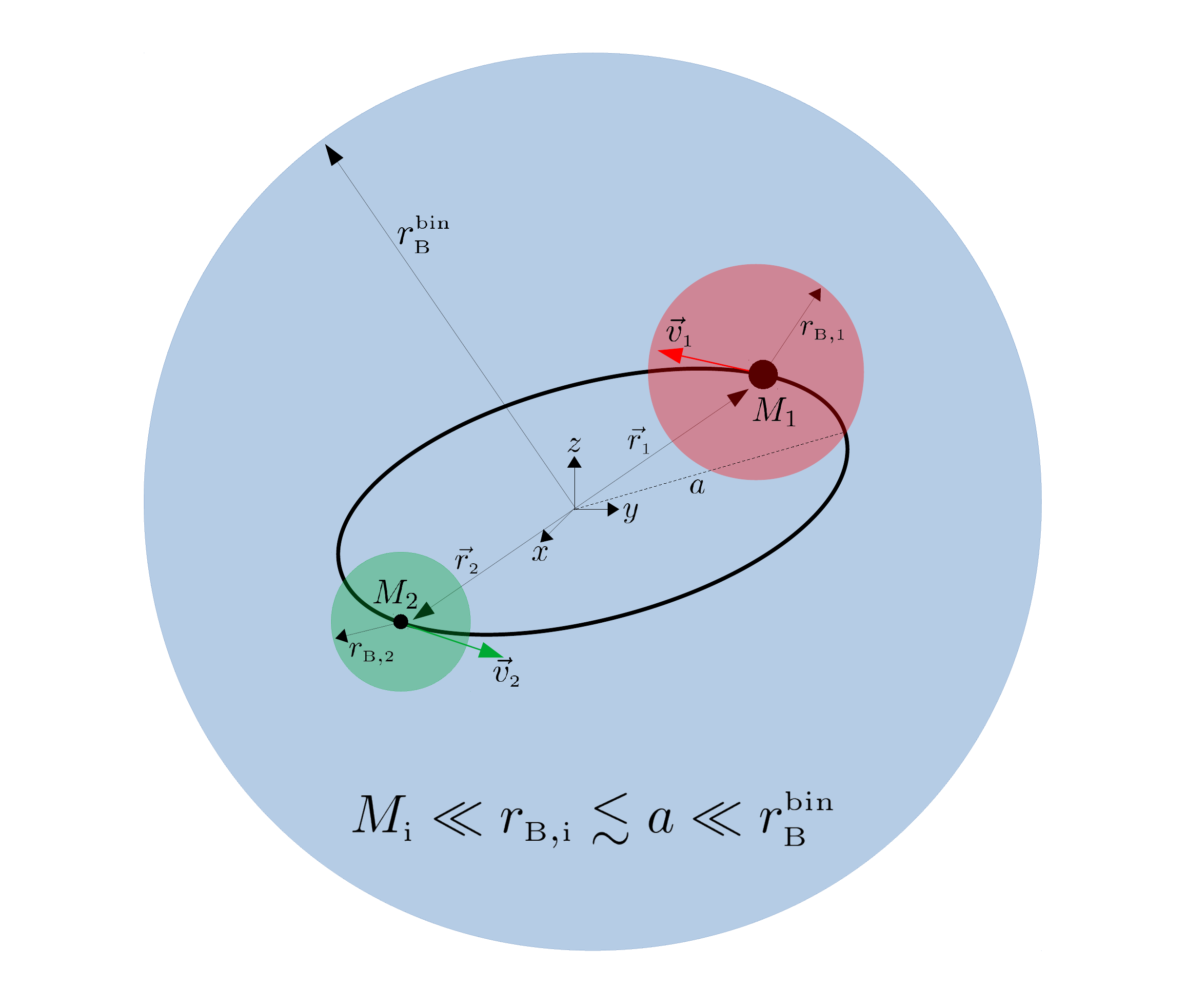}
	\caption{\it Schematic illustration of the relevant scales involved in the accretion process 
		for a PBH binary system. The Bondi radius of the binary is much bigger than the orbital separation.
	}\label{fig:draw}
\end{figure}

In order to study the accretion onto binary PBHs, one has to take into account both {\it 
	global} accretion processes (i.e., of the binary as a whole) and {\it local} accretion processes (i.e., onto 
the 
individual components of the binary). Here we extend the analysis of Ref.~\cite{paper1} to account for generic mass 
ratios and eccentric orbits.
We consider a binary of total mass $M_\text{\tiny tot} = M_1 + M_2$, reduced mass $\mu = M_1 M_2 /(M_1 + M_2)$, mass 
ratio $q=M_2/M_1\leq1$, semi-major axis $a$, and eccentricity $e$.
The situation is schematically illustrated in Fig.~\ref{fig:draw} and described below.

Two distinctive regimes occur depending on how the binary separation compares with the Bondi radius of the binary, 
\begin{equation}
	r^\text{\tiny bin}_\text{\tiny B} = \frac{M_\text{\tiny tot}}{v^2_\text{\tiny eff}}\,, \label{rBbin}
\end{equation}
where the effective velocity $v_\text{\tiny eff}= \sqrt{c_s^2 + v^2_\text{\tiny rel}}$  depends on the 
velocity $v_\text{\tiny rel}$ of the center of mass of the binary relative to the surrounding gas with mean cosmic 
density $n_\text{\tiny gas}$.
Indeed, if $a\gg r^\text{\tiny bin}_\text{\tiny B}$, accretion occurs onto the two individual PBHs independently (each 
one moving at a characteristic velocity that also depends on the orbital one), as discussed in the previous section. 
However, as the binary hardens, the orbital separation becomes much smaller than the Bondi radius of the binary ($a\ll
r^\text{\tiny bin}_\text{\tiny B}$) and, as we shall discuss, this occurs much before GW emission becomes the dominant 
driving mechanism of the inspiral.
In this case, accretion occurs on the binary as a whole at rate given by
\be \label{R1bin}
\dot M_\text{\tiny bin} = 4 \pi \lambda m_H n_\text{\tiny gas} v^{-3}_\text{\tiny eff} M^2_\text{\tiny tot}\,.
\ee
We now evaluate how the two binary components accrete in this configuration. The PBH positions and 
velocities with respect to the center of mass are given by~\cite{poisson2014gravity}
\be
r_1 = \frac{q}{1+q}r, \qquad v_1 = \frac{q}{1+q}v;  \qquad r_2 = \frac{1}{1+q}r, \qquad v_2 = \frac{1}{1+q}v  
\ee
in terms of their relative distance and velocity
\be
r = a (1-e \, {\rm cos}u), \qquad v = \sqrt{M_\text{\tiny tot} \lp \frac{2}{r}-\frac{1}{a}\rp}\,,
\ee
both expressed as a function of the semi-major axis $a$, eccentricity $e$, and angle $u$.
The time evolution of the latter is implicitly given by $\sqrt{a^3/M_\text{\tiny tot}}(u(t)-e \sin u(t))=t-T$, 
where $T$ is an integration constant~\cite{poisson2014gravity}.
The PBH effective velocities with respect to the gas are given by
\be
v_\text{\tiny eff,1} = \sqrt{v^2_\text{\tiny eff} + v^2_1 }, \qquad  v_\text{\tiny eff,2} = \sqrt{v^2_\text{\tiny 
		eff} + v^2_2}. \label{veffi}
\ee
Since the Bondi radius of the binary is much bigger than the typical semi-axis of the binary (see 
Fig.~\ref{fig:draw}), the total infalling flow of baryons towards the binary is constant, i.e.
\be
4 \pi m_H n_\text{\tiny gas}(R) v_\text{\tiny ff}(R)R^2 = {\rm const} = \dot M_\text{\tiny bin} 
\ee
where the free fall velocity of the gas, $v_\text{\tiny ff}$, is computed by assuming that at large distances, $R \sim 
r^\text{\tiny bin}_\text{\tiny B} $, it reduces to the usual effective velocity $v_\text{\tiny eff}$, i.e.
\be
v_\text{\tiny ff} (R) = \sqrt{v^2_\text{\tiny eff} + \frac{2M_\text{\tiny tot}}{R} - \frac{2M_\text{\tiny 
			tot}}{r^\text{\tiny bin}_\text{\tiny B}}}\,,
\ee
while $n_\text{\tiny gas}(R)$ identifies the density profile at a distance $R$ from the center of mass of the binary,
\be
n_\text{\tiny gas}(R) = \frac{\dot M_\text{\tiny bin}}{4\pi  m_H v_\text{\tiny ff}(R)R^2 }.
\ee
In other words, being the infalling flow of baryons constant, their density near the binary increases relative to its 
mean cosmic value at the Bondi radius of the binary.

The accretion rates for the single components of the binary are then given in terms of the effective Bondi 
radii\footnote{Note that the local velocities of the individual PBHs [Eq.~\eqref{veffi}] are of the order of the 
	orbital velocity. The latter is always much larger than $v_\text{\tiny rel}$ and $c_s$ for orbital separations 
	smaller than the Bondi-Hoyle radius of the binary. Therefore, the Bondi radii $r_\text{\tiny B,i}$ of the 
individual 
	PBHs are much smaller than the Bondi radius $r^\text{\tiny bin}_\text{\tiny B}$ of the binary.} of the binary 
components $r_\text{\tiny B,i} = M_i/ v_\text{\tiny eff,i}^2$ as
\be
\dot M_1 = 4 \pi  m_H n_\text{\tiny gas}(r_\text{\tiny B,1}) v^{-3}_\text{\tiny eff,1} M_1^2, \qquad 
\dot M_2 = 4 \pi  m_H n_\text{\tiny gas}(r_\text{\tiny B,2}) v^{-3}_\text{\tiny eff,2} M_2^2. \label{Rindiv}
\ee
Note that for the single accretion rates we considered two naked PBHs, for which the parameter $\lambda \approx 1$ at 
low redshift, 
while we have described the binary with a dark halo clothing with parameter $\lambda$ which takes 
into account the ratio 
of the Bondi radius of the binary with respect to the dark halo radius~\cite{Ricotti:2007au}, as discussed above.

Putting together the above formulas, the accretion rates~\eqref{Rindiv} can be written in terms of the orbital 
parameters in a cumbersome (albeit analytical) form, 
\begin{align}
	& \dot M_1 = \dot M_\text{\tiny bin} \sqrt{\frac{1+\zeta+(1-\zeta) \gamma ^2}{2 (1+\zeta) 
			(1+q)+(1-\zeta)(1+2 q) \gamma ^2}}\label{M1dotGEN}\,, \\
	& \dot M_2 =  \dot M_\text{\tiny bin} \sqrt{\frac{(1+\zeta) q+(1-\zeta) q^3 \gamma ^2}{2 (1+\zeta)
			(1+q)+(1-\zeta) q^2 (2+q) \gamma ^2} }\,, \label{M2dotGEN}
\end{align}
where we defined $\zeta=e\cos u$ and $\gamma^2=a v^2_\text{\tiny eff}/ \mu q$.
The $q\to0$ limit is particularly simple and does not depend on $e$ nor $u$,
\begin{align}
	\dot M_1 = \dot M_\text{\tiny bin}+{\cal O}(q)\,, \qquad \dot M_2 = \sqrt{\frac{q}{2}} \dot M_\text{\tiny bin} 
+{\cal 
		O}(q^{3/2})\,.  \label{M1M2smallq}
\end{align}
In the general case, since the orbital period 
is much smaller than the accretion time scale $\tau_\text{\tiny acc}$, we can average over the angle $u$ and eliminate 
the explicit time dependence in Eqs.~\eqref{M1dotGEN} and \eqref{M2dotGEN}. After this averaging procedure the 
dependence on the eccentricity is negligible. 
Therefore, we can finally write
\begin{align}
	& \dot M_1 = \dot M_\text{\tiny bin} \sqrt{\frac{M_1 q^2 + a (1+q) v^2_\text{\tiny eff}}{(1+q) [2 M_1 q^2 + a (1 
+ 
			2q) v^2_\text{\tiny eff}]} }\,,\nonumber \\
	& \dot M_2 =  \dot M_\text{\tiny bin} \sqrt{\frac{q[M_1 + a (1+q) v^2_\text{\tiny eff}]}{(1+q) [2 M_1 + a (2 + 
q) 
			v^2_\text{\tiny eff}]} }\,.   \label{M1M2dotGEN}
\end{align}
The above rates can be even further simplified when
\be
M_1 q^2 \gg a v^2_\text{\tiny eff},
\ee
which is satisfied for $M_1 \sim \mathcal{O}(M_\odot)$ and $a \sim {\cal O}(10^6)M_1$ for any $q > 
10^{-2}$. In this case, the previous expressions reduce to
\begin{align}
	&\dot M_1 = \dot M_\text{\tiny bin}  \frac{1}{\sqrt{2 (1+q)}}, \nonumber \\
	&\dot M_2 = \dot M_\text{\tiny bin}  \frac{\sqrt{q} }{\sqrt{2 (1+q)}}. \label{M1M2dotFIN}
\end{align}
Note that the expected behaviour $\dot M_1 = \dot M_2 =  \dot M_\text{\tiny bin}/2$ is recovered in the limit $q \to 1$.

As can be checked a posteriori, we will always be interested in a regime in which Eq.~\eqref{M1M2dotFIN} is an 
excellent approximation. In the following we shall therefore use these simplified formulas, which have the great 
advantage to be independent of the orbital parameters. In other words, knowing $\dot M_\text{\tiny bin}$ and the 
initial mass ratio, Eq.~\eqref{M1M2dotFIN} provides the time evolution of the masses of the binary components, 
regardless of the orbital evolution.

In terms of the Eddington normalised rates, $\dot m_i = \tau_\text{\tiny Salp} \dot M_i/M_i$, one gets
\begin{align}
	\dot m_1 =  \dot m_\text{\tiny bin} \sqrt{\frac{1+q}{2}}, \qquad \dot m_2 =  \dot m_\text{\tiny bin} 
	\sqrt{\frac{1+q}{2q}}. \label{m1m2dotFIN}
\end{align}
The evolution equation for the mass ratio is given by
\be
\dot q = q \lp \frac{\dot M_2}{M_2} -  \frac{\dot M_1}{M_1} \rp = \frac{q}{\tau_\text{\tiny Salp}} \lp \dot m_2 - \dot 
m_1 \rp.
\ee
The above equation shows an important point that will be relevant in the following: if $\dot m_2>\dot m_1$, the growth 
rate of the mass ratio is positive, i.e. the mass ratio grows until it reaches a stationary point when $q=1$ and $\dot 
m_1=\dot m_2$. From Eq.~\eqref{m1m2dotFIN}, it is clear that $\dot m_2>\dot m_1$ in any case.
This shows that accretion onto a binary PBH implies that the binary masses tend to balance each other on secular time 
scales.

\subsubsection{Effects on the mass function and PBH abundance}

In addition to changing the masses and spins of PBHs, accretion also affects their mass distribution function, as well 
as their mass fraction relative to that of the DM, in a redshift-dependent fashion~\cite{paper2}.

\begin{figure}[t!]
	\centering	
	\includegraphics[width=0.486\linewidth]{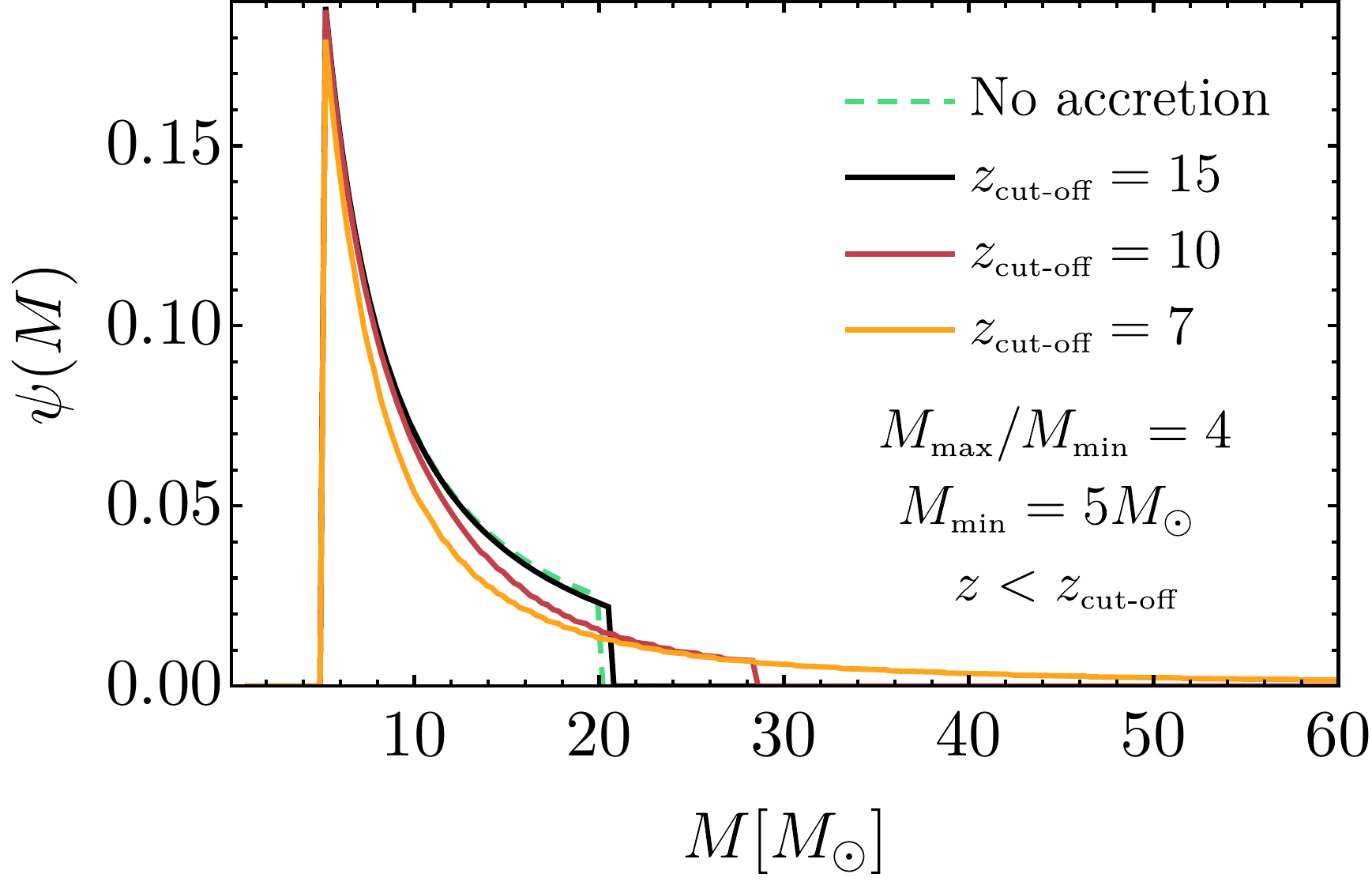}
	\includegraphics[width=0.48 \linewidth]{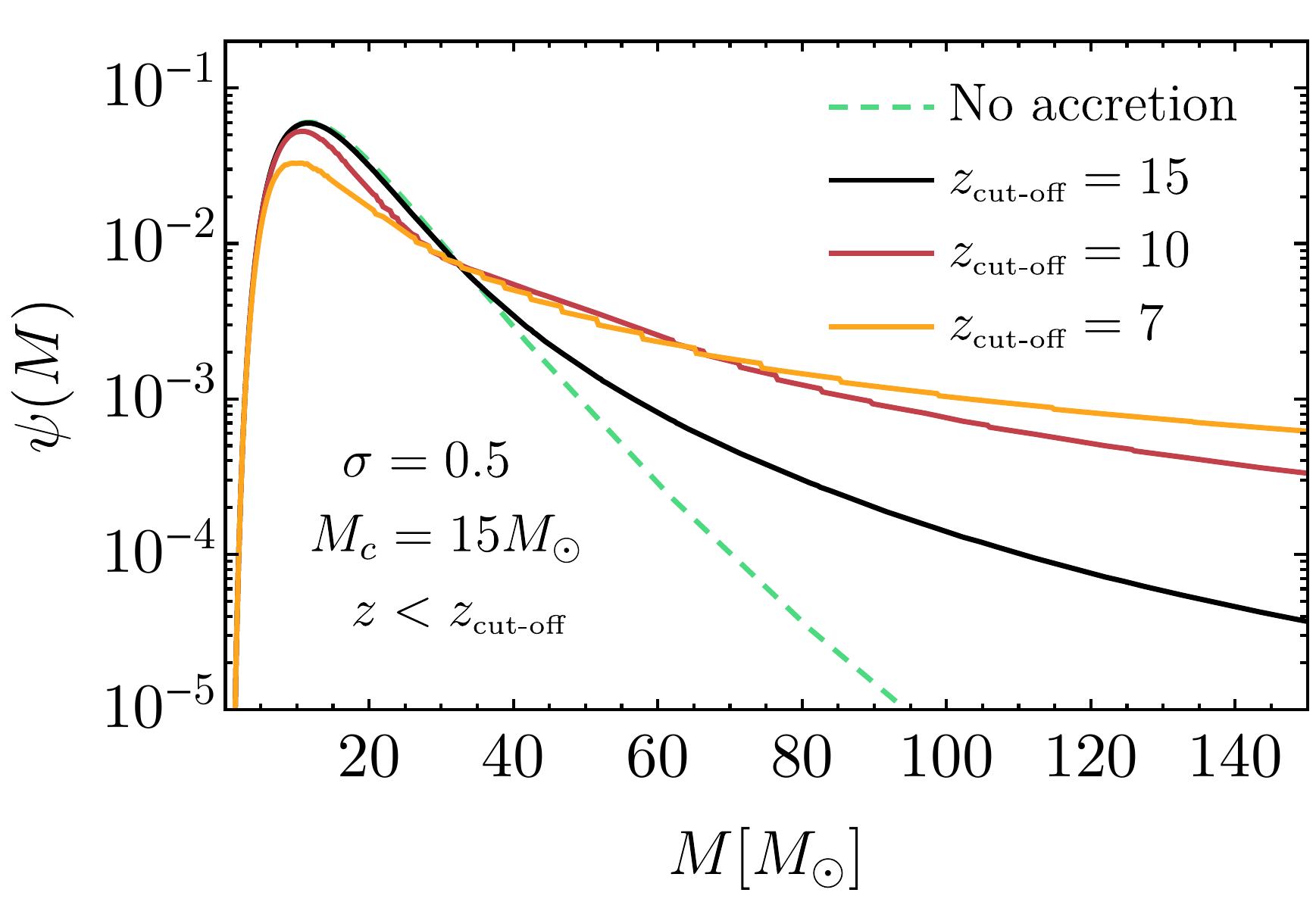}
	\caption{\it Example of evolution of the power-law (left) and lognormal (right) mass functions.
	}
	\label{mfevo}
\end{figure}

Let us define the mass function $\psi(M,z)$ as the fraction of PBHs with mass in the interval $(M, M + \d M)$ at 
redshift $z$. For an initial $\psi (M_\ii,z_\ii)$ at formation redshift $z_\ii$, its evolution is governed 
by~\cite{paper1,paper2}
\begin{equation}
	\psi(M(M_\ii,z),z) \d M = \psi(M_\ii,z_\ii) \d M_\ii \label{psiev}
\end{equation}
where $M(M_\ii,z)$ is the final mass at redshift $z$ for a PBH with mass $M_\ii$ at redshift $z_\ii$. 
We stress that --~since the evolution of the mass function is needed to re-weight existing constraints on the PBH 
abundance and since the latter are mainly due to isolated PBHs~-- for the evolution of the mass function  we have considered the 
evolution of the mass of a {\it single} BH; in this case we have to follow Sec.~\ref{sec:accretionisolated}.
The main effect of accretion on the mass distribution is to make the latter broader at high masses, producing a 
high-mass tail that can be orders of magnitude above its corresponding value at formation~\cite{paper2}.
A representative example is shown in Fig.~\ref{mfevo}, where we compare the final mass function (after the 
accretion phase) with that at formation for some choices of the initial mass distribution.

\begin{figure}[t!]
	\centering	
	\includegraphics[width=0.57\linewidth]{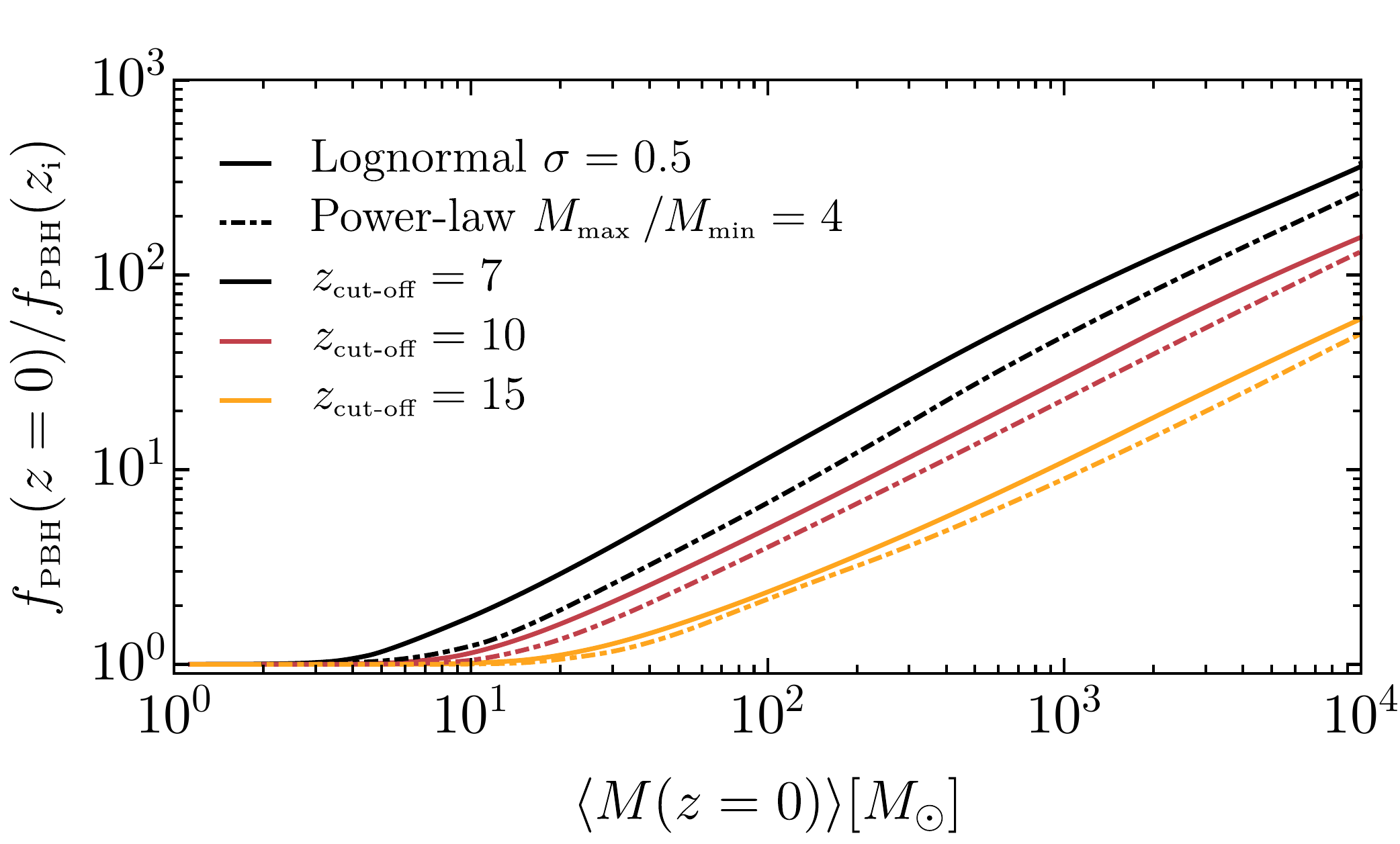}
	\caption{\it Example of evolution of the PBH abundance for both power-law and lognormal  mass functions.
	}
	\label{fevo}
\end{figure}

Finally, also the value of $f_\PBH$ is affected by accretion. Assuming for simplicity a 
non-relativistic dominant DM component (whose energy density scales as the inverse of the volume), it is easy to show 
that~\cite{paper2}
\begin{align}
	\label{fev}
	f_\PBH(z)  &
	= \frac{\langle M(z)\rangle}{\langle 
		M(z_\ii)\rangle(f^{-1}_\PBH(z_\ii)-1)+\langle M(z)\rangle},
\end{align}
where we defined the average mass 
\begin{equation}
	\langle M(z)\rangle =\int \d M M \psi (M, z).
\end{equation}
Due to the presence of accretion, $f_\PBH(z)$ can be significantly larger than $f_\PBH(z_\ii)$, see an example in Fig.~\ref{fevo}.

\subsubsection{Effects on the PBH spins}

The physics of accretion is very complex, since the accretion rate and the geometry of the accretion flow are 
intertwined, and they are both crucial in determining the evolution of the PBH mass. 
In addition, the infalling accreting gas onto a PBH can carry angular momentum which crucially determines the geometry 
of the accreting flow, and the evolution of the PBH spin \cite{paper1} (see also \cite{bv}).\footnote{The spin of PBHs might change and decrease also due to the  plasma-driven superradiant 
instabilities~\cite{Pani:2013hpa,Brito:2015oca,Conlon:2017hhi}. This   effect is dependent upon  the geometry of the 
plasma surrounding  the BH and is negligible for realistic systems~\cite{Dima:2020rzg}. For this reason we shall neglect 
plasma-driven superradiant instabilities.}

\paragraph{Conditions for a thin-disk formation in isolated and binary PBHs.}

For accretion onto an isolated BH, angular momentum transfer is relevant if the typical gas velocity (given by the 
baryon velocity variance~\cite{Ricotti:2007au}) is larger than the Keplerian velocity close to the PBH.
This implies that the minimum PBH mass for which the accreting gas flow is non-spherical is~\cite{Ricotti:2007au,paper1}
\be
\label{critM}
M \gsim 6\times 10^2 M_\odot \,D^{1.17} \xi^{4.33}(z) \frac{\lp 1+z/1000\rp^{3.35}}{\llp 1 + 0.031 
	\lp1+z/1000\rp^{-1.72} \rrp^{0.68}},
\ee
where $\xi(z)= {\rm Max}[1, \langle v_\text{\tiny eff} \rangle/c_s]$ describes the effect of a (relatively small) PBH 
proper motion in reducing the Bondi radius, and the constant $D \sim \mathcal{O}(1\div 10)$ takes into account 
relativistic corrections.

Besides the necessary condition~\eqref{critM}, the geometry of the disk also depends on the accretion rate. If $\dot 
m<1$ and accretion is non-spherical, an advection-dominated accretion flow~(ADAF) may form~\cite{NarayanYi}. When $\dot 
m\gtrsim1$, the non-spherical accretion can give rise to a geometrically thin accretion disk~\cite{Shakura:1972te}. For 
$\dot 
m\gg1$ the accretion luminosity might be strong enough that the disk ``puffs up'' and becomes thicker.
For simplicity, here we follow Ref.~\cite{Ricotti:2007au} and assume that 
a thin disk forms when Eq.~\eqref{critM} is satisfied and
\be
\label{a}
\dot m \gsim 1\,.
\ee
In the regimes we are interested in, the latter condition is always more stringent than 
condition~\eqref{critM}. Therefore, $\dot m\gsim1$ can be considered as the sufficient condition for the 
formation of a thin disk around an isolated PBH~\cite{paper1}.

If instead accretion occurs onto a PBH binary, angular momentum transfer on each PBH is much more 
efficient~\cite{paper1}. In this case each binary component has a velocity of the order of the orbital velocity $v$. 
Although the Bondi radius of the individual PBHs is much smaller than the Bondi radius of the binary, it is 
still parametrically larger than the radius of the innermost stable circular 
orbit~(ISCO, see Eq.~\eqref{ISCO} below). Therefore, the accretion onto the binary components is never spherical and a 
disk can form. Compared to the aforementioned case of an isolated PBH, in this case condition~\eqref{critM} is absent.

To summarize, for both isolated and binary PBHs we can assume that a thin accretion disk forms whenever $\dot 
m\gtrsim 1$ along the cosmic history. Furthermore, since $\dot m$ never exceeds unit significantly~\cite{paper1}, the 
thin-disk approximation should be reliable in the super-Eddington regime of PBHs.

\paragraph{Evolution of the spin.}
When the conditions for formation of a thin accretion disk are satisfied, i.e. Eq.~\eqref{a}, 
mass accretion is accompanied by an increase of the PBH spin. 
A thin accretion disk is located along the equatorial plane~\cite{Shakura:1972te,NovikovThorne} and therefore the PBH 
spin is aligned 
perpendicularly to the disk plane. In such a configuration, 
one can use a geodesic model to describe the angular-momentum accretion~\cite{Bardeen:1972fi}. 

For circular disk motion the rate of change of the magnitude $J\equiv|\vec J|\equiv\chi M^2$ of the PBH angular 
momentum 
is related to the mass accretion rate (see also Refs.~\cite{Bardeen:1972fi,thorne,Brito:2014wla,volo})
\be\label{spinchange}
\dot {J} = \frac{L(M,J)}{E(M,J)} \dot {M},
\ee
where
\be
E(M,J) = \sqrt{1- 2 \frac{M}{3 r_\text{\tiny ISCO}}}
\qquad 
\text{and}
\qquad
L(M,J) = \frac{2 M }{3 \sqrt{3} } \lp 1+ 2 \sqrt{ 3 \frac{r_\text{\tiny ISCO} }{M }-2}\rp\,,
\ee
and the ISCO radius reads
\be
r_\text{\tiny ISCO}(M,J) = M \llp 3 + Z_2 - \sqrt{\lp 3-Z_1\rp \lp 3+Z_1+2 Z_2\rp } \rrp, \label{ISCO}
\ee
with $Z_1= 1+ \lp 1- \chi^2 \rp ^{1/3} \llp \lp 1+\chi\rp^{1/3}+\lp 1-\chi\rp^{1/3} \rrp$ and $Z_2= \sqrt{3 \chi^2 + 
	Z_1^2}$.

Finally, Eq.~\eqref{spinchange} can be re-arranged to describe the time evolution of the dimensionless Kerr parameter
\be
\dot \chi = \lp {\cal F} (\chi) - 2 \chi \rp \frac{\dot {M}}{M},
\ee
where we have defined the combination ${\cal F} (\chi) \equiv L(M,J)/M E(M,J)$, which is only a function of 
$\chi$.
The above spin evolution equation predicts that the spin grows over a typical accretion time scale until it reaches 
extremality, $\chi=1$. However, radiation effects limit the actual maximum value of the spin to $\chi_{\rm 
	max}=0.998$~\cite{thorne}. Magnetohydrodynamic simulations of accretion 
disks around Kerr BHs suggest that the maximum spin might be smaller, $\chi_{\rm max} \simeq 
0.9$~\cite{Gammie:2003qi}. However, this limit may not apply to geometrically thin disks. 
The geometrically thin-disk approximation is expected to be valid for each PBH when $\dot m_i\gtrsim1$. For larger 
values of the accretion rates, the disk might be geometrically thicker and angular momentum accretion might be less 
efficient. However, the spin evolution time scale does not change significantly in more realistic accretion
models~\cite{Gammie:2003qi}.

To summarize, another key prediction of the PBH scenario is the fact that the spin of sufficiently massive PBHs (those 
that go through epochs of super-Eddington accretion during the cosmic history) should be highly spinning. Note that 
the condition~\eqref{a} is more easily fulfilled by binary PBHs than by the isolated ones, since in the former case 
accretion is enhanced by the larger total mass of the binary.

\subsection{Summary: theoretical distributions of the PBH binary parameters}\label{sec:summary}

In this section we summarize the results for the theoretical distributions of the binary parameters obtained as 
previously discussed.
For ease of notation, we shall denote by $M_j$ and $\chi_j$ the {\it final} mass and spin of the 
$j$-th binary component; likewise $q$ will be the final mass ratio of the binary. The {\it initial} mass and 
spin of the $j$-th binary component are denoted by $M^\text{\tiny i}_j$ and $\chi^\ii_j$ whereas the initial mass ratio is 
denoted by $q_\text{\tiny i}$.~\footnote{We 
remind that we conventionally define the mass ratio to be smaller than unity. If $M^\text{\tiny 
i}_2\leq M^\text{\tiny i}_1$, the ordering is preserved during the accretion-driven evolution and therefore $M_2\leq 
M_1$.} Clearly, in the absence of accretion or hierarchical mergers, the initial and final 
quantities coincide. In this case the mass and spin distributions are those at formation, see 
Secs.~\ref{sec:massfunctions} and \ref{sec:spindistribution}.

\subsubsection{Mass evolution}

\begin{figure}[t!]
	\centering
	\includegraphics[width=0.48 \linewidth]{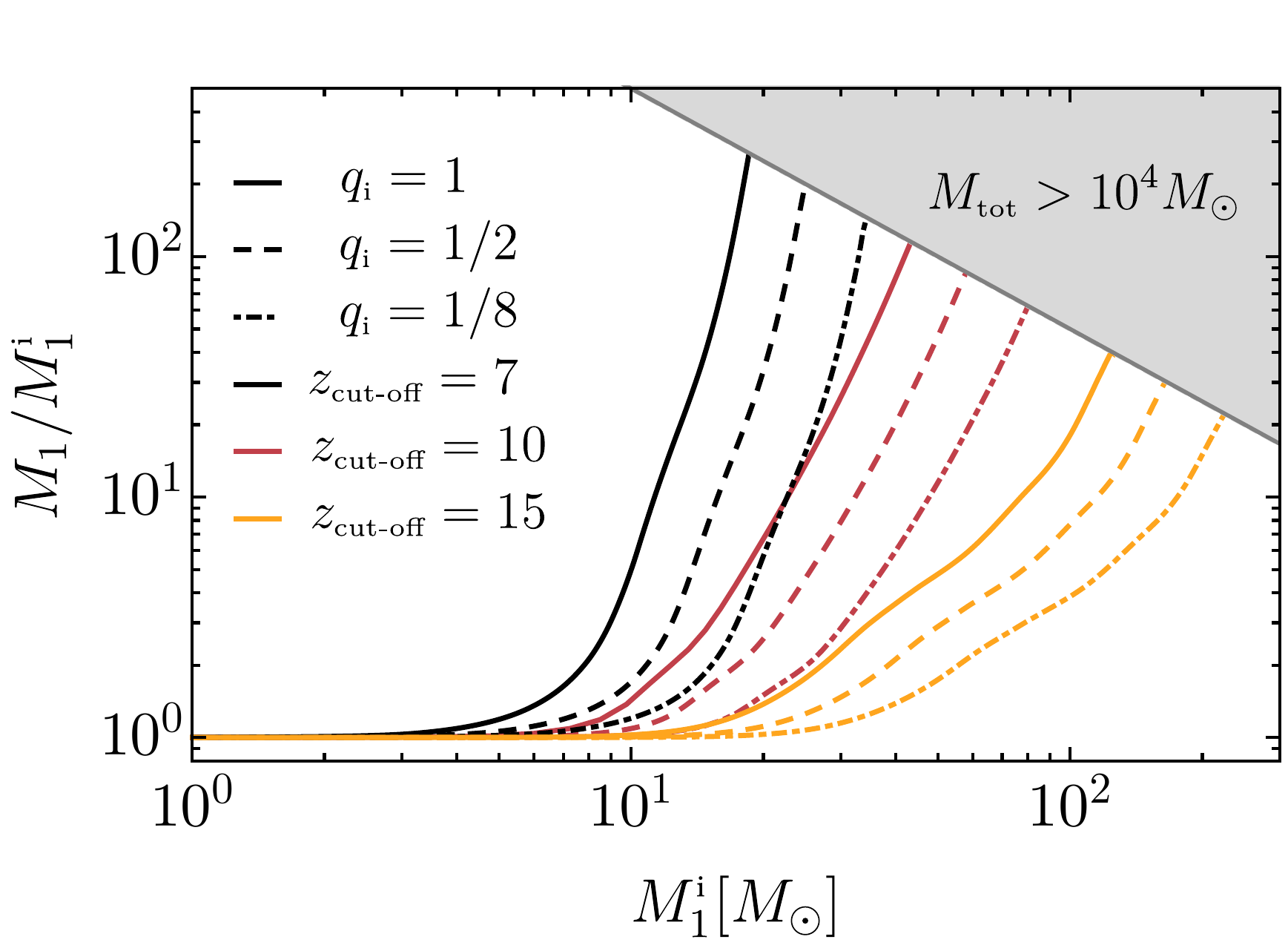}
	\includegraphics[width=0.49 \linewidth]{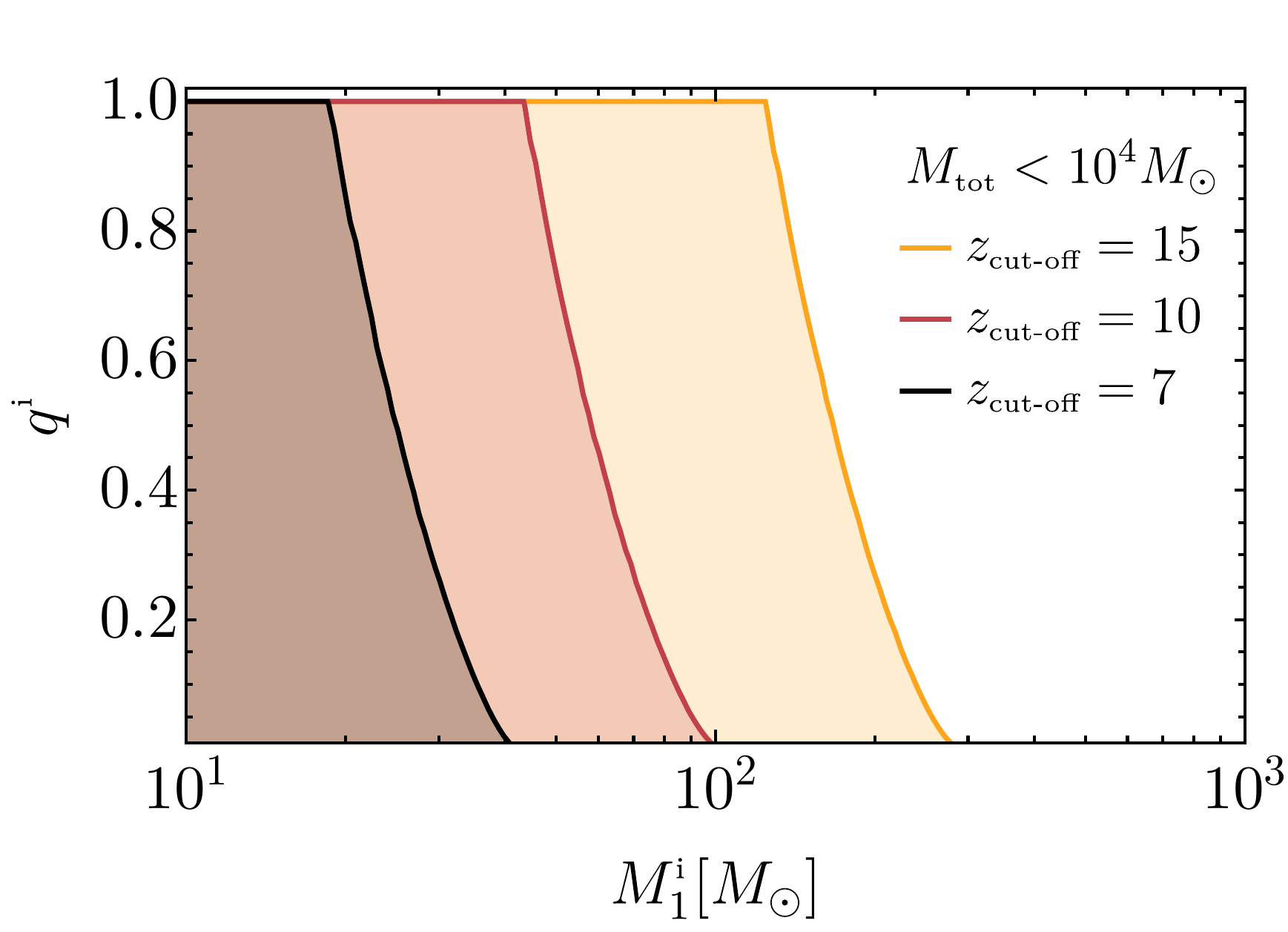}
	\caption{\it Left: Evolution of the mass of the primary for fixed values of the initial mass ratio. Right: the 
region in the $(q_\text{\rm \tiny i}-M^\text{\rm \tiny i}_1)$ plane (shaded areas) that allows the total final mass of the binary 
to be below $M_\text{\rm \tiny tot}= 10^4 M_\odot$. In both panels we considered three different choices of $z_\text{\rm \tiny 
cut-off}$.
}
	\label{qmax}
\end{figure}

As a representative example, in the left panel of Fig.~\ref{qmax} we show the evolution of the primary 
component mass of the binary for various choices of the initial mass ratio $q_\text{\tiny i}$, and for three choices of 
the cut-off redshift. One can notice that the effect of accretion becomes important above initial masses $\sim 10 
M_\odot$ and is stronger for larger masses and for (initially) nearly-equal mass binaries.
Overall, the masses can increase by one or two orders of magnitude due to accretion. Since the merger frequency of 
binaries with total mass $M_\text{\tiny tot}\gtrsim 10^4 M_\odot$ is certainly below the frequency band of current ground-based
detectors, we can safely neglect all the binaries that are pushed above that value by accretion. 
In the right panel of Fig.~\ref{qmax} we show the region in the $q_\text{\tiny i}-M^\text{\tiny 
i}_1$ plane that allows the total final mass of the binary to be below $M_\text{\tiny 
tot}= 10^4 M_\odot$. The parameter space outside the shaded areas would give rise to binaries with $M_\text{\tiny tot}> 
10^4 M_\odot$, which are irrelevant for our study.

Nonetheless, it is intriguing to note that, due to a strong accretion phase at $z\approx (10\div 30)$, PBHs formed with 
$M^\text{\tiny i}\gtrsim 20-100 M_\odot$, can have much larger masses when they are detected at small redshift. 
These objects would be natural candidates for intermediate-mass BHs, which are sources for third-generation 
ground-based detectors and especially for LISA.

\begin{figure}[t!]
	\centering
	\includegraphics[width= 0.3 \linewidth]{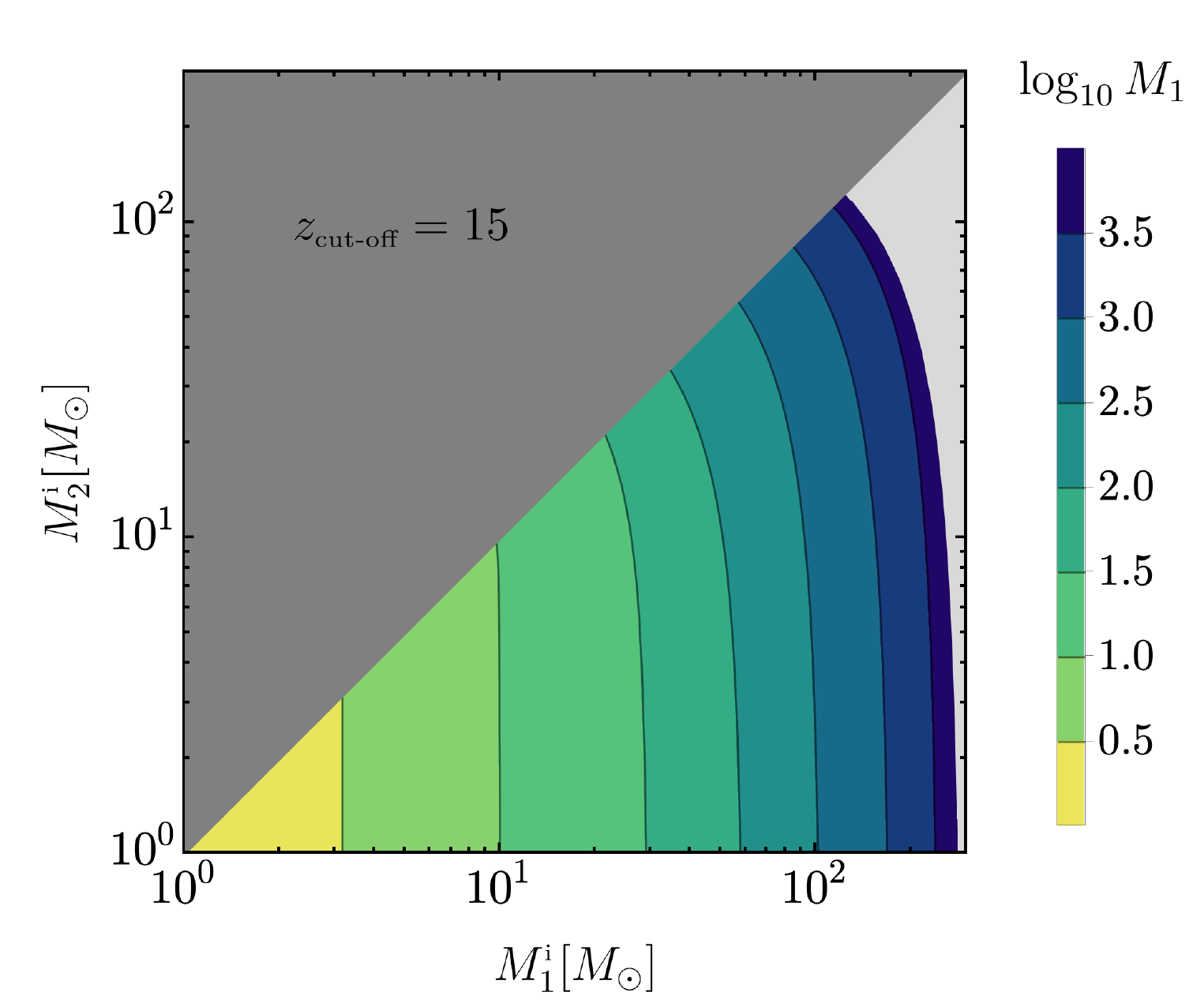}
	\includegraphics[width= 0.3 \linewidth]{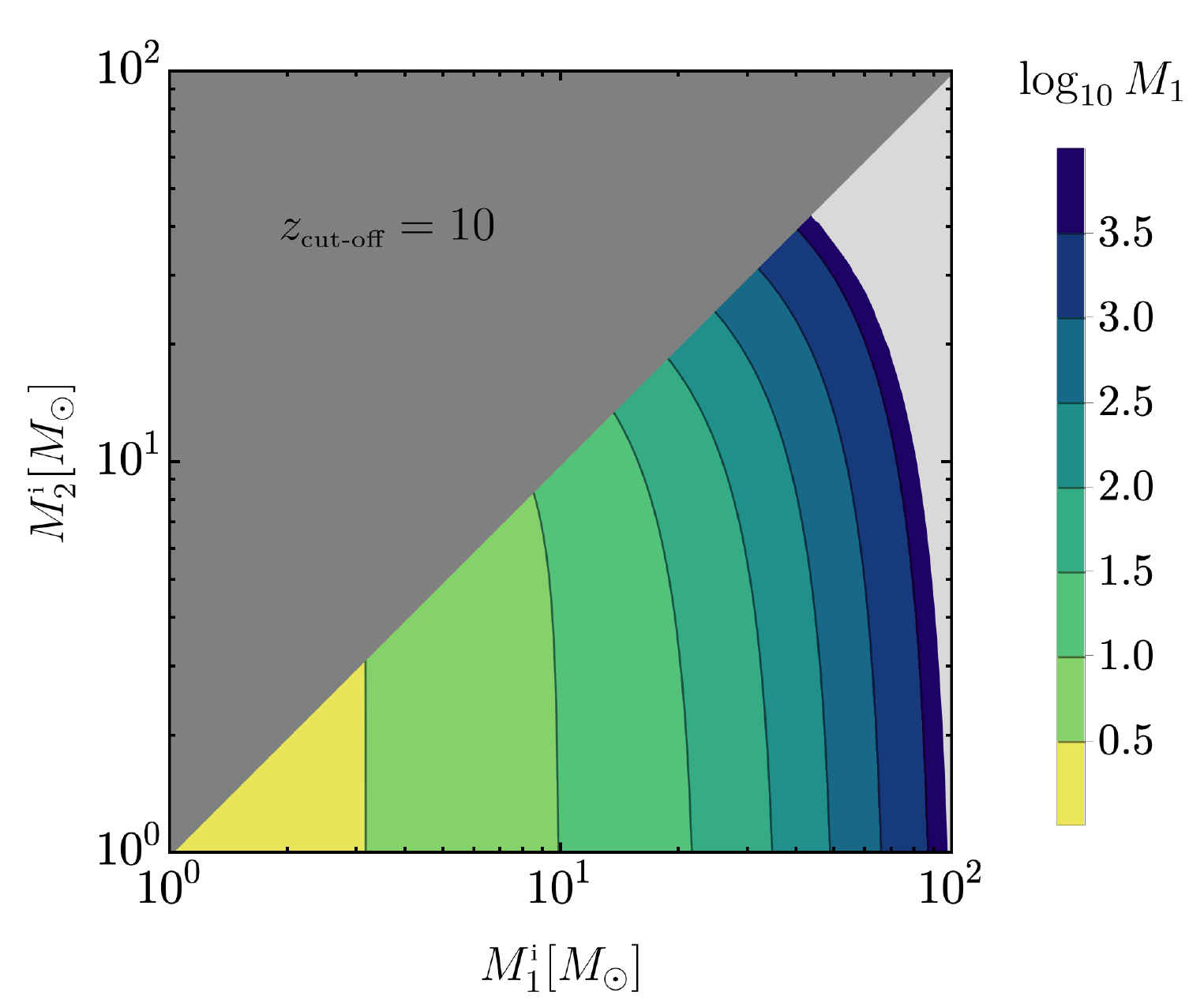}
	\includegraphics[width= 0.3 \linewidth]{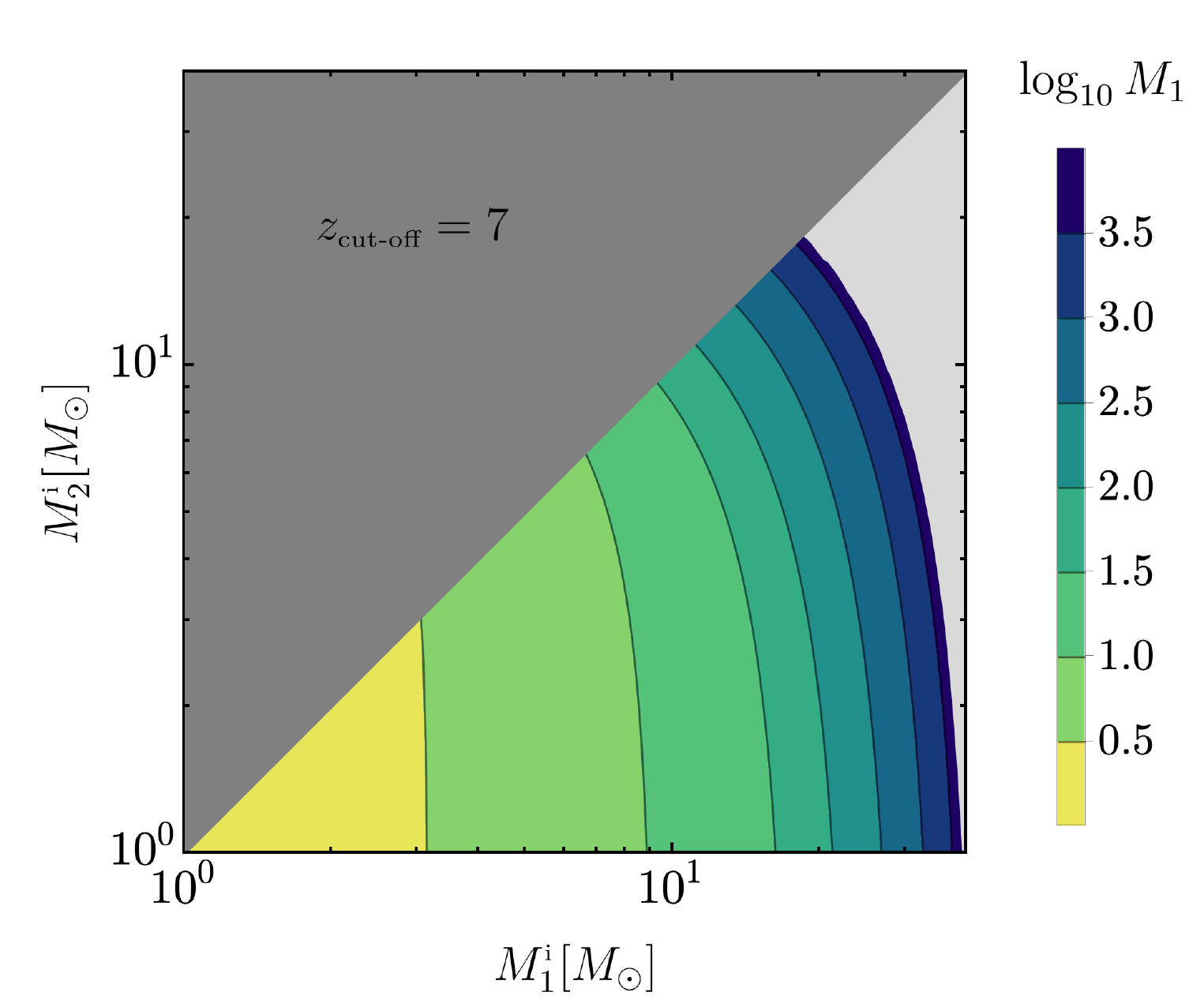}
			
	\includegraphics[width= 0.3 \linewidth]{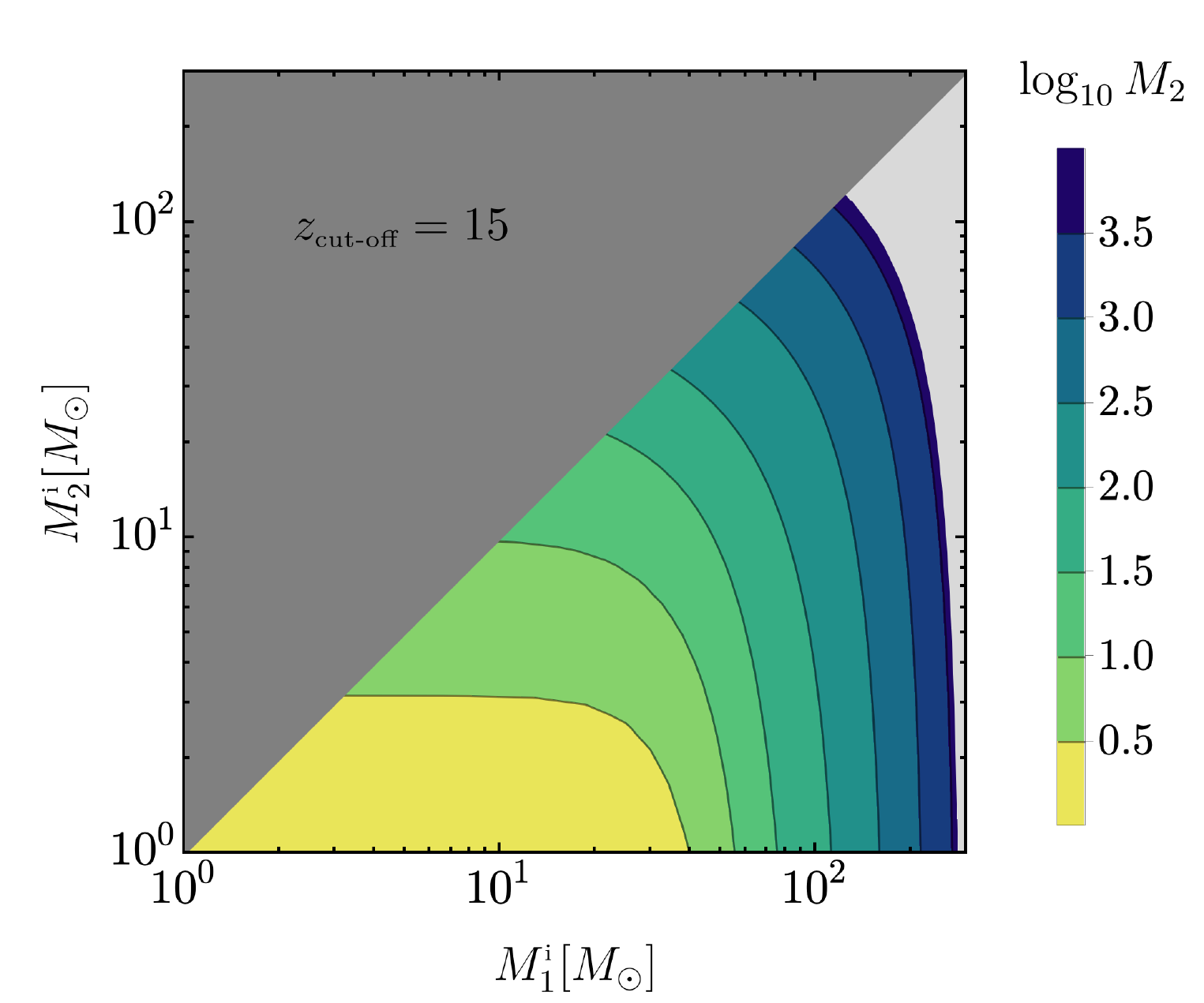}
	\includegraphics[width= 0.3 \linewidth]{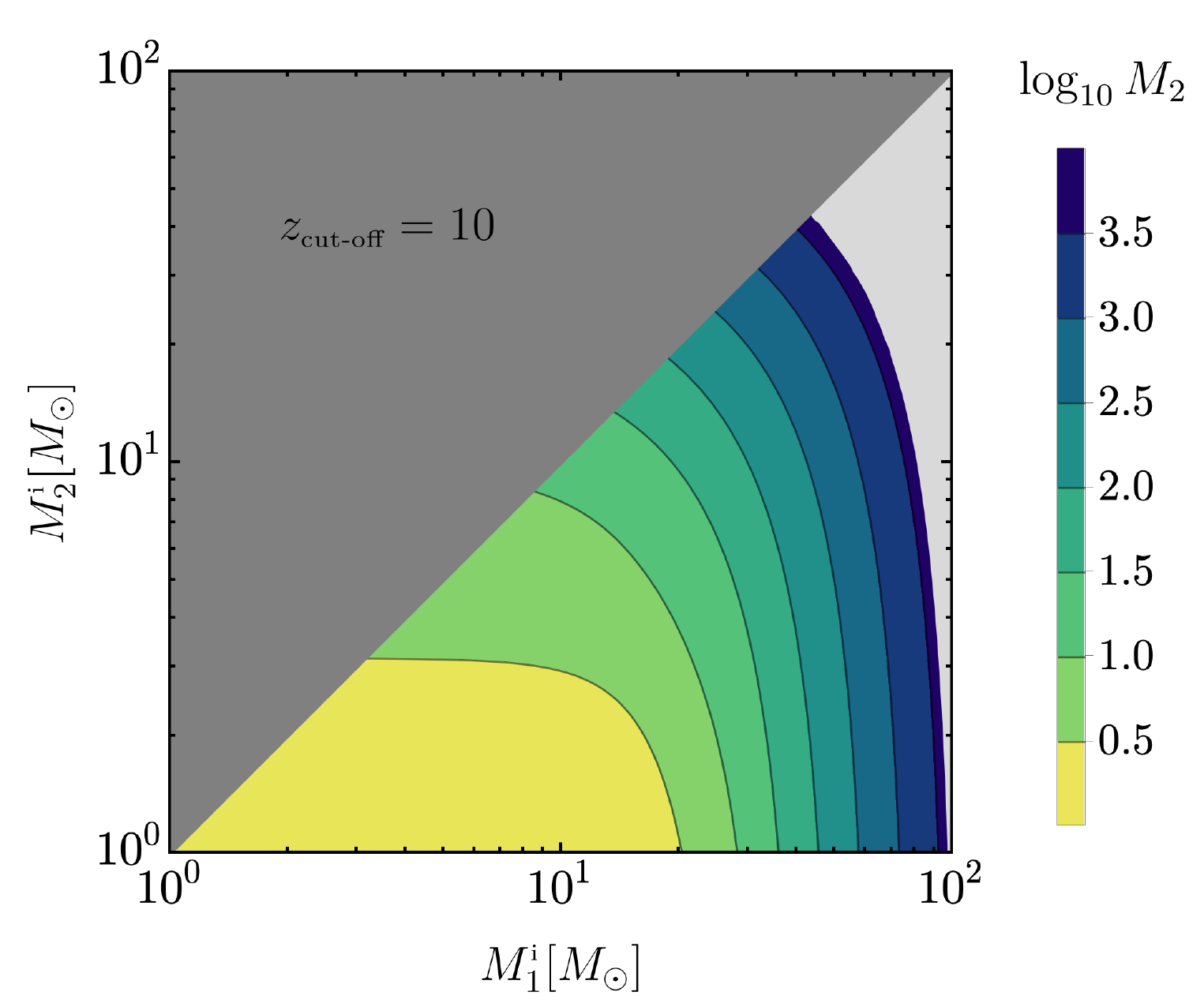}
	\includegraphics[width= 0.3 \linewidth]{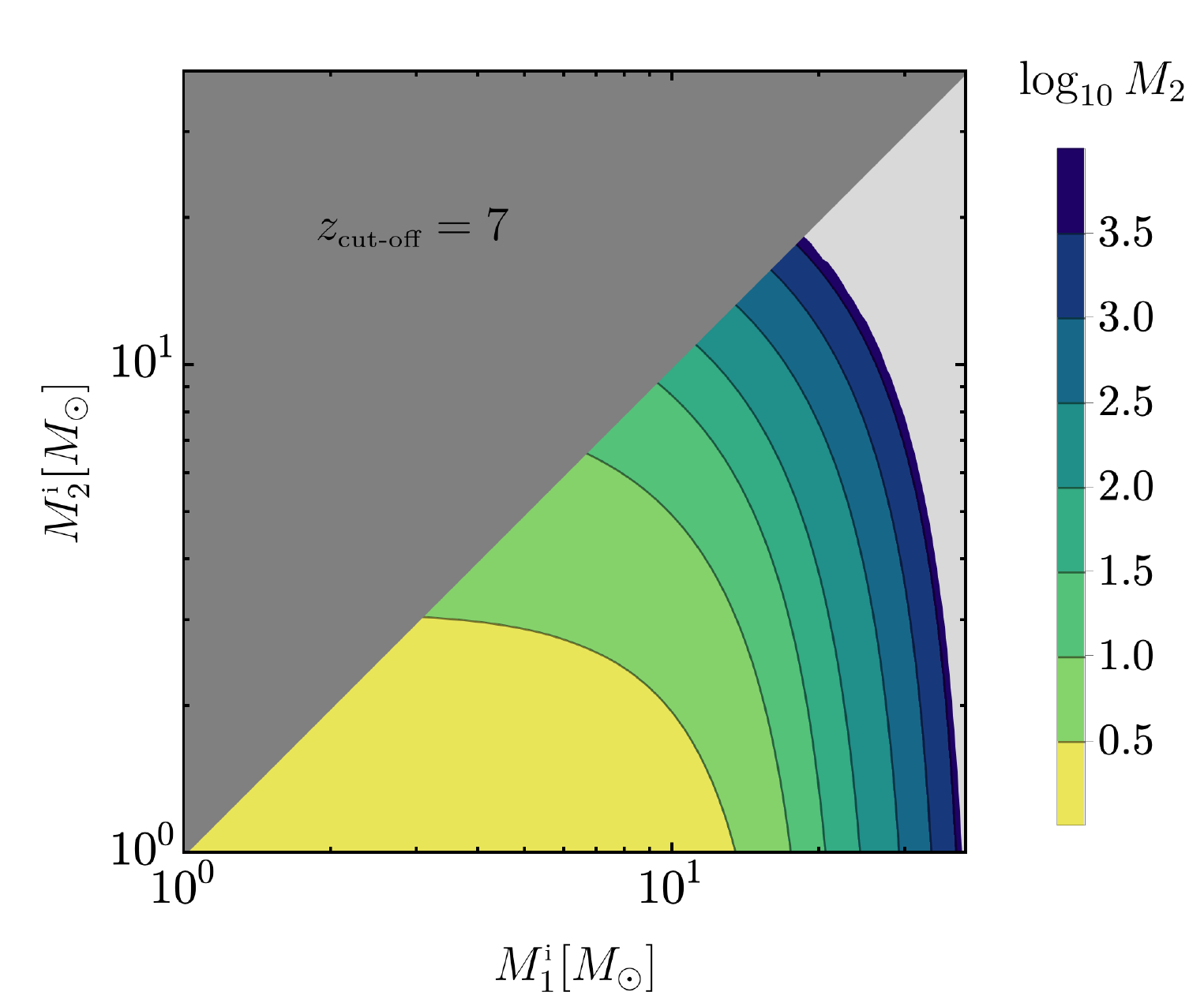}
	
	\includegraphics[width= 0.3 \linewidth]{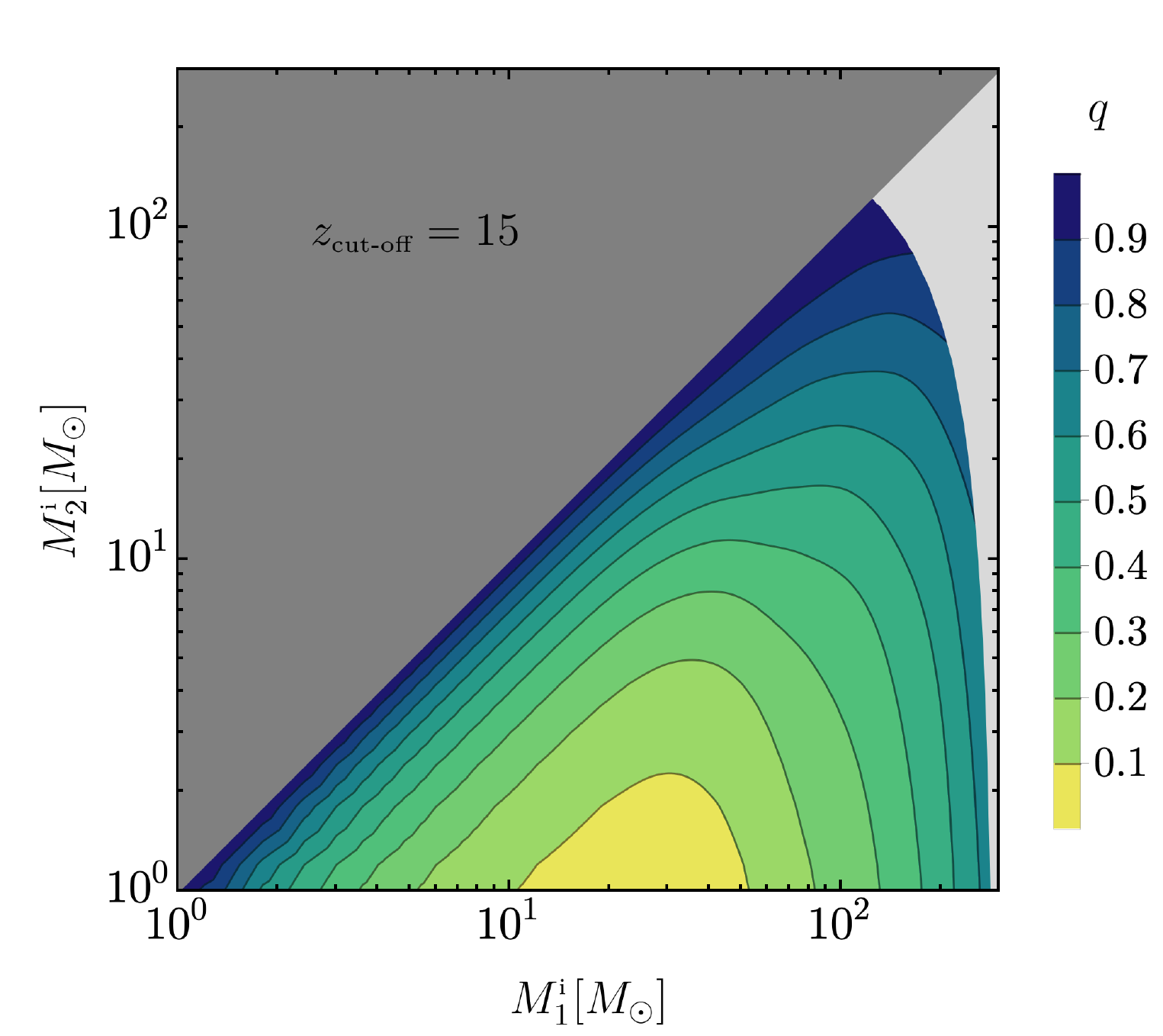}
	\includegraphics[width= 0.3 \linewidth]{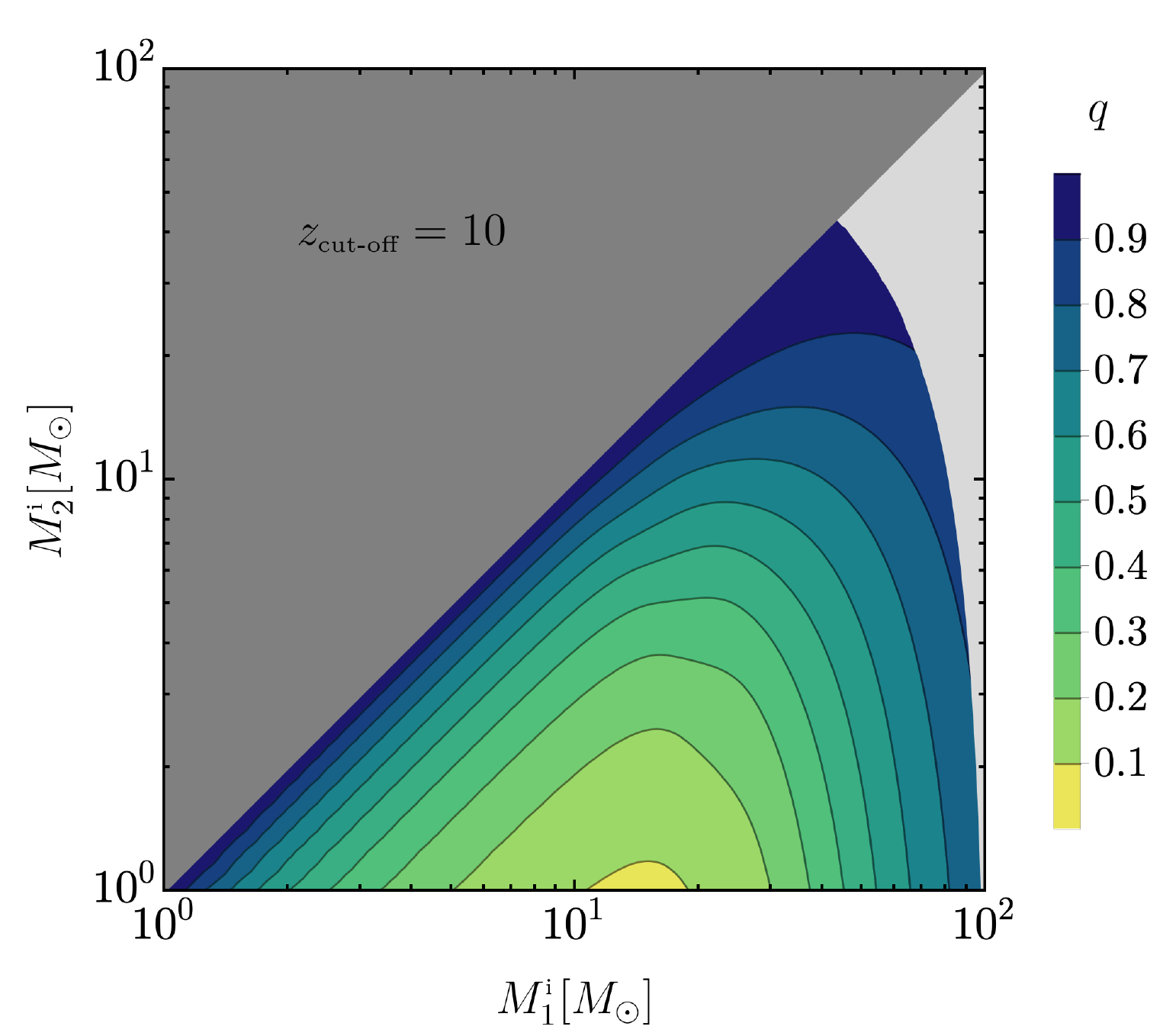}
	\includegraphics[width= 0.3 \linewidth]{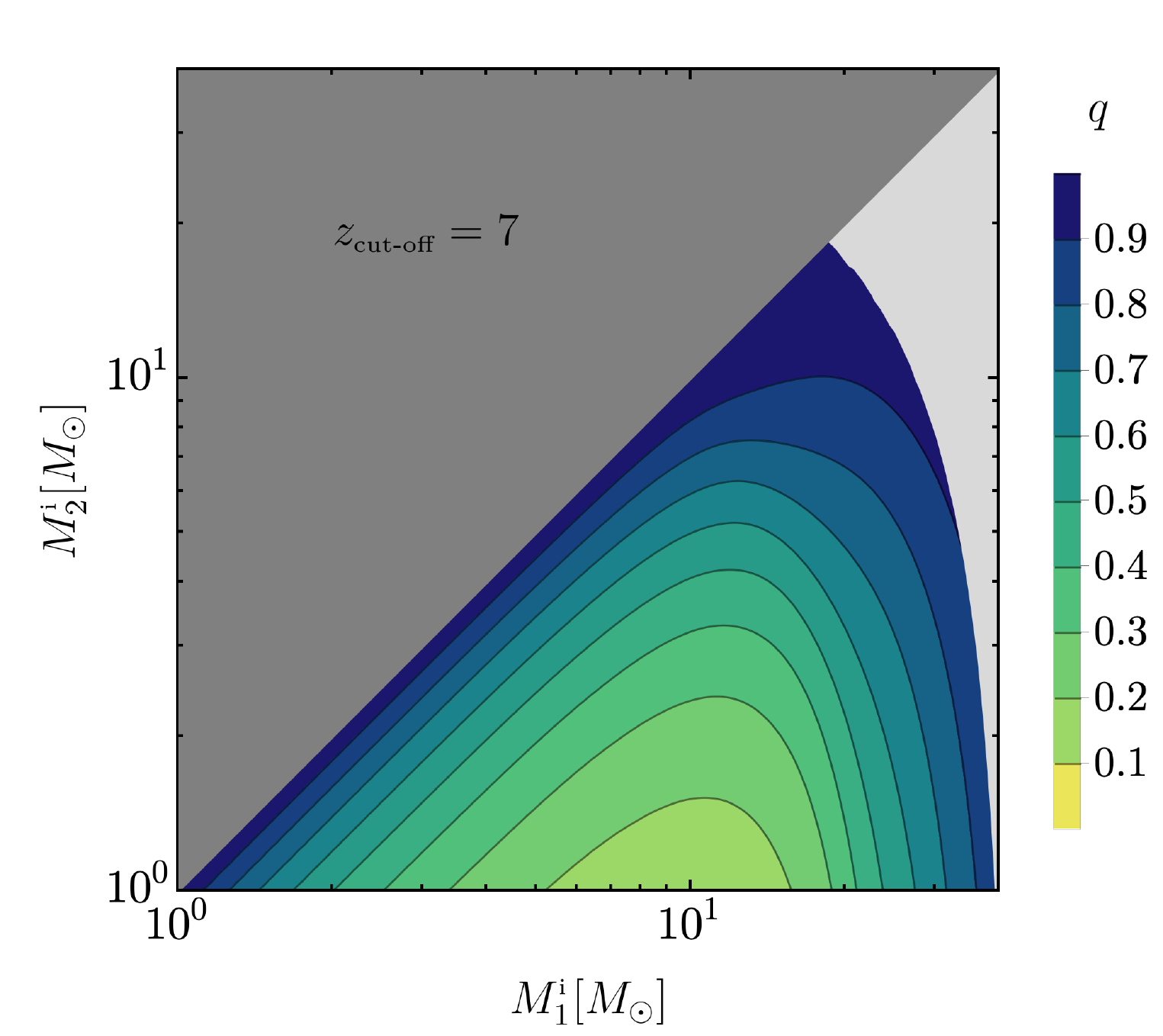}
	\caption{\it From top to bottom: final masses $M_1$, $M_2$, and final mass ratio $q$ as a function of the 
initial masses $(M^\text{\tiny\rm  i}_1, M^\text{\tiny \rm i}_2)$. We consider three accretion scenarios parametrised 
by 
$z_\text{\tiny \rm cut-off}=15$ (left panels), $z_\text{\tiny \rm cut-off}=10$ (middle panels), and $z_\text{\tiny \rm cut-off}=7$ 
(right panels).
	}
	\label{MMqplots}
\end{figure}

Since the initial binary parameters are unmeasurable, it is more relevant to analyse the dependence of the final masses 
$M_1, M_2$ and their mass ratio $q$ in terms of the initial masses $M^\text{\tiny i}_1, M^\text{\tiny i}_2$. This is 
done in Fig.~\ref{MMqplots} for an accretion evolution 
until the cut-off redshift $z_\text{\tiny cut-off}=(15,10,7)$ (from left to right columns).

\begin{figure}[t!]
	\centering	
	\includegraphics[width=0.45 \linewidth]{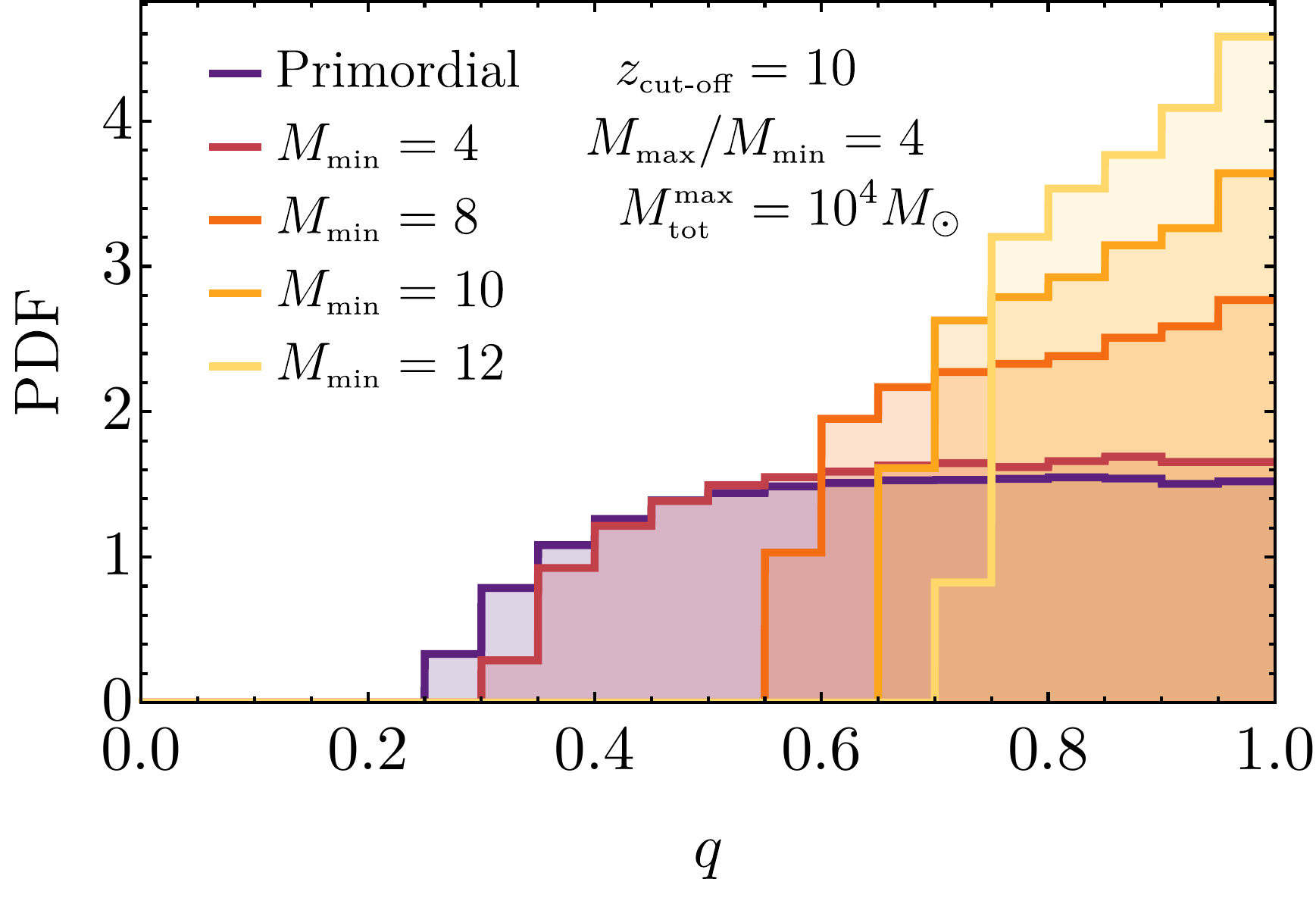}
	\includegraphics[width=0.45 \linewidth]{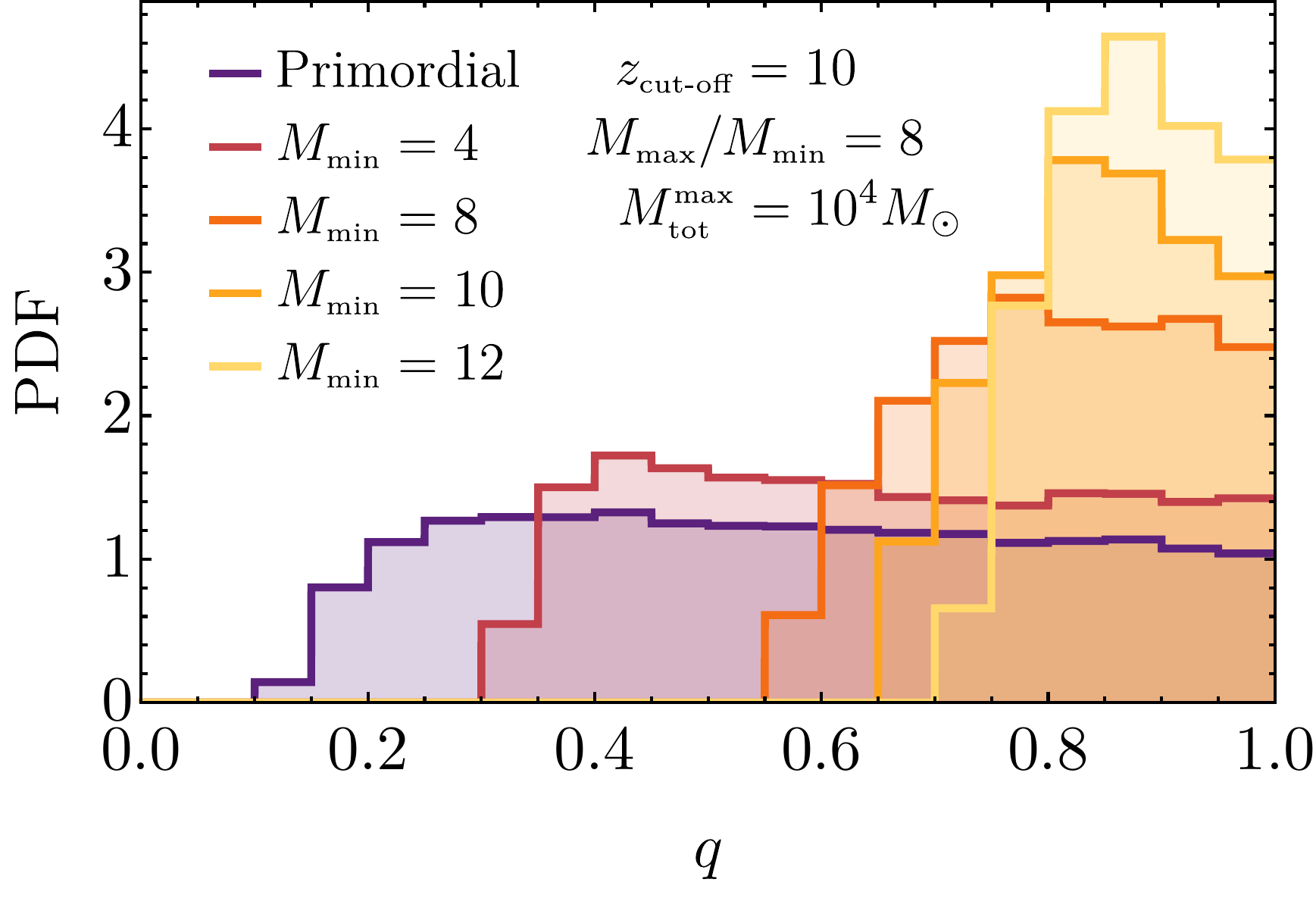}
	
	\includegraphics[width=0.45 \linewidth]{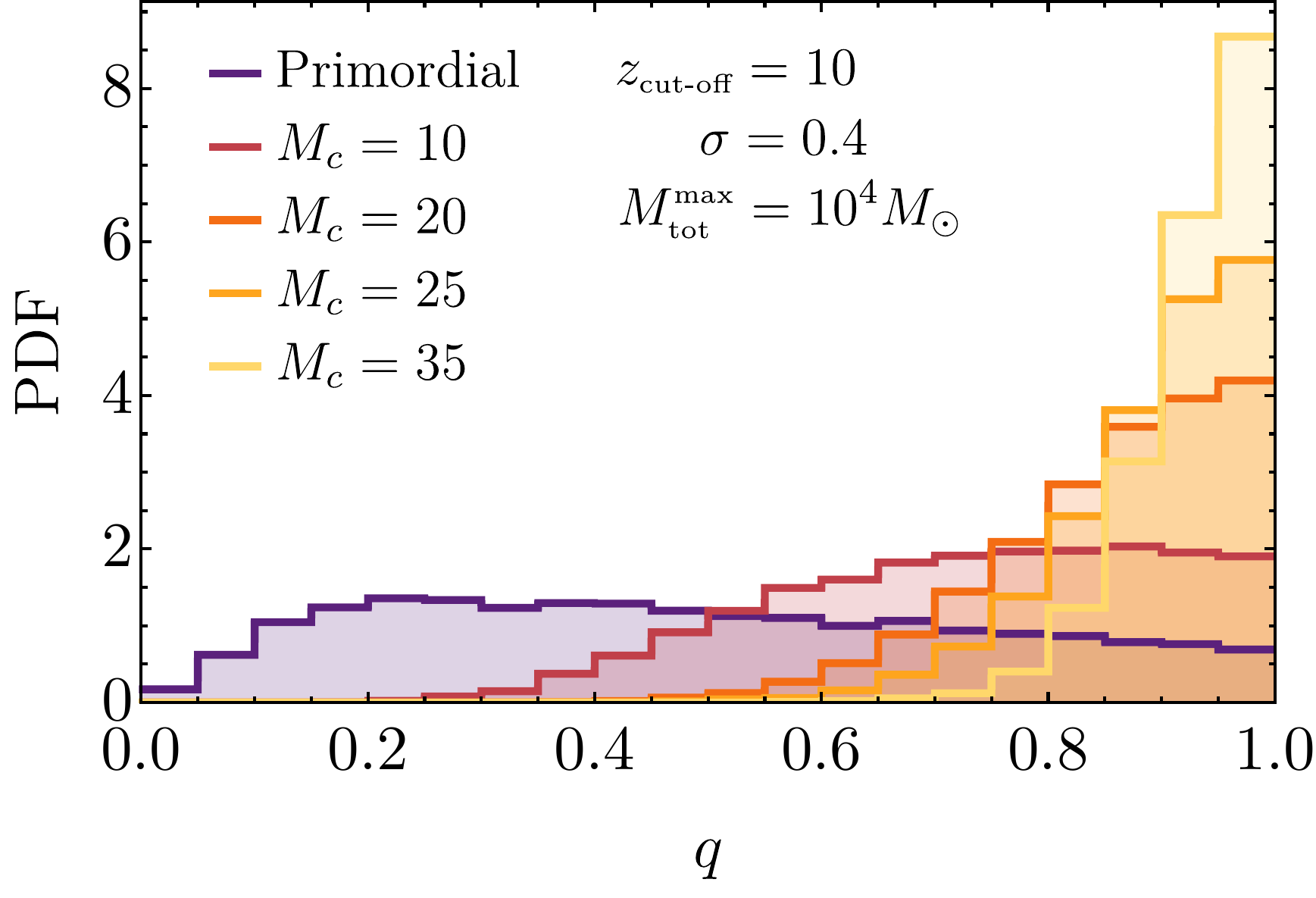}
	\includegraphics[width=0.465 \linewidth]{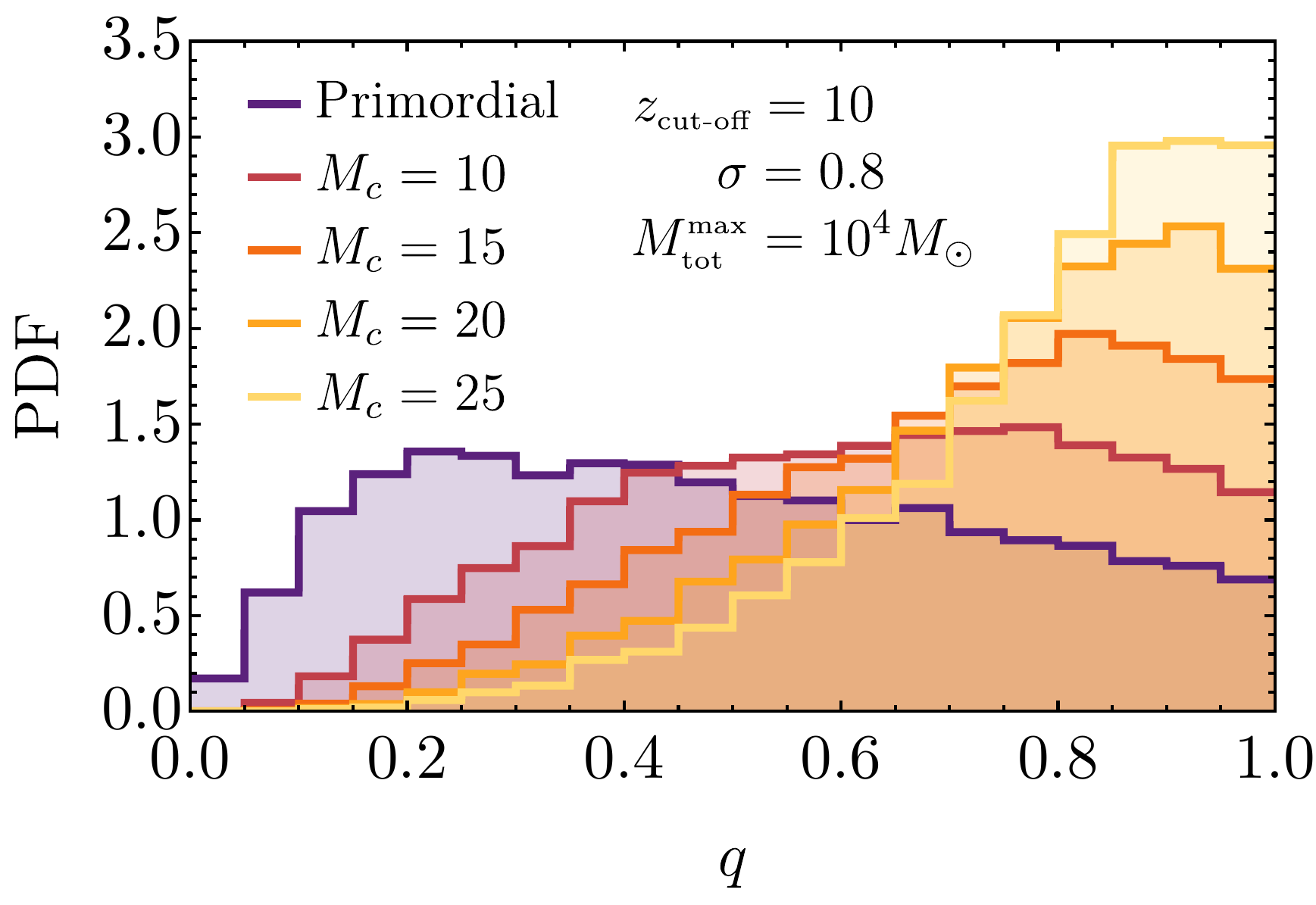}
	\caption{\it Examples of evolution of the $q$ distribution within the accretion scenario with $z_\text{\tiny \rm cut-off}=10$.	The 
distributions are found imposing a cut on the total mass equal to $M_\text{\tiny \rm tot}^\text{\tiny \rm max} =10^4 M_\odot$. 
Top and bottom panels correspond to power-law and lognormal mass functions, respectively.
	}
\label{qdistnocut}
\end{figure}

As shown in Sec.~\ref{sec:accretionbinary}, accretion onto a PBH binary system is such that the secondary body 
experiences a stronger (specific) accretion rate compared to the primary body. Since the large majority of PBH binaries 
that merge in the LIGO/Virgo band are formed before accretion is relevant~\cite{Sasaki:2018dmp}, it is expected that 
strongly-accreting binaries tend to have mass ratios close to unity. This is shown in Fig.~\ref{qdistnocut}, in which 
we 
present the distribution of the final mass ratio for an initial power-law and lognormal mass function for several 
choices of ($M_{\text{\tiny min}}$, $M_{\text{\tiny max}}$) and ($\sigma$, $M_c$), respectively. The distributions are 
constructed by drawing the values of ($M^\text{\tiny i}_1, M^\text{\tiny i}_2$) from a 
given initial mass function, subject to the constraint that $M_\text{\tiny tot}^\text{\tiny max} <10^4 M_\odot$.
Note, however, that this plot does not represent the distribution of $q$ expected in actual events, since it does not 
take into account neither how the merger rates depend on the parameters nor the sensitivity curve of LIGO/Virgo, which 
is optimal for frequencies corresponding to the merger of binary BHs  with $M_\text{\tiny tot}^\text{\tiny max} <10^2 
M_\odot$. A direct confrontation with current GW events will be presented in Sec.~\ref{sec:confrontation}.

\subsubsection{Spin evolution}\label{subsec:spinevol}
In the absence of accretion, or if mass accretion is not efficient enough, the spin of PBHs is natal.
As discussed in Sec.~\ref{sec:spindistribution} the dimensionless spin parameter $\chi_\text{\tiny i}$ in the most 
likely formation scenarios is of the order of the percent or smaller, although larger values are predicted in less 
standard scenarios, see e.g.~\cite{Harada:2017fjm,Cotner:2017tir}. 

\begin{figure}[t!]
	\centering
	\includegraphics[width=0.32 \linewidth]{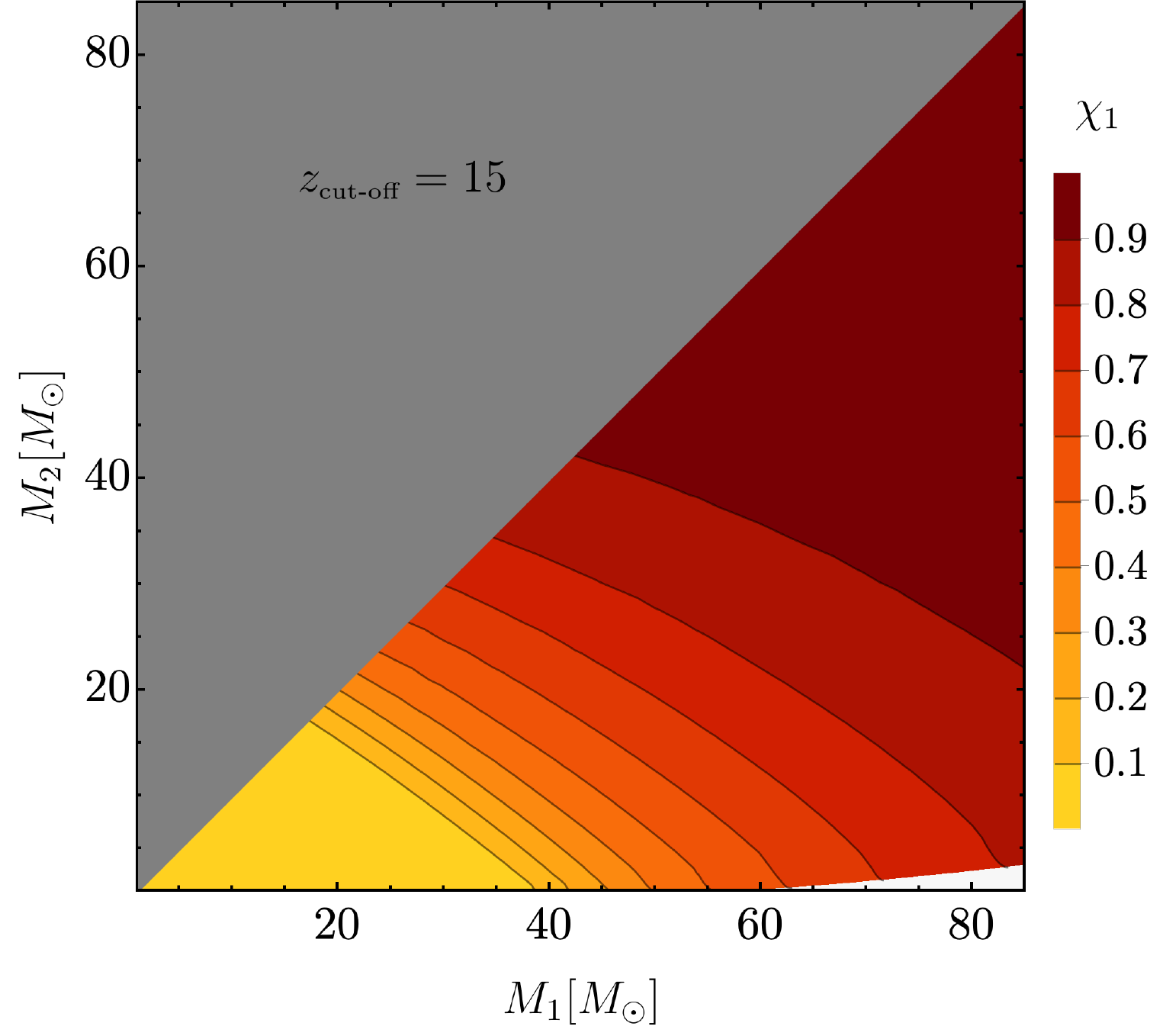}
	\includegraphics[width=0.32 \linewidth]{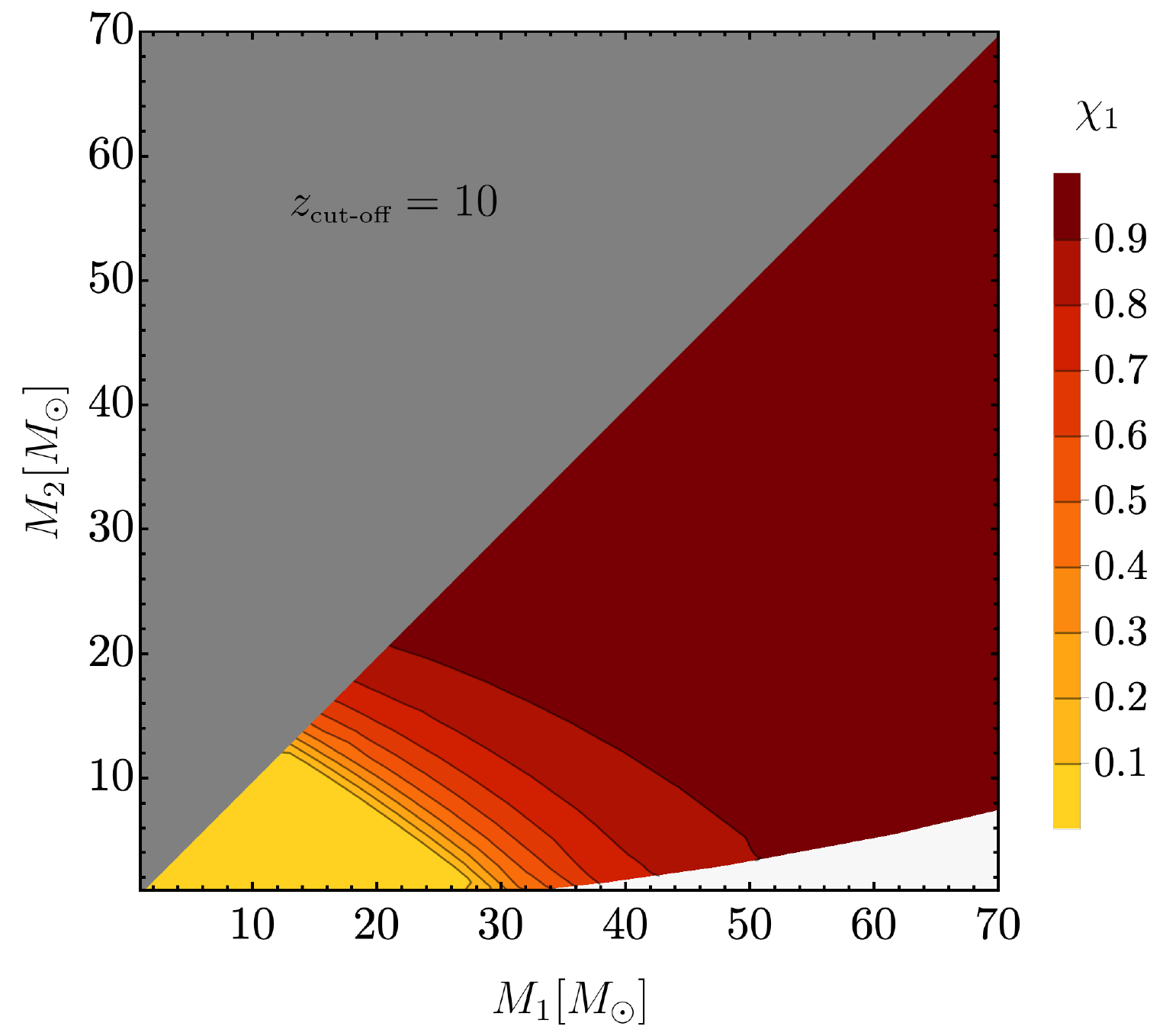}
	\includegraphics[width=0.32 \linewidth]{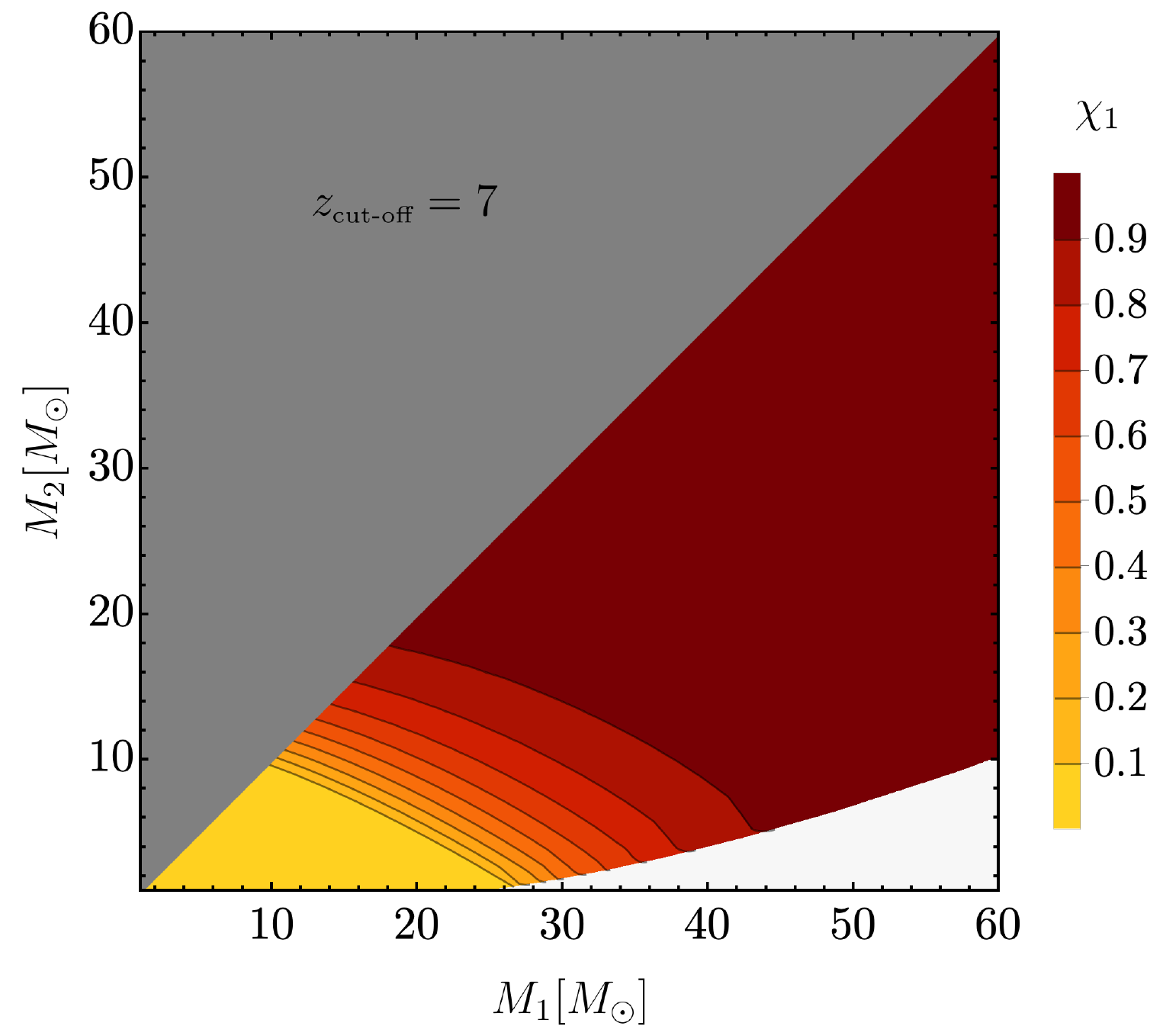}
	
	\includegraphics[width=0.32 \linewidth]{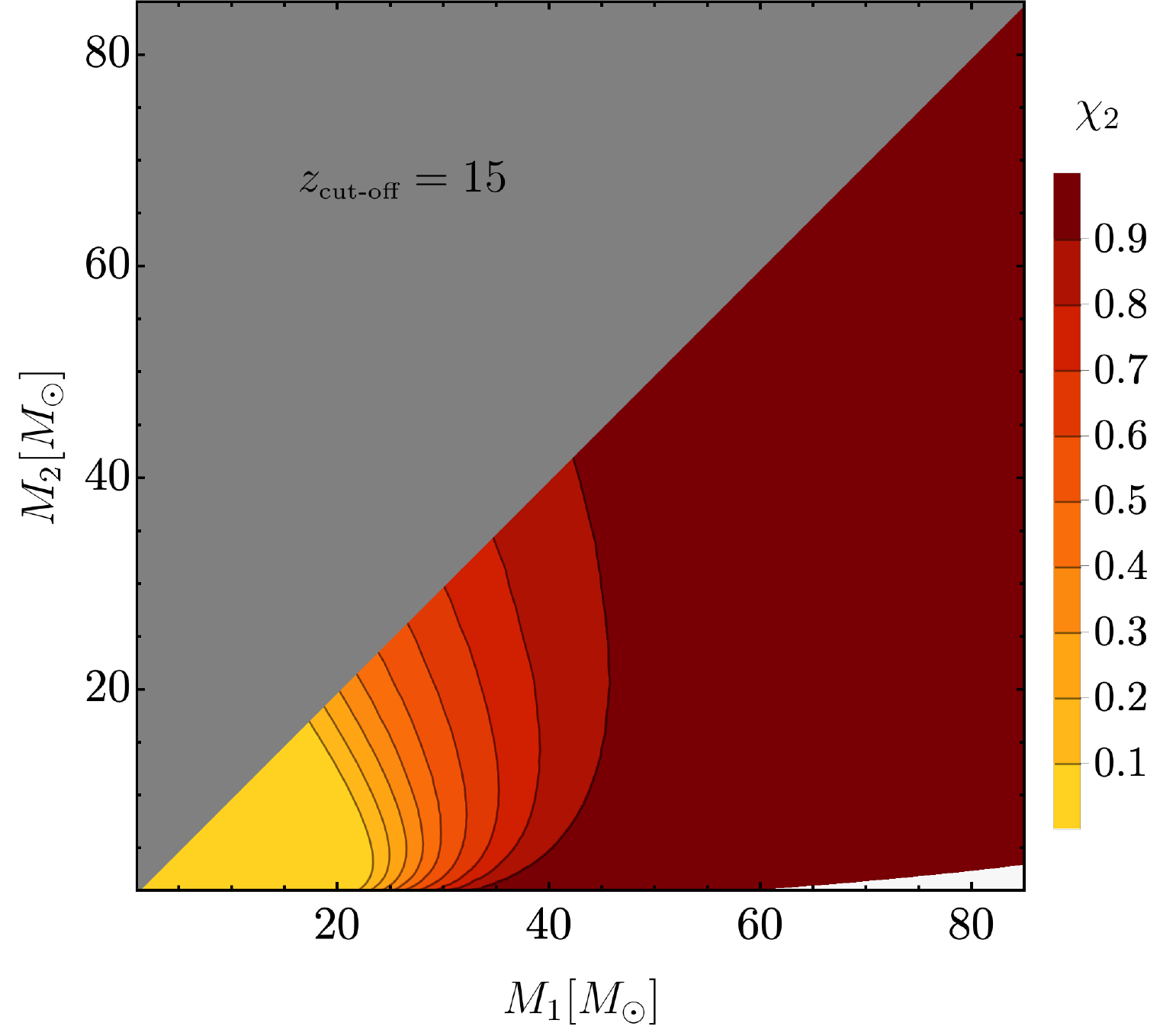}
	\includegraphics[width=0.32 \linewidth]{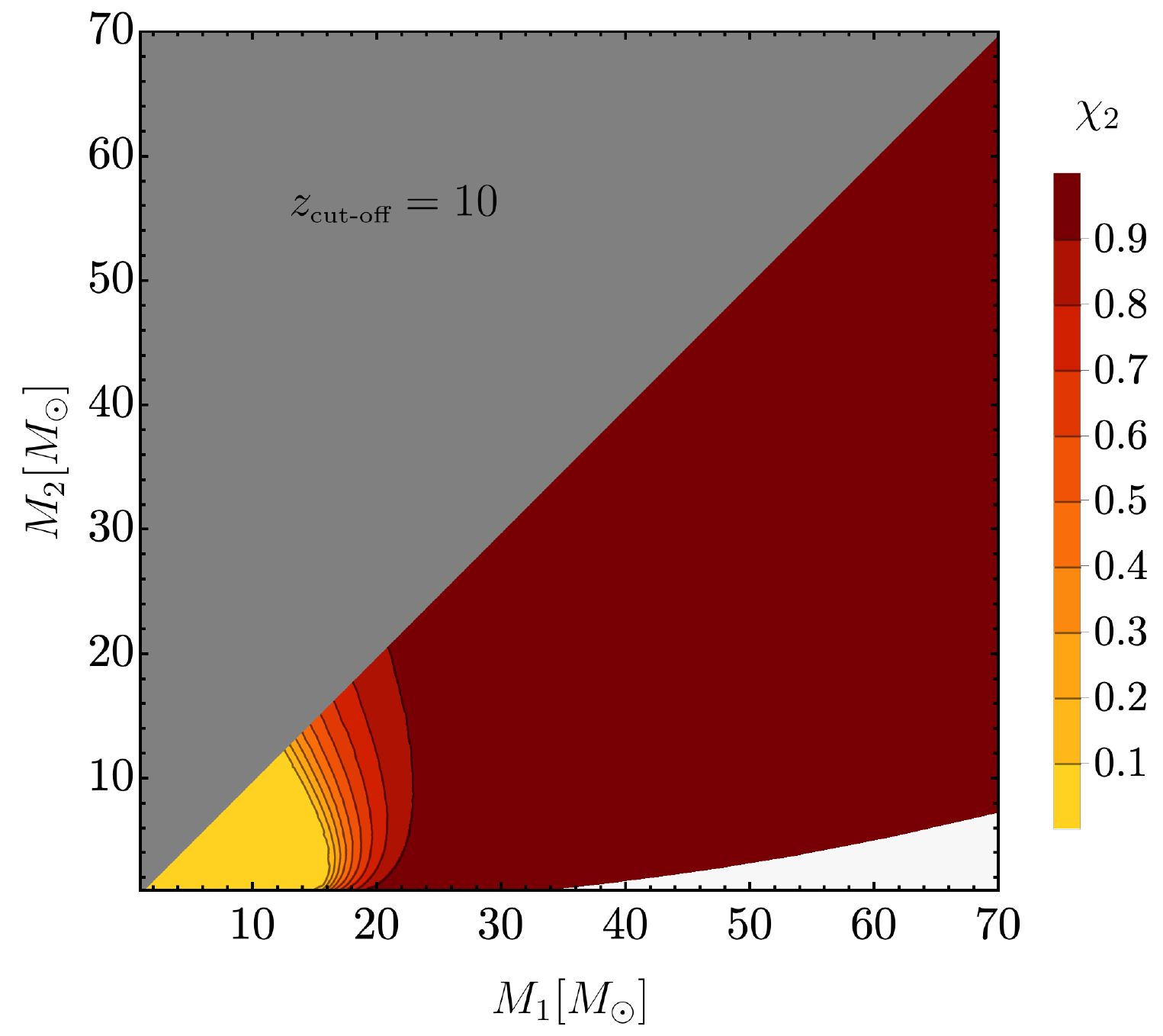}
	\includegraphics[width=0.32 \linewidth]{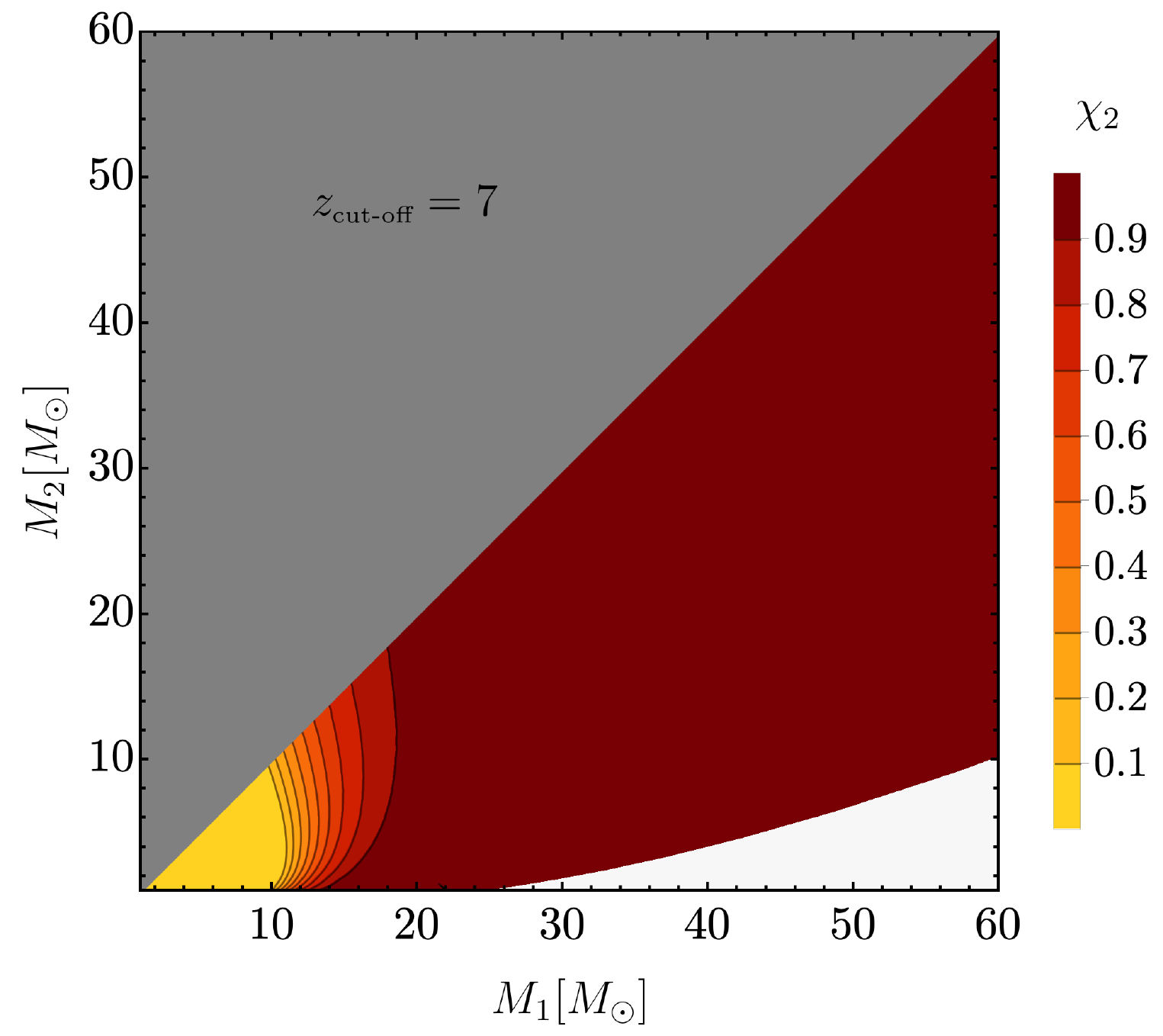}	
	\caption{\it Final spins $\chi_1$ (top panels) and $\chi_2$ (bottom panels) as functions of the {\it final} 
masses $(M_1, M_2)$. As in Fig.~\ref{MMqplots}, from left to right we consider three cut-off
redshifts below which accretion is suppressed.}
\label{spinevo}
\end{figure}

The situation changes drastically in the case of efficient accretion. In
Fig.~\ref{spinevo} we show the final spins of the PBH binary components as a function of their final masses for three 
different choices of $z_\text{\tiny cut-off}$. 
Besides the quantitative difference between different choices of 
$z_\text{\tiny cut-off}$, the qualitative trend is the same. Namely, the final 
mass and final spin of the PBHs are correlated: low-mass PBHs are slowly spinning or non-spinning, whereas high-mass 
PBHs are rapidly spinning. The mass scale at which this continuous transition occurs depends on the cut-off redshift 
and it is always around $(10\div 40)\,M_\odot$. In particular, smaller cut-offs favour a lower-mass transition, as expected by 
the fact that in this case the accreting phase lasts longer during the cosmic evolution.%

\begin{figure}[h!]
	\centering
	\includegraphics[width=0.24 \linewidth]{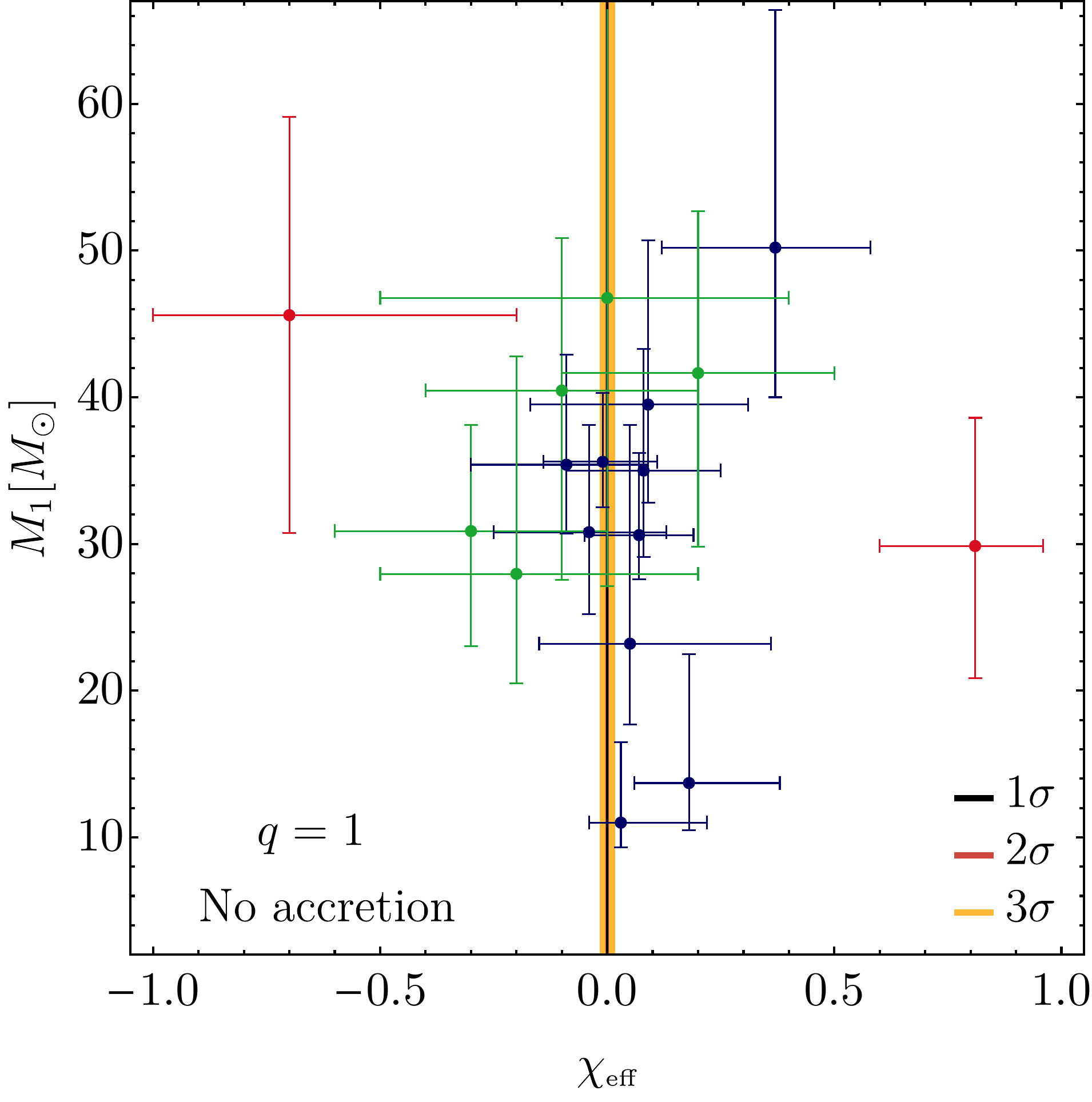}
	\includegraphics[width=0.24 \linewidth]{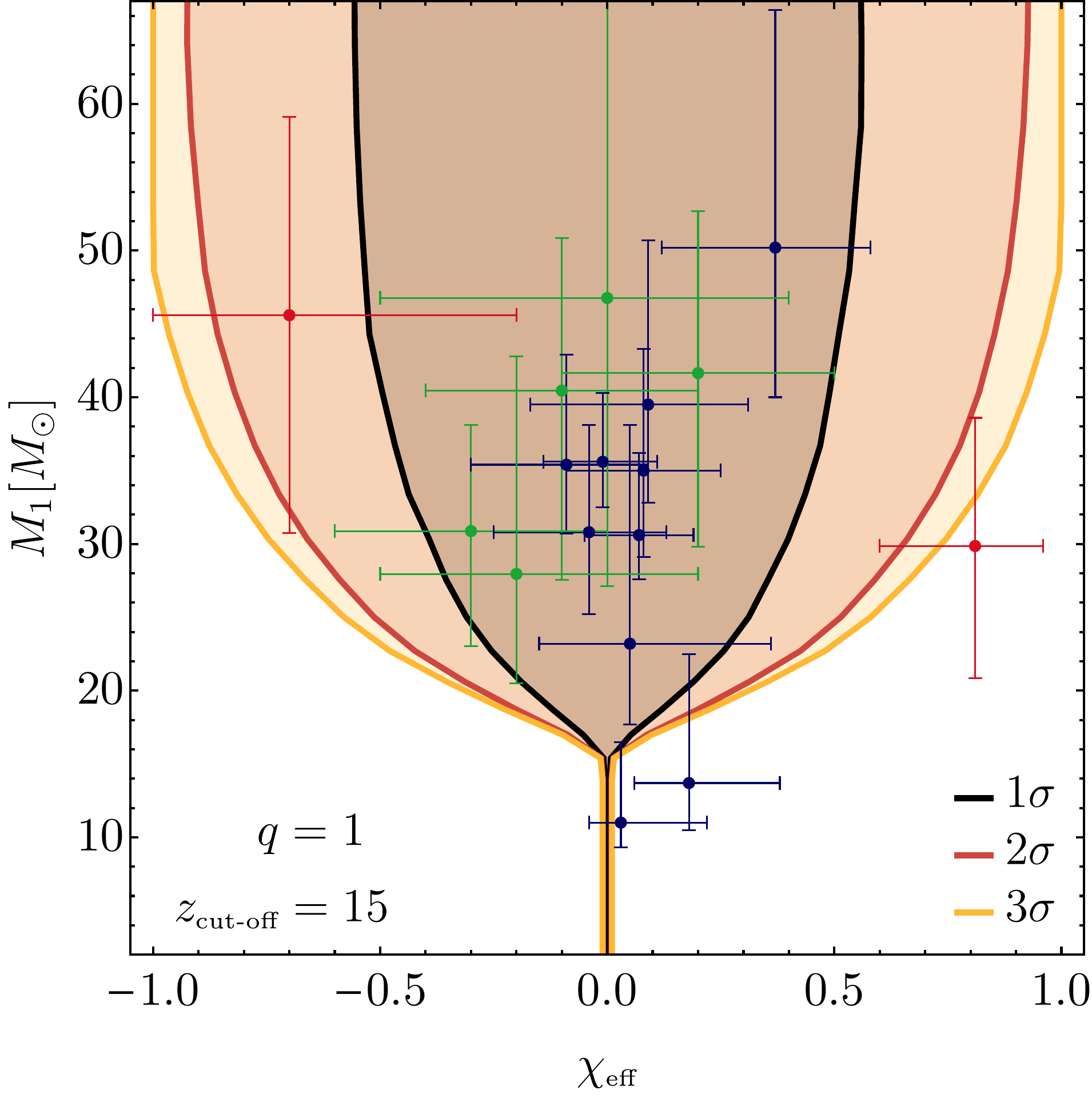}
	\includegraphics[width=0.24 \linewidth]{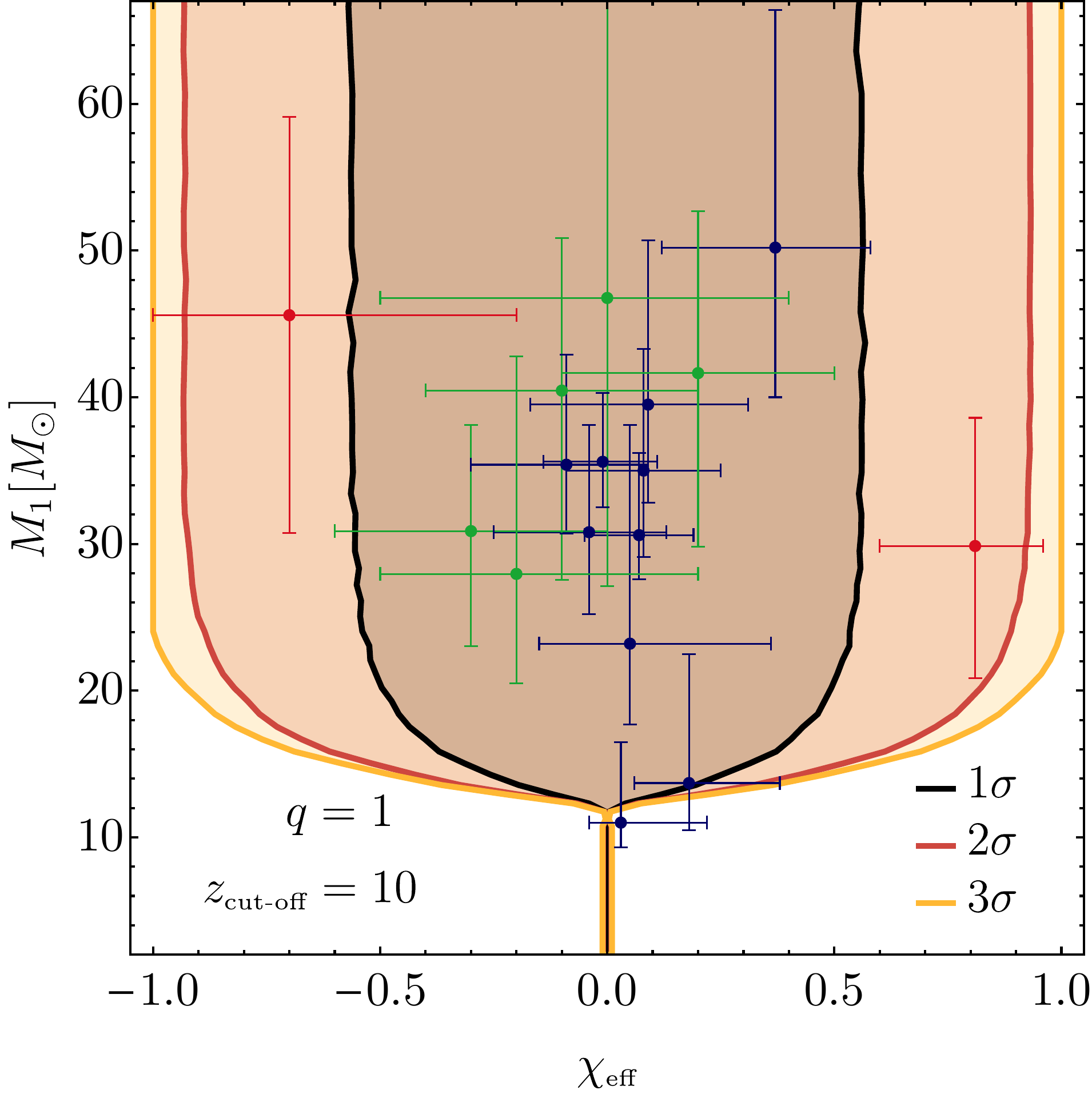}
	\includegraphics[width=0.24 \linewidth]{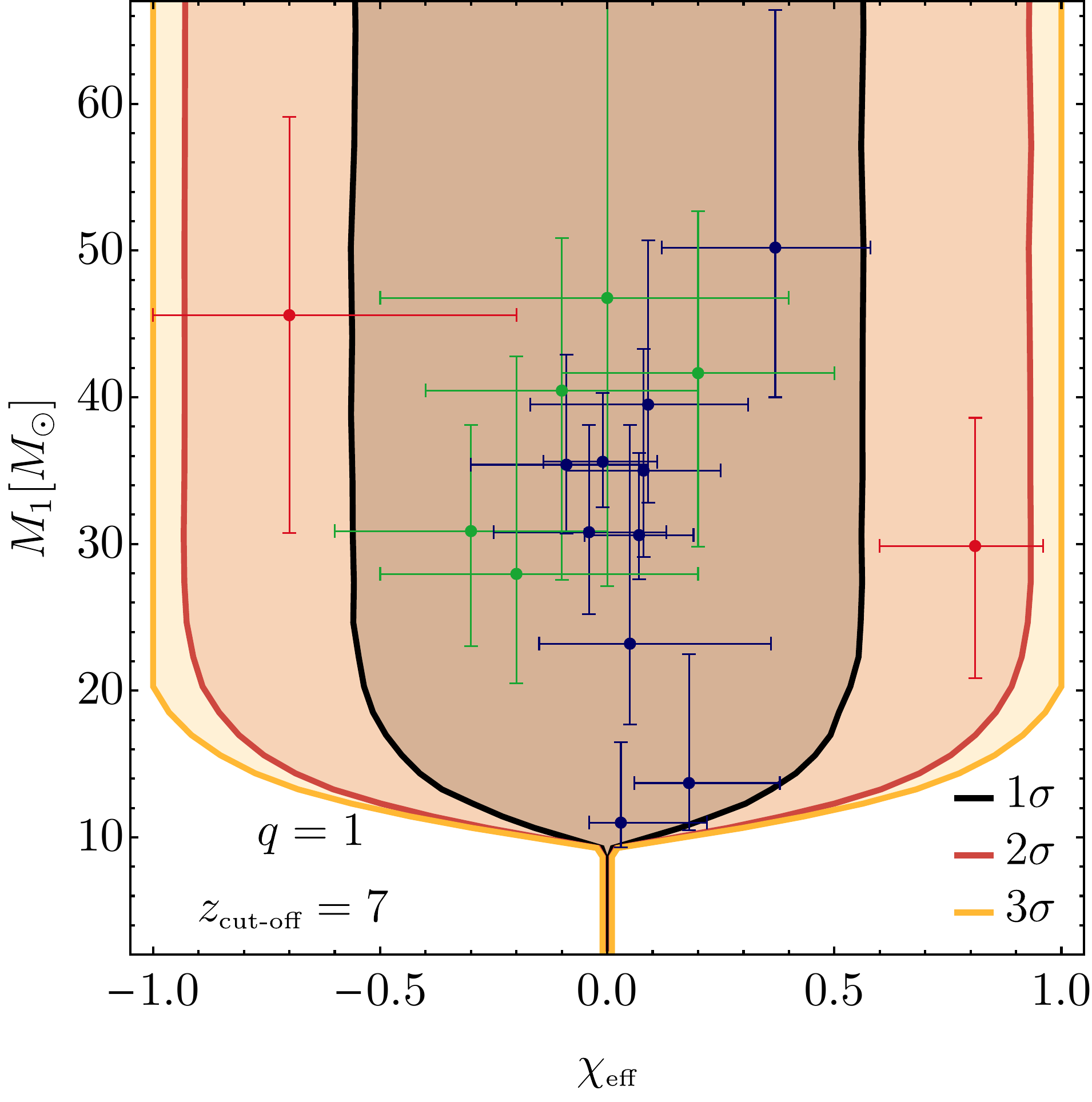}
	
	\includegraphics[width=0.24 \linewidth]{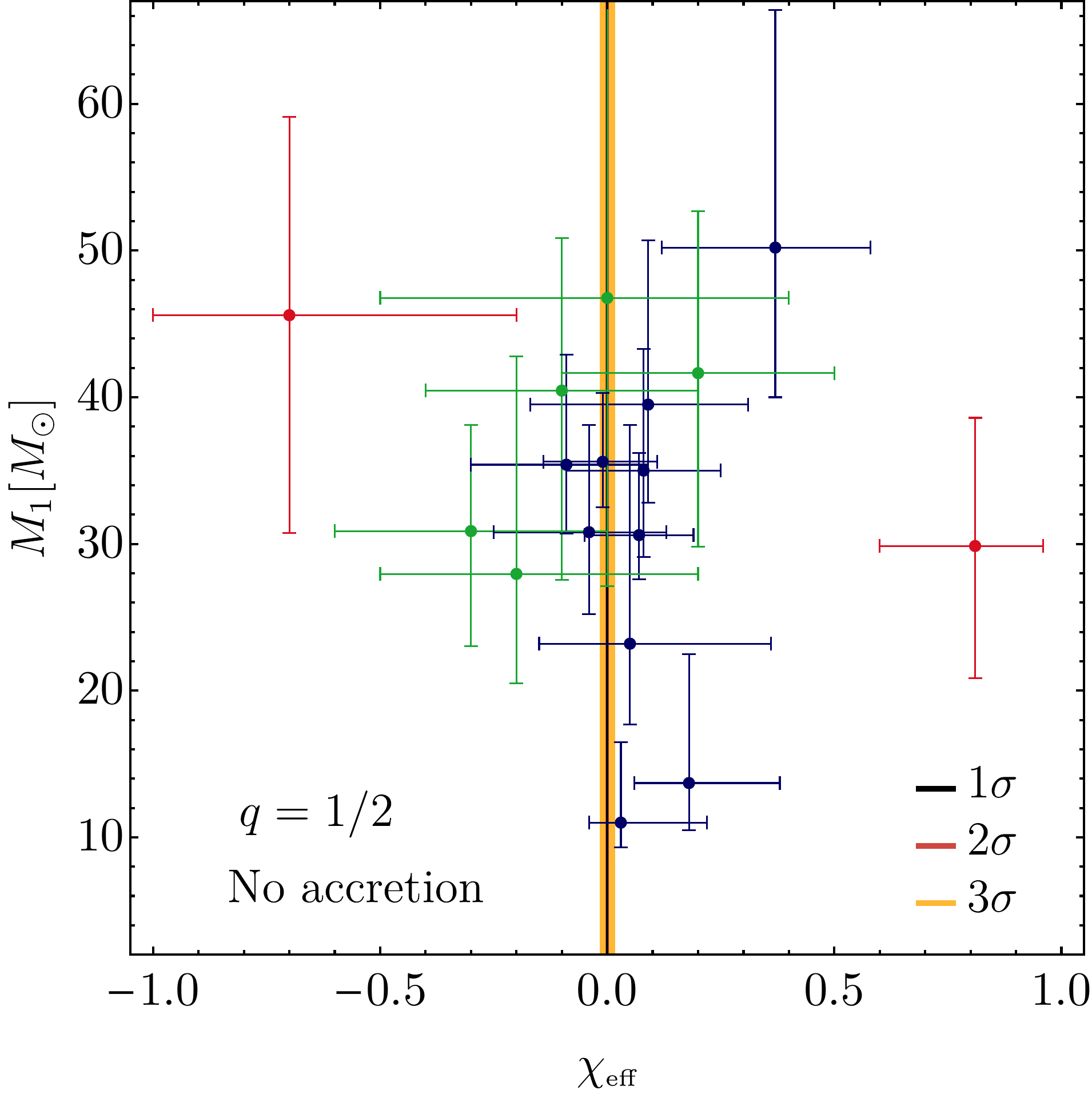}
	\includegraphics[width=0.24 \linewidth]{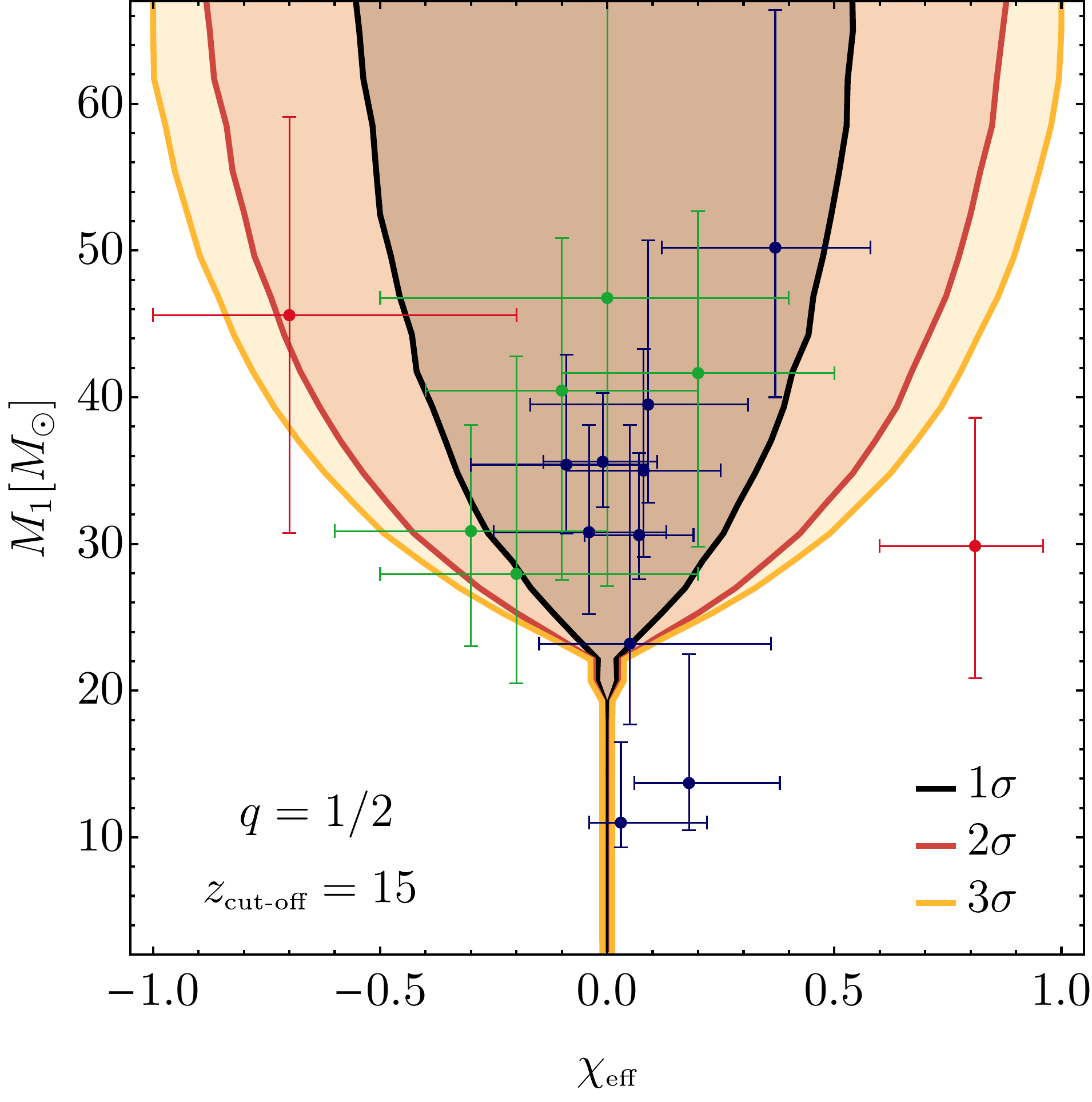}
	\includegraphics[width=0.24 \linewidth]{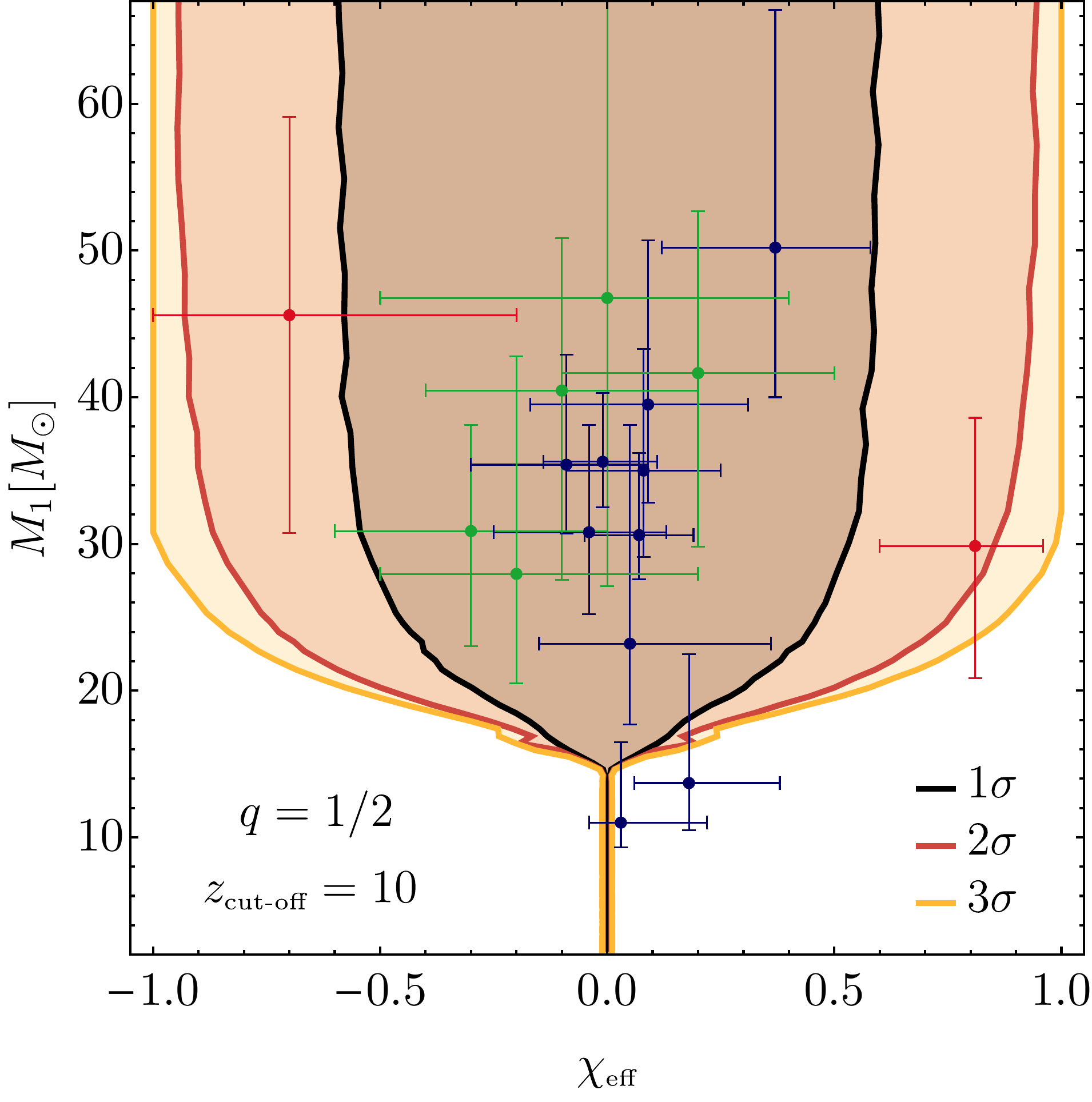}
	\includegraphics[width=0.24 \linewidth]{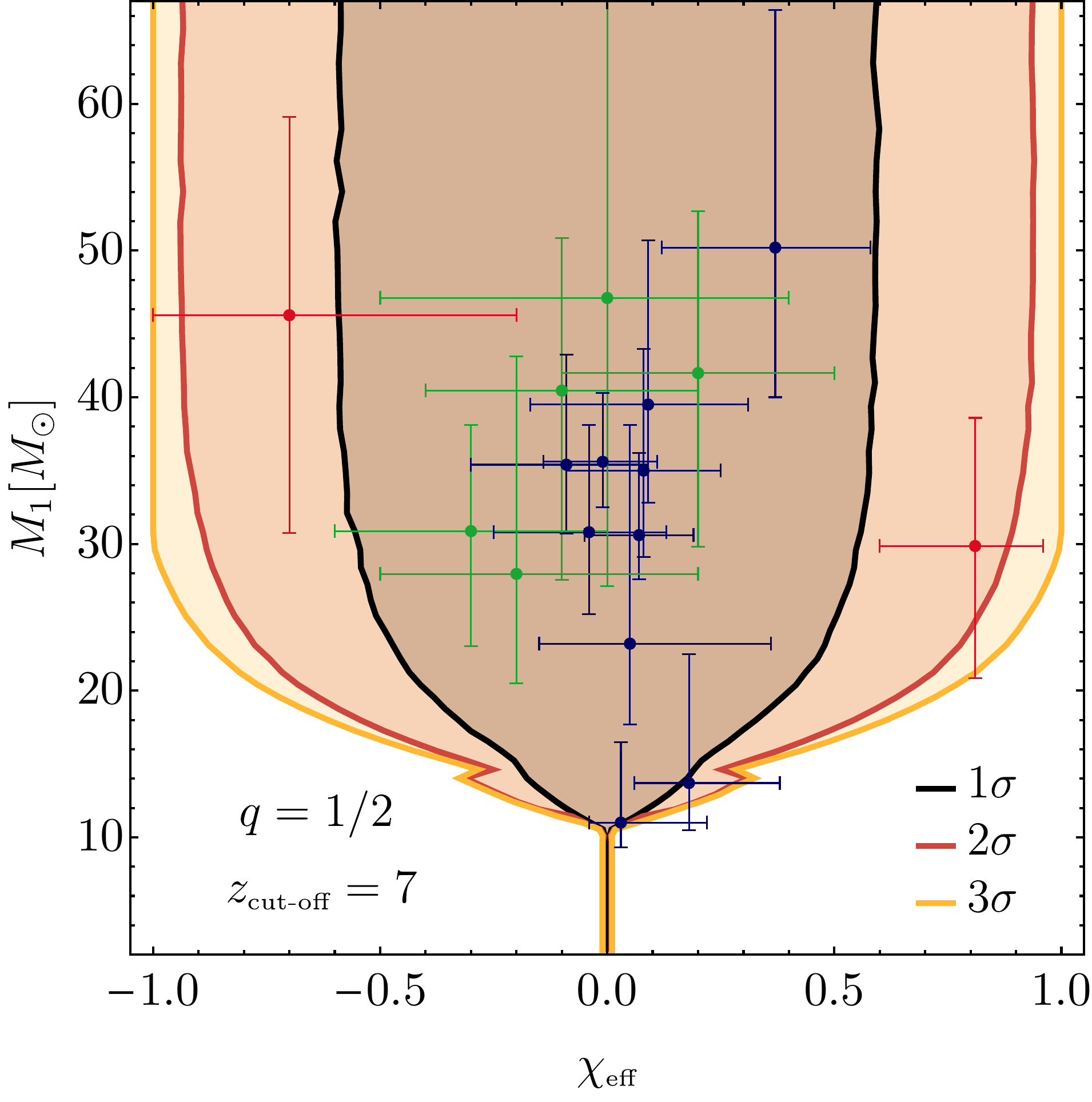}
	
	\includegraphics[width=0.24 \linewidth]{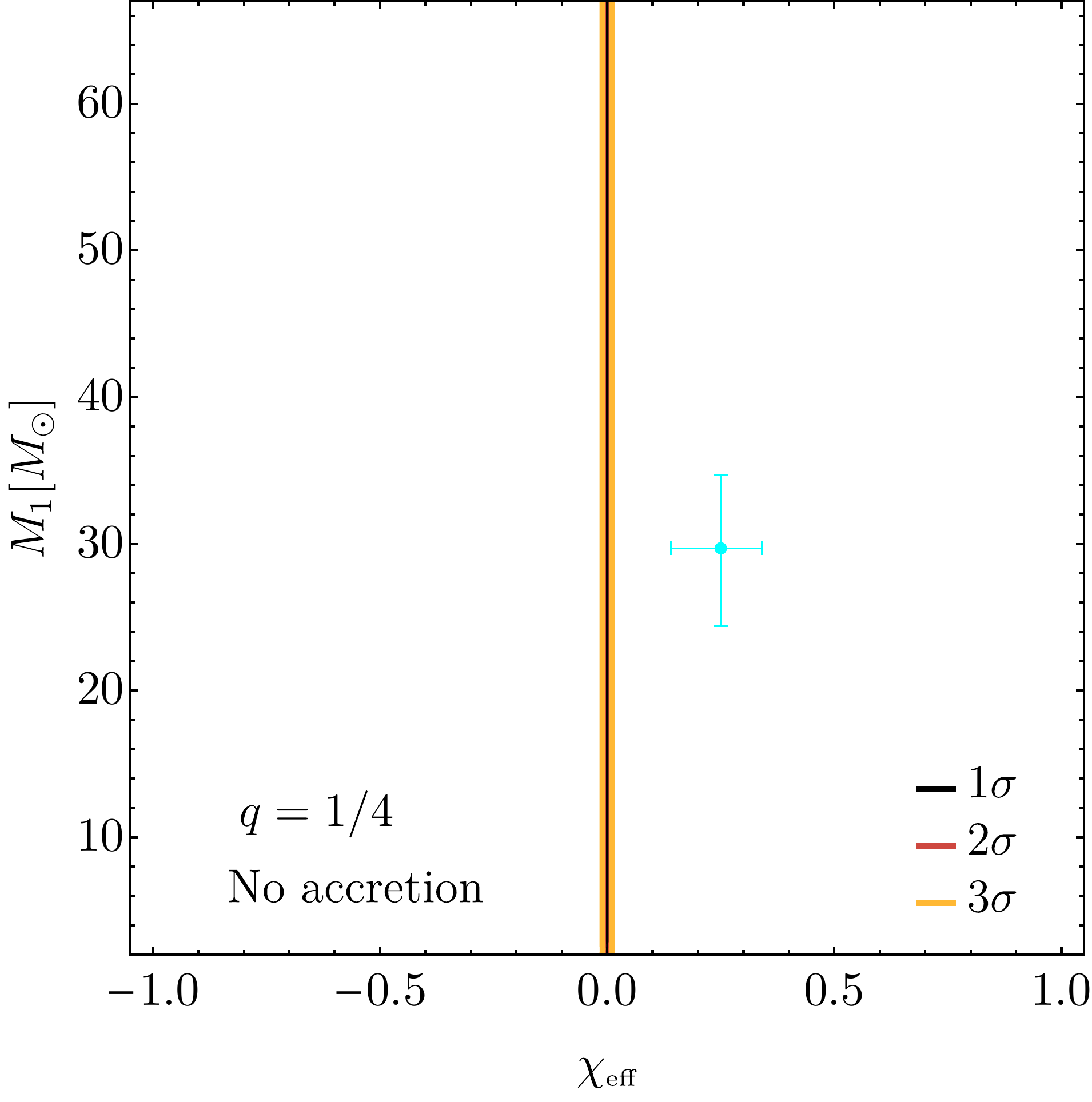}
	\includegraphics[width=0.24 \linewidth]{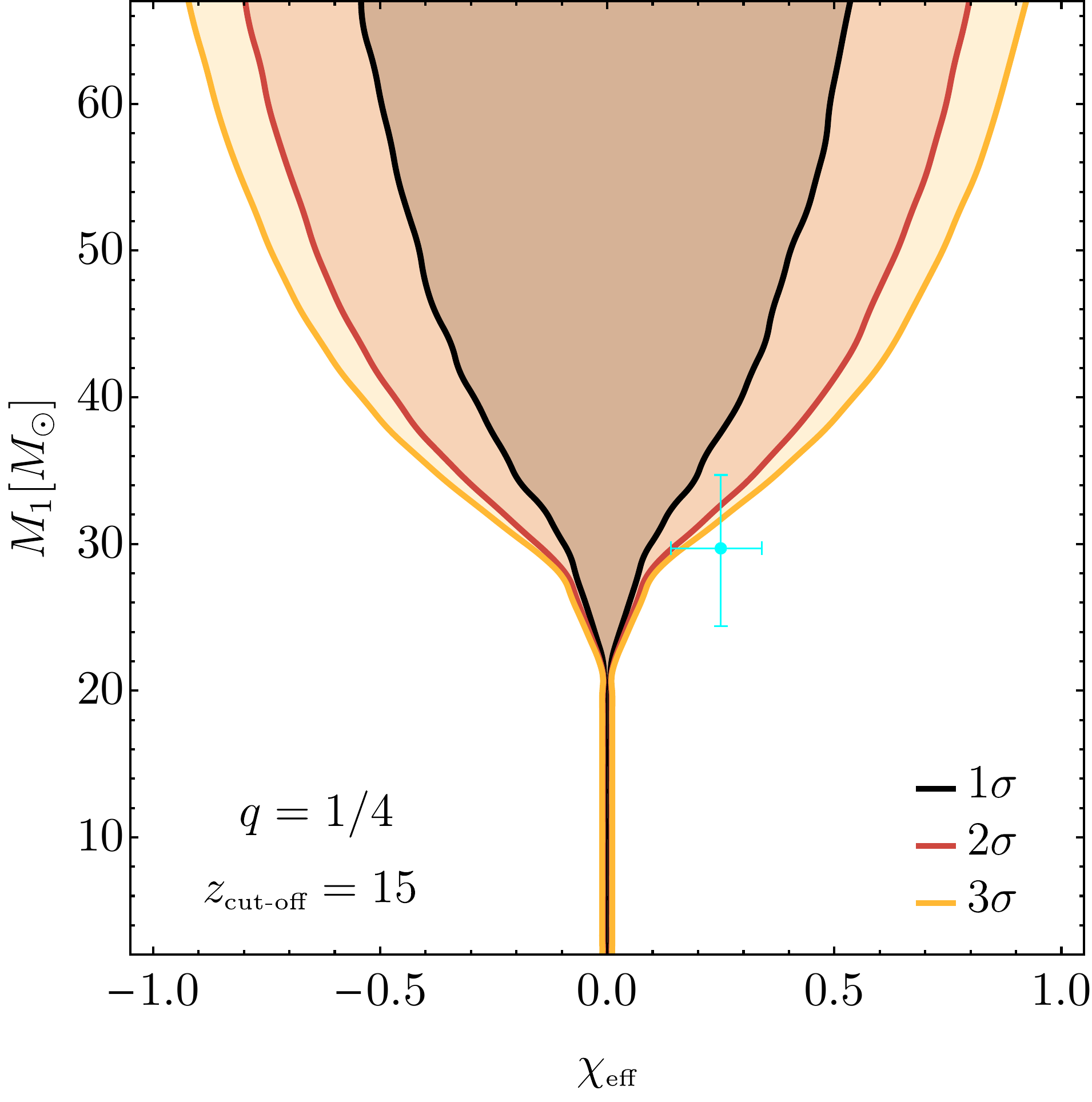}
	\includegraphics[width=0.24 \linewidth]{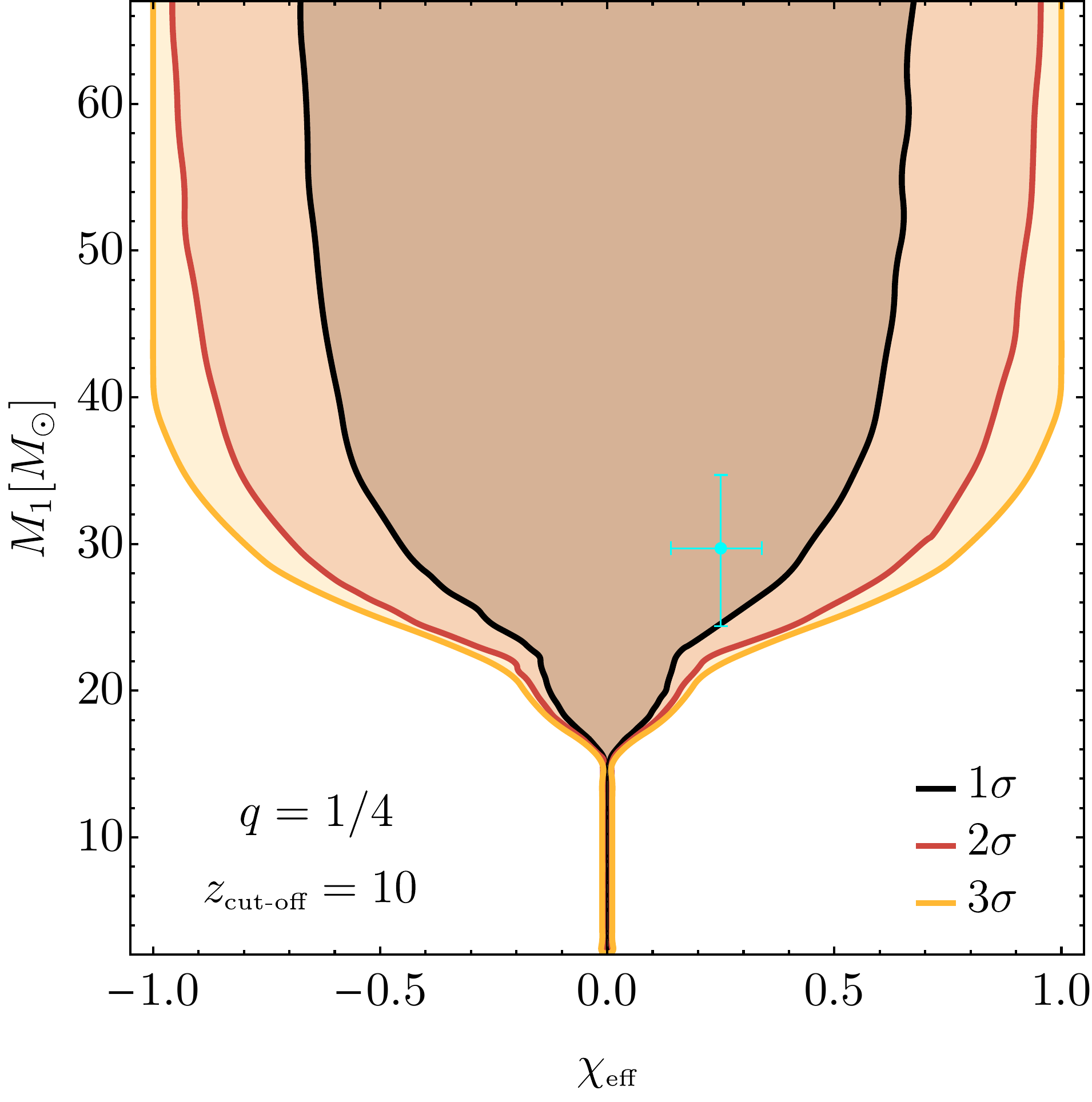}
	\includegraphics[width=0.24 \linewidth]{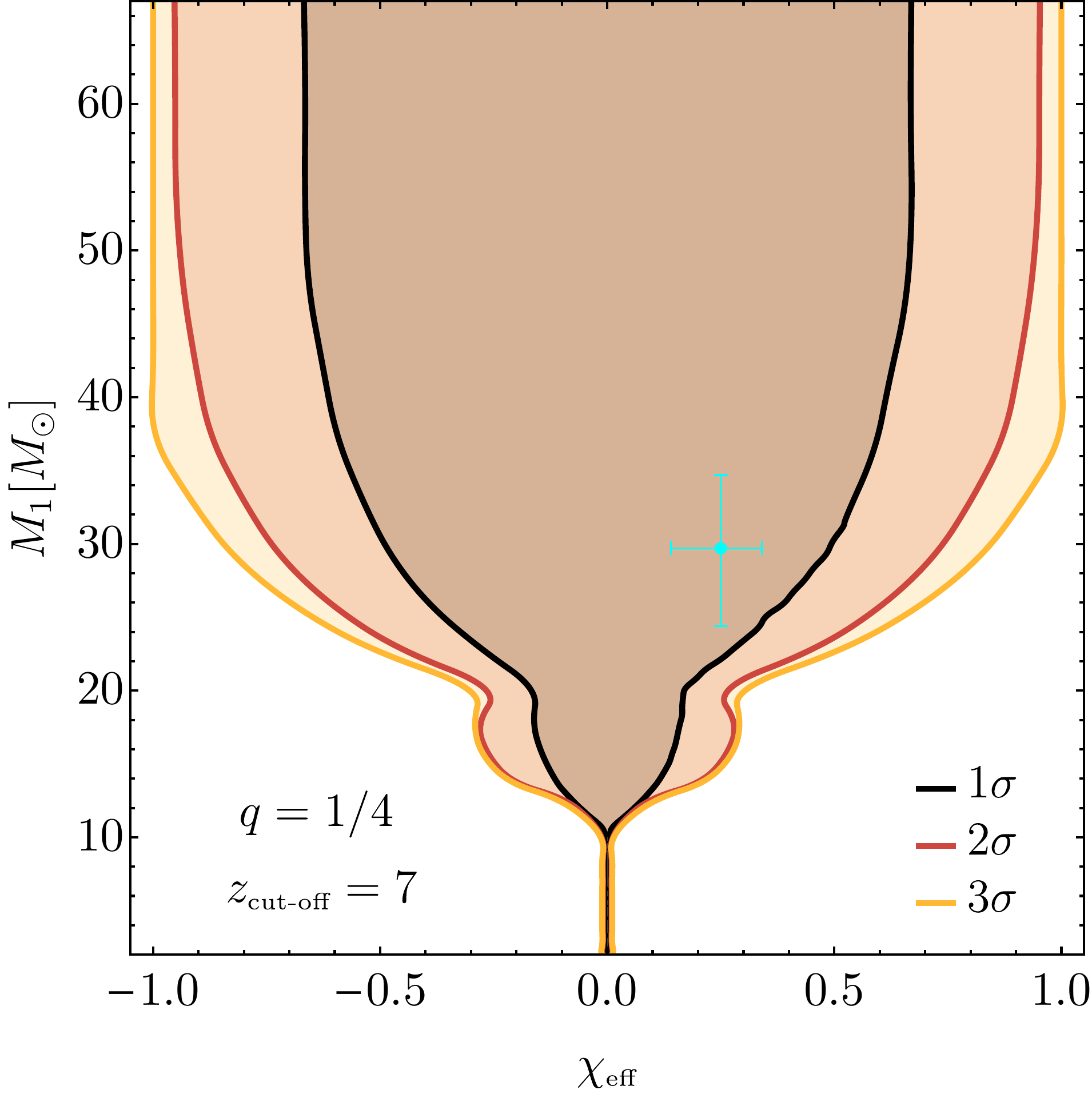}

	\includegraphics[width=0.24 \linewidth]{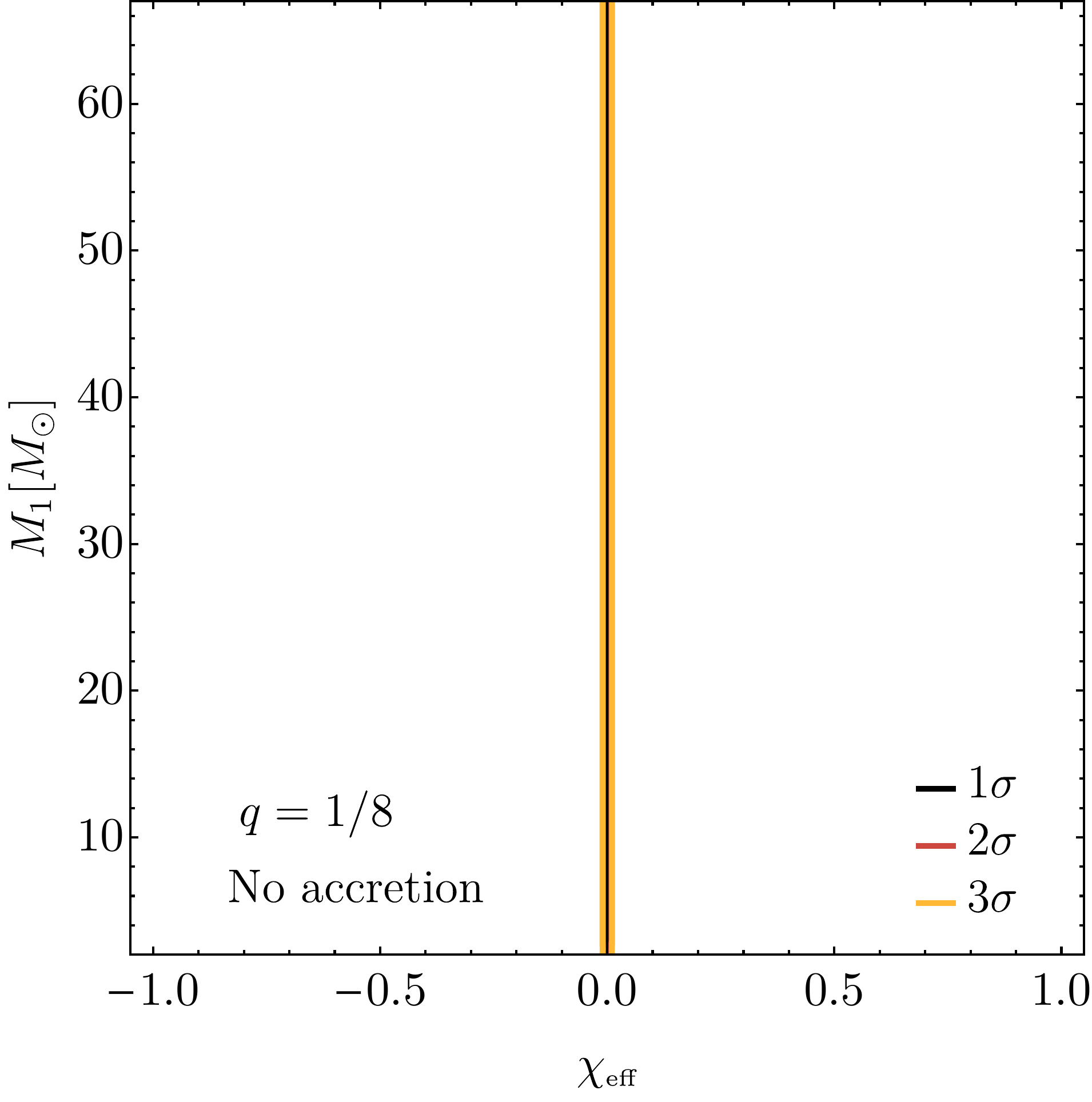}	
	\includegraphics[width=0.24 \linewidth]{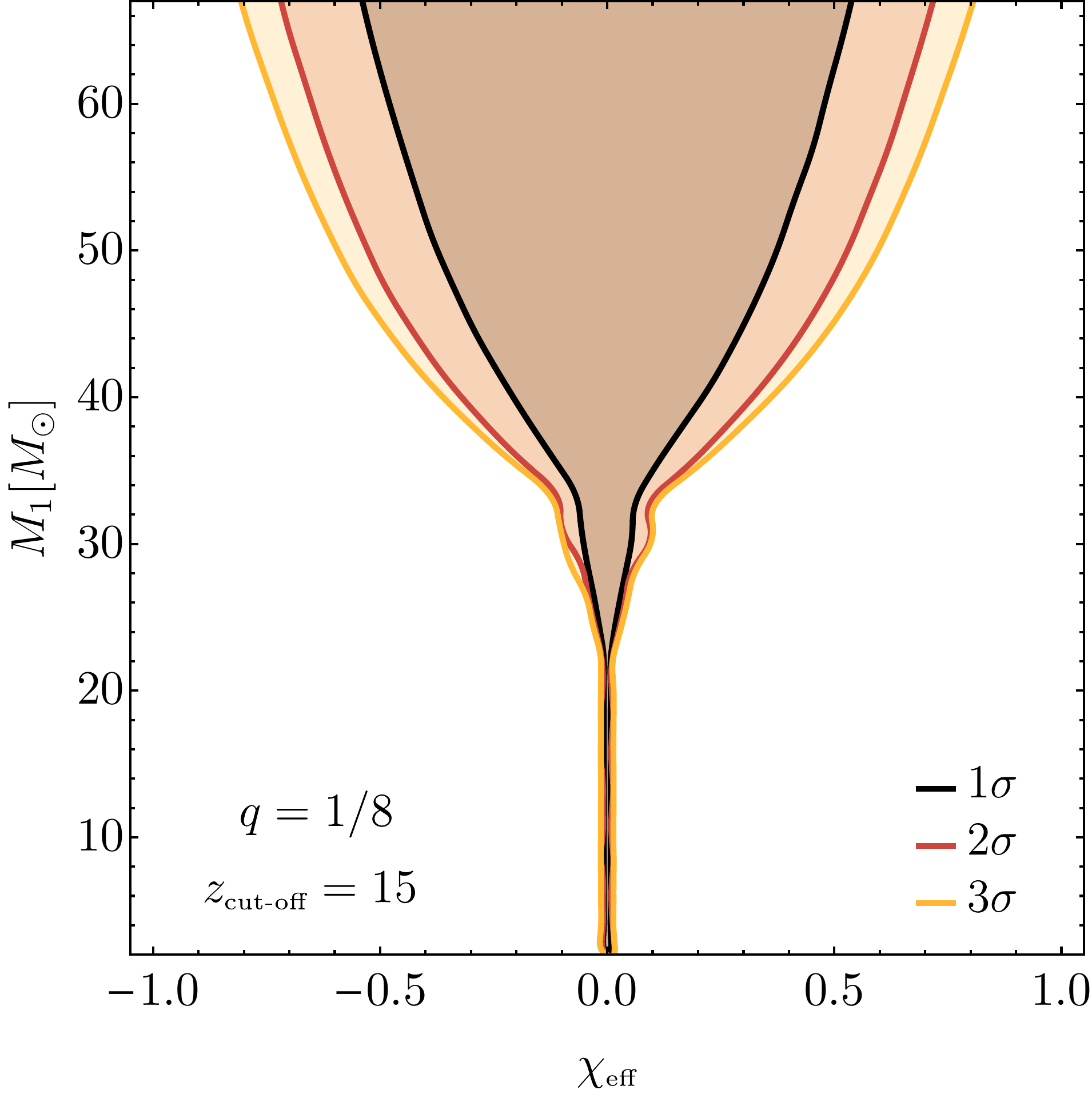}
	\includegraphics[width=0.24 \linewidth]{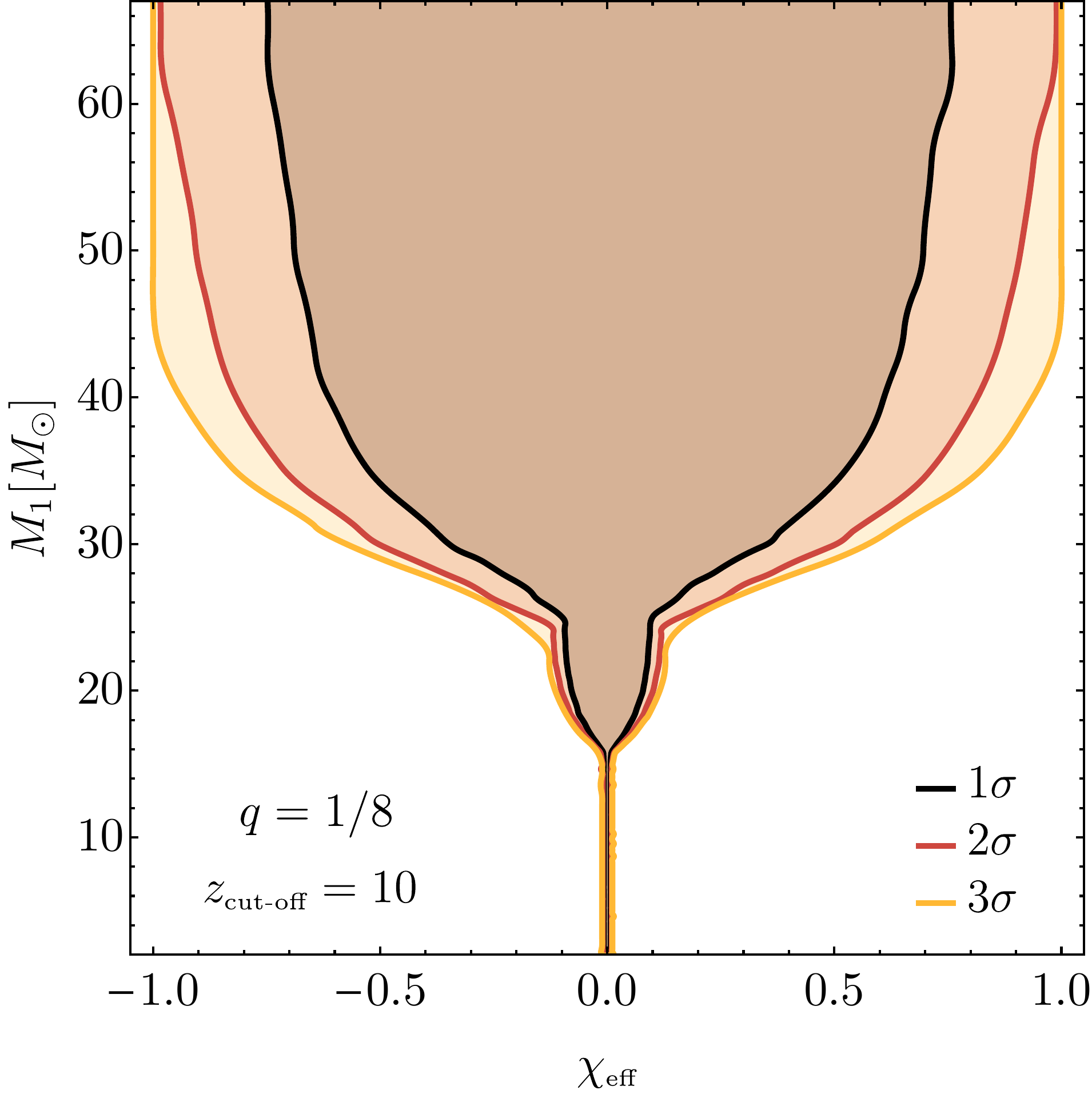}
	\includegraphics[width=0.24 \linewidth]{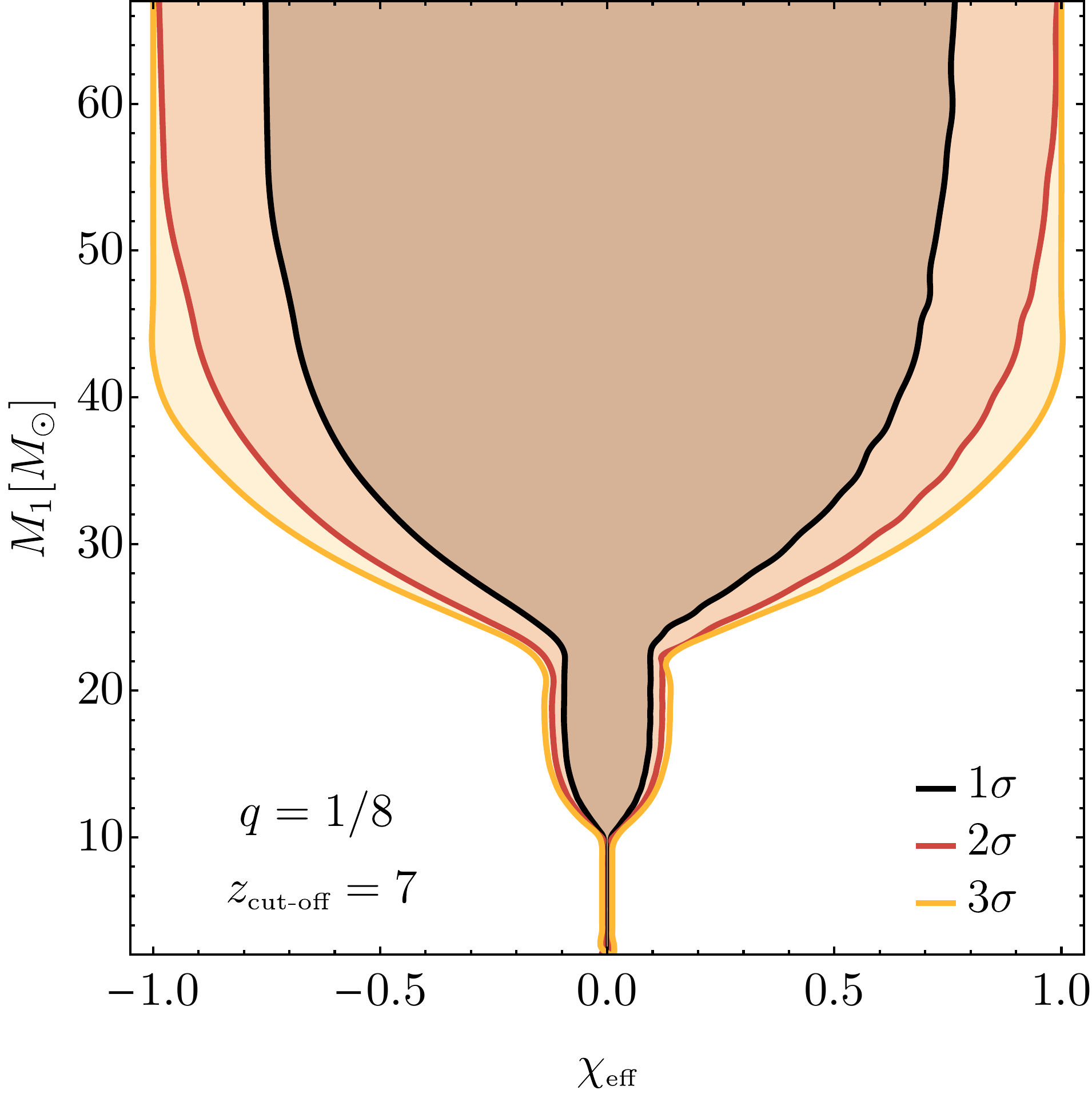}
	\caption{\it The distribution of $\chi_{\text{\tiny \rm eff}}$ as a function of the PBH mass $M_1$ for 
selected values of the parameter $q$. Blue data points refer to the events listed in 
Ref.~\cite{LIGOScientific:2018mvr}, whereas 
green and red data points refer to the events discovered in Refs.~\cite{Zackay:2019tzo,Venumadhav:2019lyq}; the 
red data points refer to GW151216 and GW170403, for which the measured value of $\chi_\text{\tiny \rm eff}$ is 
significantly affected by the prior on the spin angles~\cite{Huang:2020ysn}. The cyan data point refers to GW190412 
recently reported in Ref.~\cite{LIGOScientific:2020stg}.
	}
\label{chieffdistsf}
\end{figure}

In Fig.~\ref{chieffdistsf} we show the distribution of the effective spin parameter  defined as 
\be
\chi_{\text{\tiny eff}}=\frac{\vec J_1 /M_1  + \vec J_2 /M_2 }{M_1+M_2}\cdot \hat L, \label{chieffdef}
\ee
where $M_1$ and $M_2$ are the individual BH masses, $\vec J_1$ and $\vec J_2$  are the corresponding 
angular-momentum vectors, and $\hat L$ is the direction of the orbital angular momentum. Following Ref.~\cite{paper1} we 
have averaged over the angles between the
total angular momentum and the individual PBH spins. We have plotted $\chi_{\text{\tiny eff}}$ as a function of the 
final PBH mass $M_1$ for different fixed values of the mass ratio parameter $q$. When accretion is present, such 
distributions reflect the transition from initially vanishing values of $\chi_{\text{\tiny eff}}$ to large ones. 
In the top two rows of Fig.~\ref{chieffdistsf} (for $q=1$ and $q=1/2$), we have shown only the 
current data which are compatible (within their reported errors) with the corresponding chosen value of $q$. In the 
third row ($q=1/4$) we report only GW190412 as a reference.

\subsection{Limitations of the accretion model} \label{sec:limitations}

Accretion onto compact objects through the cosmic history is a complex phenomenon and the accretion 
rate relies on a set of assumptions, above all through the dependence on the velocity. 
While from the discussion of the previous sections it seems clear that accretion should play a relevant role, it is 
also important to spell out the main uncertainties in the 
accretion modelling~\cite{Ricotti:2007au,Ricotti:2007jk,Ali-Haimoud:2016mbv}.

\begin{itemize}
	\item[a)] {\it Local feedback}: our analysis neglects the effect of feedback on the accretion flow. 
	The effect of local heating for the PBH masses of interest for LIGO/Virgo can be safely 
	neglected~\cite{Ricotti:2007jk,Ali-Haimoud:2016mbv}.
	\item[b)] {\it Global feedback \& X-ray pre-heating}: this effect was estimated in~\cite{Ricotti:2007jk}, 
	including the X-ray heating of the gas and the extra contribution due to PBH 
	accretion, but neglecting other possible sources of X-ray heating~\cite{Oh:2003pm}. However, since then, the 
	analysis 
	of the cosmic ionisation provided in Ref.~\cite{Ricotti:2007jk} has been strongly revisited by later work, in 
	particular after GW150914 and the suggestion that those BHs might be of primordial origin~\cite{Bird:2016dcv}. 
	Indeed, 
	the detailed analysis of Ref.~\cite{Ali-Haimoud:2016mbv} shows that global feedback is much less important for 
	LIGO/Virgo BHs. In particular, taking all relevant effects into account, Ref.~\cite{Ali-Haimoud:2016mbv} found 
	an accretion rate which is consistent with that estimated by Ref.~\cite{Ricotti:2007jk} {\it without} the 
	effect of global heating. 
	Modelling the temperature of the intergalactic medium at redshift $10\lesssim z \lesssim 30$ is particularly 
	relevant, since an increase in the temperature is followed by an increase in the sound speed and, in turn, 
	by a reduction of the accretion rate.
	\item[c)] {\it DM halo}: the accretion rate computed in our model is consistent with that of 
	Ref.~\cite{Ali-Haimoud:2016mbv}, with the important inclusion of a DM halo, which seems 
	inevitable if PBHs form a small fraction of the DM, at least if the latter is of particle origin.
	%
	\item[d)] {\it Structure formation}: part of the population of PBHs starts falling in the gravitational 
	potential well of large-scale structures after redshift around $z\simeq 10$, experiencing an increase of the 
	relative 
	velocity up to one order of magnitude, see for example Ref.~\cite{Hasinger:2020ptw} for a recent analysis.  
	This will result in a consequent suppression of the accretion 
	rate~\cite{Ricotti:2007au,Ali-Haimoud:2017rtz,raidalsm}. While this motivates our 
	choice $z_\text{\tiny cut-off}=10$, the precise dynamics depends on the complex modelling of the 
	global thermal feedback and the change in the relative velocity due to the structure formation. Furthermore, it 
	is hard to estimate the fraction of PBHs that stop accreting efficiently enough due to this effect. For 
	example, the captured PBHs might settle at the center of the halo within a Hubble time, due e.g. to dynamical 
	friction effects, and might keep accreting efficiently.
	\item[e)] {\it Spherical accretion \& disk geometry}: most of the semi-analytical studies on accretion onto 
	compact objects necessarily assume a quasi-spherical flow. However, this approximation might break down, 
	e.g. in the case of outflows~\cite{Bosch-Ramon:2020pcz}. The efficiency of the latter in reducing the accretion 
	rate depends on the geometry of the accretion flow and on the relative direction of the outflow. 
	\item[f)] {\it Angular momentum transfer}: As discussed above, when $\dot m\sim1$ 
	the disk is geometrically thin and can be described with a geodesic model. When $\dot m\ll1$ or $\dot m\gg1$, 
	the geometry of the disk is different, and this impacts on the accretion luminosity and feedback. In the 
	super-Eddington regime the disk is expected to ``puff up'', becoming geometrically thicker. Angular momentum 
	accretion in this case is more complex, although numerical simulations suggest that the spin evolution time 
	scale does not change significantly~\cite{Gammie:2003qi}. 
\end{itemize}
While the above details require complex and model-dependent simulations, the uncertainties in certain aspects of the 
accretion model can be parametrised in an agnostic way through a cut-off 
redshift $z_\text{\tiny cut-off}$.  For instance, as far as the X-ray pre-heating effect~\cite{Oh:2003pm} on the 
accretion rate is concerned, the suppression factor in $\dot m$ with respect to the case of no X-ray 
pre-heating may be at most an order of magnitude or smaller.
Most importantly, this effect can be caught in our model by increasing the value of $z_\text{\tiny cut-off}$ without 
including the X-ray pre-heating\footnote{It is also worth mentioning that the analysis of Ref.~\cite{Oh:2003pm} is 
based on WMAP data, which estimated the reionisation redshift as $z_\text{\tiny re} = 	17 \pm 5$~\cite{Spergel:2003cb}. 
The current value is $z_\text{\tiny re} = 7.8 \pm 0.8$, as measured by Planck~\cite{Aghanim:2018eyx}, well within 
$z_\text{\tiny cut-off}=10$ which we consider the most realistic choice for the cut-off. Thus, it seems likely that also 
the pre-reionisation era might be shifted to smaller redshifts, in which case its effect is much smaller, possibly not 
affecting the accretion evolution up to $z<7$.}.

Based on the above discussion, we consider $z_\text{\tiny cut-off}=10$ as the most reasonable choice for the cut-off 
redshift; the value $z_\text{\tiny cut-off}=7$ advocated in Ref.~\cite{Ricotti:2007au} might be considered as an 
optimistic choice that assumes a non-negligible fraction of PBHs which keeps accreting significantly even 
after structure formation, whereas larger values of the cut-off, e.g. $z_\text{\tiny cut-off}=15$ would suppress the 
effect of accretion and correspond to a scenario in which the temperature of the intergalactic medium is high enough 
even before reionisation.

\section{Binary evolution}\label{sec:binaryevolution}
Having discussed the masses and spins of isolated and binaries PBHs, we now turn our attention to the evolution of PBH 
binaries. First we consider the case in which accretion is absent or inefficient, in which case the binary evolves only 
through GW radiation-reaction, which we review following Sec.~4.1 of 
Ref.~\cite{maggiore}. Then, we consider the case in which baryonic mass accretion is 
modelled as discussed in the previous sections, extending the results of Ref.~\cite{Caputo:2020irr} to the case 
of accretion-driven inspiral for eccentric binaries with generic mass ratio.

\subsection{GW-driven evolution}
\begin{figure}[t!]
	\centering
	\includegraphics[width=0.32 \linewidth]{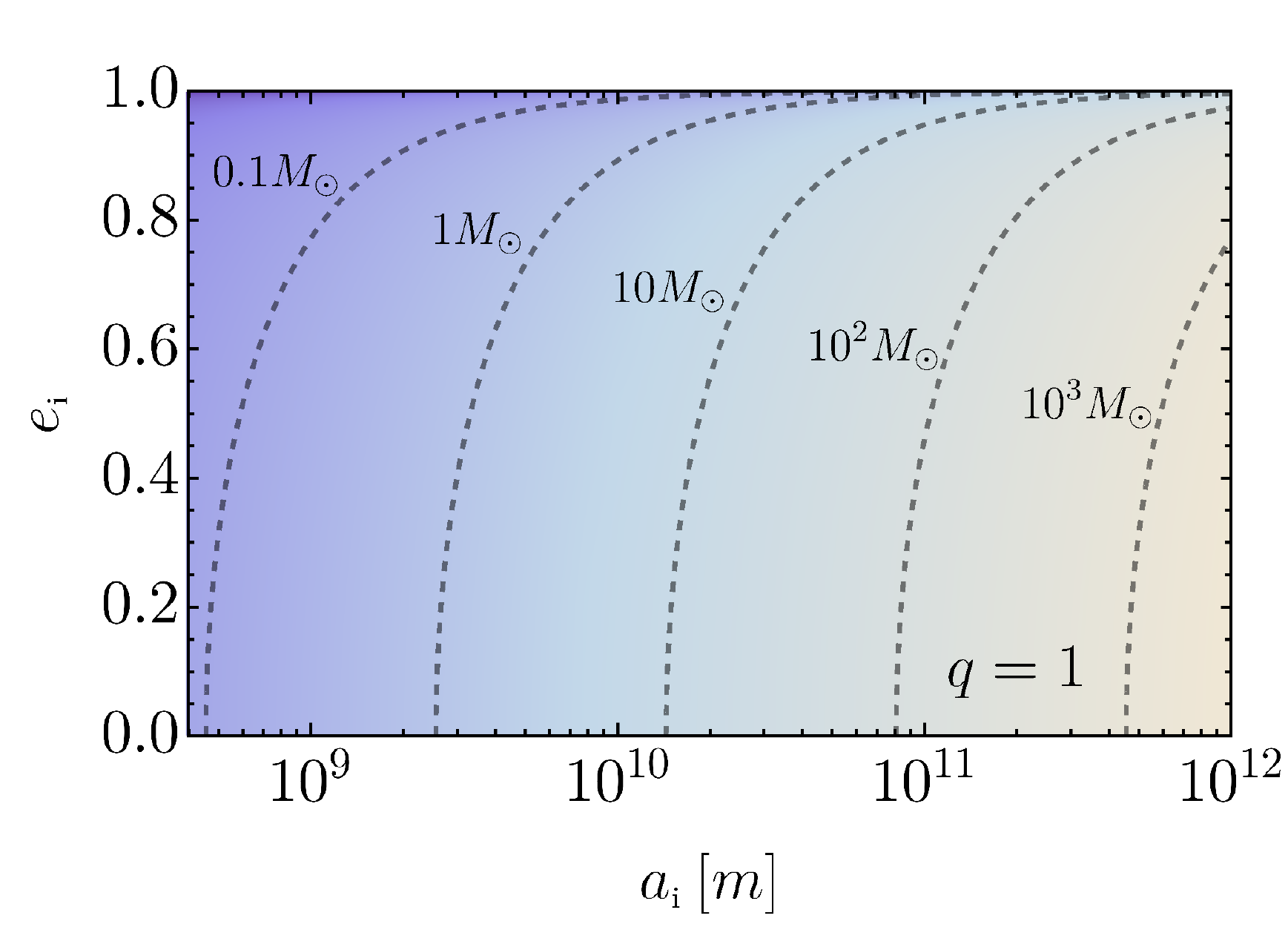}
	\includegraphics[width=0.32 \linewidth]{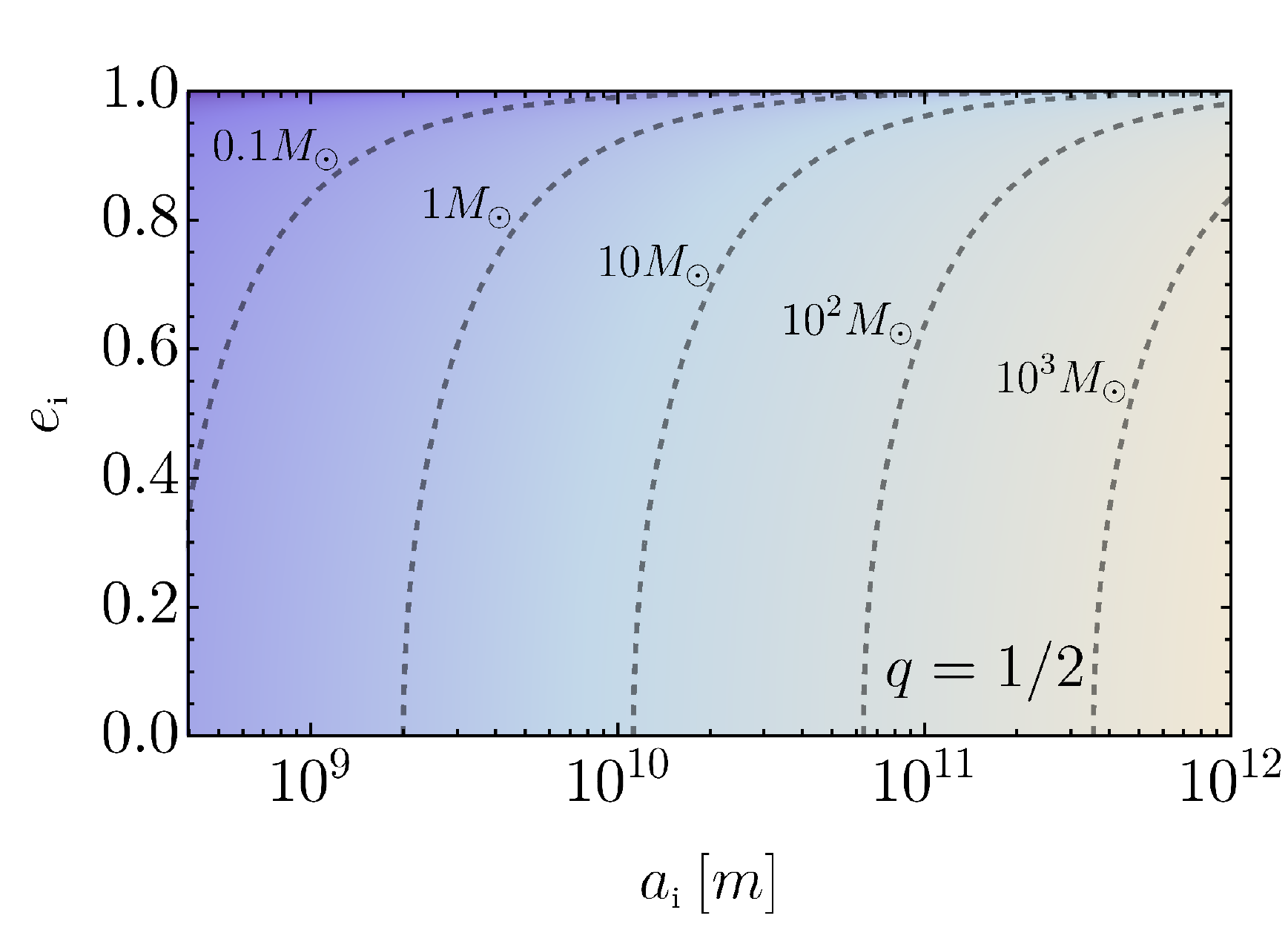}
	\includegraphics[width=0.32 \linewidth]{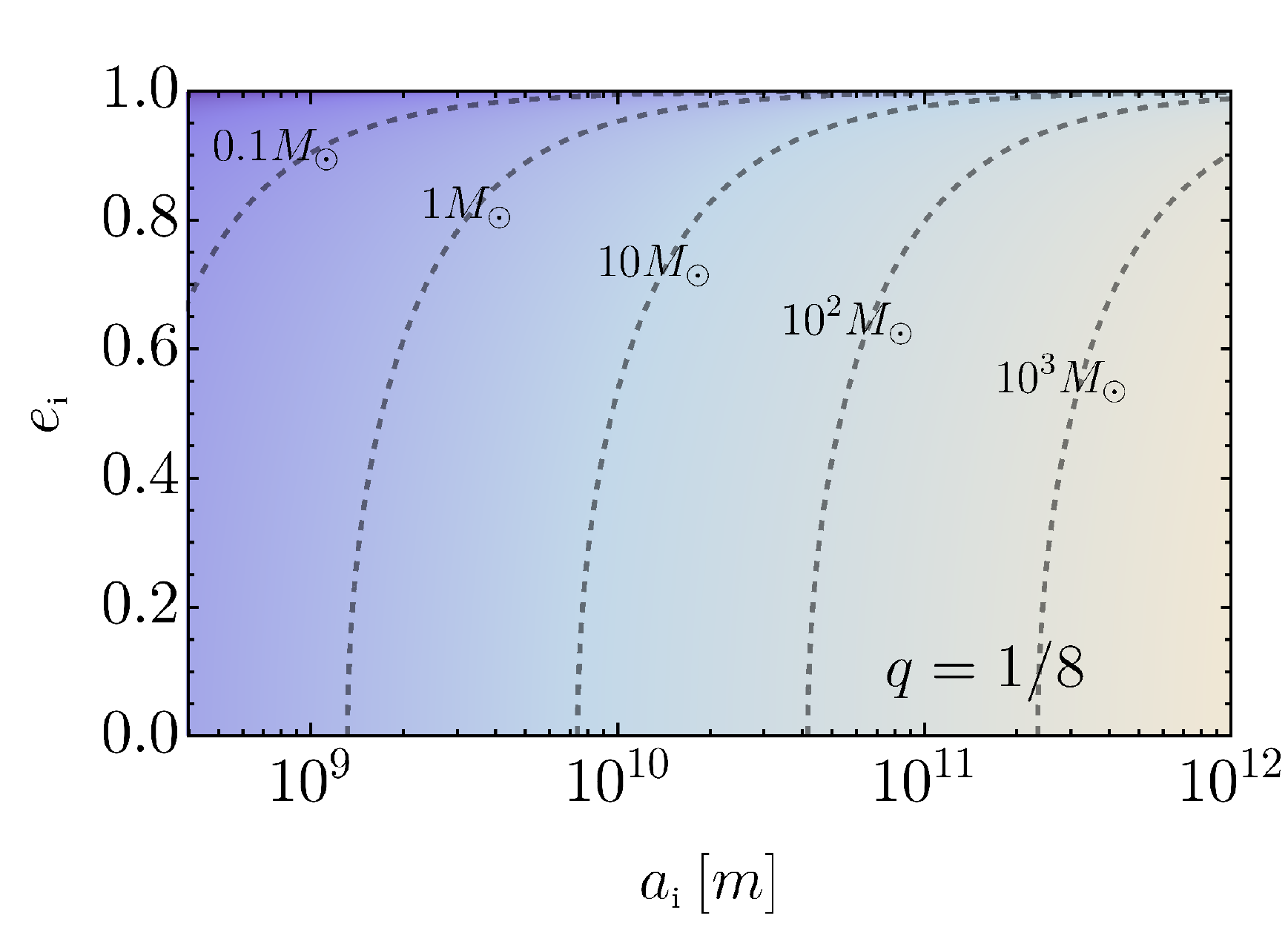}
	\caption{\it The contour lines indicate the combination of the parameters $e_\text{\tiny\rm i}$ and $a_\text{\tiny\rm i}$ giving a coalescence 
		time equal to the age of the universe $t_c(a_\text{\tiny\rm i},e_\text{\tiny\rm i}) = t_0$. The label indicates $M_1$ and the mass 
ratio considered is 
		$q=1$, $1/2$ and $1/8$ respectively.}
	\label{co}
\end{figure}

In the absence of GW radiation-reaction, the eccentricity $e$ and semi-major axis  $a$ are constants of motion and can 
be 
expressed in terms of the angular momentum and energy as
\begin{equation}
	e^2 = 1+ \frac{2 E L ^2}{M_\text{\tiny tot}^2 \mu^3} \,,\qquad a = \frac{G M_\text{\tiny tot} \mu}{2 |E|}\,.
\end{equation}
To the leading order in the weak-field/slow-motion approximation, one can use the quadrupole formula to evaluate the 
energy and angular momentum losses through the GW emission. The energy and angular momentum of the binary evolves 
as~\cite{Peters:1963ux, Peters:1964zz}
\begin{align}
	\frac{\d E}{\d t} &= -\frac{32}{5} \frac{\mu^2 M_\text{\tiny tot}^3}{a^5} \frac{1}{(1-e^2)^{7/2}}\lp 1+ 
	\frac{73}{24} e^2 + \frac{37}{96} e^4\rp,
	\nonumber \\
	\frac{\d L}{\d t} &= -\frac{32}{5} \frac{\mu^2 M_\text{\tiny tot}^{5/2}}{a^{7/2}} \frac{1}{(1-e^2)^{2}}\lp 
	1+ \frac{7}{8}  e^2\rp.
\end{align}
In terms of the adiabatic evolution of $e$ and $a$, one can recast the system of equations in the form
\begin{align}
	\frac{\d a}{\d t} &= -\frac{64}{5} \frac{\mu M_\text{\tiny tot}^2}{a^3} \frac{1}{(1-e^2)^{7/2}}\lp 1+ 
	\frac{73}{24} e^2 + \frac{37}{96} e^4\rp,
	\nonumber \\
	\frac{\d e}{\d t} &= -\frac{304}{15} \frac{\mu M_\text{\tiny tot}^{2}}{a^{4}} \frac{e}{(1-e^2)^{5/2}}\lp 1+ 
	\frac{121}{304}e^2\rp.
\end{align}
%
This system can be solved to find an estimate for the merging time $t_c$, here defined as $a(t_c)=0$.
For an initial orbit with $e(t_\text{\tiny i})=e_\text{\tiny i}$ and $a(t_\text{\tiny i})=a_\text{\tiny i}$ one finds
\begin{equation}
	t_c(a_\text{\tiny i},e_\text{\tiny i}) = t_c(a_\text{\tiny i}) \frac{48}{19} \frac{1}{g^4(e_\text{\tiny i})} \int_0^{e_\text{\tiny i}} \d e \frac{g^4(e)(1-e^2)^{5/2}}{e (1+121 
e^2 
		/304)}\,, \label{tauGWGEN}
\end{equation}
where we defined the function
\begin{equation}
	g(e) =\frac{ e^{12/19}}{1-e^2} \lp 1+ \frac{121}{304} e^2\rp^{870/2299}\,.
\end{equation}
For a circular orbit, Eq.~\eqref{tauGWGEN} reduces to 
\begin{equation}
	t_c(a_\text{\tiny i},e_\text{\tiny i}=0)\equiv t_c(a_\text{\tiny i}) = \frac{5}{256} \frac{a_\text{\tiny i}^4}{M_\text{\tiny tot}^2 \mu}.\label{tauGWCIRC}
\end{equation}
Conversely, in the limit $e_\ii \to 1$, one finds that
\begin{equation}
	t_c(a_\text{\tiny i},e_\text{\tiny i} \to 1) \simeq t_c(a_\text{\tiny i}) \frac{768}{429} (1-e_\text{\tiny i}^2)^{7/2}.
\end{equation}
The parameters leading to a coalescence time equal to the age of the universe, $t_0 = 13.7\, {\rm Gyr}$, are shown in 
Fig.~\ref{co}. 
As one can notice, the initial value of $a_\text{\tiny i}$ diverges as $e_\text{\tiny i}$ tends towards unity, since 
the coalescence 
time tends to shrink rapidly in that limit.

The above relation is used by \cite{Sasaki:2016jop,Sasaki:2018dmp} to compute an estimate of the merger rate, as we 
discuss in Sec.~\ref{sec:rates}.

\subsection{Accretion-driven evolution}\label{sec:accretion-driven}
Accretion introduces a further secular change to the orbital parameters, in addition to GW 
radiation-reaction~\cite{Barausse:2014tra,Macedo:2013qea,Caputo:2020irr}.
As we are going to discuss, due to the time scales involved in the problem, we can study the accretion-driven phase and 
the GW-driven phase separately. 
Indeed, since for $z<z_\text{\tiny cut-off}$ accretion is drastically suppressed, the evolution of the binary from 
$z_\text{\tiny cut-off}$ to the redshift of detection (which we shall assume to be $z\approx0$ having in mind the 
current horizons of LIGO and Virgo) is governed by GW emission only.
From Eq.~\eqref{tauGWGEN}, a merger occurring at $z\approx0$ corresponds to a binary with orbital separation $a={\cal 
O}(10^{11}\,{\rm m})$ at $z_\text{\tiny cut-off}=10$. In this configuration the time scale of variation of the 
semi-major axis due to GW
emission is
\begin{align}
	T_\GW &\sim \left .\frac{a}{\dot a} \right |_\GW = 4  \times 10^{17} \lp \frac{a}{1.4 \times 10^{11} {\rm 
m}}\rp ^{4} \lp\frac{M}{30 M_\odot} \rp ^{-3} \lp \frac{1-e^2}{0.1} \rp^{7/2}  \,{\rm s}
\end{align}
where the normalisation $a=1.4 \times 10^{11} {\rm m}$ has been chosen as the one giving a merger in a time equal to 
the age of the universe for binary components of equal mass $M=30 M_\odot$ and $e=0.95$. As can be directly checked, 
owing to the strong dependence $T_\GW\propto a^4$, for $z>z_\text{\tiny cut-off}$ the typical accretion time 
scale~\eqref{tauACC} is much smaller than the one governing GW radiation-reaction. 
In other words, we can assume\footnote{The case in which the detected events is at $z>z_\text{\tiny cut-off}$ requires 
	a different analysis, which will be relevant for third-generation ground-based detectors such as the Einstein 
	Telescope~\cite{Maggiore:2019uih} and for the space mission LISA~\cite{Audley:2017drz}.} that the inspiral is 
driven 
solely by accretion when 
$z>z_\text{\tiny cut-off}$ and solely by GW radiation-reaction when $z<z_\text{\tiny cut-off}$.

It is also  worth noting that, for the range of parameters we are 
interested in, the mass accretion time scale~\eqref{tauACC} is always much larger than the characteristic orbital time 
scale,
\begin{equation}
	T_\text{\tiny orbital} \sim \lp \frac{M_\text{\tiny tot} }{a^3} \rp ^{-1/2} \sim 8\times  10^5 \lp \frac{M}{30 
		M_\odot}\rp^{-1/2} \lp\frac{a}{1.4 \cdot 10^{11}\, {\rm m}} \rp^{3/2}\,{\rm s} \,.
\end{equation}
Therefore one can treat the mass accretion as an adiabatic process keeping constant the adiabatic invariants of the 
elliptical motion, as we now discuss.

If the masses vary adiabatically, one can compute the action variables $I_k = \int p \d k /(2\pi)$ for the elliptic 
motion (where $k=r,\phi$ are the polar coordinates) and then employ the fact that they are adiabatic invariants. 
Specifically, the adiabatic invariants for the Keplerian two-body problem are~\cite{landau1976mechanics}
\begin{align}
	I_\phi &= \frac{1}{2\pi} \int_0^{2\pi} p_\phi \d \phi = L_z,  \\
	I_r &=  \frac{1}{2\pi} \int_{r_{\rm min}}^{r_{\rm max}} p_r \d r  = - L_z + \sqrt{M_\text{\tiny tot} \mu^2 a},
\end{align}
where we recall the definition of energy and angular momentum of the binary
\begin{align}
	\label{ellip}
	E &= -\frac{\mu M_\text{\tiny tot}}{2a}, \qquad  L_z = \sqrt{ \frac{M_\text{\tiny tot}^2 \mu^3 (e^2-1)}{2E}} = 
\sqrt{1-e^2} \sqrt{ M_\text{\tiny tot} \mu^2 a}.
\end{align}
The invariance of $I_\phi$ and $I_r$ implies that 
\begin{align}
	\frac{\d I_\phi}{\d t} &= \frac{\partial L_z}{\partial e}  \frac{\partial e}{\partial t}  + \frac{\partial 
		L_z}{\partial a}  \frac{\partial a}{\partial t}  + \frac{\partial L_z}{\partial \mu} \frac{\partial 
\mu}{\partial t} +  
	\frac{\partial L_z}{\partial M_\text{\tiny tot}} \frac{\partial M_\text{\tiny tot}}{\partial t} = 0\,, 
\nonumber 
\\
	\frac{\d I_r}{\d t} &= \frac{\partial I_r}{\partial e}  \frac{\partial e}{\partial t}  + \frac{\partial 
I_r}{\partial a} 
	\frac{\partial a}{\partial t}  + \frac{\partial I_r}{\partial \mu} \frac{\partial \mu}{\partial t} +  
\frac{\partial 
		I_r}{\partial M_\text{\tiny tot}} \frac{\partial M_\text{\tiny tot}}{\partial t}   =0 .
\end{align}
It is easy to prove that the system of equations can be recast as
\begin{align}
	\frac{\partial a}{\partial t}  &=-  \frac{ 2 a}{\mu} \frac{\partial \mu}{\partial t} -  \frac{a}{M_\text{\tiny 
tot}} \frac{\partial 
		M_\text{\tiny tot}}{\partial t} ,
	\nonumber \\
	\frac{\partial e}{\partial t}  &= 0\,,
\end{align}
which shows that the eccentricity is a constant of motion for an accretion-driven inspiral. This does not 
come as a surprise as the eccentricity can be expressed in terms of the adiabatic invariants 
as~\cite{landau1976mechanics}
\be
e = \sqrt{1 - \lp \frac{I_\phi}{I_\phi + I_r} \rp^2}.
\ee
The only non-trivial dynamical equation can be written as
\be\label{eqa}
\frac{\dot a}{a}  + 2\frac{\dot \mu}{\mu} +  \frac{\dot M_\text{\tiny tot}}{M_\text{\tiny tot}}  =0.
\ee
This results recovers the one found in Ref.~\cite{Caputo:2020irr} in the absence of GW emission and for circular 
binaries, and extends it for a generic eccentricity, which remains a constant of motion.

Finally, it is convenient to write Eq.~\eqref{eqa} in terms of the mass accretion rates of the isolated objects
\be\label{eqabis}
\frac{\dot a}{a}  +\frac{M_2(M_1+2 M_2) \dot M_1+M_1 (2M_1+M_2) \dot M_2}{M_1 M_2(M_1+M_2)} =0\,.
\ee
In general, this equation is coupled to the evolution equations for $M_1$ and $M_2$, namely Eqs.~\eqref{M1M2dotGEN}. 
However, within the aforementioned approximations, one can evolve $M_i$ using Eq.~\eqref{M1M2dotFIN} and finally 
evolve the semi-axis major using Eq.~\eqref{eqabis}.
In the limit $q\to 1$, i.e. $M_1=M_2$, Eq.~\eqref{eqabis} simplifies to the more familiar form 
\be
\frac{\dot a}{a} + 3\frac{\dot M_1}{M_1} =0\,.
\ee

\section{PBH merger rates and phenomenology} \label{sec:rates}
After the discussion about the evolution of the PBH binary, we move to the computation of the PBH merger rate both 
without and including accretion, and we will then discuss its impact on the PBH phenomenology.\footnote{We assume that 
PBHs are   not initially clustered \cite{cl1,cl2,cl3,dizgah} which is a good approximation in  the absence of  
primordial non-Gaussianity.}

\subsection{Merger rates without accretion}
Following the notation of Ref.~\cite{raidal}, one can write down the differential merger rate at the time of coalescence $t$ in the form
\be
\label{mergerrate}
\d R = \frac{1.6 \times 10^6}{{\rm Gpc^3 \, yr}} f_\PBH^{\frac{53}{37}} (z_\ii)   \lp \frac{t}{t_0} 
\rp^{-\frac{34}{37}}  
\eta_\ii^{-\frac{34}{37}} \lp \frac{M^\ii_\text{\tiny tot}}{M_\odot} \rp^{-\frac{32}{37}}  
S\lp M^\ii_\text{\tiny tot}, f_\PBH (z_\ii)  \rp
\psi(M^\ii_1, z_\ii) \psi (M^\ii_2, z_\ii)\d M^\ii_1 \d M^\ii_2,
\ee
where $f_\PBH (z_\ii) $ is the initial fraction of DM in the form of PBHs, 
$\eta_\ii=\mu^\ii/M^\ii_\text{\tiny tot}$ is the symmetric mass ratio, defined in terms of the reduced mass $\mu^\ii = M^\ii_1 
M^\ii_2/M^\ii_\text{\tiny tot}$ and total mass $M^\ii_\text{\tiny tot}=M^\ii_1+M^\ii_2$ of the binary components at the formation time, and $\psi 
(M^\ii,z_\ii)$ identifies the PBH mass function at the formation time $z_\ii$, normalised to unity.

The suppression factor $S$ is introduced in order to keep into account the effect of the matter density perturbations
and possible modifications due to the size of the empty region around the binary. Its expression is given by 
\cite{raidal}
\be
S \lp M^\ii_\text{\tiny tot}, f_\PBH (z_\ii)  \rp = \frac{e^{-\bar{N}(y)}}{\Gamma(21/37)} \int \d v v^{-\frac{16}{37}} {\rm exp} \llp - \bar{N}(y)  \left \langle m \right \rangle \int \frac{\d m}{m} \psi (m, z_\ii) F \lp \frac{m}{\left \langle m \right \rangle} \frac{v}{\bar{N}(y)} \rp - \frac{3 \sigma_M^2 v^2}{10 f_\PBH^2 (z_\ii)} \rrp
\ee
in terms of the generalised hypergeometric function
\be
F(z) = \tensor[_1]{F}{_2} \lp-\frac{1}{2}; \frac{3}{4};\frac{5}{4};-\frac{9z^2}{16}\rp-1\,,
\ee
the rescaled variance of matter density perturbations 
$\sigma_M^2 = (\Omega_{M}/\Omega_{\rm DM})^2 \left \langle \delta_M^2 \right \rangle \simeq 3.6 \cdot 10^{-5}$
 at the time at which the binary is formed, and 
\be
\left \langle m \right \rangle = \lp \int \frac{1}{m} \psi (m,z_\ii) \d m \rp^{-1}.
\ee
The number  $\bar{N}(y)$ of PBHs in a spherical volume of radius $y$ is chosen such that the binaries do not get 
destroyed by other PBHs, i.e.
\be
\bar{N}(y) =  \frac{M^\ii_\text{\tiny tot}}{\left \langle m \right \rangle}  \frac{f_\PBH (z_\ii)}{f_\PBH (z_\ii)+\sigma_M}.
\ee
The integral over the masses in the suppression factor gives an estimate of the typical mass of the perturber PBHs 
responsible for the torque, which prevents two PBHs to collide directly and is responsible for forming the binary 
itself. 

\begin{figure}[t!]
	\centering
	\includegraphics[width=0.495\linewidth]{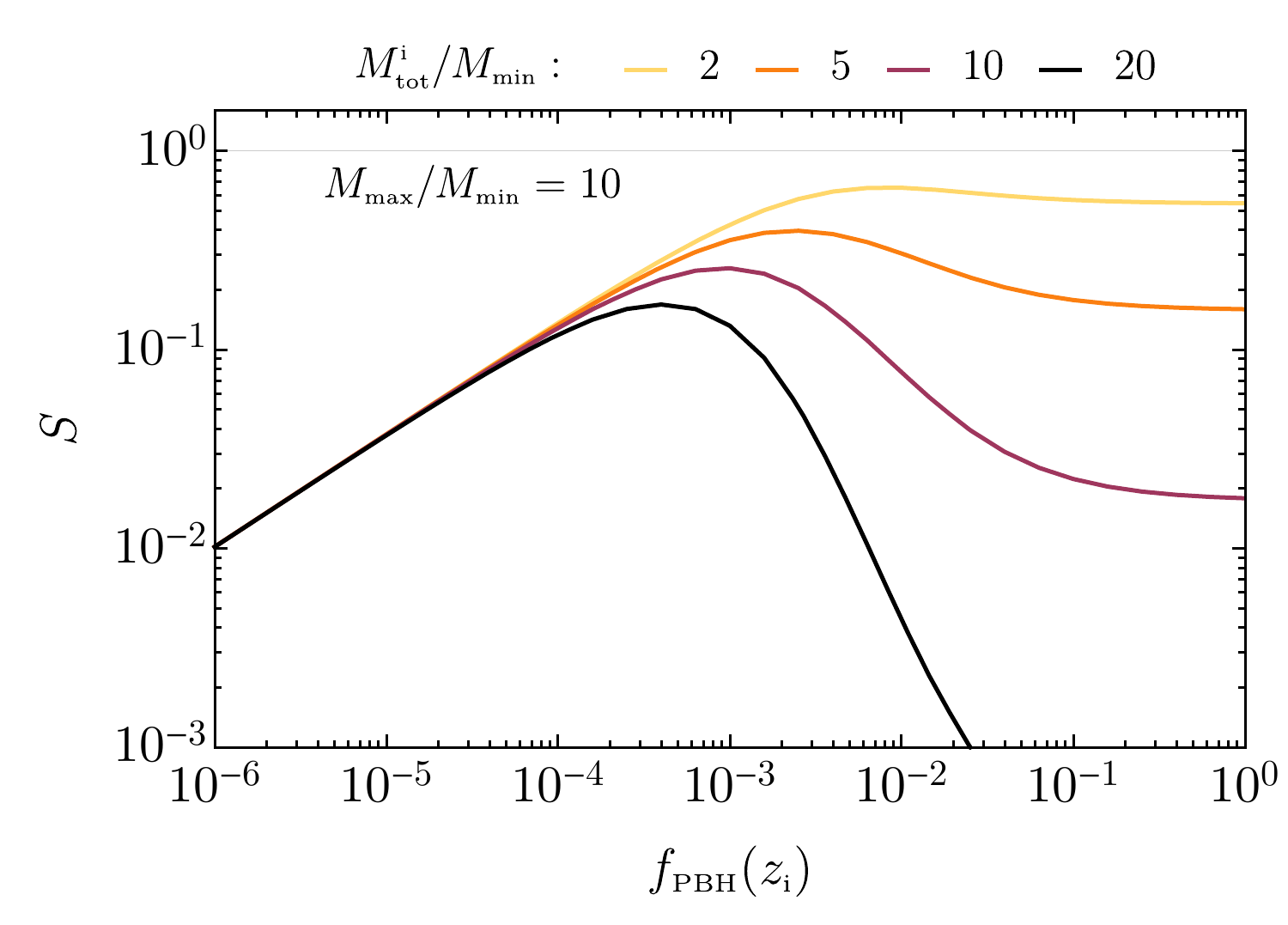}
	\includegraphics[width=0.495\linewidth]{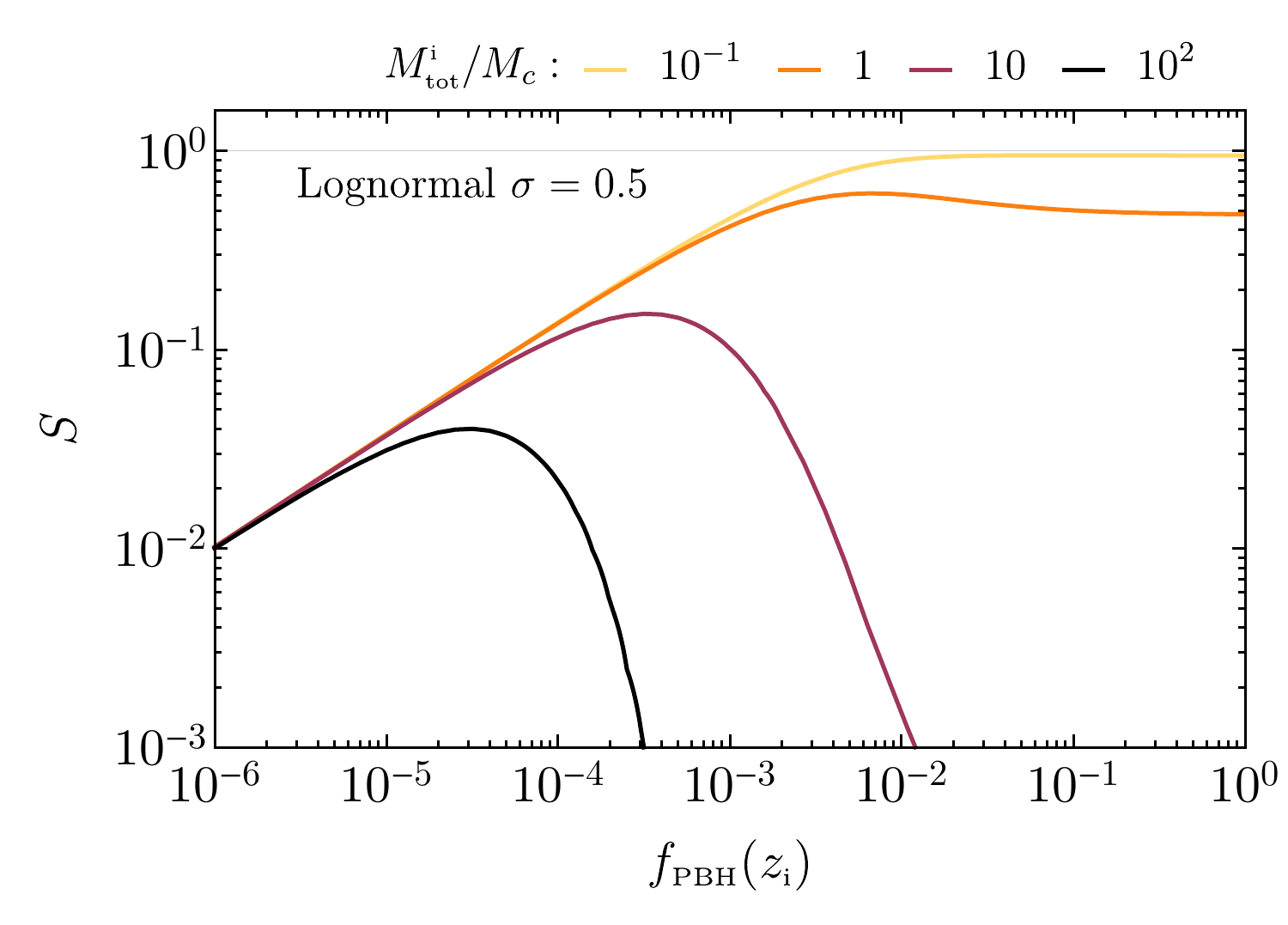}
	\caption{\it Left:  Suppression factor for 
a power-law mass function with $M_\text{\tiny \rm max}=10 M_\text{\tiny \rm min}$ and various $M^\ii_\text{\tiny \rm tot}$. 
 Right:
 Suppression factor for a lognormal mass function with $\sigma=0.5$.
}
	\label{supp}
\end{figure}

In Fig.~\ref{supp} we show the behaviour of the suppression factor for the cases of a lognormal and power-law mass 
function.

\subsection{Merger rates with accretion}
In this subsection we will describe the main impact of accretion on the PBH merger rate.

As already discussed in Sec.~\ref{sec:binaryevolution}, due to the time scales involved in the problem, one can study 
the accretion-driven and the GW-driven binary evolution separately. Indeed, accretion 
dominates the evolution of the binary up to redshift $z_\text{\tiny cut-off}$, while GW radiation-reaction is dominant 
from that redshift up to the detection time.

To understand the effect of accretion on the merger rate we remind that, using the formalism of Ref.~\cite{raidal}, the latter is defined by integrating the  probability distribution of the orbital parameters
\be
\d R (t) = \int \d a \d e \frac{\d P}{\d e} \delta\lp t - t_c (\mu, M_\text{\tiny tot} , a ,e)\rp
\ee
at the coalescence time of the binary, which has been computed in Sec.~\ref{sec:binaryevolution} without accretion and 
which we report here for convenience in a more compact form
\begin{equation}
	t_c=\frac{3}{85}  \frac{a_\ii^4 {(1-e_\ii^2)}^{7/2} }{\eta (z_\ii) M_\text{\tiny tot}^3(z_\ii)},
\end{equation}
where all the quantities are set at the formation time $z_\ii$.

In the presence of accretion the masses change in time. Even though the ellipticity is a constant of motion, the semi-major axis $a$ does evolve with time due to the masses evolution, as one can see from Eq.~\eqref{eqa}.
This will have an impact on the coalescence time of the binary. Since the accretion-driven phase occurs earlier and 
independently of GW emission, once accretion is included the coalescence time becomes
\begin{equation}
	t^\text{\tiny  acc}_c=\frac{3}{85}  \frac{ {\cal N} ^4a_\ii^4 {(1-e_\ii^2)}^{7/2} }{\eta (z_\text{\tiny cut-off}) M_\text{\tiny tot}^3(z_\text{\tiny cut-off})}
	\equiv \frac{{\cal N}^4}{\mathcal{S}} 	t_c( M^\text{\tiny i}_j),
\end{equation}
where we have defined the factor
\begin{equation}
	{\cal S}=\frac{\eta (z_\text{\tiny cut-off}) M_\text{\tiny tot}^3(z_\text{\tiny cut-off})}{\eta (z_\text{\tiny i}) M_\text{\tiny tot}^3 (z_\text{\tiny i}) }
\end{equation}
which keeps into account the masses evolution from initial time $z_\ii$ to the cut-off redshift $z_\text{\tiny cut-off}$, and the shrinking factor of the orbit
\be
{\cal N} \equiv \frac{a (z_\text{\tiny cut-off})   }{a_\ii} = 
\exp\llp -\int_{t_\ii} ^{t_\text{\tiny cut-off}} \d t \lp \frac{\dot M_\text{\tiny tot}}{M_\text{\tiny tot}} + 2 \frac{\dot \mu}{\mu} \rp
\rrp,
\ee 
which properly considers the semi-major axis evolution. As we 
discussed, for $z<z_\text{\tiny cut-off}$ accretion is negligible, so the binary proceeds in the standard GW 
radiation-reaction scenario, but with different masses with respect to the no accretion case.

Implementing the fact that the suppression factor does not depend on time and using Eq.~\eqref{psiev} for the mass function evolution, one can rescale the coalescence time to compute the final differential merger rate as
\begin{align}
	\label{diffaccrate}
	\d R_\text{\tiny  acc} (t, M_j, f_\PBH(z_\ii)) &=
	\frac{\mathcal{S}}{{\cal N}^4}
	\d R \left(t \mathcal{S} / {\cal N}^4, M^\text{\tiny i}_j ,f_\PBH (z_\ii) \right) 
	\nonumber \\
	& = \mathcal{N}^{-12/37} \mathcal{S}^{3/37}  \d R \left(t, M^\text{\tiny i}_j  ,f_\PBH (z_\ii) \right) 
\end{align}
where with $M^\text{\tiny i}_j$ and $M_j$ we identify respectively the couple $(M_1,M_2)$ at the formation and final time.
In the last line the merger rate in terms of 
the initial quantities is explicitly given by Eq.~\eqref{mergerrate}.  As such, the bounds coming from merger rates do not depend upon the evolution
of the mass functions discussed in subsection 2.3.3.

\subsection{Phenomenology of PBH mergers without accretion} 
In this subsection we will discuss the implications of the physics of the PBH mergers on their phenomenology, by 
focusing on the constraints on the PBH abundance both from the total number of observed binary BH merger events by the 
LIGO/Virgo collaboration, and from the absence of the stochastic GW background produced by unresolved sources. We will 
follow the procedure outlined in Ref.~\cite{raidal}, without including the effects of accretion. The latter will be 
discussed in Sec.~\ref{sec:mergeraccr}.

\subsubsection{Likelihood analysis for GW observations without accretion}
One can start by performing a maximum-likelihood analysis, considering all the BH merger events observed by the 
LIGO/Virgo collaboration to date \cite{LIGOScientific:2018mvr,LIGOScientific:2020stg} and assuming that they all have a primordial origin, to find out the best-fit values 
for the parameters of the PBH mass function. For concreteness, we shall assume either a lognormal or a 
power-law distribution. 

The log-likelihood function is given by 
\be
{\cal L} = \sum_j {\rm ln} \frac{\int \d R (M_1, M_2, z) /(1+z) \d V_c (z) p_j (M_{j,1}| M_1)  p_j (M_{j,2}| M_2) p_j (z_j| 
z) \Theta (\rho(M_1,M_2,z) - \rho_c)}{ \int \d R (M_1, M_2, z) /(1+z) \d V_c (z) \Theta (\rho(M_1,M_2,z) - \rho_c)}\,, 
\label{eq:likelihood}
\ee
where the experimental uncertainties of the detected events are assumed to be described by Gaussian probabilities $p_j 
(M_{j}| M)$ to observe a BH mass $M_j$ given that the BH has mass $M$, and by Gaussian probabilities  $p_j (z_j| z)$ to 
observe a merger at redshift $z_j$ given that it happens at redshift $z$.

The integral is performed over the masses and redshift, in terms of the PBH merger rate $\d R (M_1, M_2, z)$ shown in Eq.~\eqref{mergerrate}, and of the comoving volume per unit redshift 
\be
\frac{\d V_c(z)}{\d z} = \frac{4\pi}{H_0} \frac{D_c^2(z)}{E(z)} = \frac{4 \pi}{H_0^2} \frac{1}{E(z)} \lp \int_0^z \frac{\d z'}{E(z')} \rp^2,
\ee
where the comoving distance $D_c$ is
\be
D_c (z) =\frac{1}{H_0} \int_0^z \frac{\d z'}{E(z')}
\ee
with
\begin{equation}
	E(z') =\sqrt{\Omega_r (1+z')^4+\Omega_m (1+z')^3 +\Omega_\text{\tiny K} (1+z')^2+ \Omega_\Lambda},
\end{equation}
and $\Omega_\text{\tiny K}= 0.0007$, $\Omega_r = 5.38 \times 10^{-5}$, $\Omega_\Lambda = 0.685$, $\Omega_m=0.315$, 
$h=0.674$, $H_0 = 1.0227 \times 10^{-10} h\, {\rm yr}^{-1}$. 
The additional factor of redshift $1/(1+z)$ is introduced to account for the difference in the clock rates at the time 
of merger and detection.
The Heaviside function $\Theta$ is introduced in Eq.~\eqref{eq:likelihood} to implement a detectability threshold based 
on the signal-to-noise ratio (SNR) of the GW events,  $\rho_c=8$. The optimal signal-to-noise ratio $\rho_\text{\tiny 
opt}$  of individual GW events for a source with masses $M_1,M_2$ at a  redshift $z$ is given by \cite{Ajith:2007kx} 
\be
\rho_\text{\tiny opt}^2 (M_1,M_2,z)\equiv \int_0^\infty \frac{4 |\widetilde h (\nu)|^2}{S_n (\nu)} \d \nu
\ee
where the strain noise for the O2 run has been taken from Ref.~\cite{Aasi:2013wya}
 and its analytical fit in the frequency range $\nu \in [10,5000] \, {\rm Hz}$ is given by
\be
S_{n, {\rm O2}}^{1/2} (\nu)= \exp \llp \sum _{i=0}^6 c_i \log^i(\nu) \rrp,
\ee
with 
\begin{align}
	c_0&= 33.3329,
	\quad
	c_1= -75.7393,
	\quad  
	c_2 = 27.1742,
	\quad
	c_3 = -5.10534,
	\nonumber \\ 
	c_4&= 0.524229,
	\quad 
	c_5= -0.0273956,
	\quad
	c_6= 0.000557901.
\end{align}
The GW strain signal $\widetilde h$ is given in Fourier space by \cite{Ajith:2007kx} 
\be
\widetilde h (\nu) = \sqrt{\frac{5}{8}} \frac{1}{ D_\text{\tiny L} \pi} \frac{1}{\nu} \lp \frac{\d E_\text{\tiny GW} (\nu)}{\d \nu} \rp^{1/2} e^{i \phi (\nu)} 
\ee
where $\phi (\nu)$ is the phase of the waveform, not relevant for the SNR, and $D_\text{\tiny L} $ identifies the luminosity distance from the source in terms of the comoving distance $D_c$ as
\be
D_\text{\tiny L} (z) = (1+z) D_c (z) = (1+z) \frac{1}{H_0} \int_0^z \frac{\d z'}{E(z')}.
\ee
For the GW energy spectrum $\d E_\text{\tiny GW}$ with frequency between $(\nu, \nu+ \d \nu)$ we use a 
phenomenological expression which, in the non-spinning limit, is given by \cite{Ajith:2007kx,Ajith:2009bn}
\footnote{In our analysis we neglect the impact of the BH spins onto the emitted GWs energy described in 
Refs.~\cite{Ajith:2009bn,Hemberger:2013hsa}.
}
\be
\frac{\d E_\text{\tiny GW} (\nu)}{\d \nu} =  \frac{\pi^{2/3}}{3} M_\text{\tiny tot}^{5/3} \eta \times
\left \{ \begin{array}{rl}
	&\nu^{-1/3} \quad \text{for} \quad \nu < \nu_1,  \\
	&\frac{\nu}{\nu_1}\nu^{-1/3} \quad \text{for} \quad \nu_1 \leq \nu < \nu_2, \\
	& \frac{\nu^2}{\nu_1 \nu_2^{4/3}} \frac{\sigma^4}{(4 (\nu- \nu_2)^2 + \sigma^2)^2} \quad \text{for}  \quad \nu_2 \leq \nu < \nu_3 , 
\end{array}
\right .
\ee
where 
\begin{align}
	&\pi M_\text{\tiny tot} \nu_1  = (1-4.455+3.521)+0.6437\eta-0.05822\eta^2-7.092\eta^3,  \nonumber \\
	&\pi M_\text{\tiny tot} \nu_2  = (1-0.63)/2+0.1469\eta-0.0249\eta^2+2.325\eta^3,  \nonumber \\
	&\pi M_\text{\tiny tot} \sigma = (1-0.63)/4 -0.4098\eta +1.829\eta^2-2.87\eta^3,  \nonumber \\
	& \pi M_\text{\tiny tot} \nu_3 = 0.3236 -0.1331\eta -0.2714\eta^2 +4.922\eta^3.
\end{align}
The final SNR can be then obtained by performing an average over the isotropic sky locations and orientations, finding 
that \cite{Finn:1992xs,Berti:2004bd}
\be
\rho^2(M_1,M_2,z) =  \frac{1}{5}   \rho_\text{\tiny opt}^2 (M_1,M_2,z)\,,
\ee
which should be compared with the detectability threshold assumed to be $\rho_c=8$.

From all these ingredients one can perform a maximum-likelihood analysis and find the best-fit values for the PBH mass function parameters, which can then be used as benchmark values to show the constraints on the PBH abundance from GWs events.

\subsubsection{Number of events and stochastic GW background without accretion}
The predicted number of PBH merger detections in a time interval $\Delta t$ is given by 
\cite{raidal,domi,chen}
\be
N = \Delta t \int \d z \d M_1 \d M_2 \frac{1}{1+z} \frac{\d V_c(z)}{\d z} \frac{\d R(M_1,M_2,z)}{\d M_1 \d M_2} \Theta \lp \rho (M_1,M_2,z)- \rho_c \rp
\ee
in terms of the PBH merger rate, including a redshift factor $1/(1+z)$ to account for the difference in the clock rates 
at the time of merger and detection. The errors $N-N_\text{\tiny min}$ and $N_\text{\tiny max}-N$ on $N$ can be 
estimated using a Poisson statistics by computing the number of events with the PBH mass function parameters and 
$f_\PBH$ at which the likelihood is maximum, and the reference masses at the  2$\sigma$ confidence level.
One can use the range $N_\text{\tiny min} < N < N_\text{\tiny max}$ to constrain the fraction of DM as PBHs assuming 
that all the observed BH merger events are primordial, and setting an upper bound on $f_\PBH$. We will show the results 
of this procedure in Sec.~\ref{sec:confrontation}.

PBHs mergers which are not individually resolved (i.e. $\rho<\rho_c=8$) contribute to a stochastic GW 
background, which in turn can be used to constrain the PBHs abundance, see 
Ref.~\cite{Zhu:2011bd,wang,raidal0,raidal,
Chen:2018rzo}.
From the differential merger rate $\d R (z)$ at redshift $z$, one can compute the spectrum of the stochastic GW background of frequency $\nu$ 
as
\be
\Omega_\text{\tiny GW} (\nu)=  \frac{\nu}{\rho_0} \int_0^{\frac{\nu_3}{\nu}-1} \d z \d M_1 \d M_2 \,\frac{1}{(1+z)H(z)}  \frac{\d R(M_1,M_2,z)}{\d M_1 \d M_2}  \frac{\d E_\text{\tiny GW} (\nu_s)}{\d \nu_s}  \Theta(\rho_c - \rho (M_1,M_2,z)),
\ee
where $\rho_0 = 3 H_0^2/8\pi$, $\nu_s =\nu (1+z)$ is the redshifted source frequency, and now the Heaviside function 
is introduced to subtract the contribution from events which can be observed individually.

By calculating the stochastic GW 
background arising from the coalescences of PBH binaries and comparing its strength to the sensitivity of LIGO \cite{ligoSGWB,Thrane:2013oya}, one can 
constrain the fraction of DM in PBHs. The result of this procedure will be shown in Sec.~\ref{sec:confrontation}.

\subsection{Phenomenology of PBH mergers with accretion} \label{sec:mergeraccr}
In this subsection we include the effect of accretion on the likelihood analysis as well as on the estimates of the 
number of BH merger events and of the stochastic GW background from unresolved sources. 
\subsubsection{Likelihood analysis for GWs observations with accretion}

Following the same procedure of the previous subsection, one can start by performing a maximum-likelihood analysis to find out the preferred values for the parameters of the PBH initial mass function which best-fit the data.
The log-likelihood function is given by
\be
{\cal L}_\text{\tiny  acc} = \sum_j {\rm ln} \frac{\int \d R_\text{\tiny  acc} (M_1, M_2, z) /(1+z) \d V_c (z) p_j (M_{j,1}| M_1)  p_j (M_{j,2}| M_2) p_j (z_j| z) \Theta (\rho(M_1,M_2,z) - \rho_c)}{ \int \d R_\text{\tiny  acc} (M_1, M_2, z) /(1+z) \d V_c (z) \Theta (\rho(M_1,M_2,z) - \rho_c)}
\ee
in terms of the merger rate including the effect of accretion $\d R_\text{\tiny  acc}(z)$ given in Eq.~\eqref{diffaccrate}.
We stress that, once accretion is included, the final masses enter both in the Gaussian probabilities for the 
experimental uncertainties of the detected events as well as in the SNR.

\subsubsection{Number of events and GWs abundance with accretion}
Including the effect of accretion, the predicted number of PBH mergers detected in a time $\Delta t$ is given by
\be
N_\text{\tiny  acc} = \Delta t \int \d z \d M_1 \d M_2 \frac{1}{1+z} \frac{\d V_c(z)}{\d z} \frac{\d R_\text{\tiny  acc}(M_1,M_2,z)}{\d M_1 \d M_2} \Theta\left(\rho (M_1, M_2,z) - \rho_c \right)
\ee
in terms of the accretion-included merger rate $\d R_\text{\tiny  acc}(M_1, M_2,z)$ given by Eq.~\eqref{diffaccrate}.

Likewise, also the GW background gets modified by accretion as
\be
\Omega^\text{\tiny  acc}_\text{\tiny GW} (\nu)=  \frac{\nu}{\rho_0} \int_0^{\frac{\nu_3}{\nu}-1} \d z \d M_1 \d M_2 \,  \frac{1}{(1+z)H(z)} \frac{\d R_\text{\tiny  acc} (M_1,M_2,z)}{\d M_1 \d M_2} \frac{\d E_\text{\tiny GW} (\nu_s)}{\d \nu_s}  \Theta(\rho_c - \rho (M_1, M_2,z)).
\ee
The results for the constraints with the inclusion of accretion will be shown in the next section.

\section{Confrontation with LIGO/Virgo O1, O2, and GW190412}\label{sec:confrontation}

In this section we present the main results of this work. We confront the theoretical predictions discussed above with 
current observations, assuming that the BHs involved in the merger events are of primordial origin. We do so by assuming 
two mass functions (power-law and lognormal) and various scenarios, namely no accretion and accretion with three 
different cut-off redshifts ($z_\text{\tiny cut-off}=15,10,7$). As for the data, we include the LIGO/Virgo observation 
runs O1 and O2~\cite{LIGOScientific:2018mvr} and the recent GW190412 \cite{LIGOScientific:2020stg}, which 
overall comprise ${\rm N}_\text{\tiny \rm obs}=11$ events. \footnote{In the case without accretion, an analysis 
regarding the chirp mass and the mass ratio for Advanced-LIGO was done in Ref.~\cite{b}, whereas the best-fit values 
for a toy-model initial spin distribution were computed in Ref.~\cite{Fernandez:2019kyb} using O1-O2 events.}

We have proceeded by assuming that all the events seen by LIGO/Virgo are due to a first-generation merger of PBHs 
(hierarchical mergers~\cite{Gerosa:2017kvu} will be briefly discussed later on). This is admittedly a strong assumption 
as of course a fraction of these events (if not all) might be due to BHs of astrophysical 
origin.
Our goal is therefore to analyse whether motivated PBH mass functions (power-law and lognormal) are compatible with 
current GW data. 

\subsection{Best-fit parameters for the PBH mass function}
\begin{figure}[t!]
	\centering
	\includegraphics[width=0.235 \linewidth]{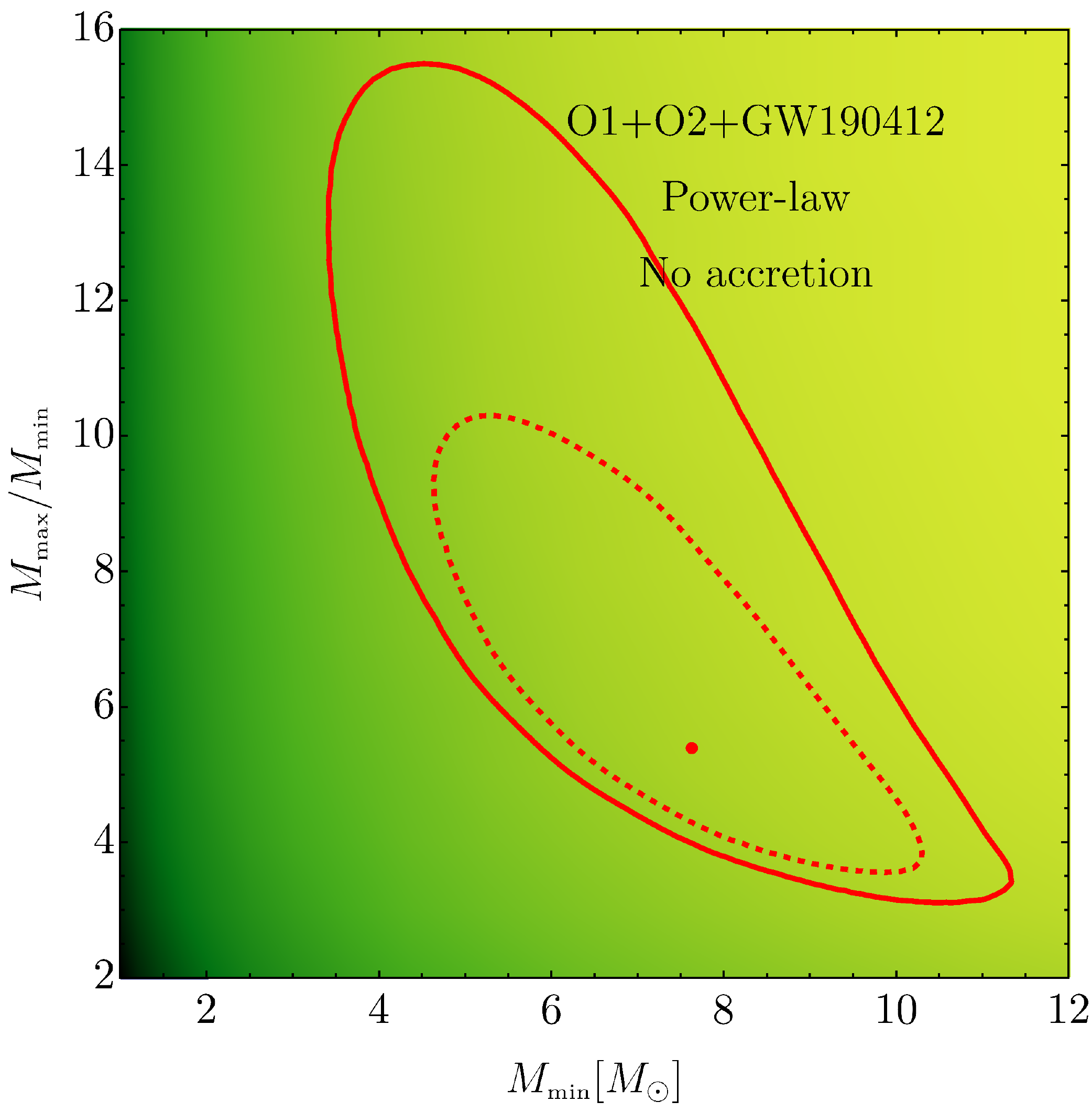}
	\includegraphics[width=0.235 \linewidth]{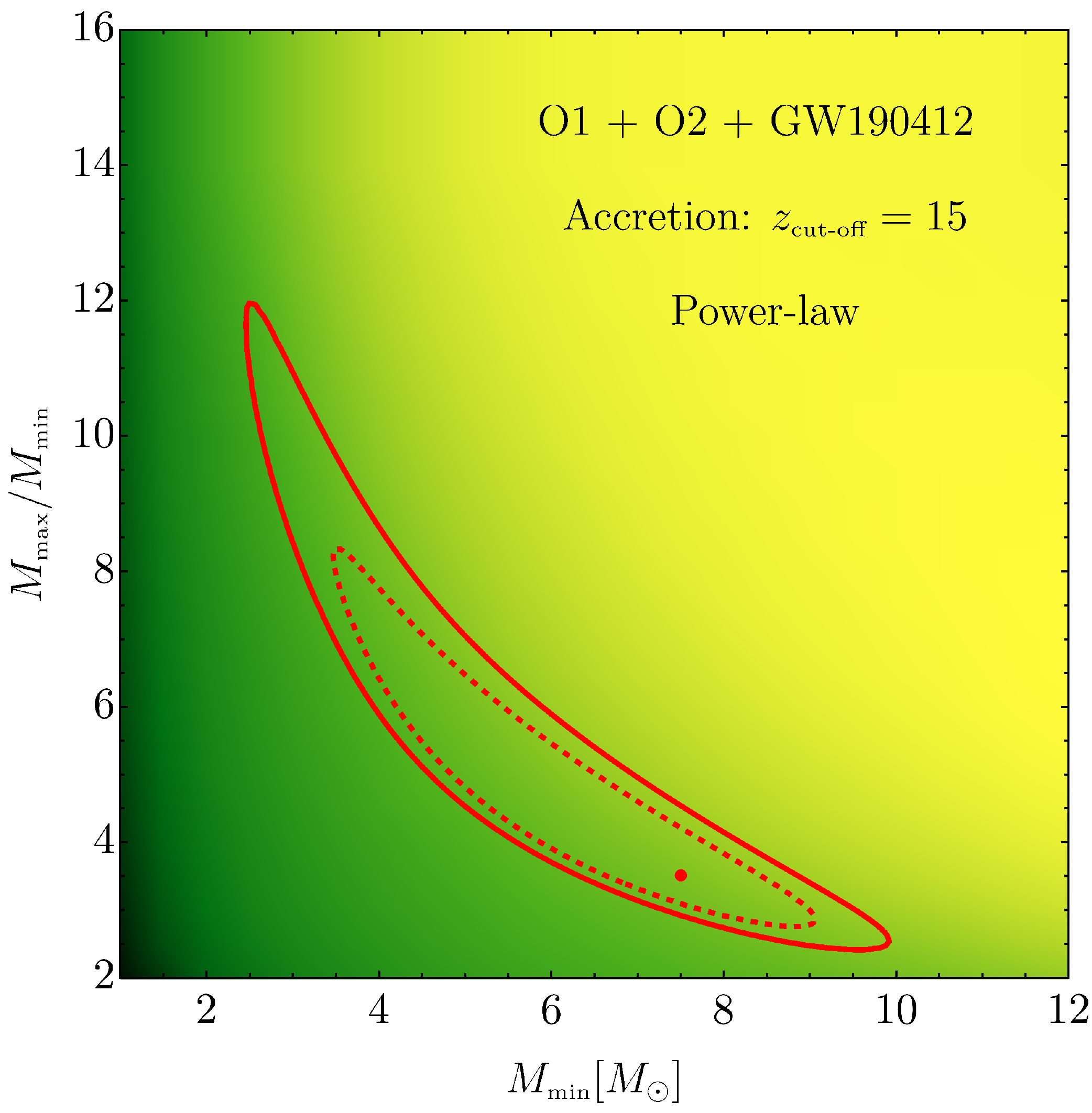}
	\includegraphics[width=0.232 \linewidth]{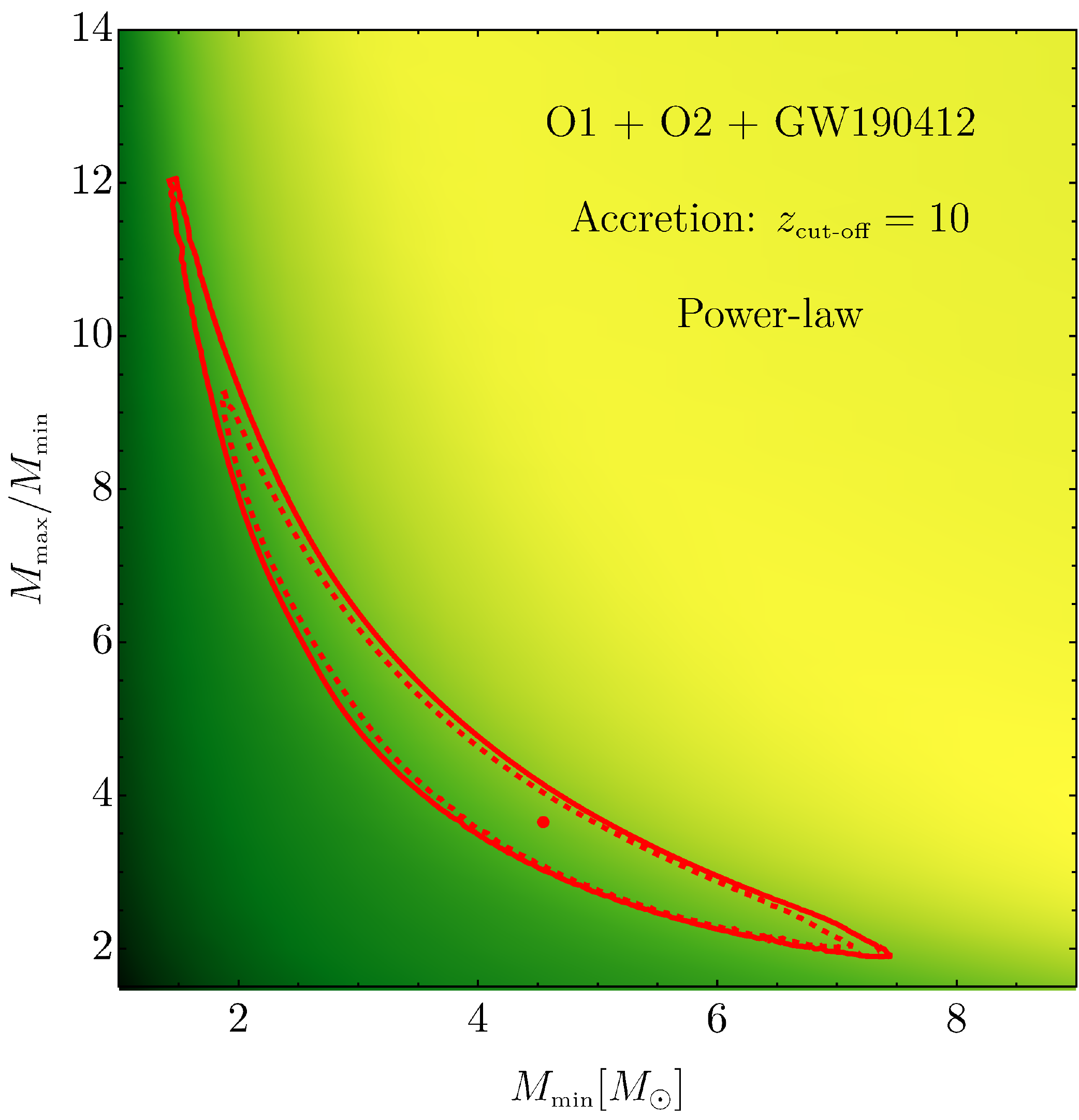}
	\includegraphics[width=0.273 \linewidth]{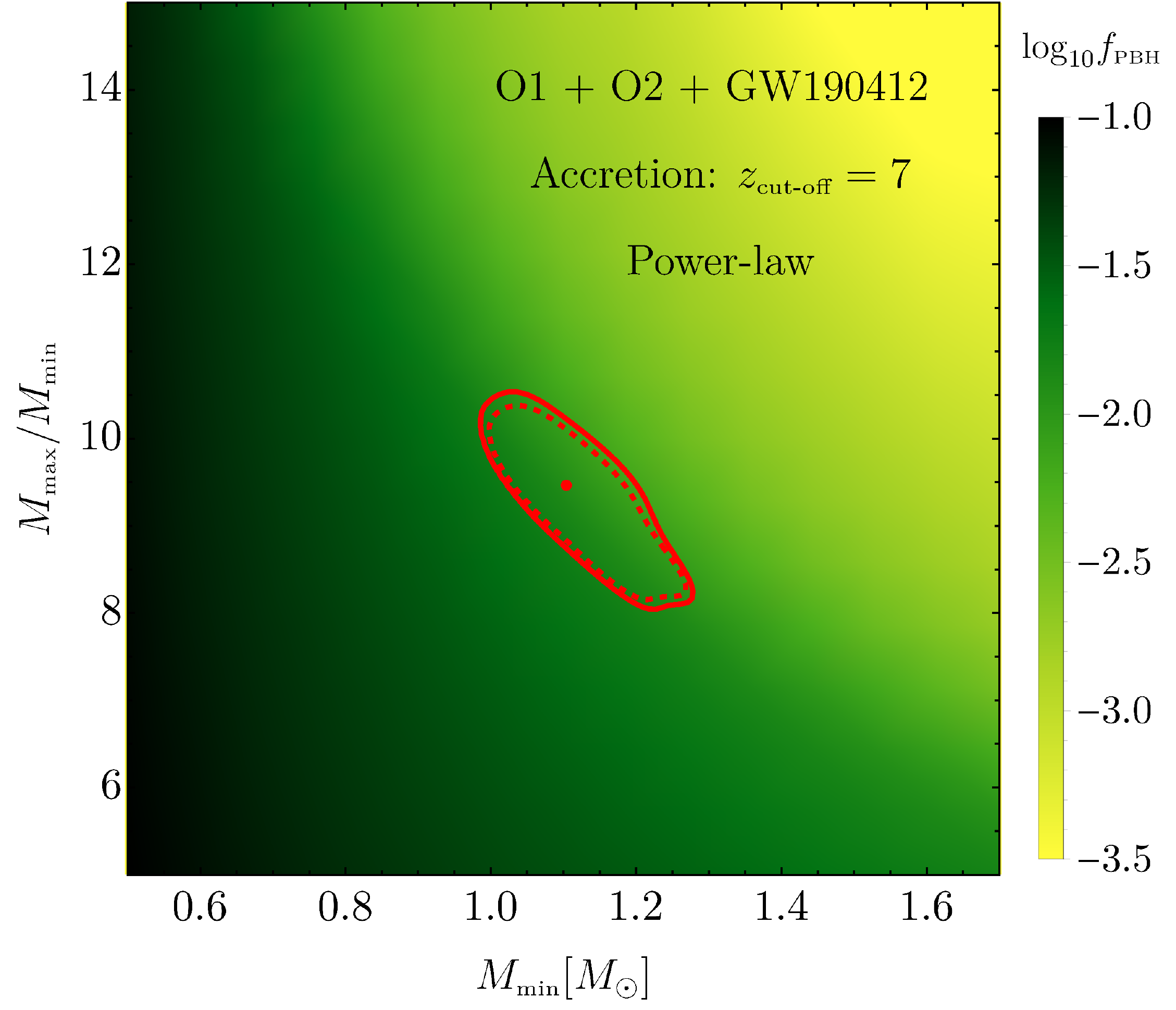}
	\includegraphics[width=0.235 \linewidth]{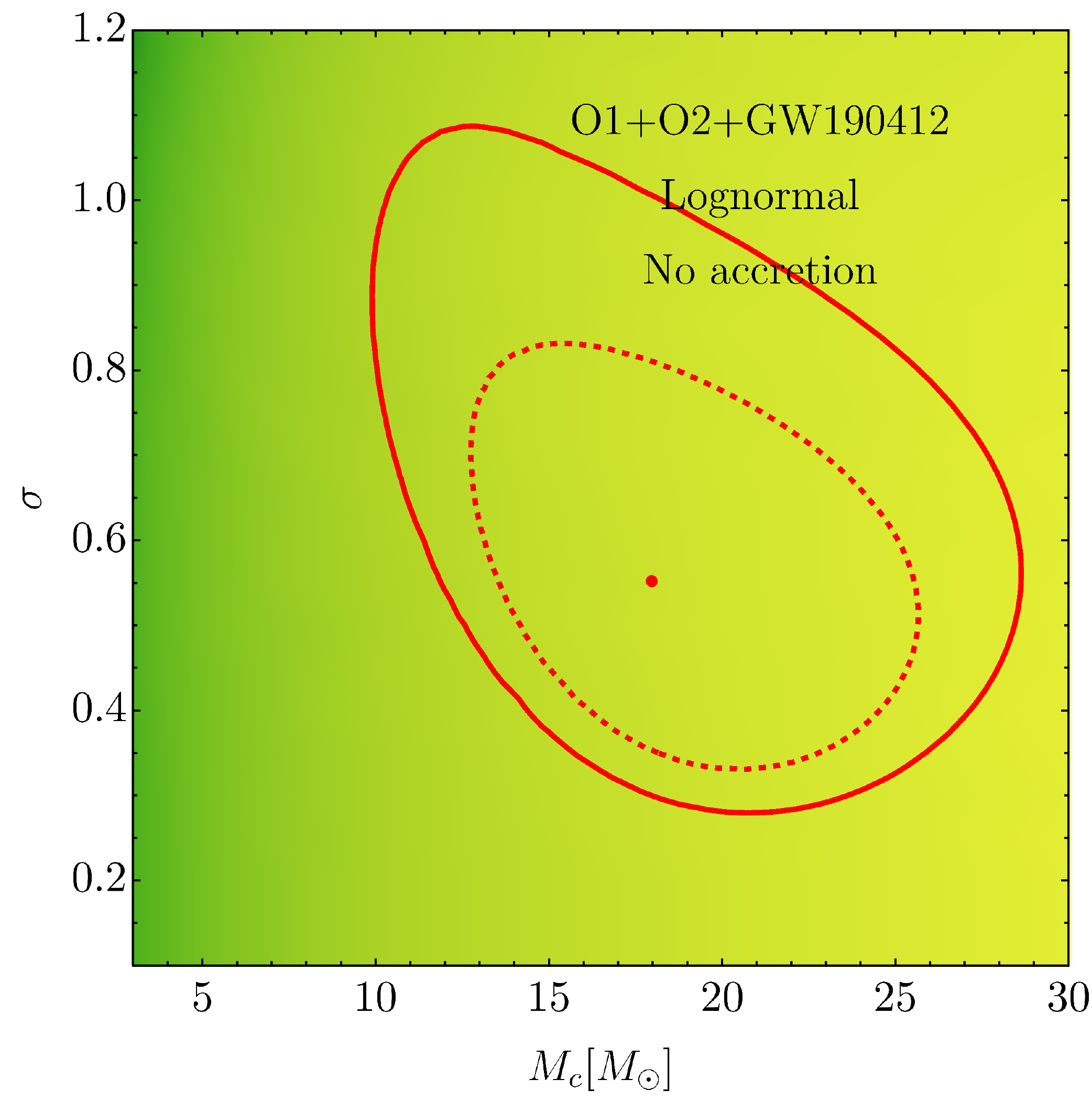}
	\includegraphics[width=0.235 \linewidth]{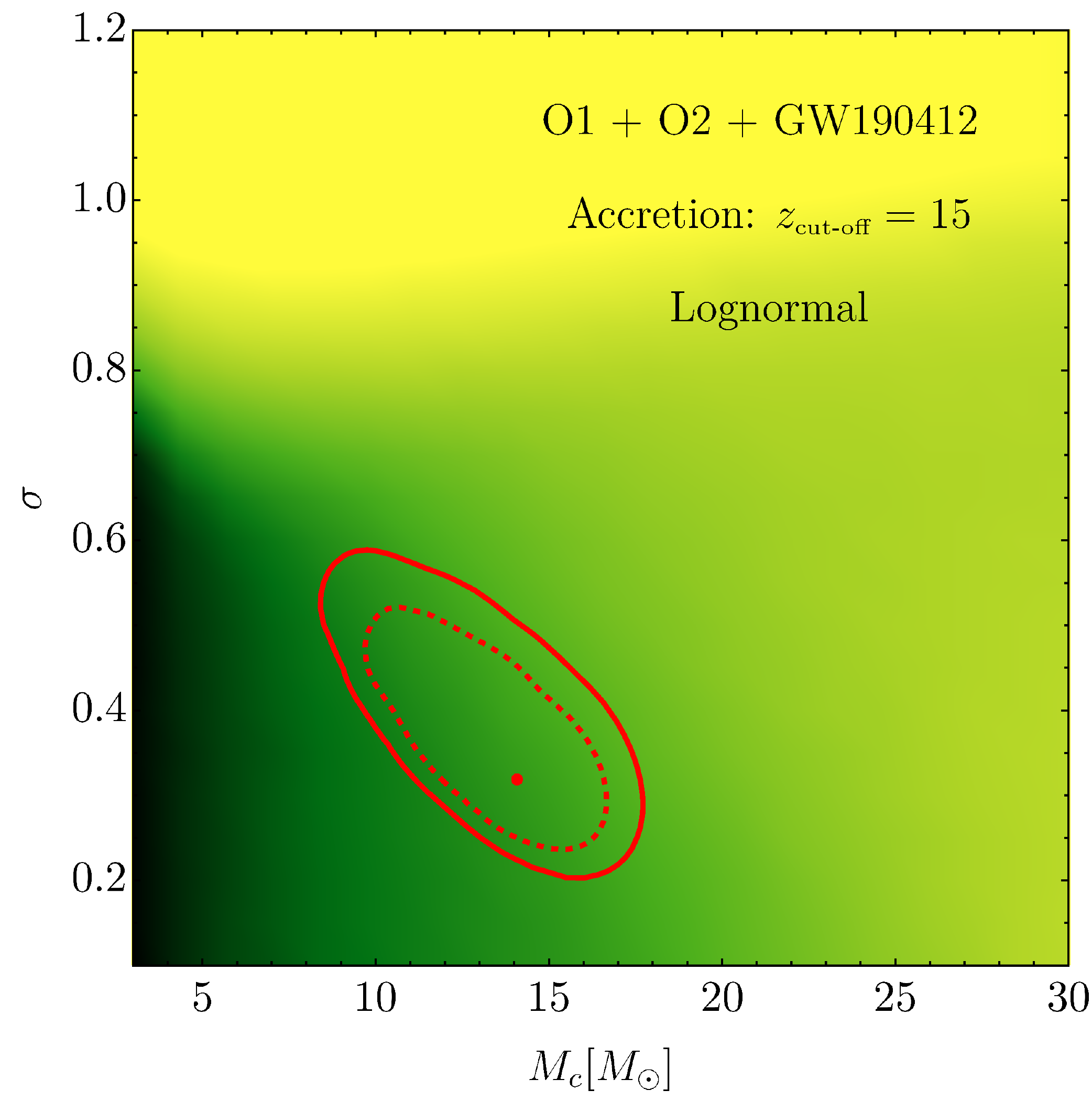}
	\includegraphics[width=0.235 \linewidth]{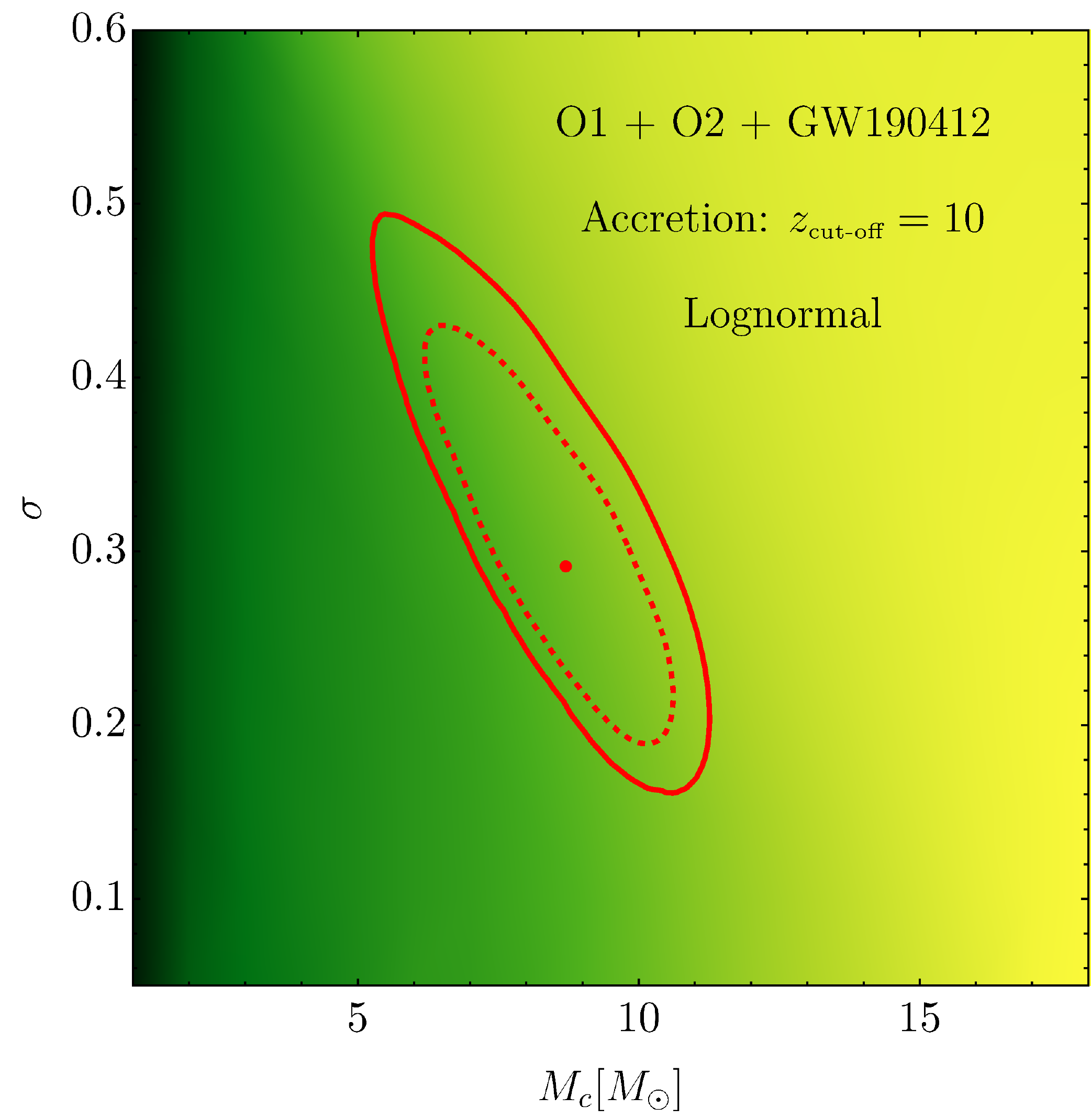}
	\includegraphics[width=0.273 \linewidth]{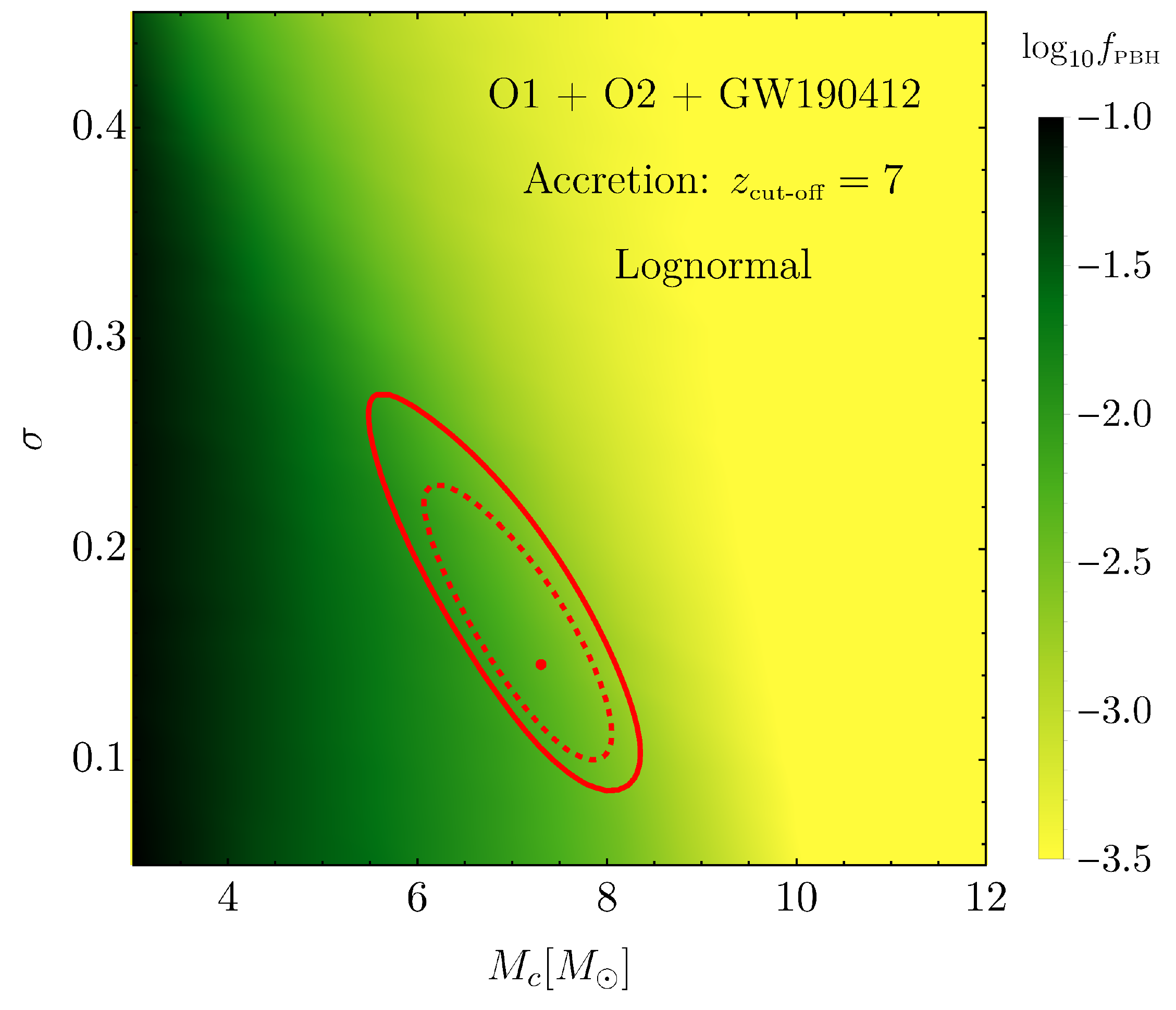}
	\caption{\it Likelihood on the parameter space for the power-law (top) and lognormal (bottom) mass functions 
requiring ${\rm N}_\text{\tiny \rm obs}=11$. The red dashed (solid) contours corresponds to $2\sigma$ ($ 3 \sigma$) 
respectively. The leftmost panels correspond to the case in which accretion is negligible, whereas the second to fourth 
columns correspond to accretion suppressed at $z_\text{\tiny \rm cut-off}=(15,10,7)$, respectively.
	}
\label{likelihood}
\end{figure}

In Fig.~\ref{likelihood} we present the likelihood on the parameter space for a power-law (top panels) and lognormal 
(bottom panels) mass function. A few comments are in order: first, the best-fit values in all 
cases end up providing a similar value of the PBH fraction in DM, namely $f_\PBH\approx {\rm few}\times 10^{-3}$; this 
upper bound becomes less stringent in the case of strong accretion (small $z_\co$); secondly, increasing 
the accretion effect (i.e., decreasing $z_\co$) makes the 
$2\sigma$ and $ 3 \sigma$ contours to shrink.
This is due to the fact that the best-fit values correspond to a narrower initial distributions and  smaller 
initial central masses; 
finally, we note that, for the case of no accretion, our best-fit values for the lognormal distribution agree with 
those recently computed in Ref.~\cite{Dolgov:2020xzo}. The latter reported $M_c=17 M_\odot$ and $\sigma=0.75$, which 
agree with the values shown in the bottom left panel of Fig.~\ref{likelihood} within $1\sigma$. This shows the 
robustness of these values, given the fact that our analysis and that of Ref.~\cite{Dolgov:2020xzo} are 
different; for example Ref.~\cite{Dolgov:2020xzo} did not include the suppression factor for the merger rate and fitted 
only the chirp masses of the events.

\begin{figure}[t!]
	\centering
	\includegraphics[width=0.24 \linewidth]{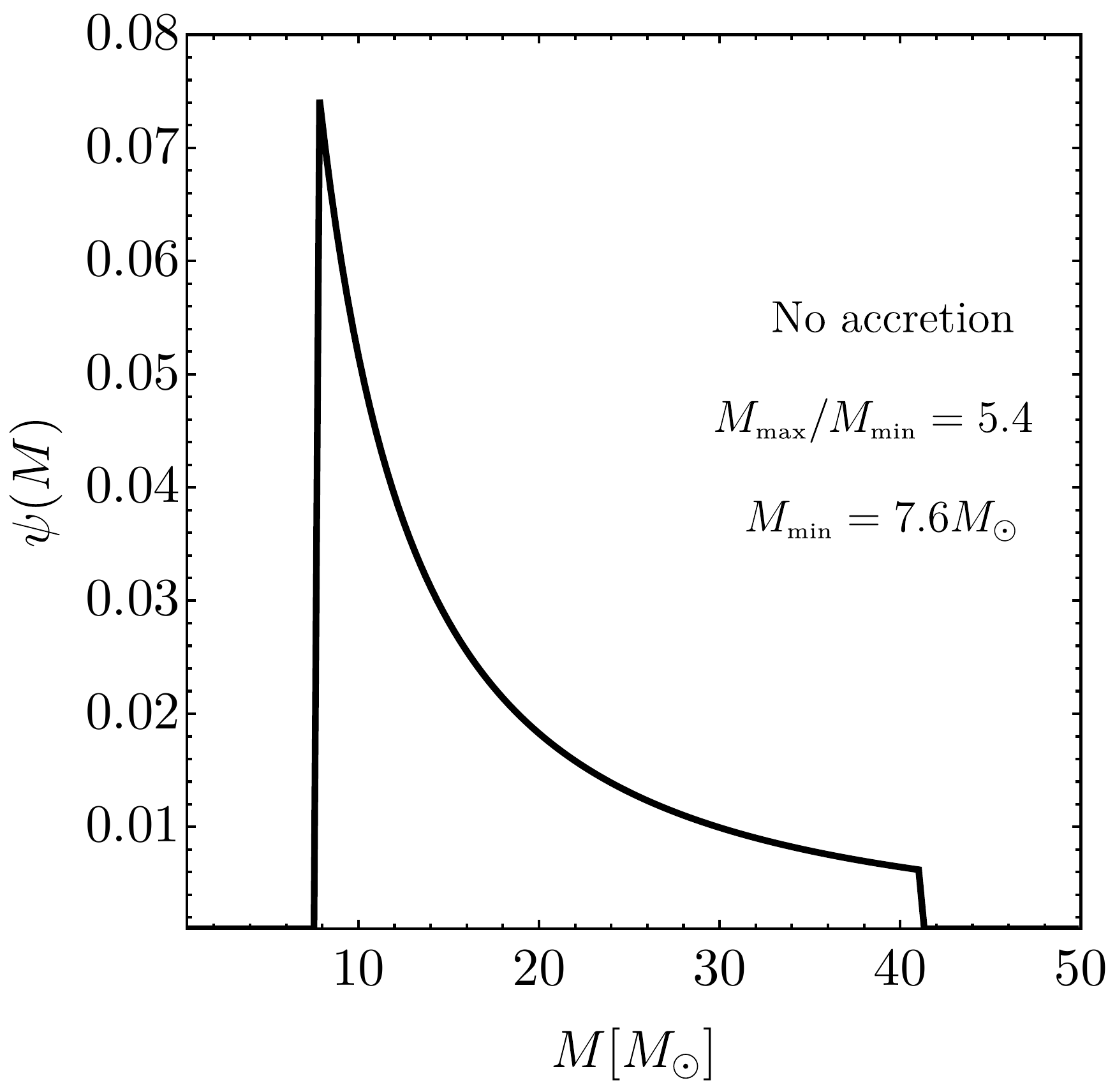}
	\includegraphics[width=0.24 \linewidth]{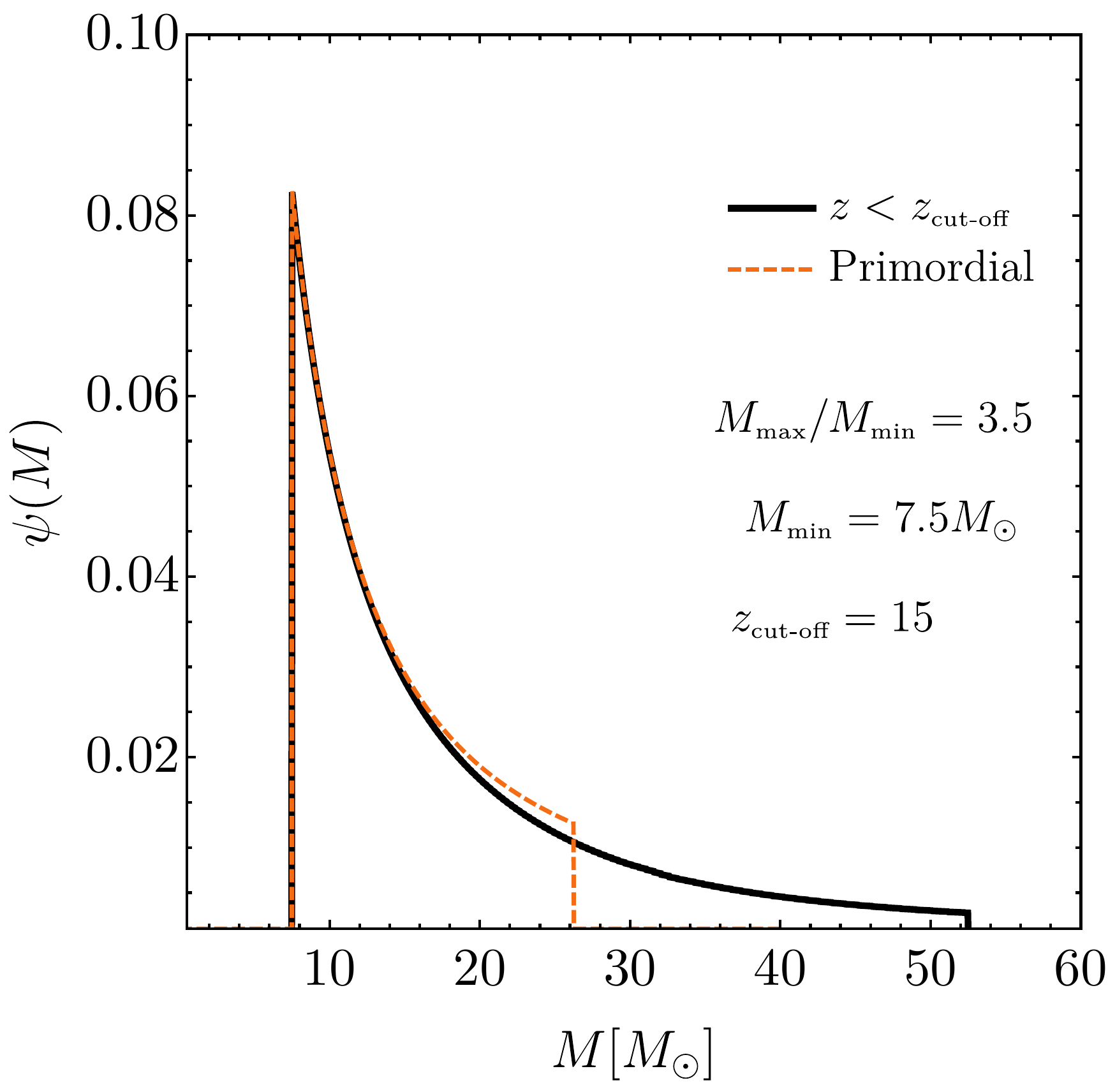}
	\includegraphics[width=0.242 \linewidth]{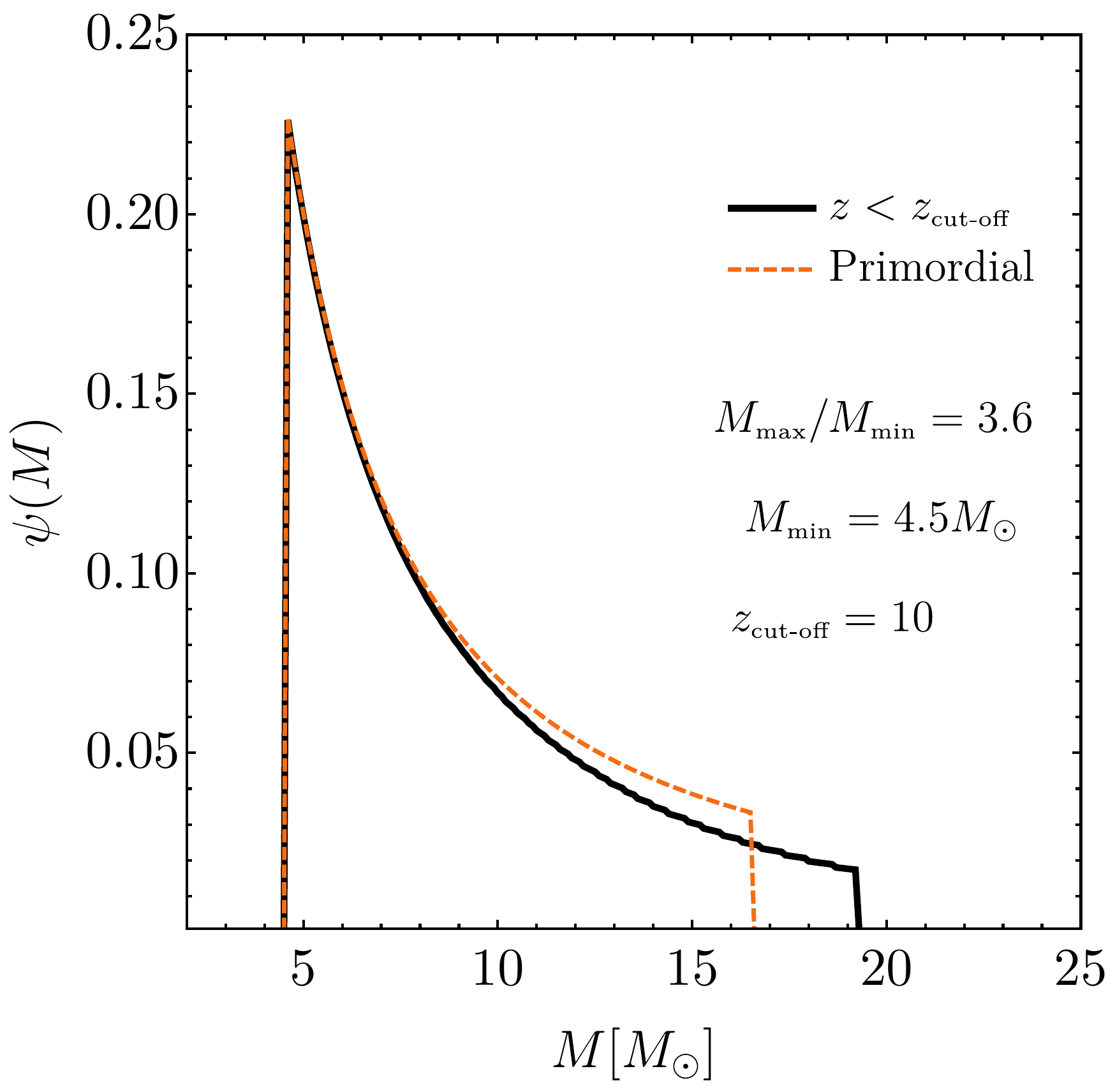}
	\includegraphics[width=0.24 \linewidth]{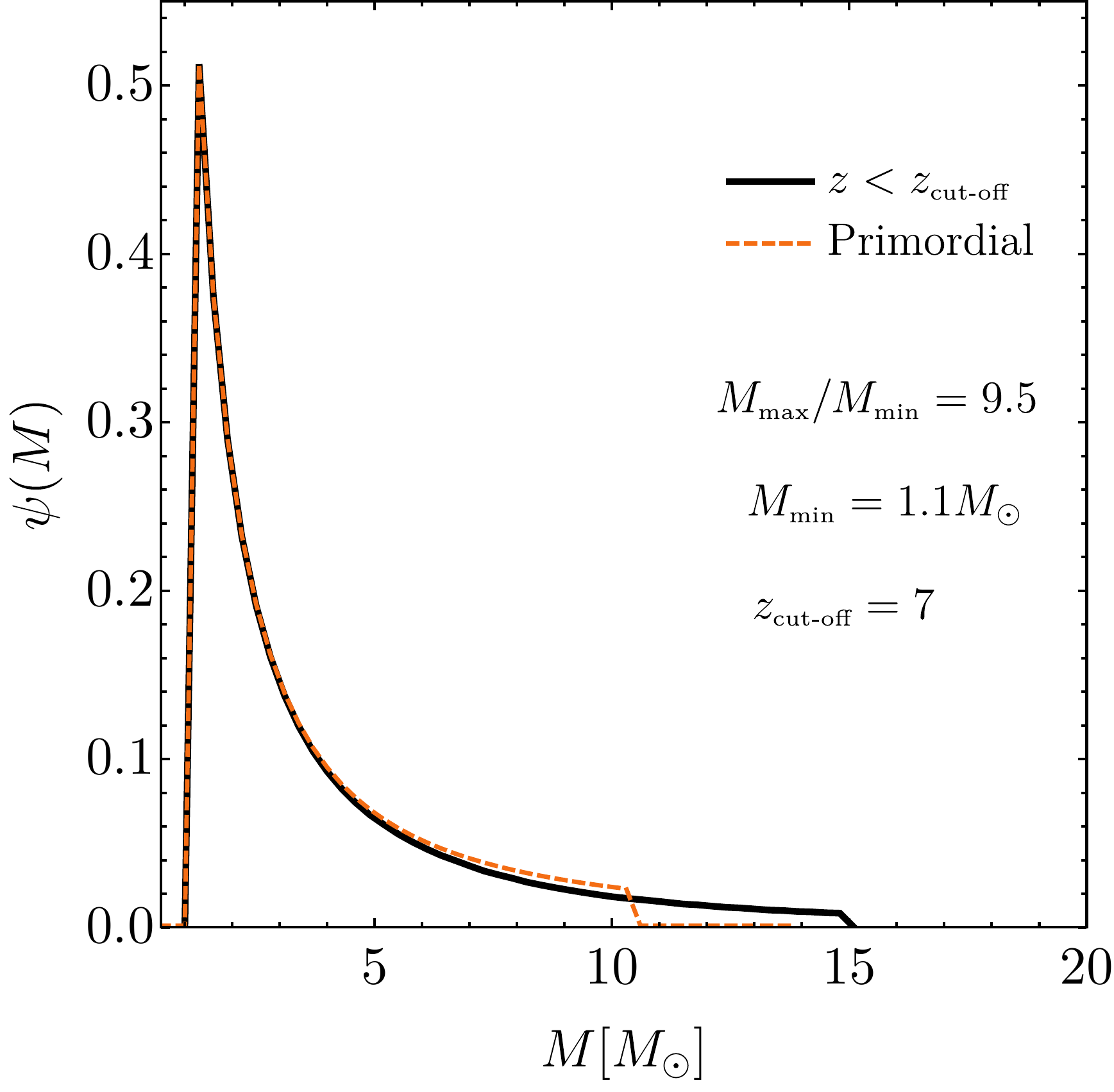}
	
	\includegraphics[width=0.24 \linewidth]{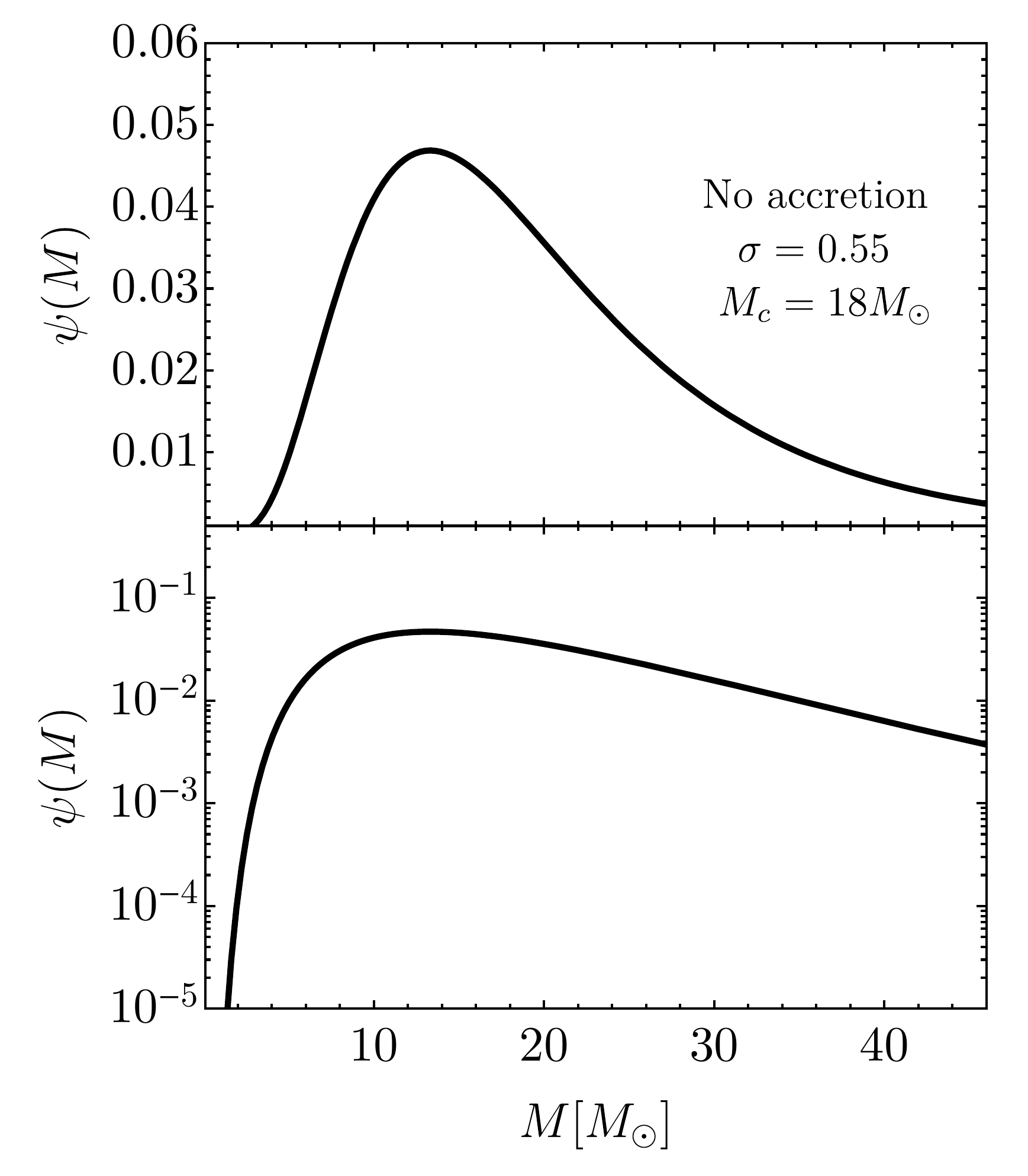}
	\includegraphics[width=0.24 \linewidth]{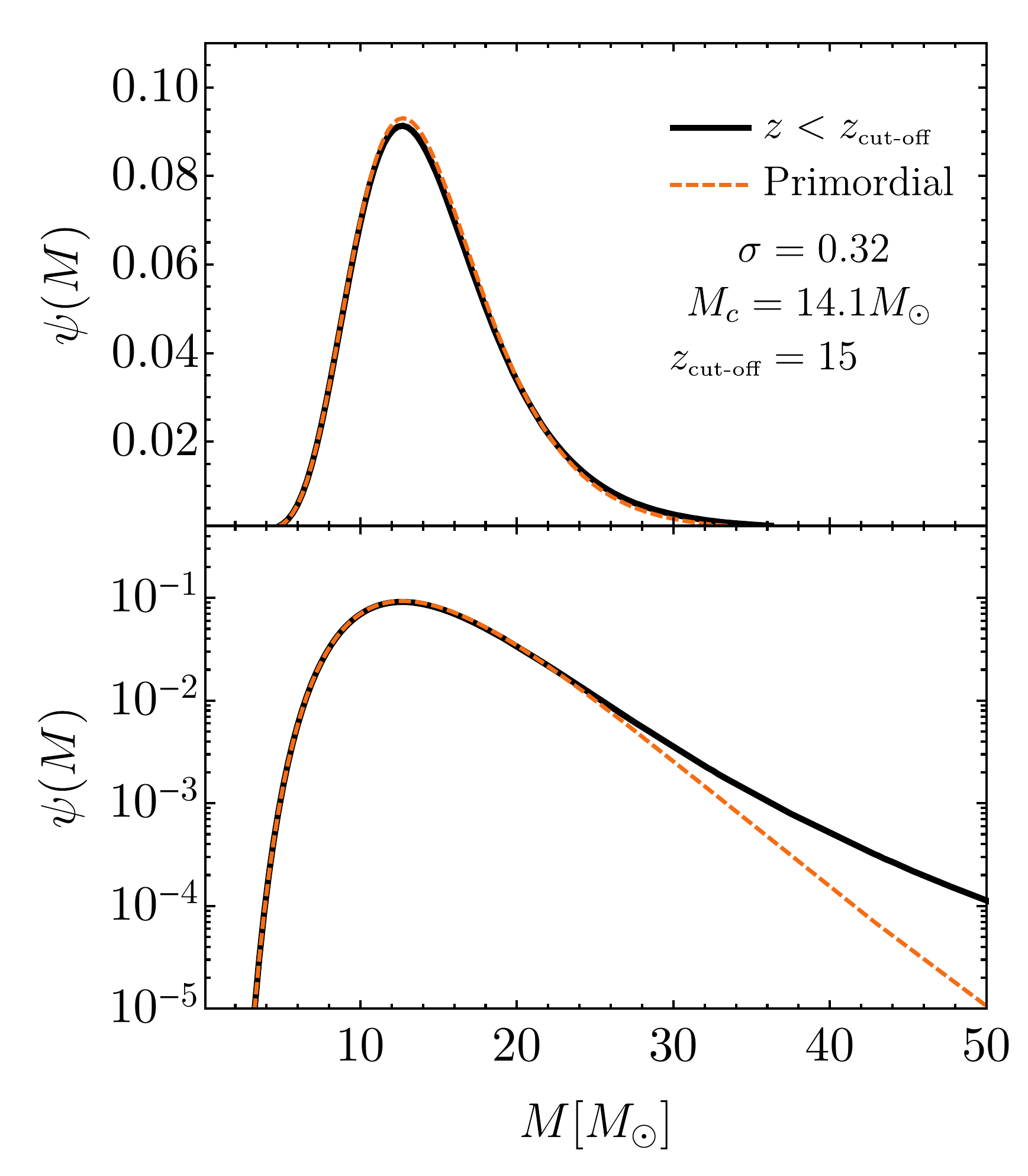}
	\includegraphics[width=0.242 \linewidth]{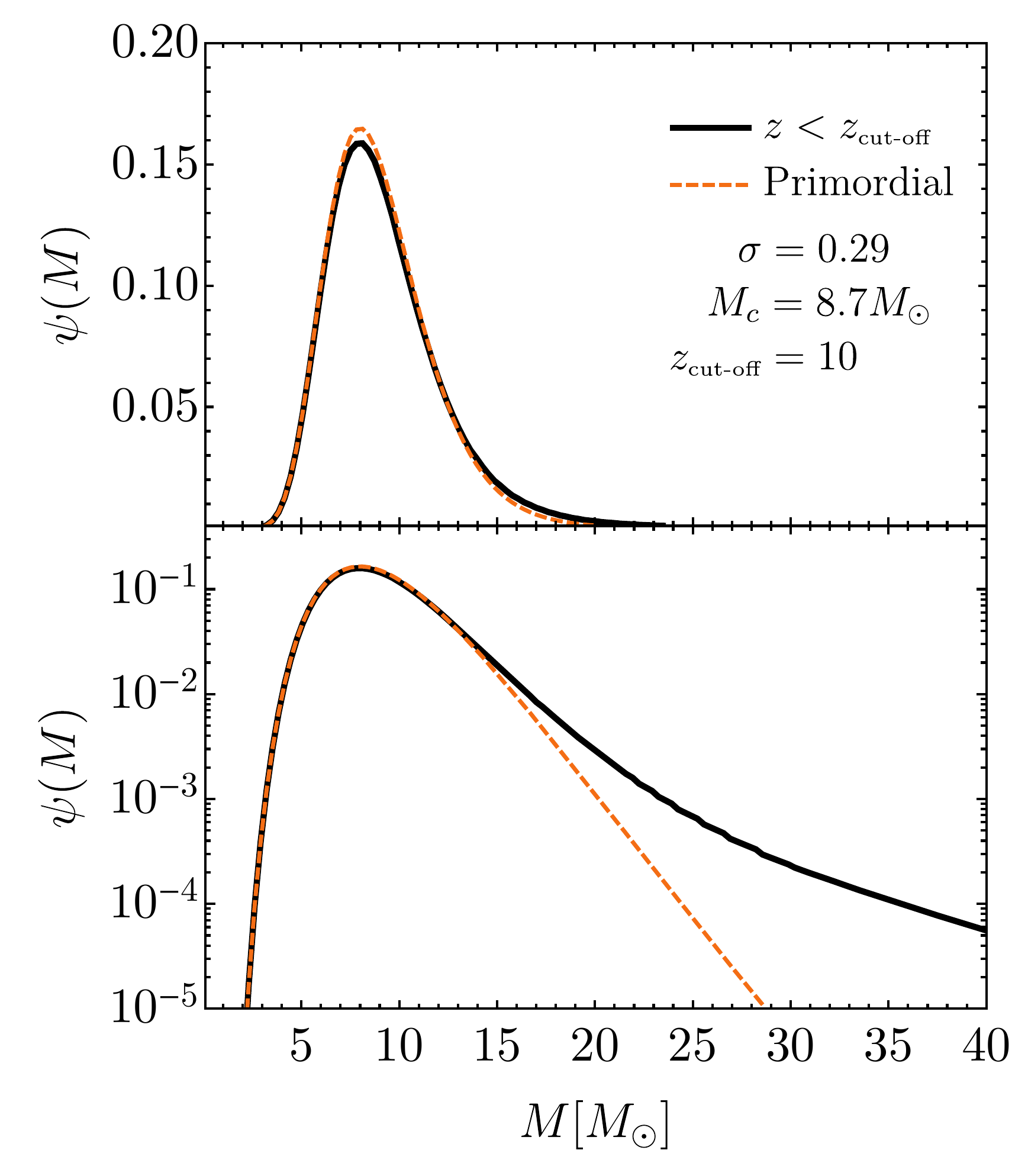}
	\includegraphics[width=0.24 \linewidth]{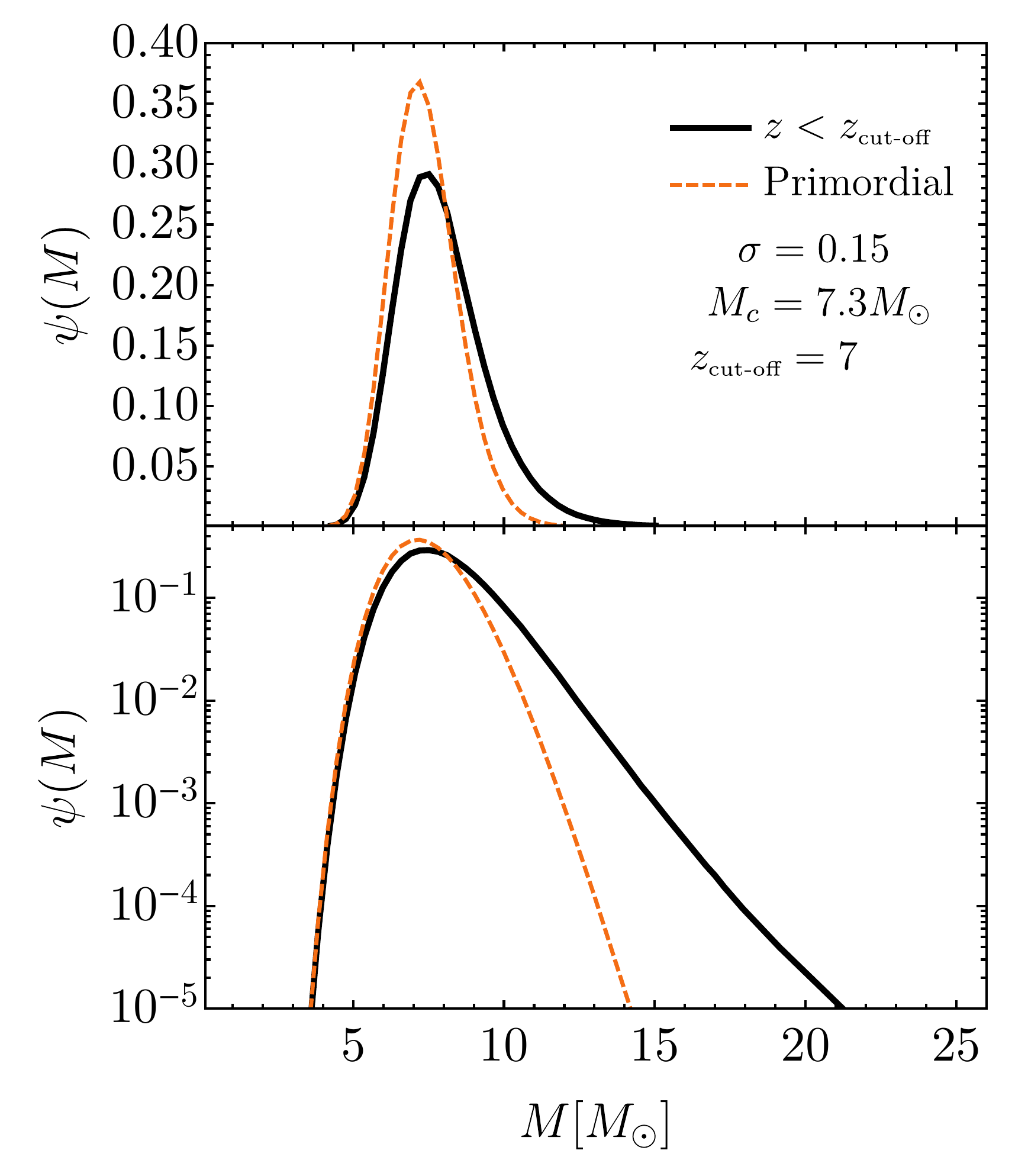}
	\caption{\it Primordial mass function and its evolution if accretion is present. In this plot we use the 
best-fit values for the parameters of the PBH mass distributions at formation, as 
obtained from the previous likelihood analysis. The organization of the panels is the same as in Fig.~\ref{likelihood}.
	}
\label{massfunction}
\end{figure}

In Fig. \ref {massfunction} we present the comparison between the initial and the final mass functions for the best-fit 
values obtained from the previous likelihood analysis. To highlight the differences at large masses, for the lognormal 
case we also show the same results in a log-linear scale. The effect of accretion is to shift the tail of the mass 
function to larger PBH masses, this shift being more pronounced when the accretion is stronger, with a consequent 
decrease of the amplitude of the peak. In other words, accretion tends to make the mass distribution 
broader~\cite{paper2}. Note, however, that the effect on the mass functions for the specific values of the parameters 
selected by the likelihood is much less pronounced than in the example shown in Fig.~\ref{mfevo}. This happens because 
the best-fit distributions peak at relatively low mass.
We also notice that this evolution is for isolated PBHs, which is the relevant case for the constraints inferred from 
the observations other than those from GW observations. However, one should take into account that for PBHs in binaries 
the effect of accretion is larger, thus allowing PBH masses larger than those indicated in Fig. \ref {massfunction}.

\subsection{Updated constraints on PBH abundance}
\begin{figure}[t!]
	\centering
	\includegraphics[width=0.22 \linewidth]{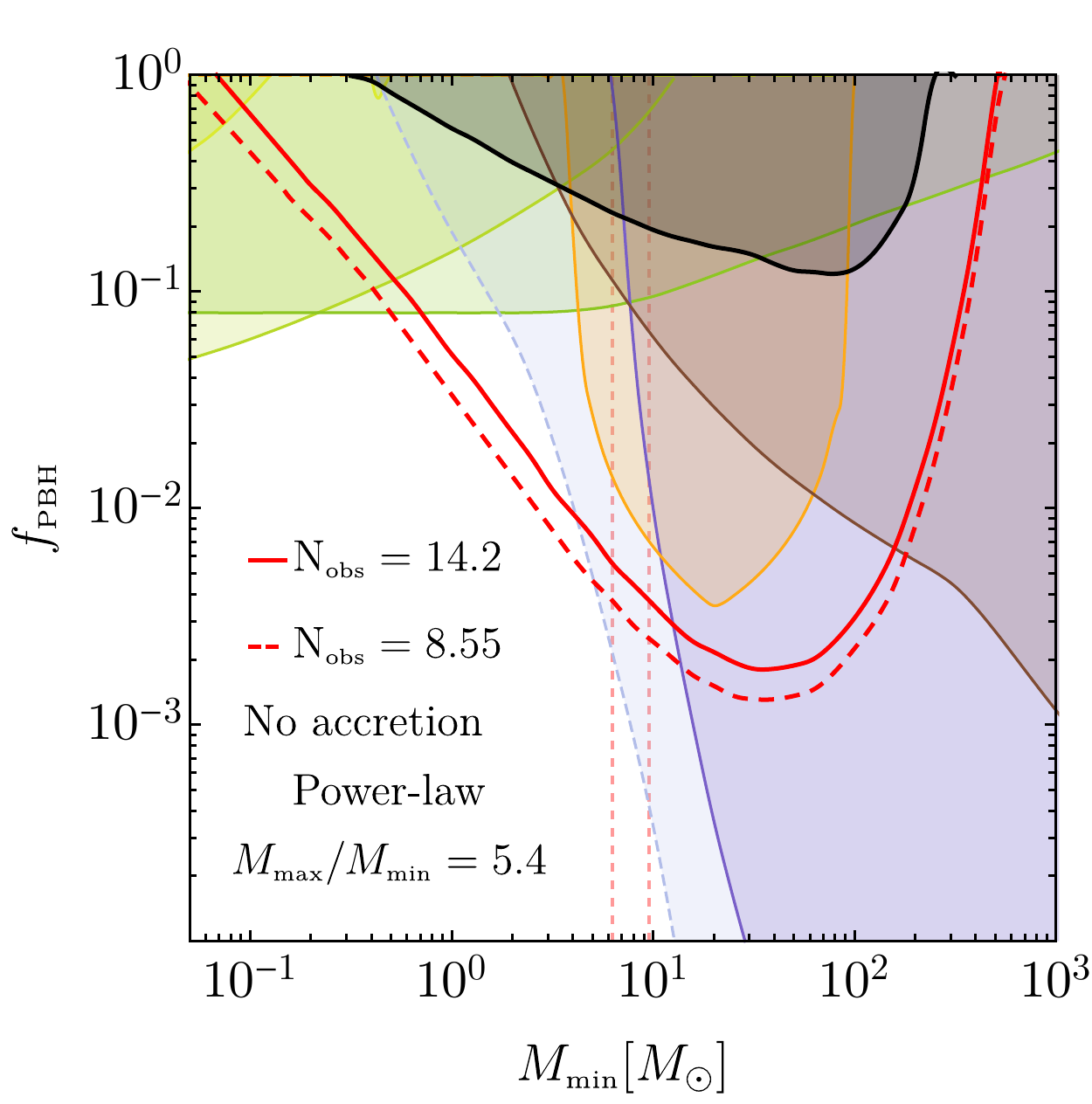}
	\includegraphics[width=0.22 \linewidth]{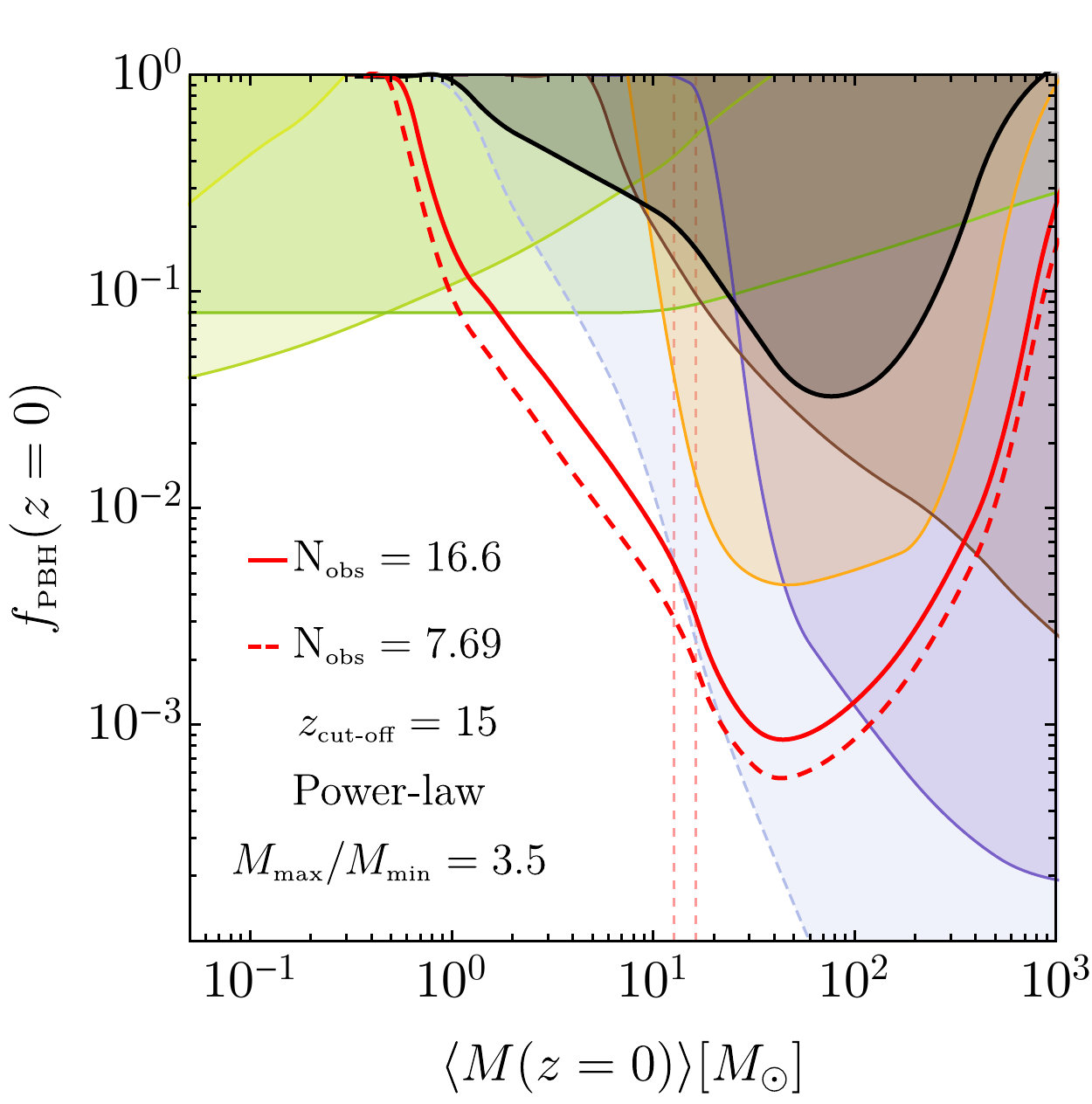}
	\includegraphics[width=0.22 \linewidth]{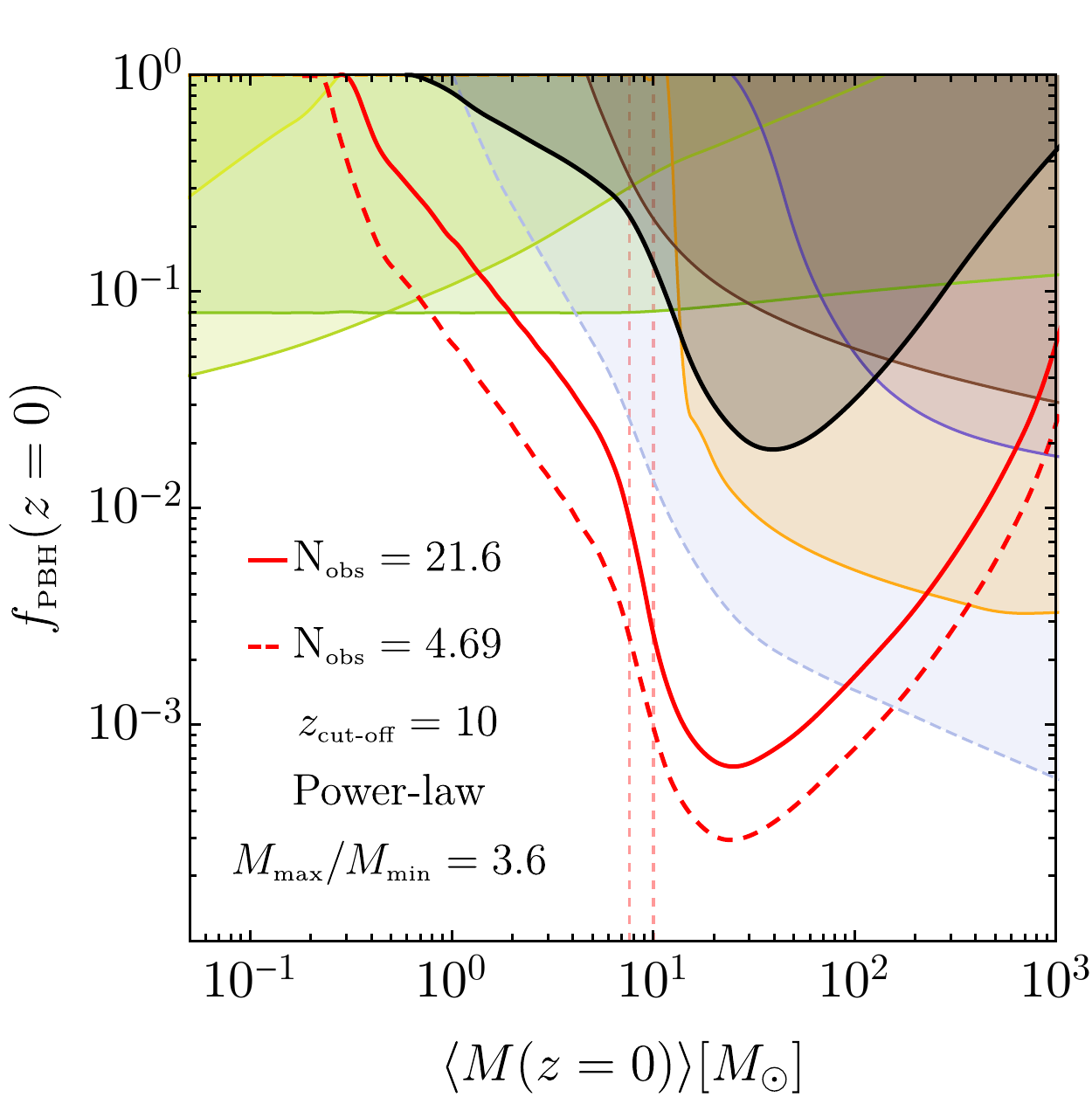}
	\includegraphics[width=0.29 \linewidth]{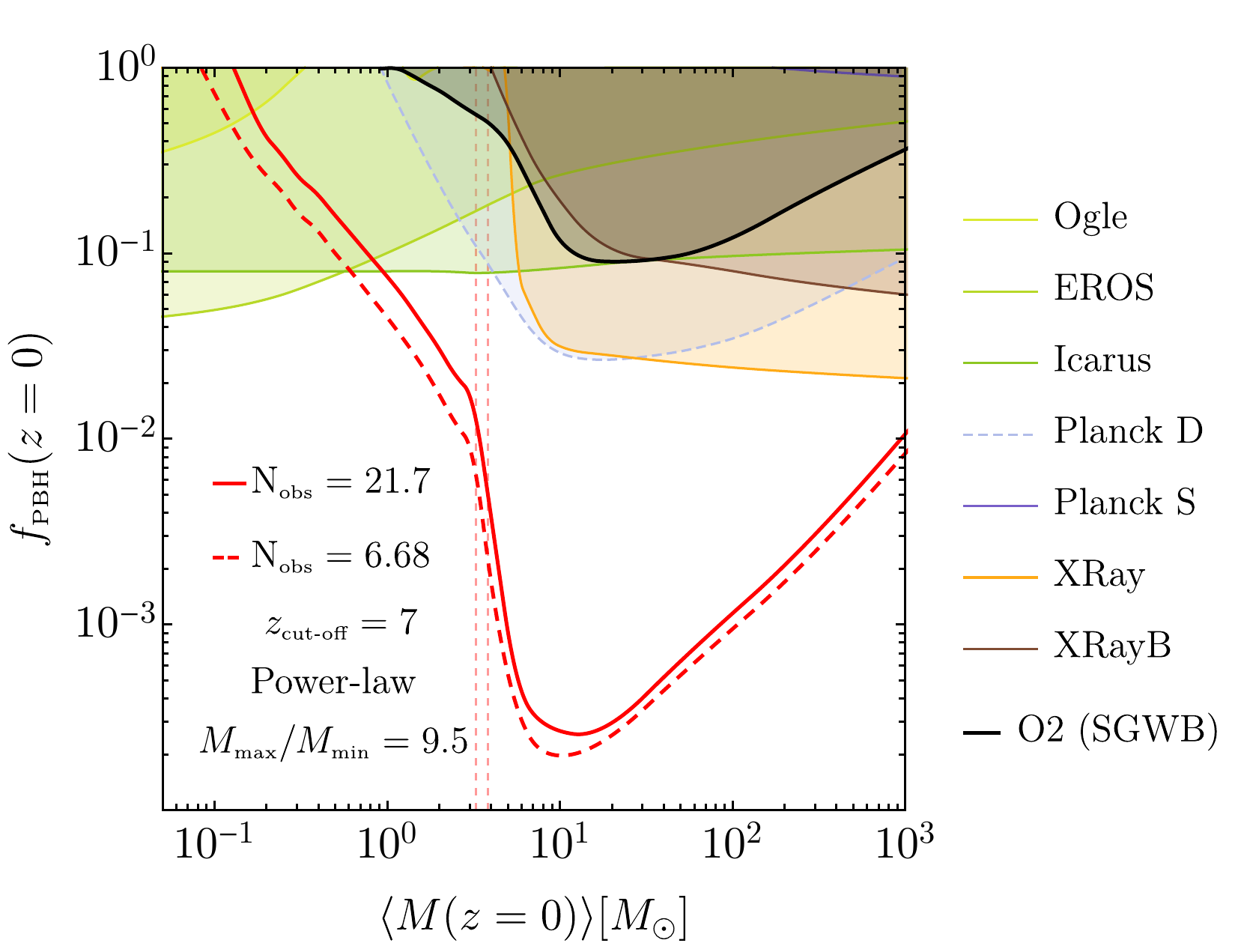}
	
	\includegraphics[width=0.22 \linewidth]{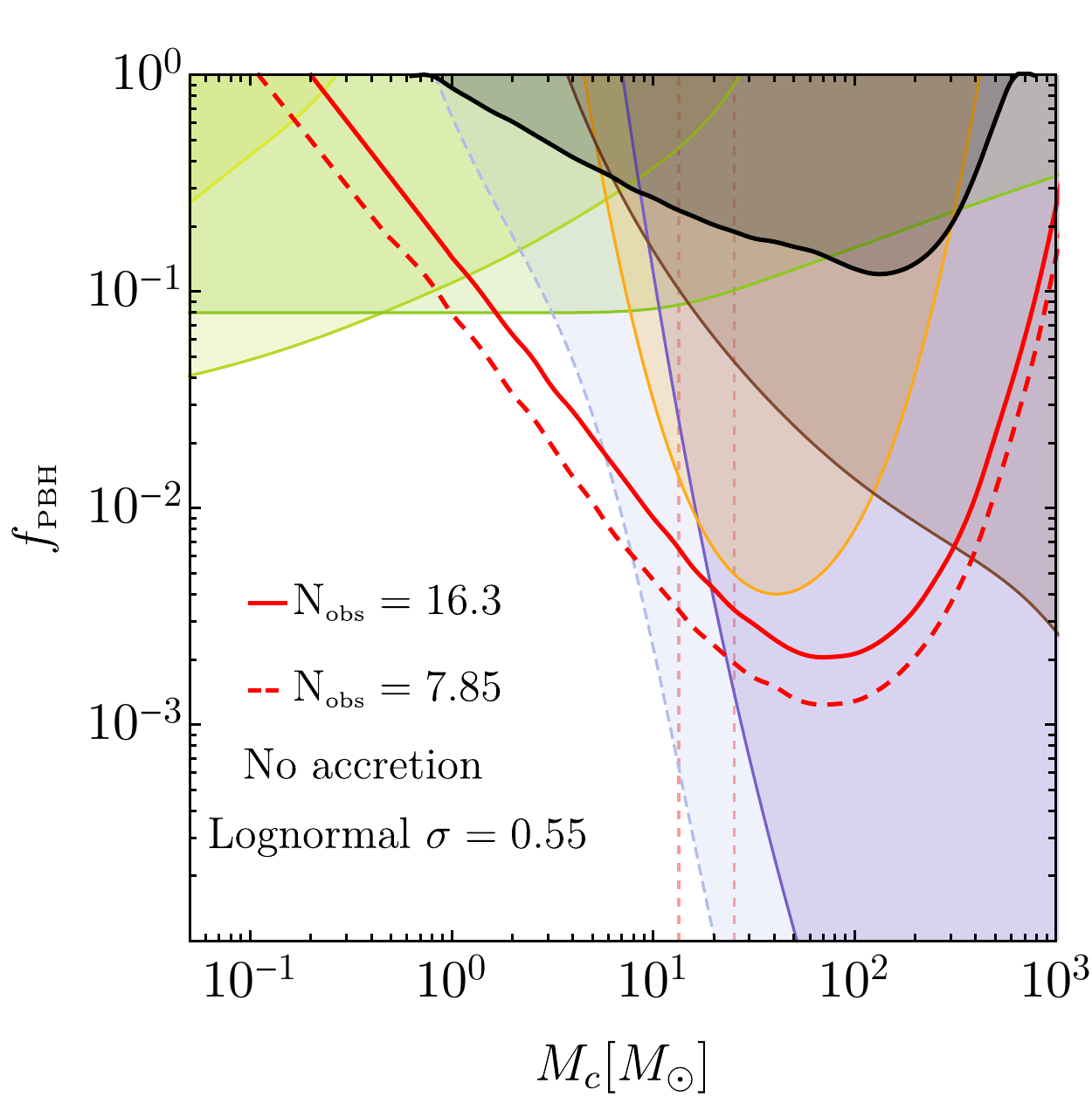}
	\includegraphics[width=0.22 \linewidth]{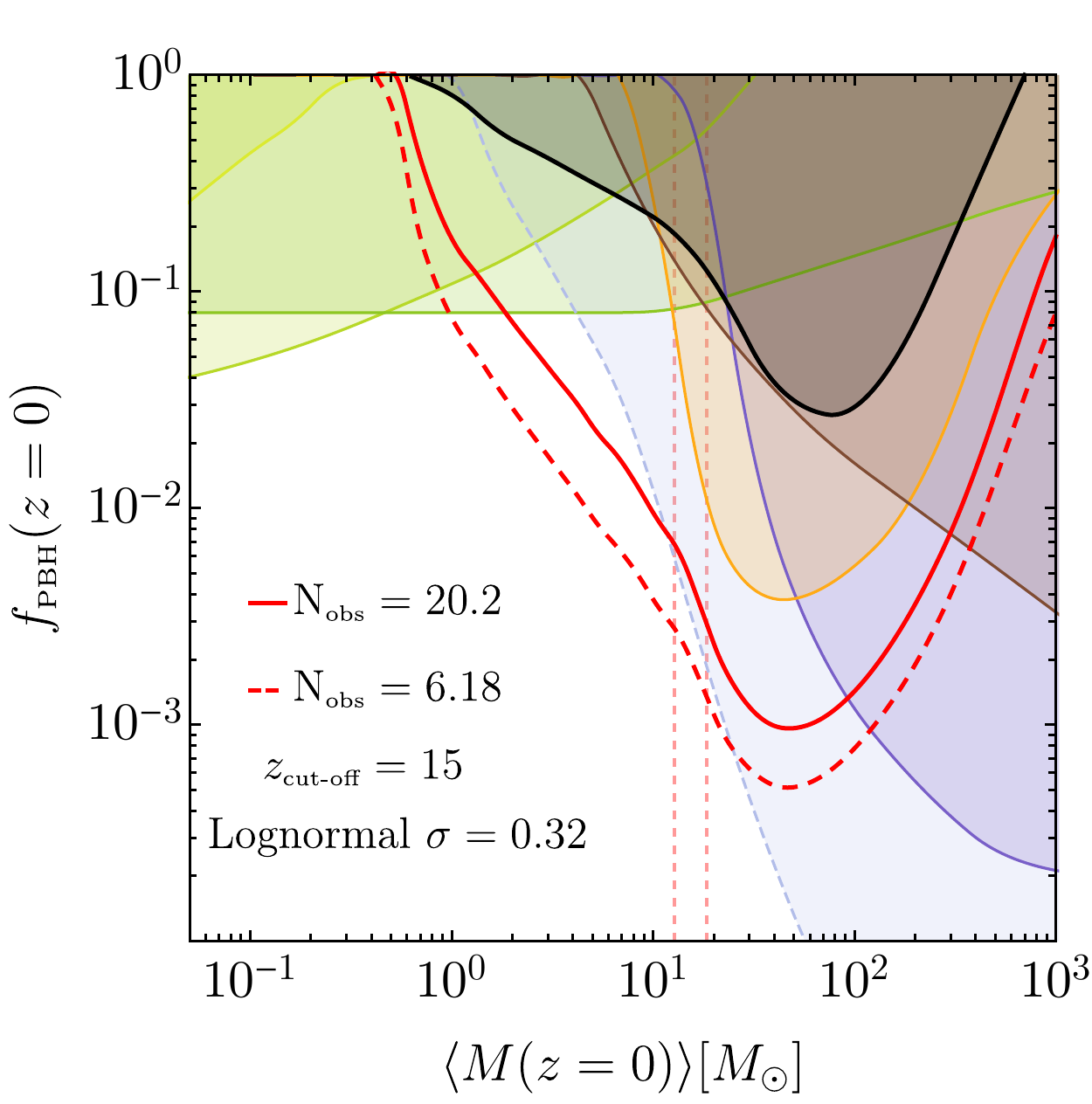}
	\includegraphics[width=0.22 \linewidth]{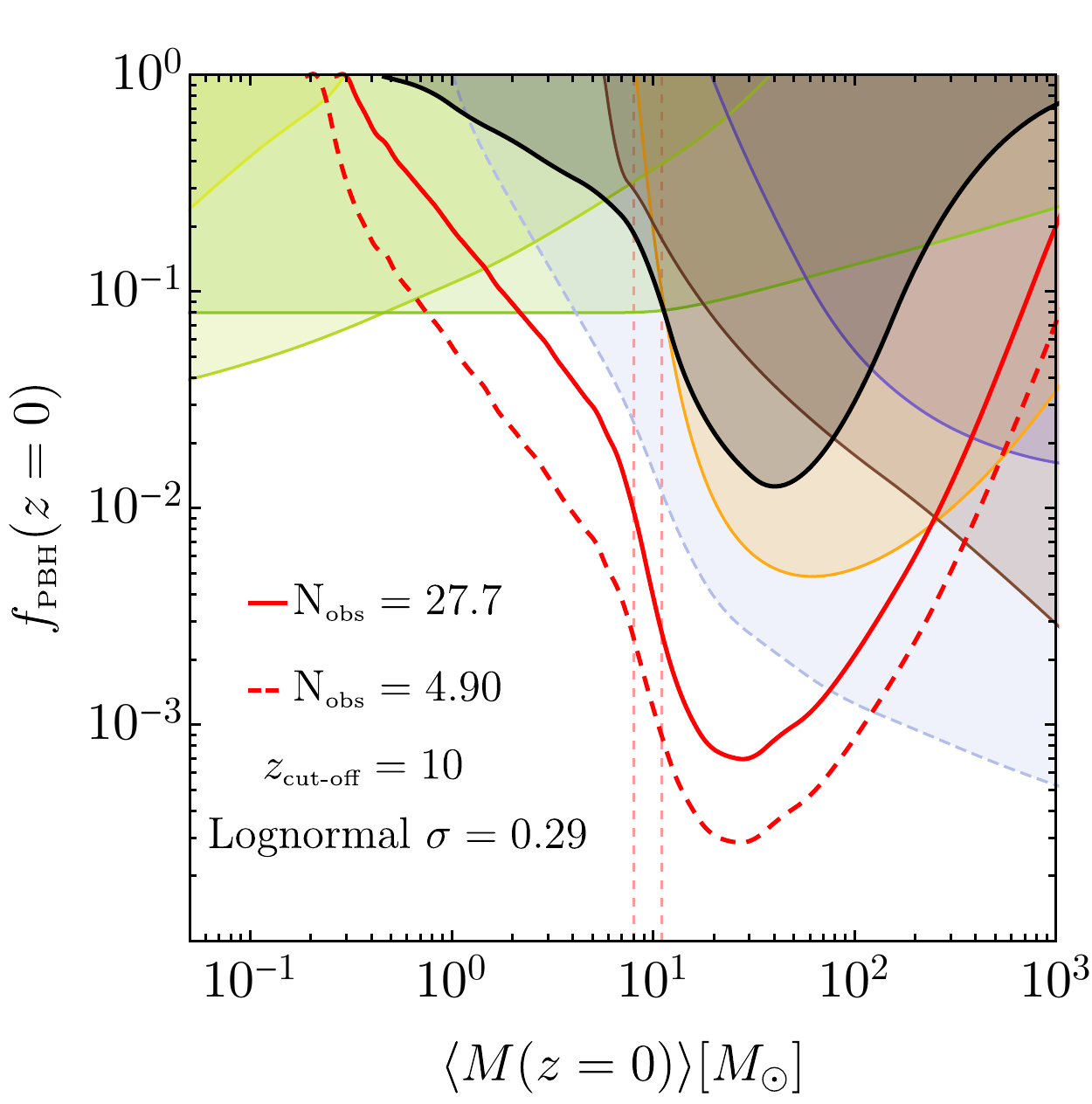}
	\includegraphics[width=0.29 \linewidth]{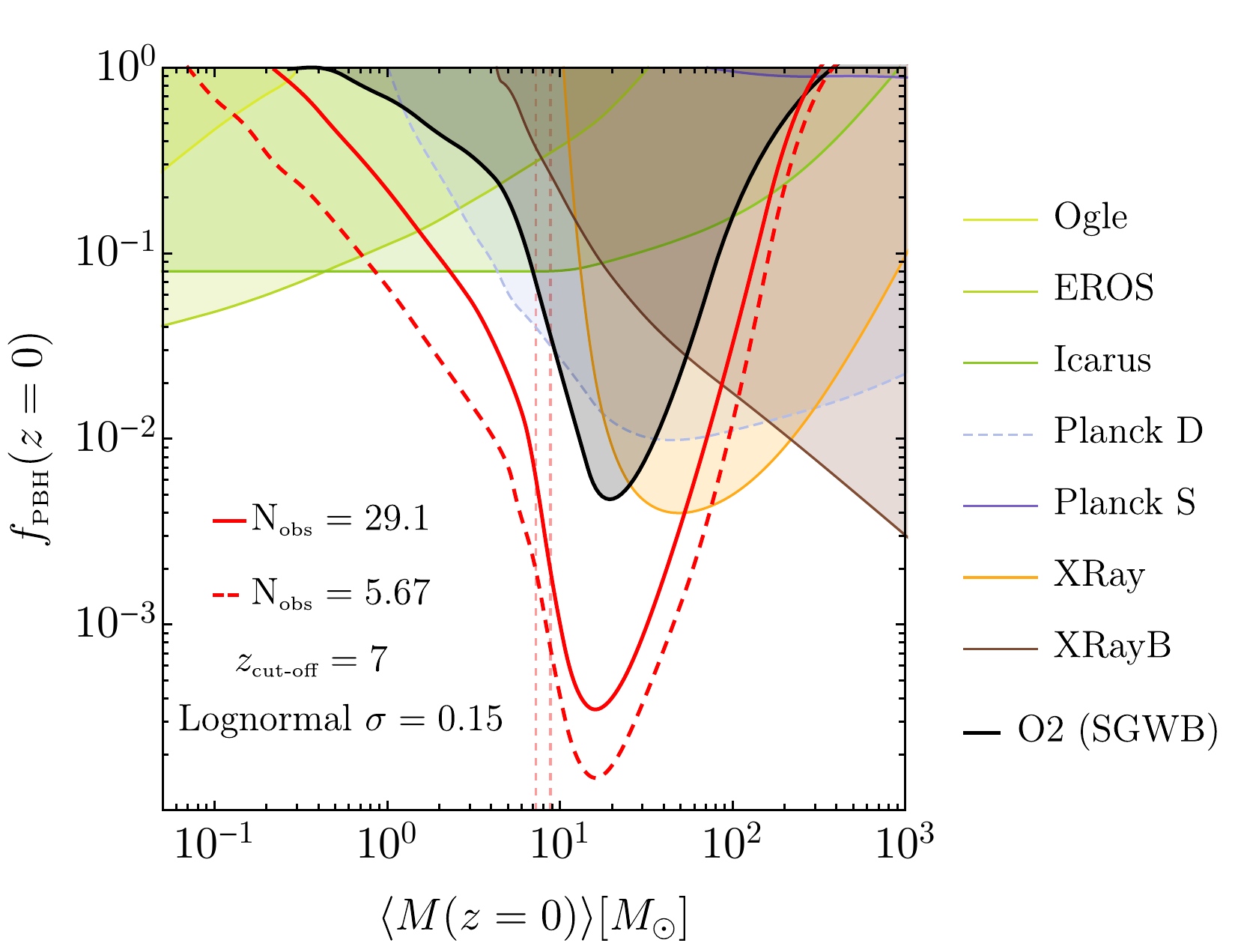}
	\caption{\it Constraints on $f_\PBH$ from experiments not related to GWs (see legend and main text), absence of 
stochastic GW background in LIGO O2 (black-solid), and upper (red-solid indicating the 2$\sigma$ exclusion region) and lower (red-dashed, obtained assuming that all BH merger events detected so far are primordial) bounds coming from the 
observed merger rates. %
As in Figs.~\ref{likelihood} and \ref{massfunction}, the leftmost panels correspond to the case in which accretion is 
negligible, whereas the second to fourth columns correspond to accretion suppressed at $z_\text{\tiny \rm
cut-off}=(15,10,7)$, respectively. 
	}
\label{constraints}
\end{figure}

In Fig.~\ref{constraints} we present the constraints on the PBH abundance as a function of the mass, again 
for the best-fit values obtained from the previous likelihood analysis. In the range of masses of interest for 
LIGO/Virgo, the most important constraints come from lensing, dynamical processes, formation of structures, and 
accretion related phenomena. The lensing bounds include those from supernovae~\cite{Zumalacarregui:2017qqd}, the 
MACHO and EROS experiments~\cite{Alcock:2000kd, Allsman:2000kg}, ICARUS~\cite{Oguri:2017ock} and radio observations, 
such as \cite{Wilkinson:2001vv} and~\cite{Niikura:2019kqi}~(Ogle). 
They all consider lensing sources at low redshift $z \ll z_\co$. Dynamical 
constraints involve disruption of wide binaries~\cite{Quinn:2009zg}, and survival of star clusters in 
Eridanus~II~\cite{Brandt:2016aco} and Segue~I~\cite{Koushiappas:2017chw} at small redshifts. Bounds also arise by 
observations of the Lyman-$\alpha$ forest at redshift before $z \approx 4$~\cite{Murgia:2019duy}. Other 
constraints involve bounds from Planck data on the CMB anisotropies induced by X-rays emitted by spherical or 
disk (Planck~S and Planck~D, respectively)~\cite{Ali-Haimoud:2016mbv,serpico} accretion at high redshifts or bounds on 
the observed 
number of X-ray~(XRay) \cite{Gaggero:2016dpq,Manshanden:2018tze} and X-ray binaries~(XRayB) at low redshifts~\cite{Inoue:2017csr}. 


In Fig.~\ref{constraints} we show a selection of the above constraints, identified by the nickname in parenthesis in 
the list above, and computed as discussed in Ref.~\cite{paper2}. In addition, we show the bounds coming from the absence 
of stochastic GW background in LIGO/Virgo data (black line) and those from the merger rate (red lines), computed as 
discussed in the previous section.  \footnote{Notice that the merger rate may be additionally suppressed by the interaction of the binaries with other PBHs in early sub-structures if $f_\PBH\gsim 0.1$ \cite{raidal,ver}, affecting  the constraint inferred from LIGO/Virgo observations.  We decided to neglect this effect as those values of $f_\PBH$ are ruled out by other constraints.}

The red dashed and continuous lines correspond to the $2\sigma$ values for the expected number of events. 
In other words, the red continuous line corresponds to the {\it upper} bound from the observed merger 
rate, since larger values of $f_\PBH(z=0)$ would yield a merger rate higher than observed at a given mass.
On the other hand, the red dashed line corresponds to a {\it lower} bound on $f_\PBH(z=0)$, assuming all the observed 
events are of primordial origin, since smaller values of $f_\PBH(z=0)$ would yield a merger rate lower than observed at 
a given mass.  Clearly, this lower bound can be made less stringent by assuming that only a fraction of 
events is of primordial origin. 
The vertical dashed red lines indicate the $2\sigma$ interval around the best-fit value of the mass parameter, as 
obtained by the likelihood analysis.

When accretion gets stronger (for masses around $\approx 10 M_\odot$), we observe two main effects: 
\begin{itemize}
 \item[i)] In the leftmost part of the red curve (i.e. on the left of the minimum) the GW bounds on $f_\PBH$ become 
{\it more stringent}, since accretion enhances the merger rate in that region. The opposite is true on the right of the 
minimum, because in that region accretion pushes the masses to large values, outside the optimal sensitivity range of 
the detectors.
 \item[ii)] the non-GW bounds become {\it weaker}, due to the broadening of the mass function and the evolution of 
$f_\PBH(z)$. The interested reader can find a more detailed discussion of this phenomenon in Ref.~\cite{paper2}.
\end{itemize}
The net result is that, if accretion is negligible, in the range of the mass-function parameters selected by the 
likelihood analysis, non-GW constraints (in particular Planck~D) would already marginally exclude the possibility that 
all BH merger events detected so far are of primordial origin. However, when accretion is significant 
the opposite is true: LIGO/Virgo constraints are the most stringent ones in the relevant mass range.
Overall, the upper bound on the PBH abundance coming from LIGO/Virgo rates is $f_\PBH(z)\lesssim {\rm few}\times 
10^{-3}$ in the relevant mass range, with the upper bound becoming more stringent in the case of strong accretion.
Moreover, in the relevant mass range existing constraints seem to exclude the possibility to detect the GW 
stochastic background from PBH mergers, because detecting the latter would require a PBH abundance which is already 
excluded.

\subsection{Confrontation of the predicted distributions of the binary parameters with observations}
\begin{figure}[t!]
	\centering
	\includegraphics[width=0.24 \linewidth]{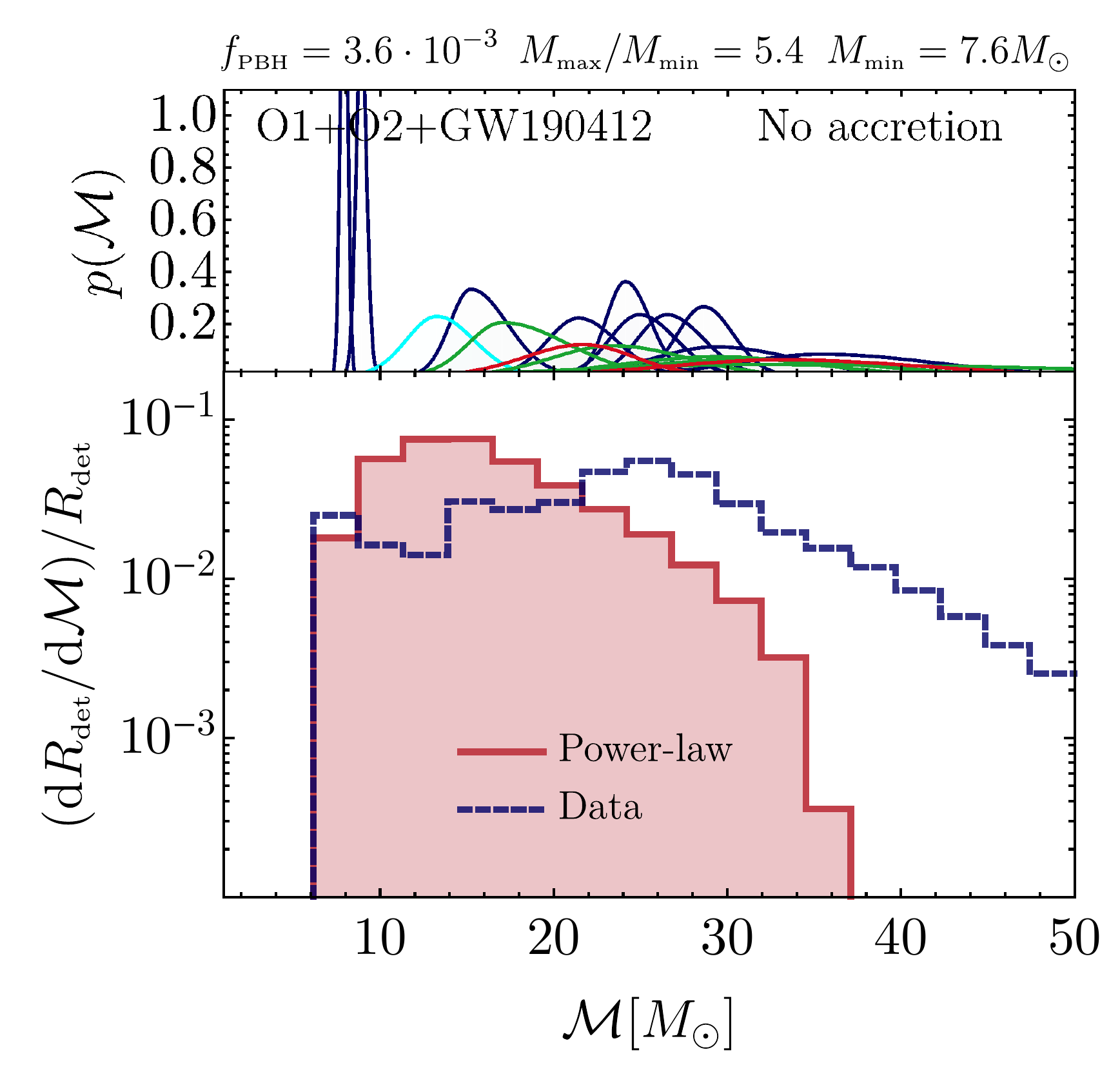}
	\includegraphics[width=0.24 \linewidth]{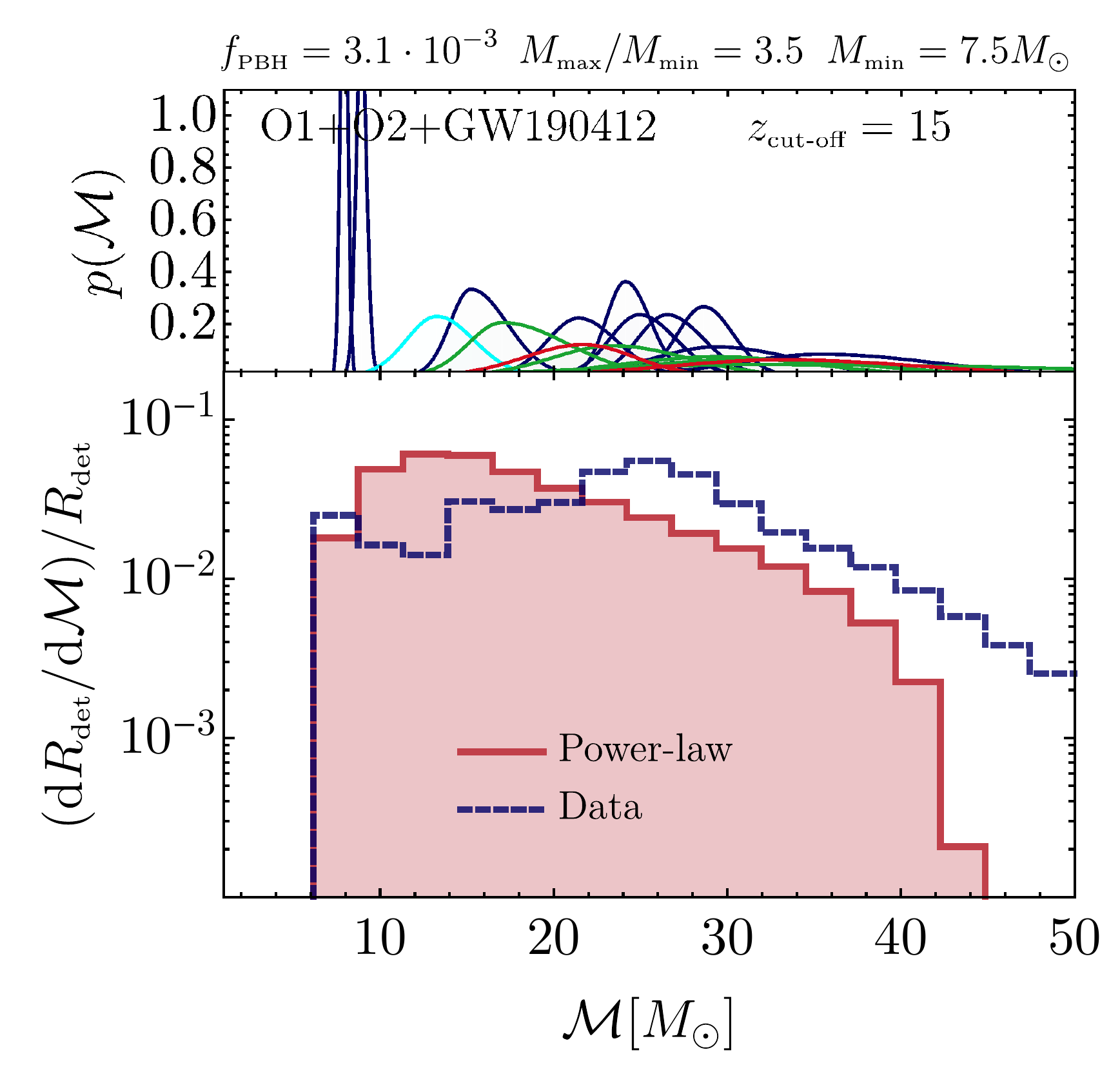}
	\includegraphics[width=0.24 \linewidth]{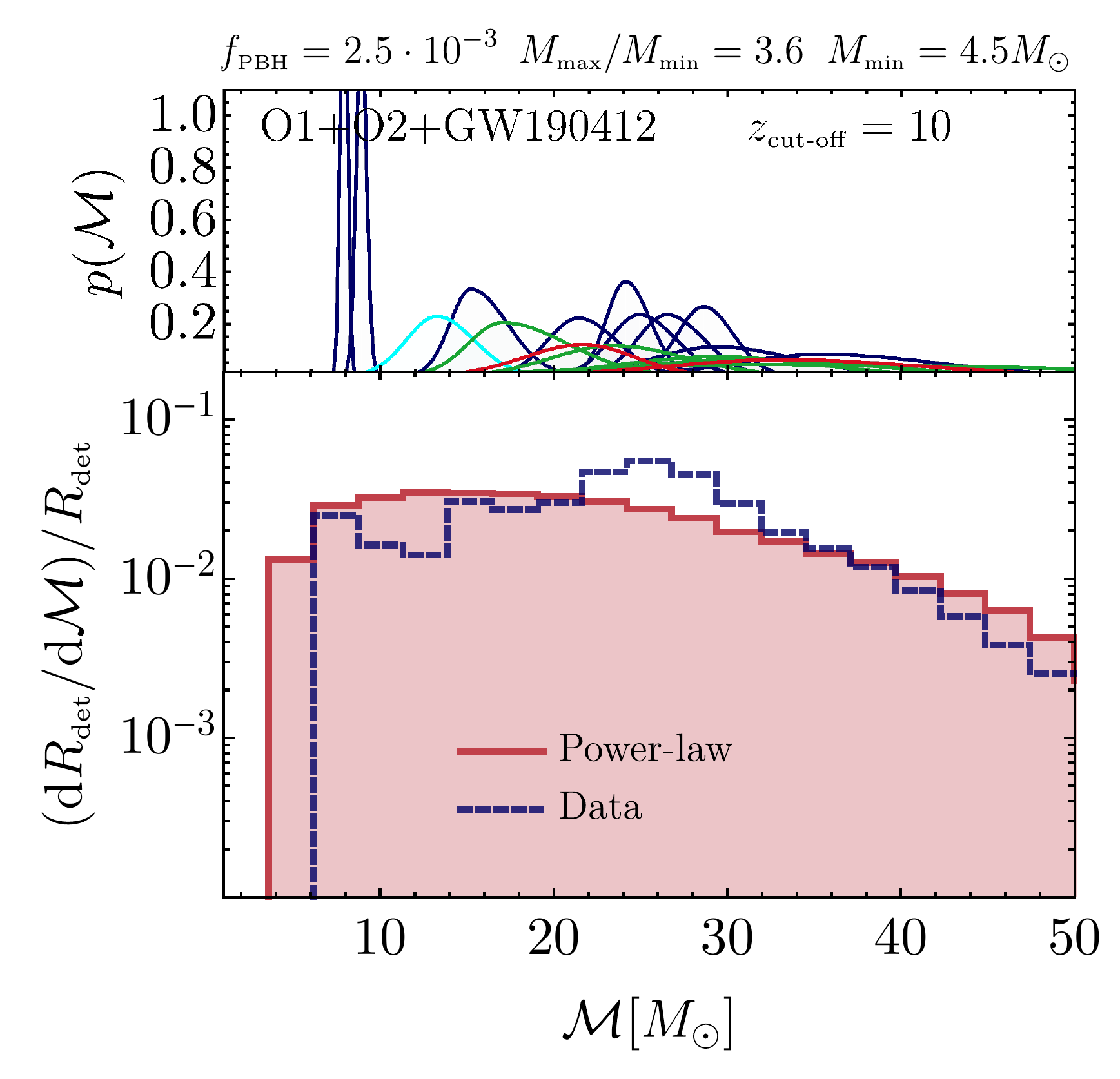}
	\includegraphics[width=0.24 \linewidth]{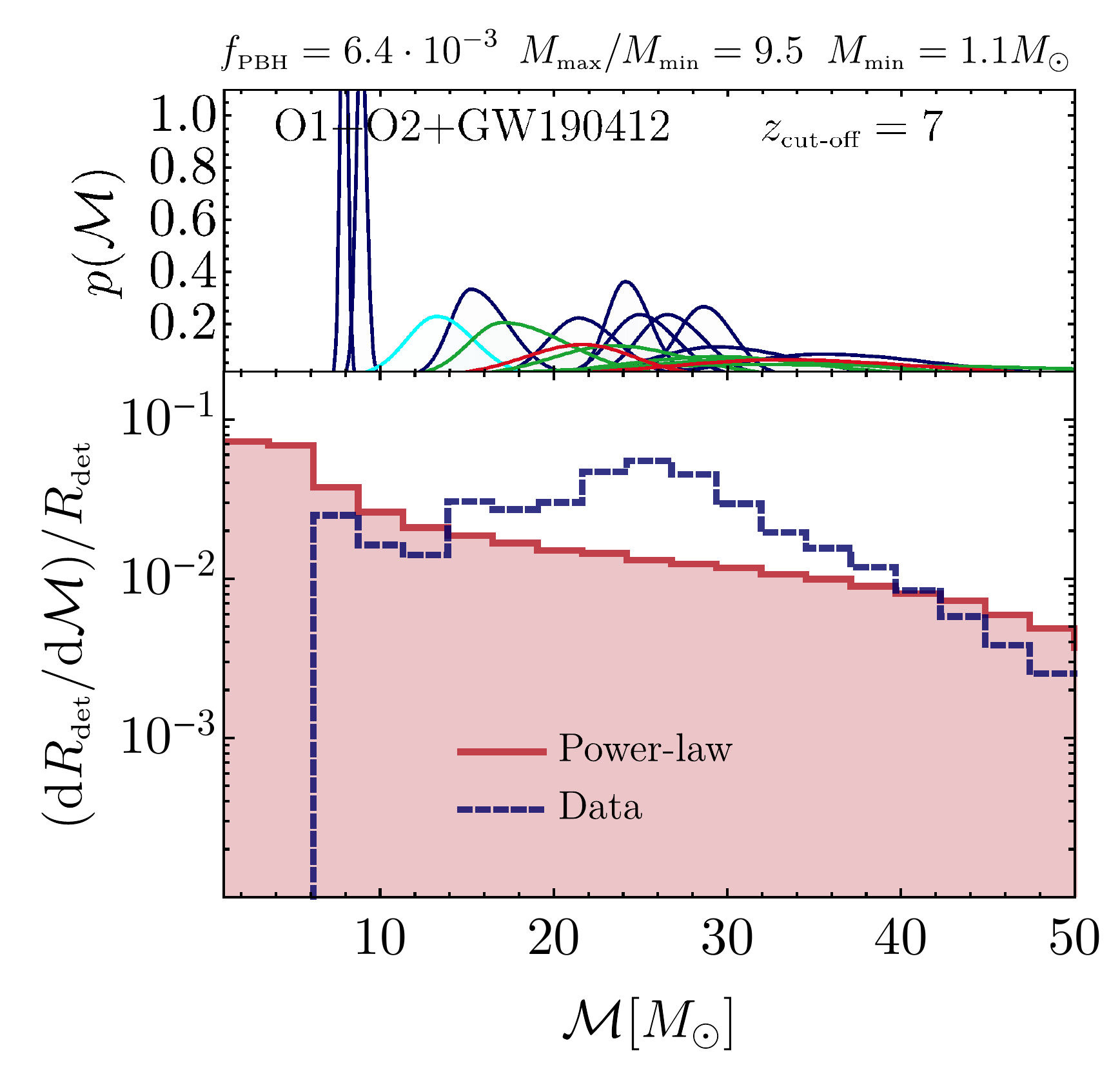}
	
	\includegraphics[width=0.24 \linewidth]{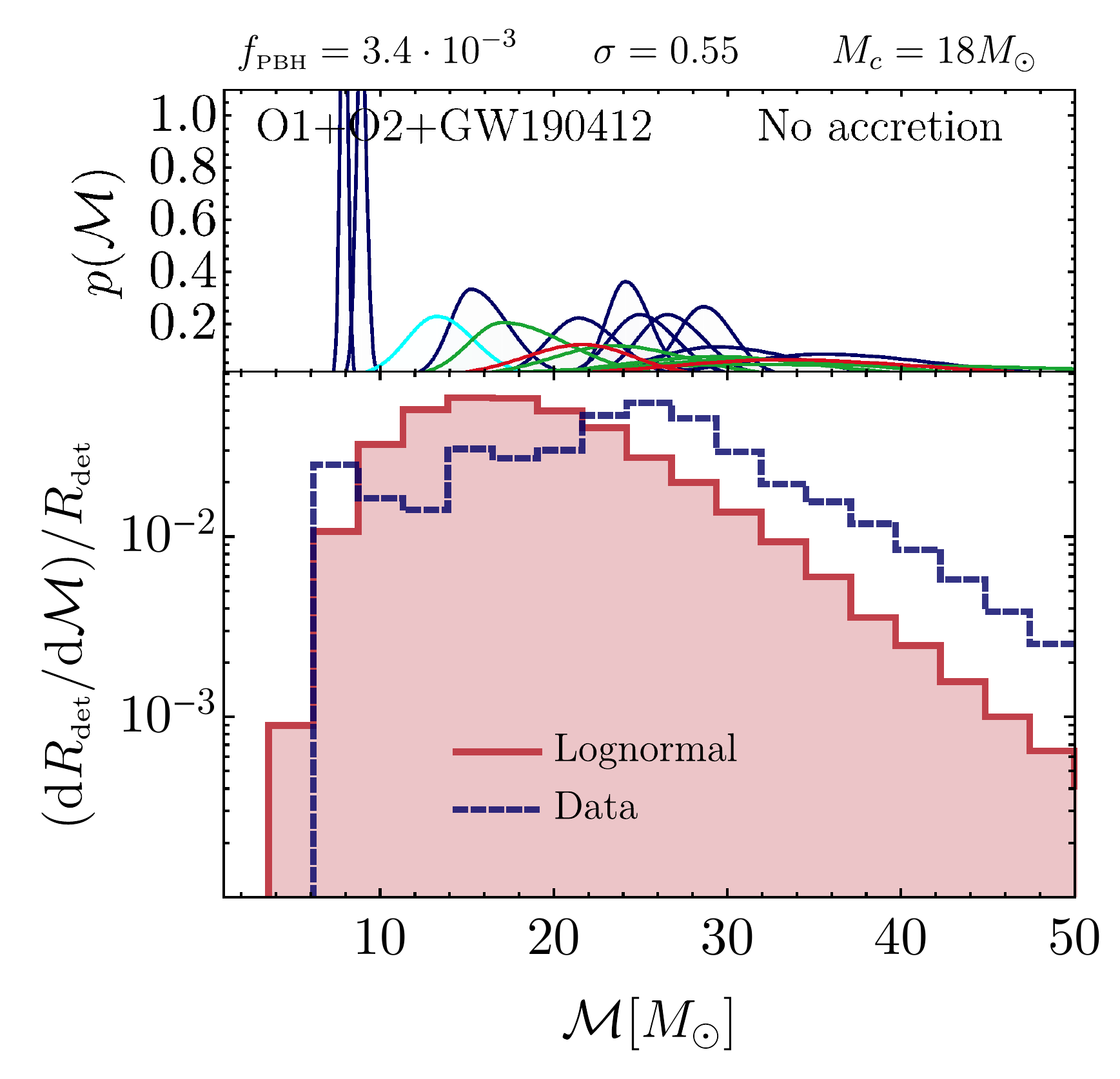}
	\includegraphics[width=0.24 \linewidth]{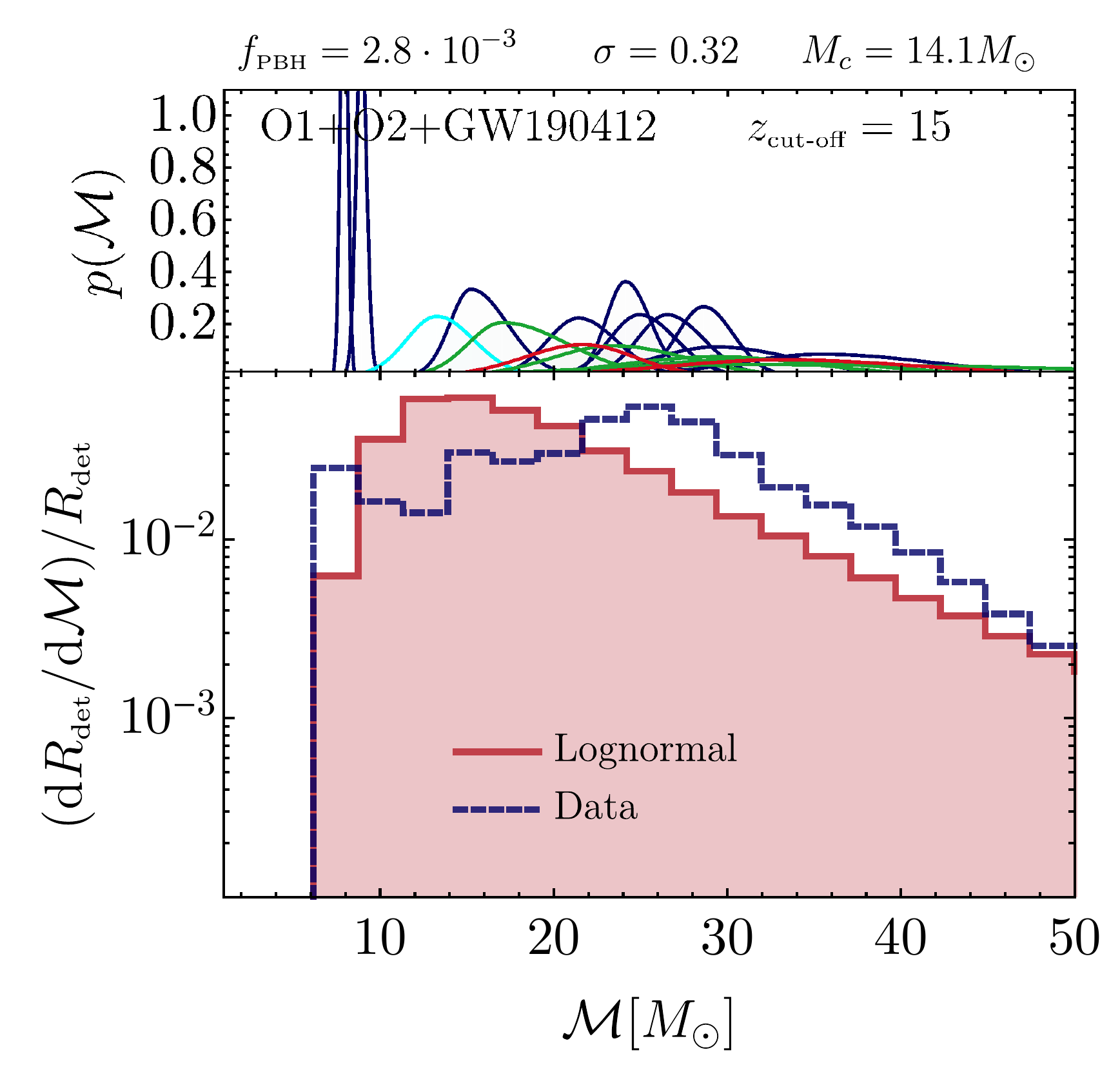}
	\includegraphics[width=0.24 \linewidth]{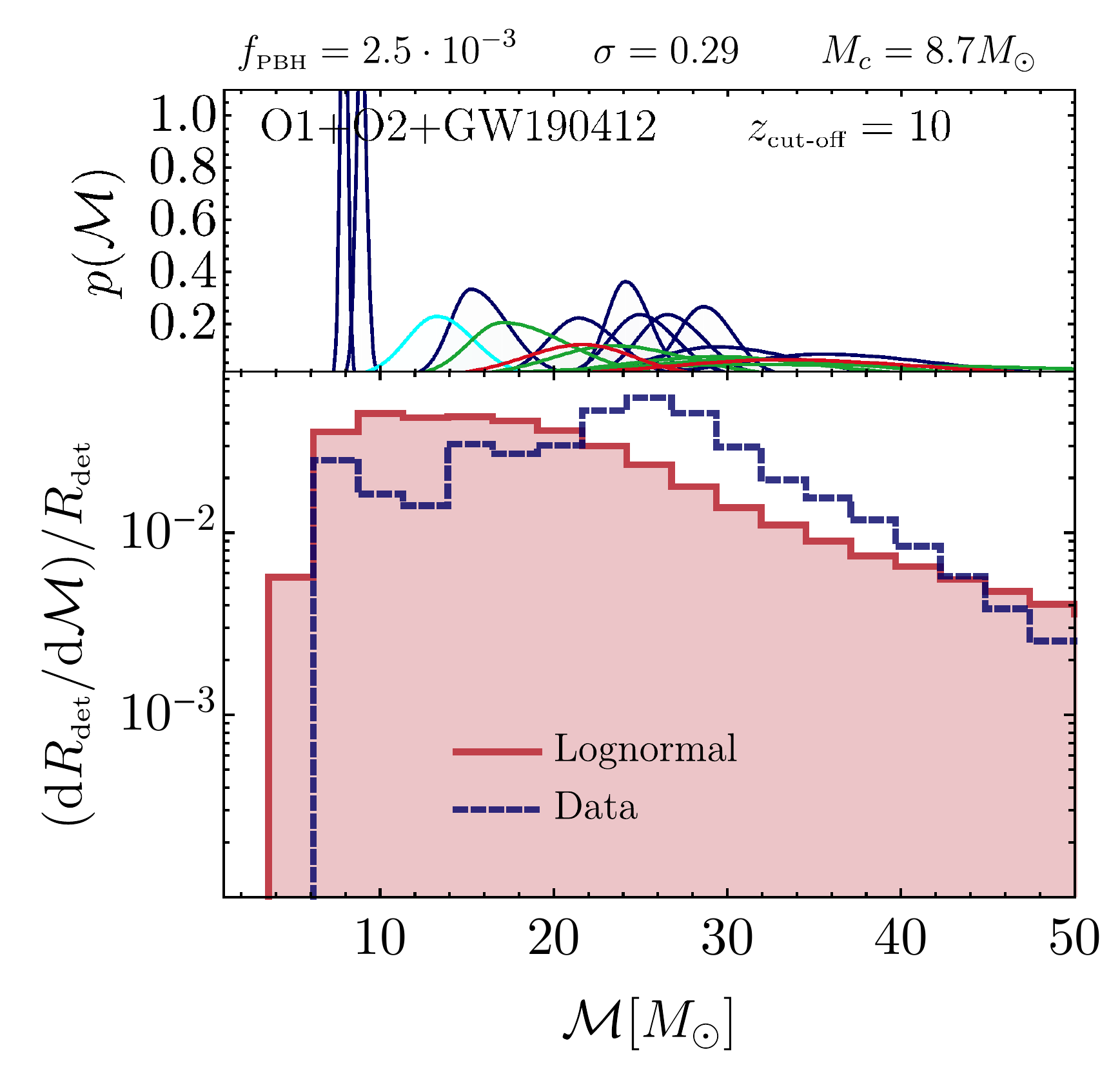}
	\includegraphics[width=0.24 \linewidth]{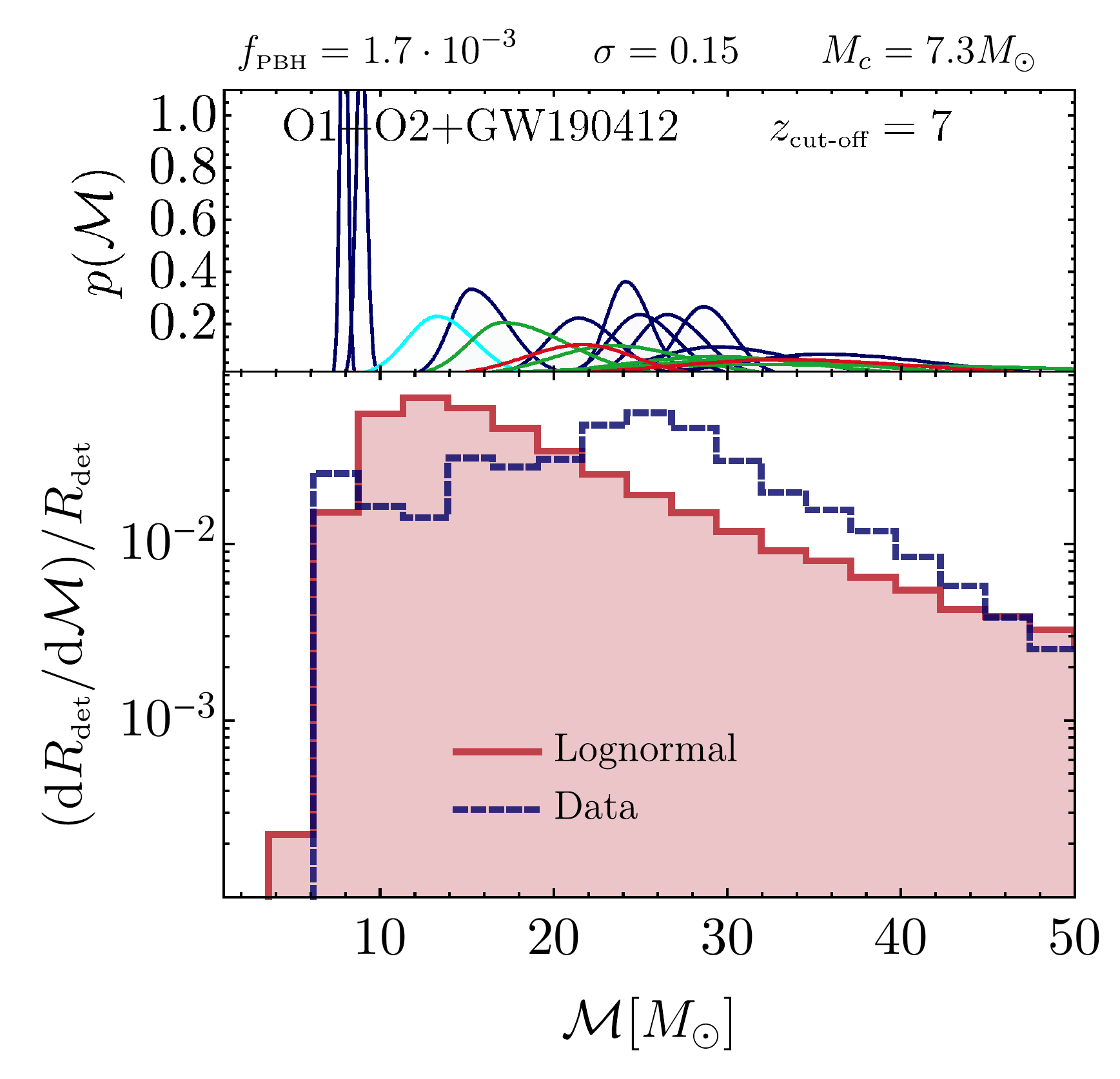}
	\caption{\it Observable distribution of chirp mass ${\cal M}$ (red histogram) compared to the data available 
(blue histogram). The organization of the panels is the same as in Fig.~\ref{constraints}. 
	}
\label{chirp}
\end{figure}

In Fig.~\ref{chirp} we show the observable distribution of the chirp mass ${\cal M}\equiv \mu^{3/5}M_{\text{\tiny 
tot}}^{2/5}$ obtained using the the best-fit values calculated from the previous likelihood analysis and compared to 
LIGO/Virgo data. 
We have plotted the differential $R_{\text{\tiny det}}$, i.e. the number of events in the chirp mass bin normalised with respect to the total number of events (red 
lines). The blue histograms indicate the distribution inferred from the data. They have been obtained by plotting, in each chirp mass bin, the integral of the posterior probability
summed over all measurements in that bin. 
On the top of the figures we have shown 
the individual posterior probability for each measured event. One can appreciate that the larger the accretion, the 
wider the distribution becomes because the high-mass tail gets more spread. 
For the power-law mass function, accretion shifts the peak of 
the distribution to smaller masses because the likelihood prefers smaller $M_{\text{\tiny min}}$.
For the same reason, for very strong accretion the predicted distribution has also support for small chirp masses. For 
the extreme case $z_\co=7$ and a power-law distribution, the predicted merger rate at small chirp mass seems in tension 
with current observations.

\begin{figure}[t!]
	\centering
	\includegraphics[width=0.24 \linewidth]{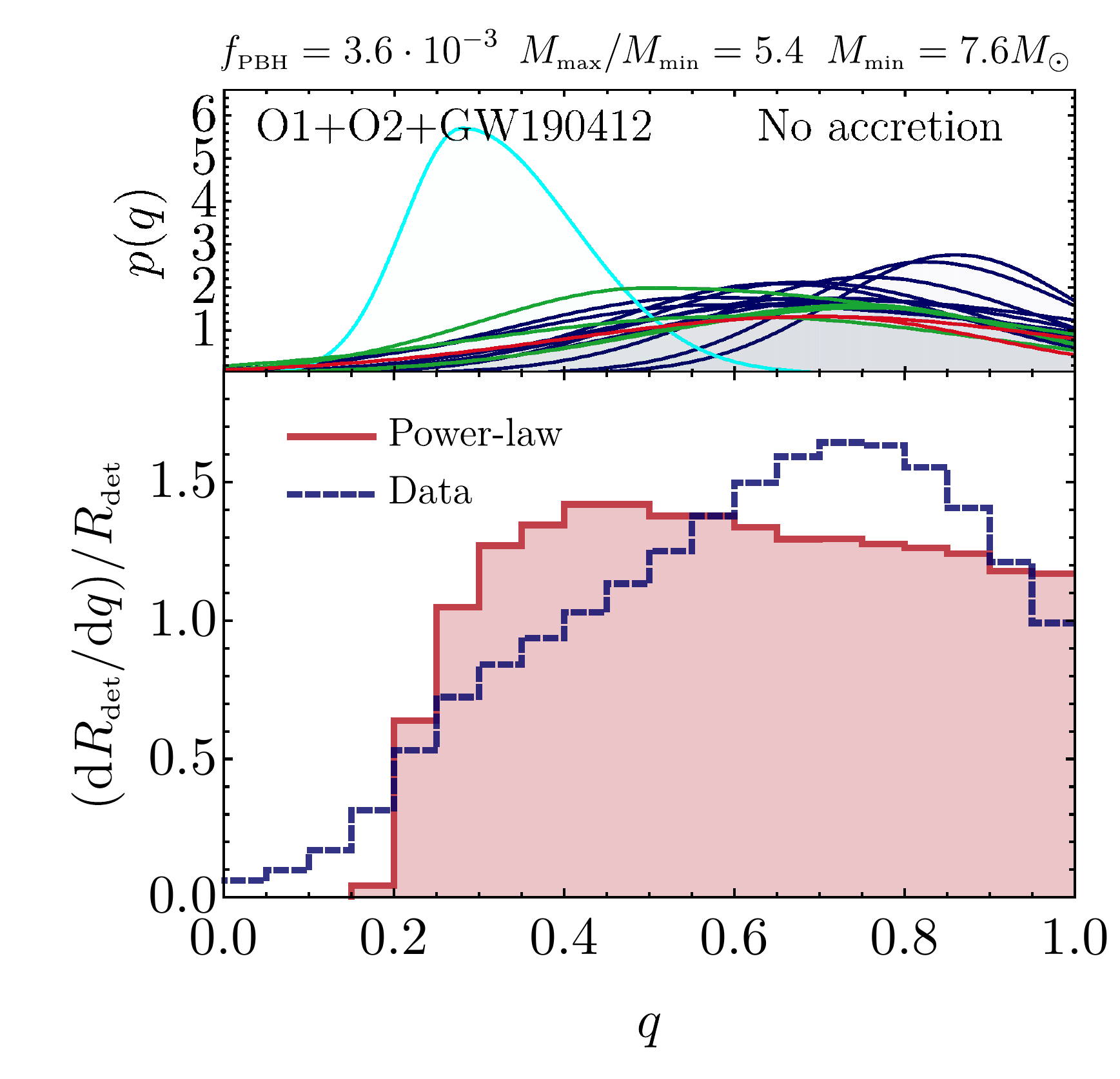}
	\includegraphics[width=0.24 \linewidth]{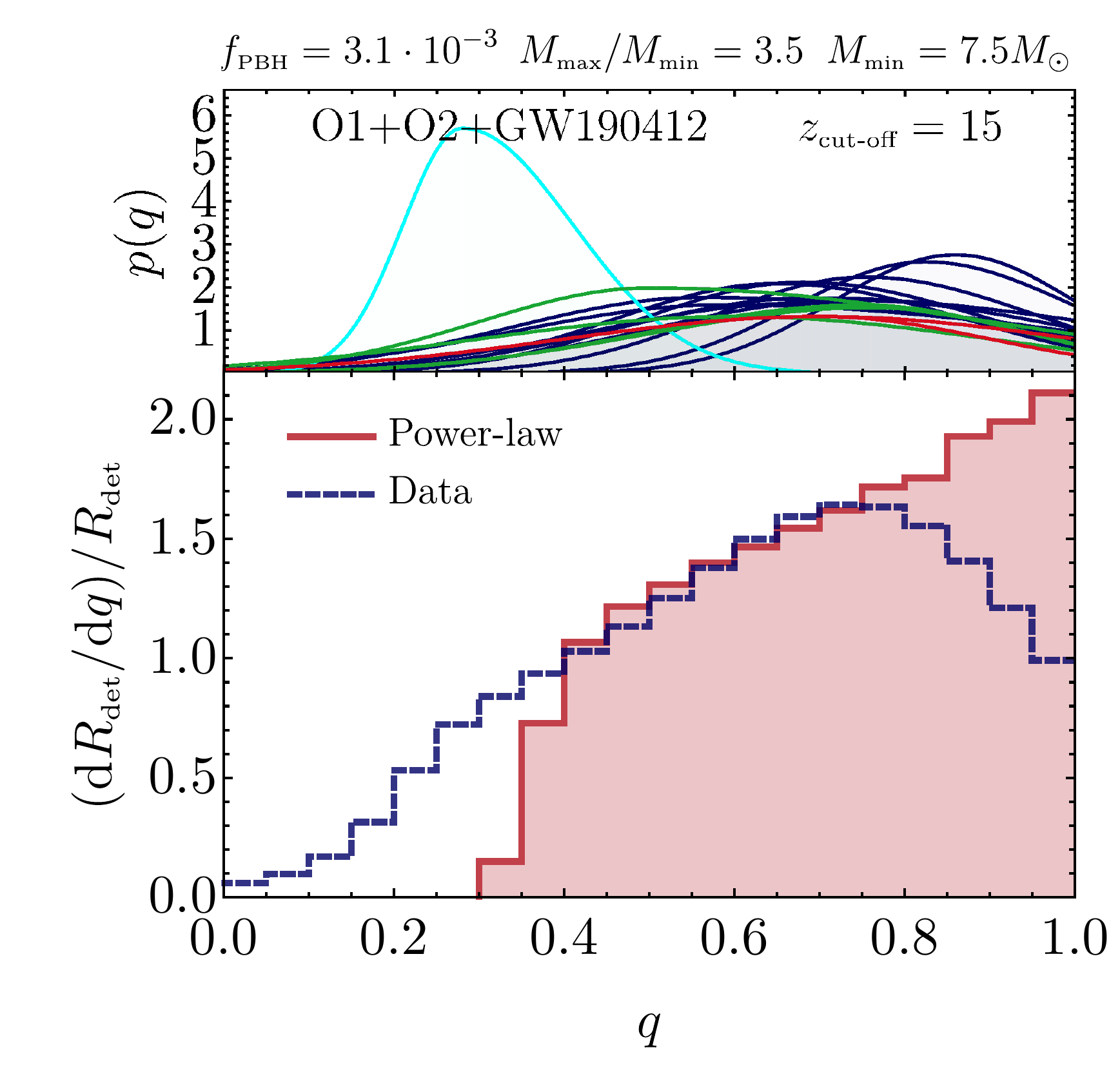}
	\includegraphics[width=0.24 \linewidth]{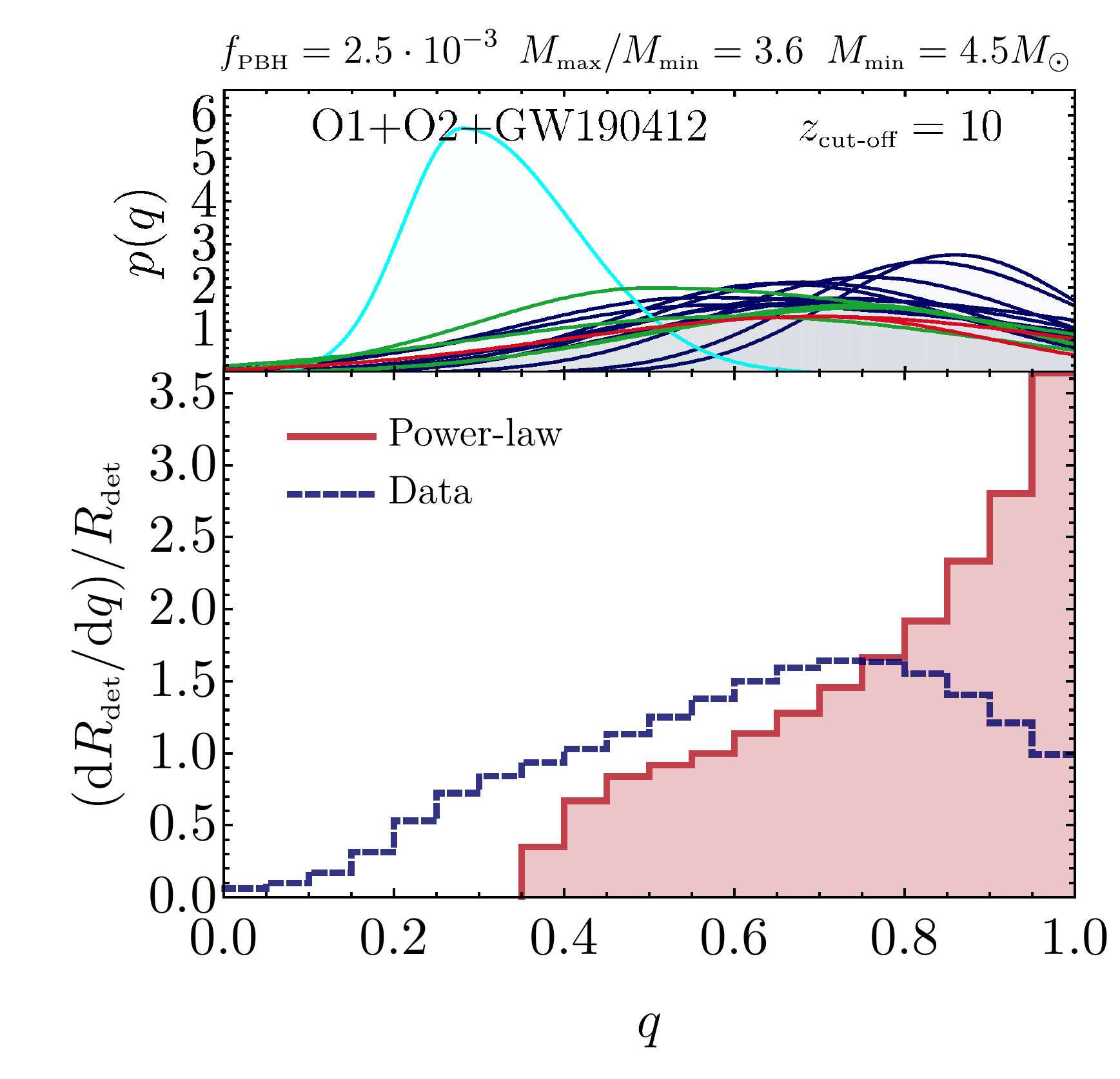}
	\includegraphics[width=0.24 \linewidth]{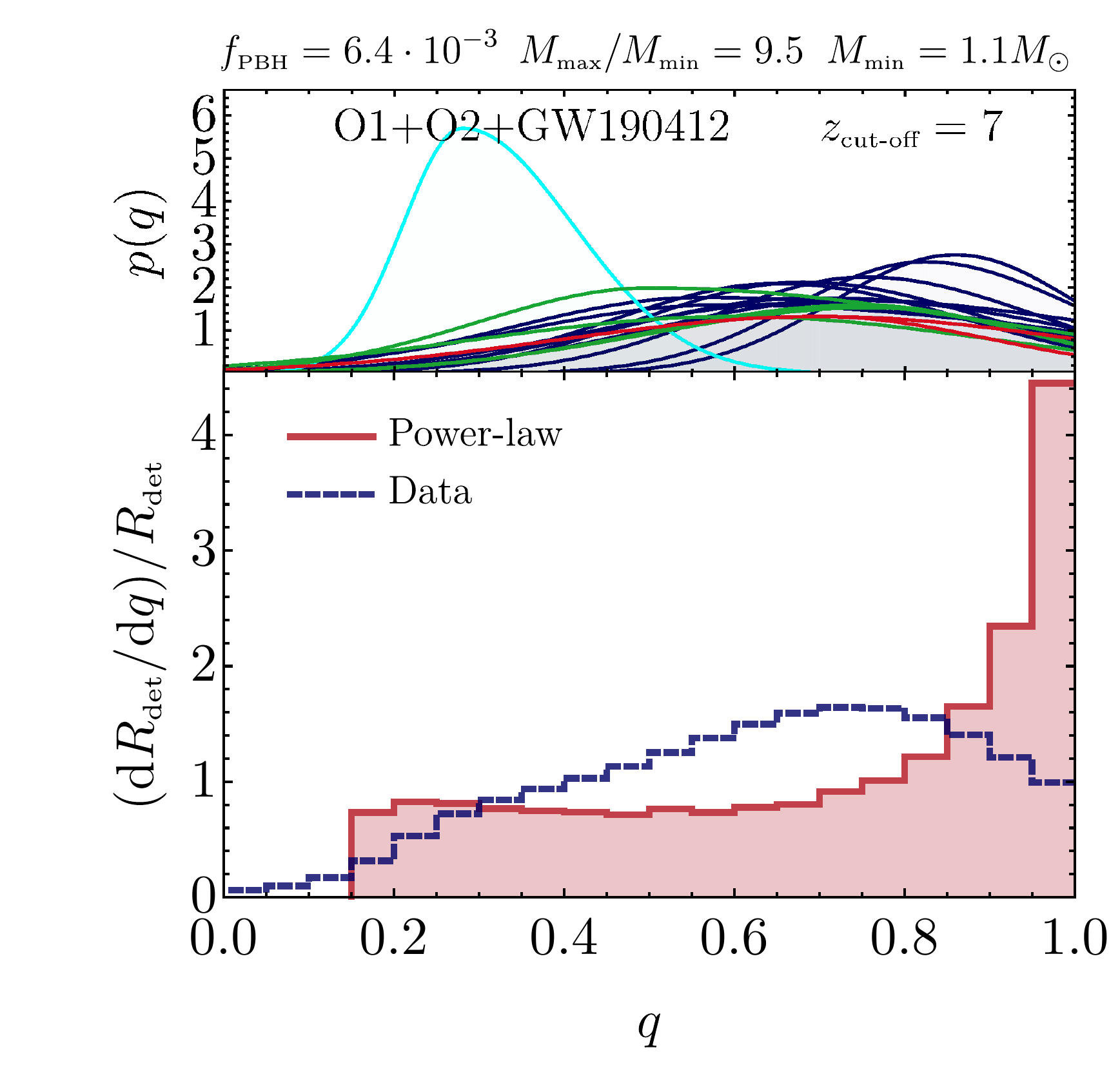}
	
	\includegraphics[width=0.24 \linewidth]{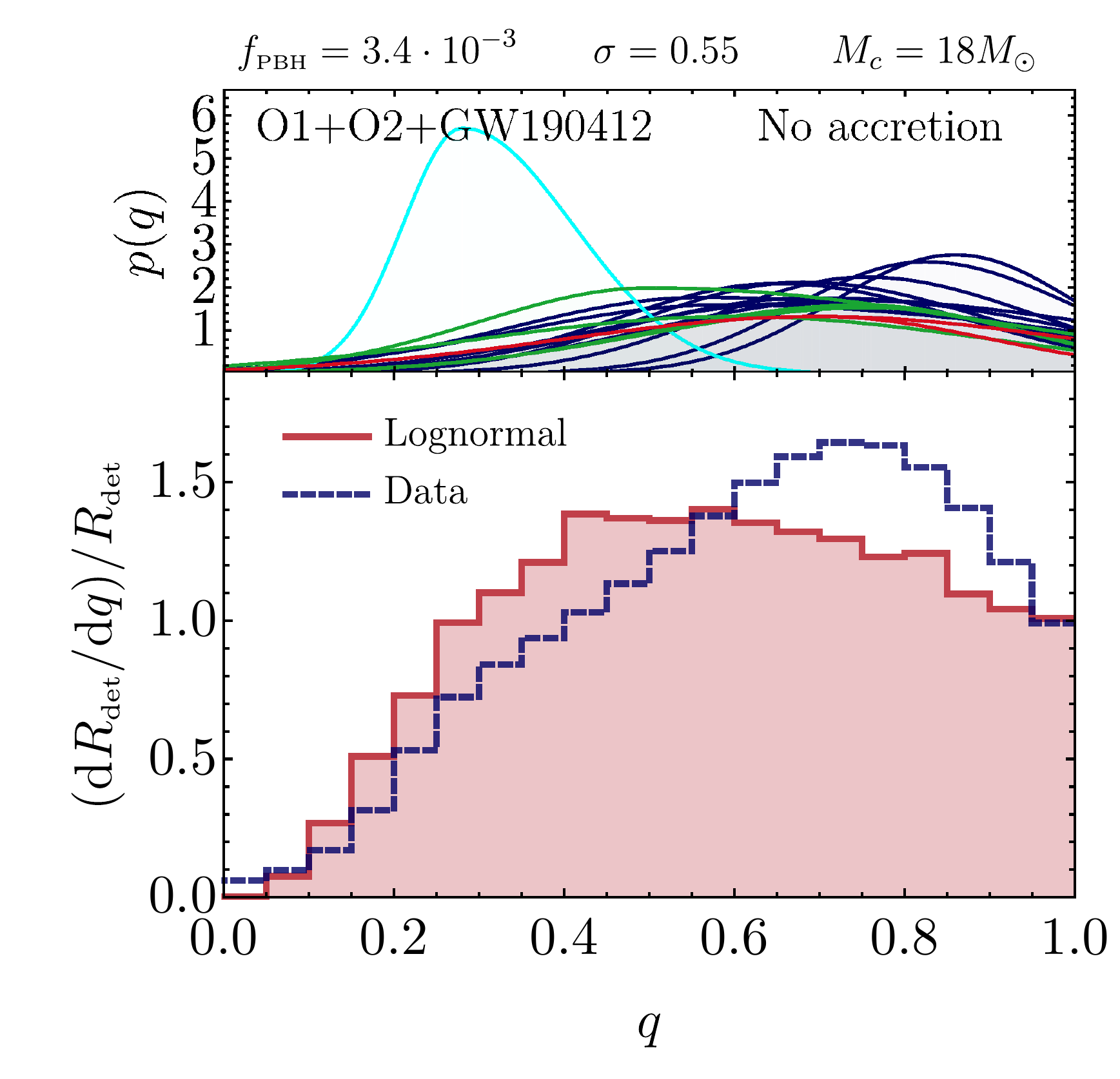}
	\includegraphics[width=0.24 \linewidth]{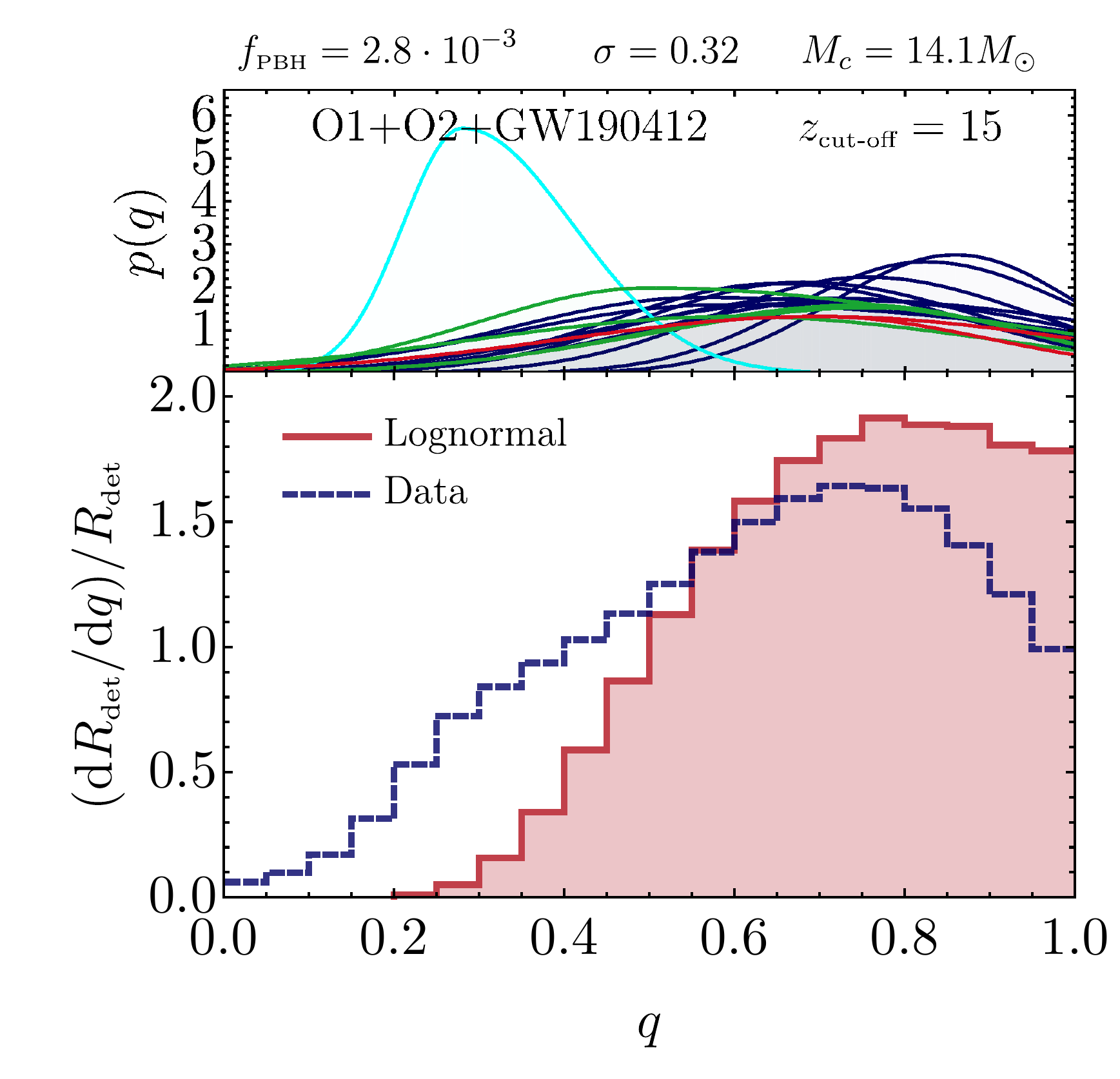}
	\includegraphics[width=0.24 \linewidth]{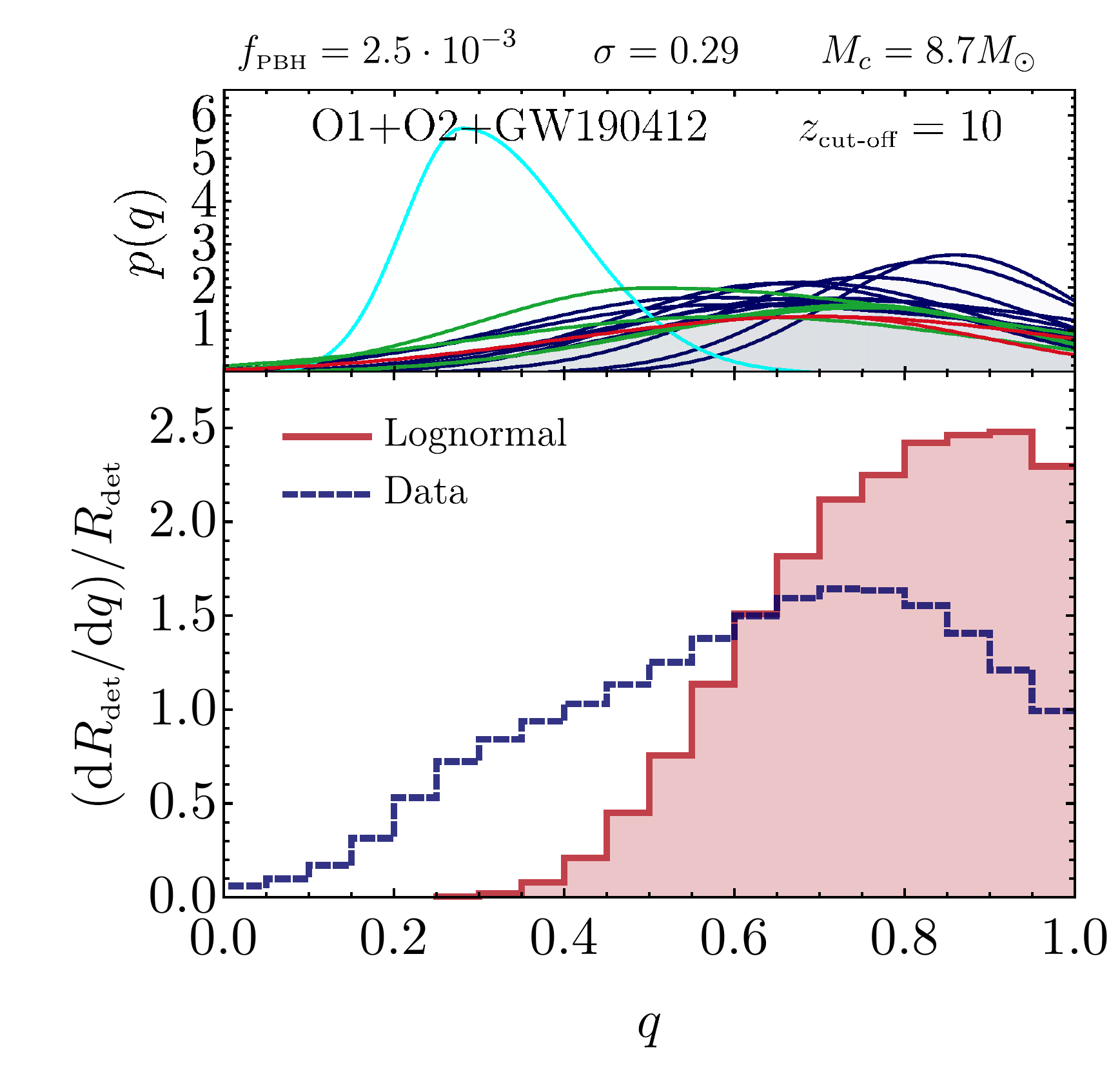}
	\includegraphics[width=0.24 \linewidth]{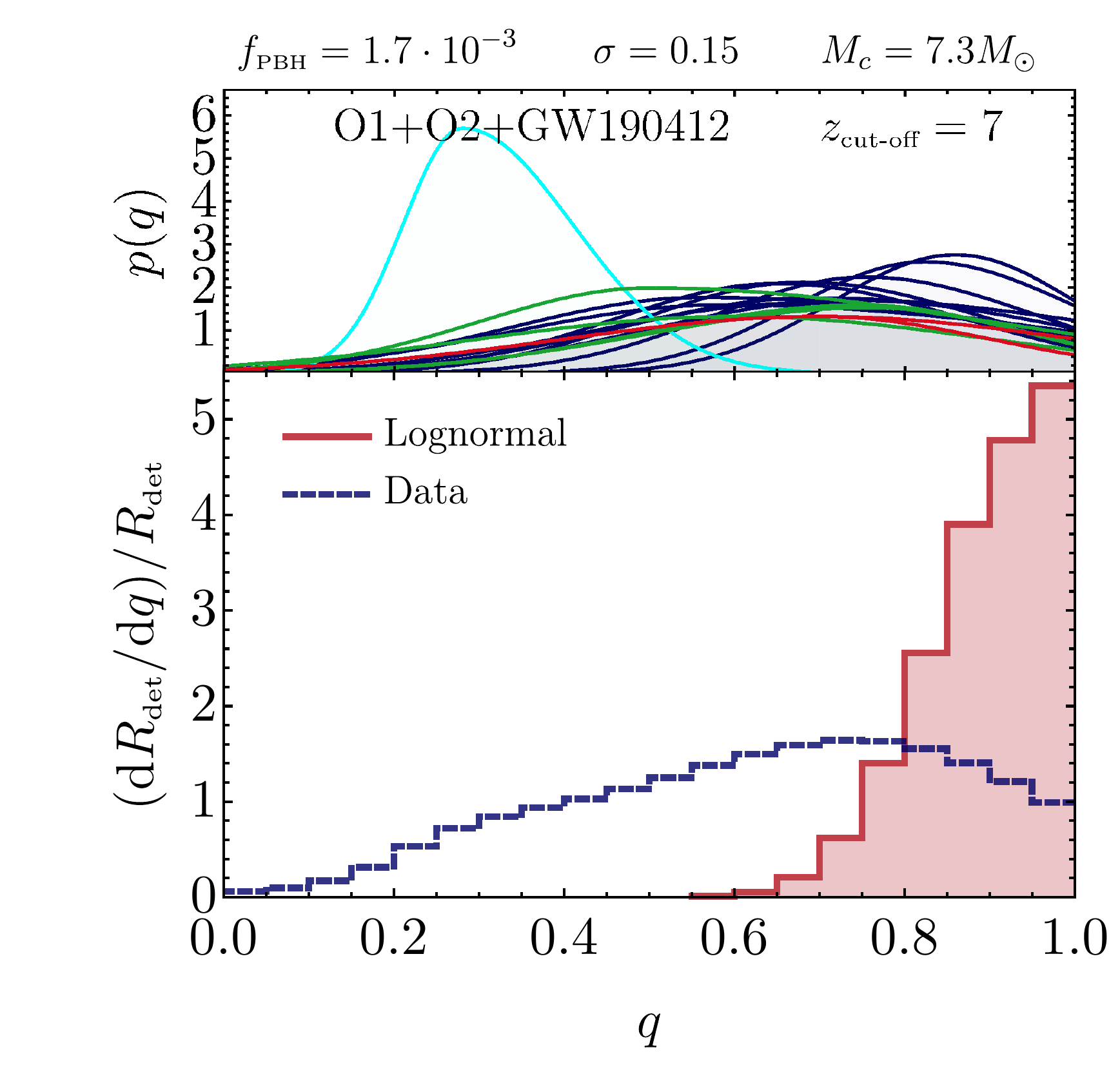}
	\caption{\it Same as Fig.~\ref{chirp}, but for the observable distribution of mass ratio $q$ compared to the 
data available.
	}
\label{q}
\end{figure}

Fig.~\ref{q} is analogous to Fig.~\ref{chirp} but for the observable distribution of the mass ratio 
parameter $q$.  Again, the histograms  (blue lines) inferred from the data  have been obtained by summing up, in each $q$ bin, the data  weighted by the corresponding posterior probability.
 
 Here we notice that the distributions, when accretion is included, move towards 
$q=1$. This is predicted by the the discussion in Sec.~\ref{sec:summary}, where we showed 
that $q=1$ is a fixed-point of the binary evolution. 

Current observed rates have a peak at about $q\approx0.7$. However, it is important to note that current errors on the 
mass ratio (especially for LIGO/Virgo O1 and O2 events~\cite{LIGOScientific:2018mvr}) are quite large. In fact, all O1 and O2 
events are compatible with $q\approx1$; GW190412 is the only BH merger detected by LIGO/Virgo to date which has a mass 
ratio significantly different from unity, as also shown in the posterior distributions on the small top panels of 
Fig.~\ref{q}. Overall, one might expect that the distributions of $q$ will change significantly as 
new events in O3 become available.
Nonetheless, a general qualitative result can be drawn: extreme accretion scenarios (like 
$z_\co\lesssim7$) are in tension with GW190412, since they would predict that the vast majority of events should have 
$q\sim1$. Furthermore, the cases of a power-law and a lognormal mass functions with $z_\co\simeq 10$, see Fig. \ref{chirp}, which seemed to be in good agreement as far as the chirp mass distribution is concerned, seem to be in tension with the corresponding mass-ratio distributions. This is of course only  a preliminary  result which should wait for
confirmation or disproval when the new flow of data will be available.

\begin{figure}[t!]
	\centering
	\includegraphics[width=0.23 \linewidth]{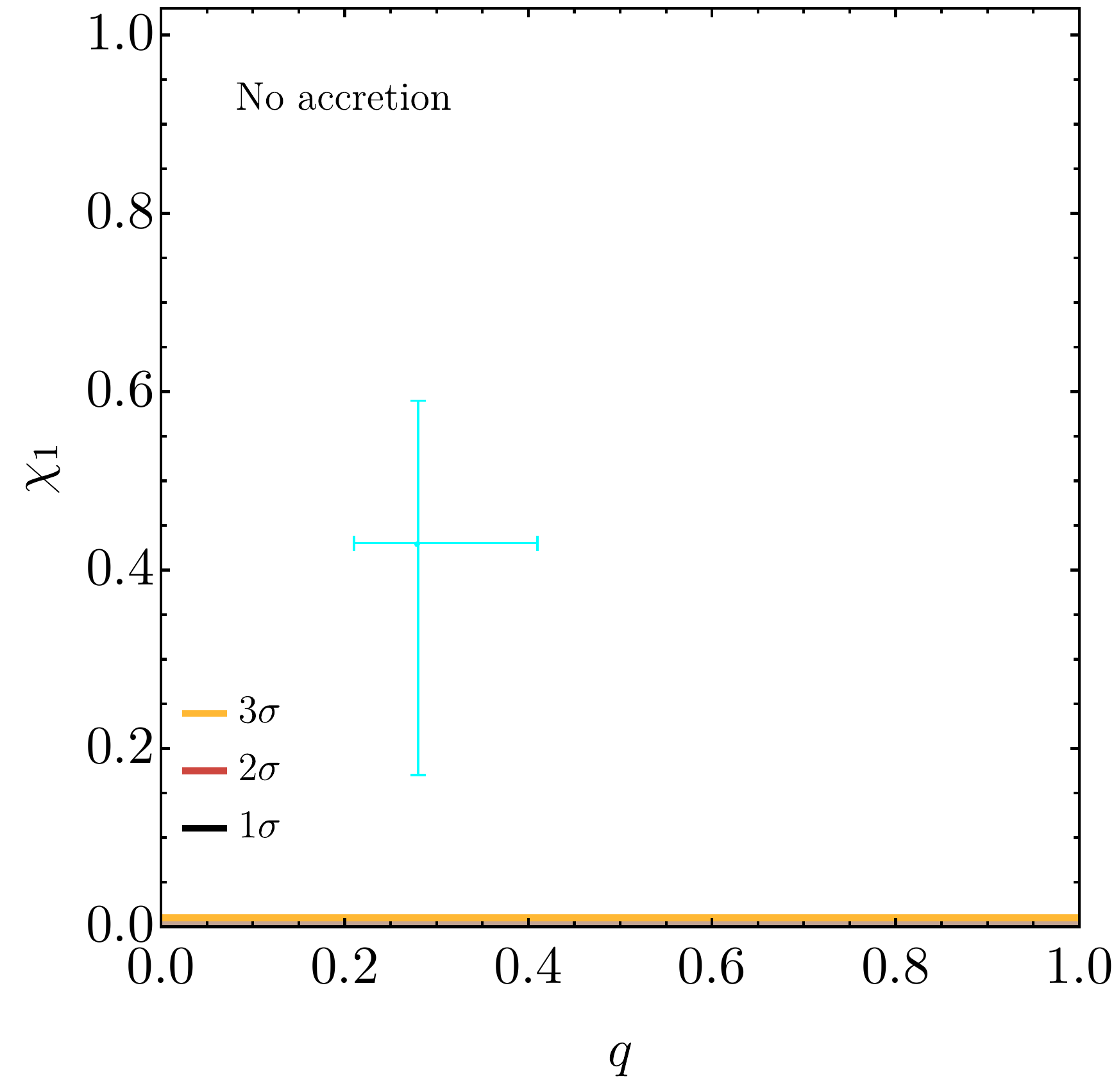}
	\includegraphics[width=0.24 \linewidth]{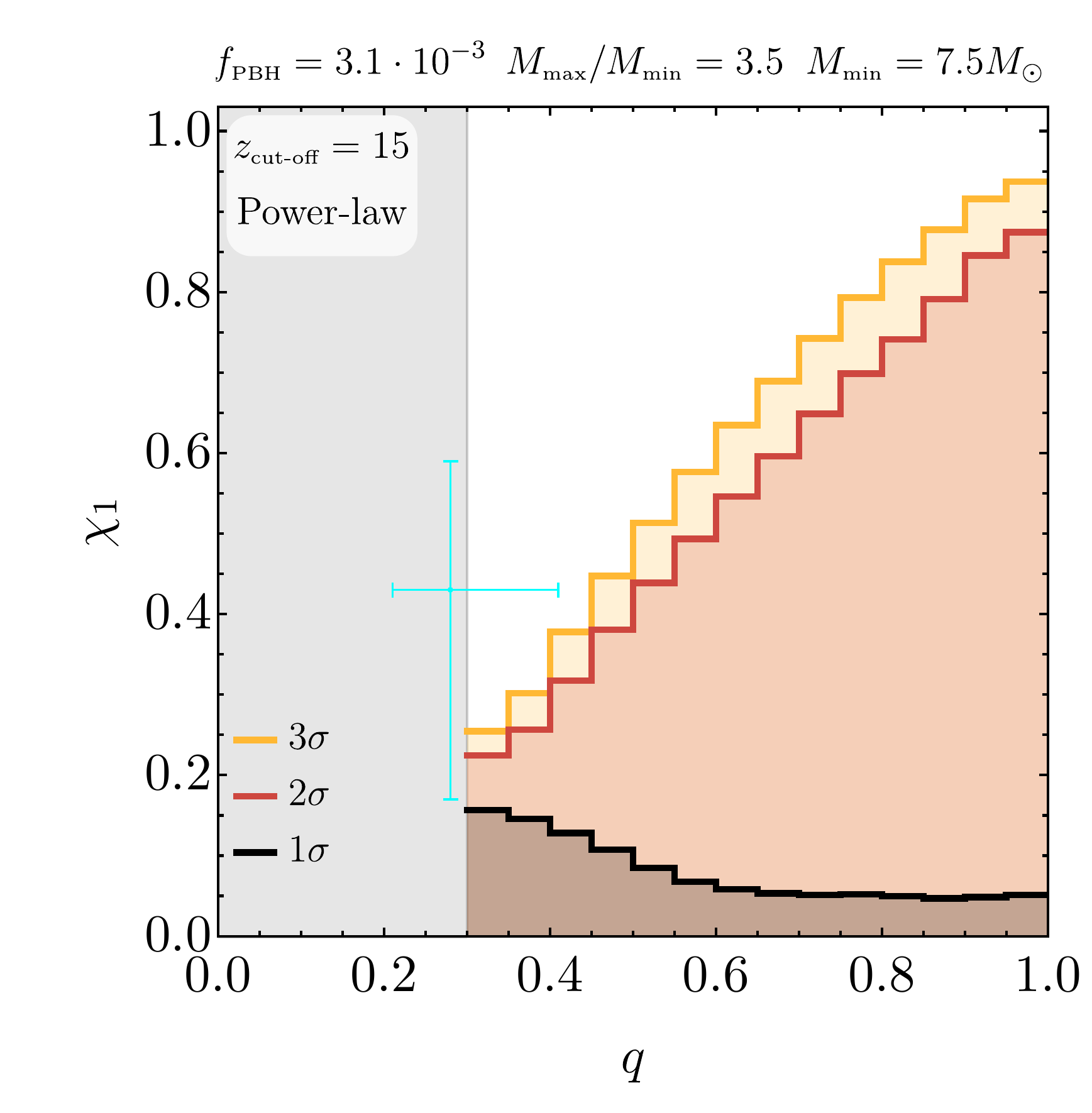}
	\includegraphics[width=0.24 \linewidth]{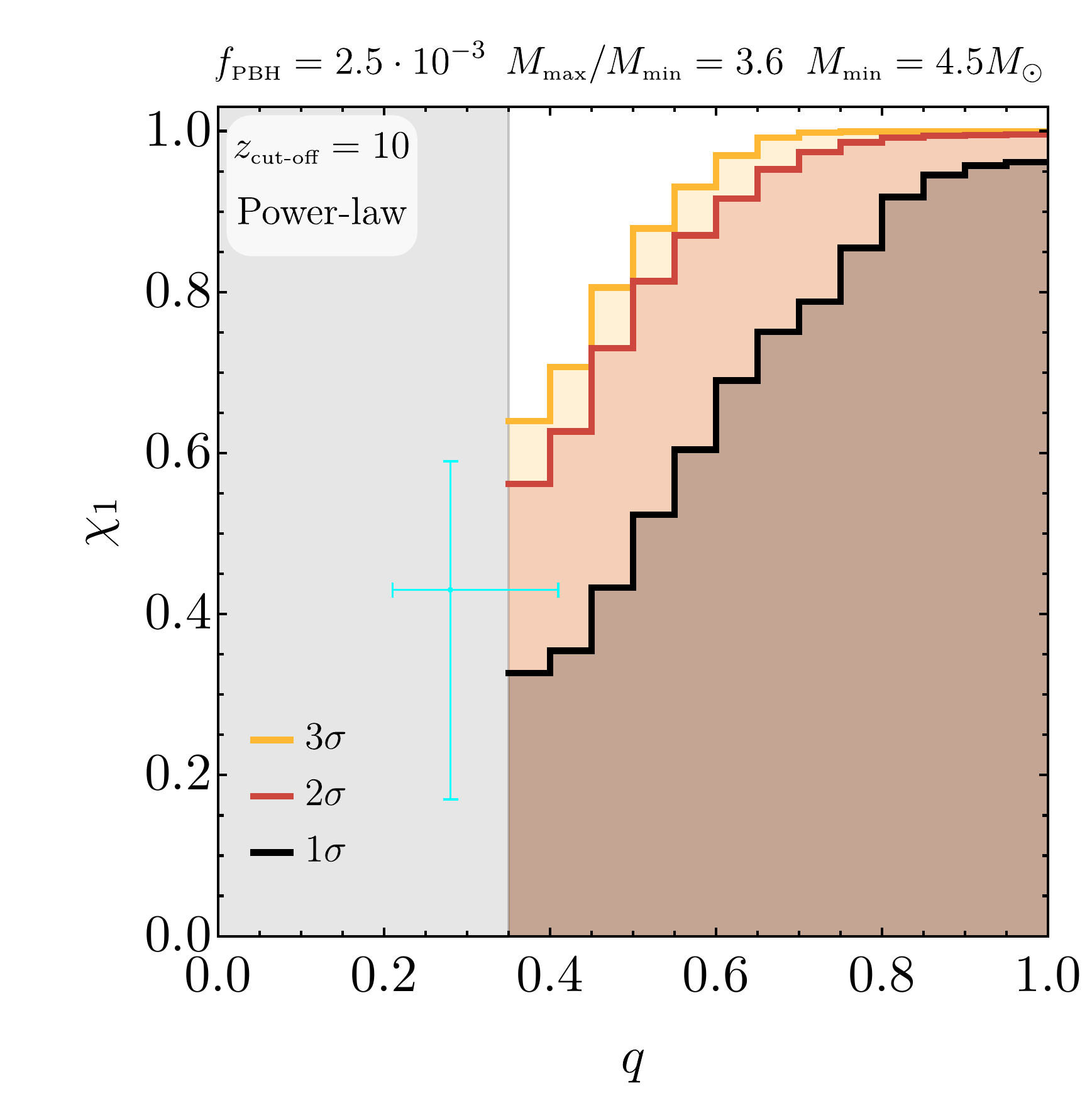}
	\includegraphics[width=0.24 \linewidth]{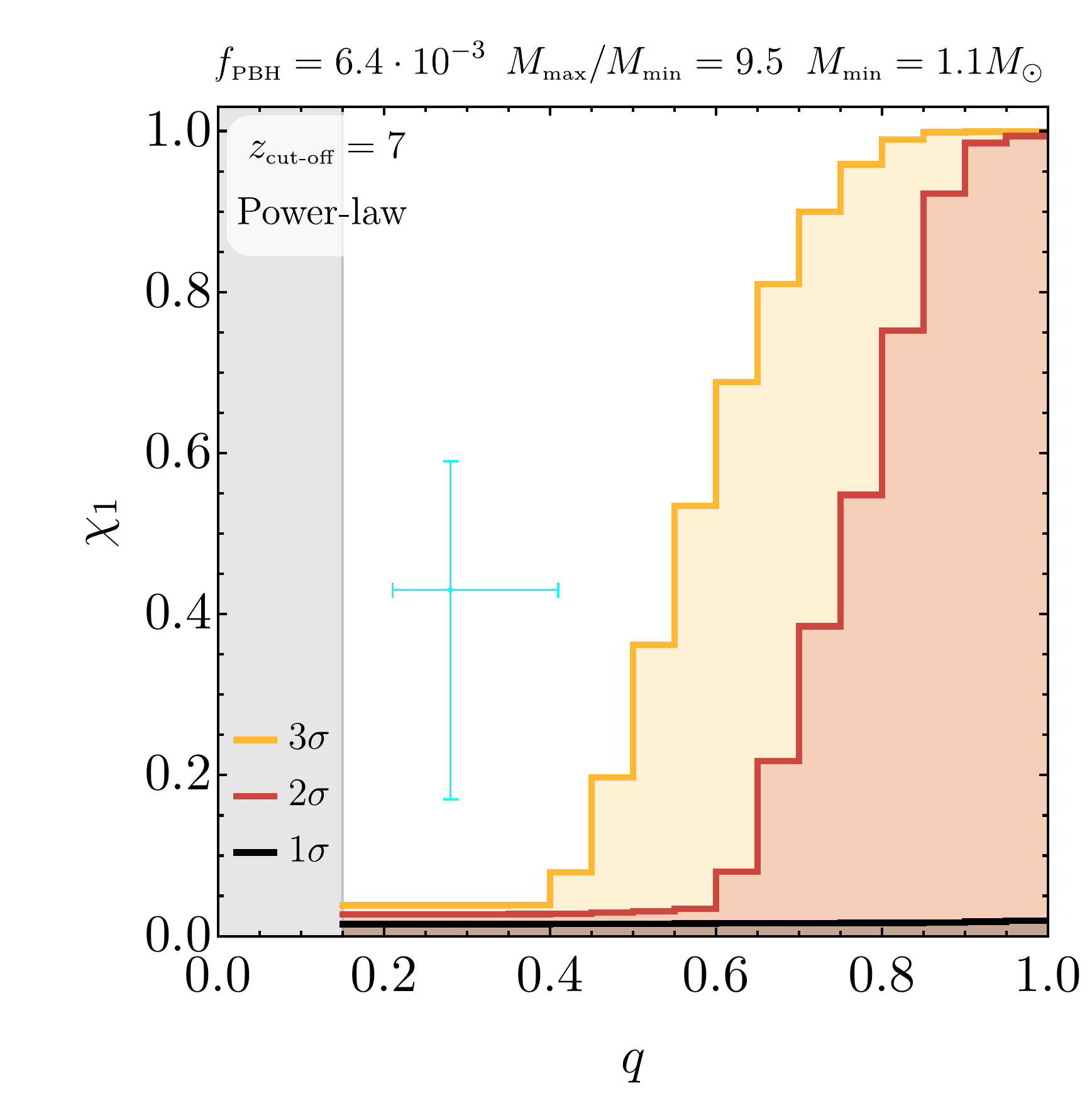}
	
	\includegraphics[width=0.23 \linewidth]{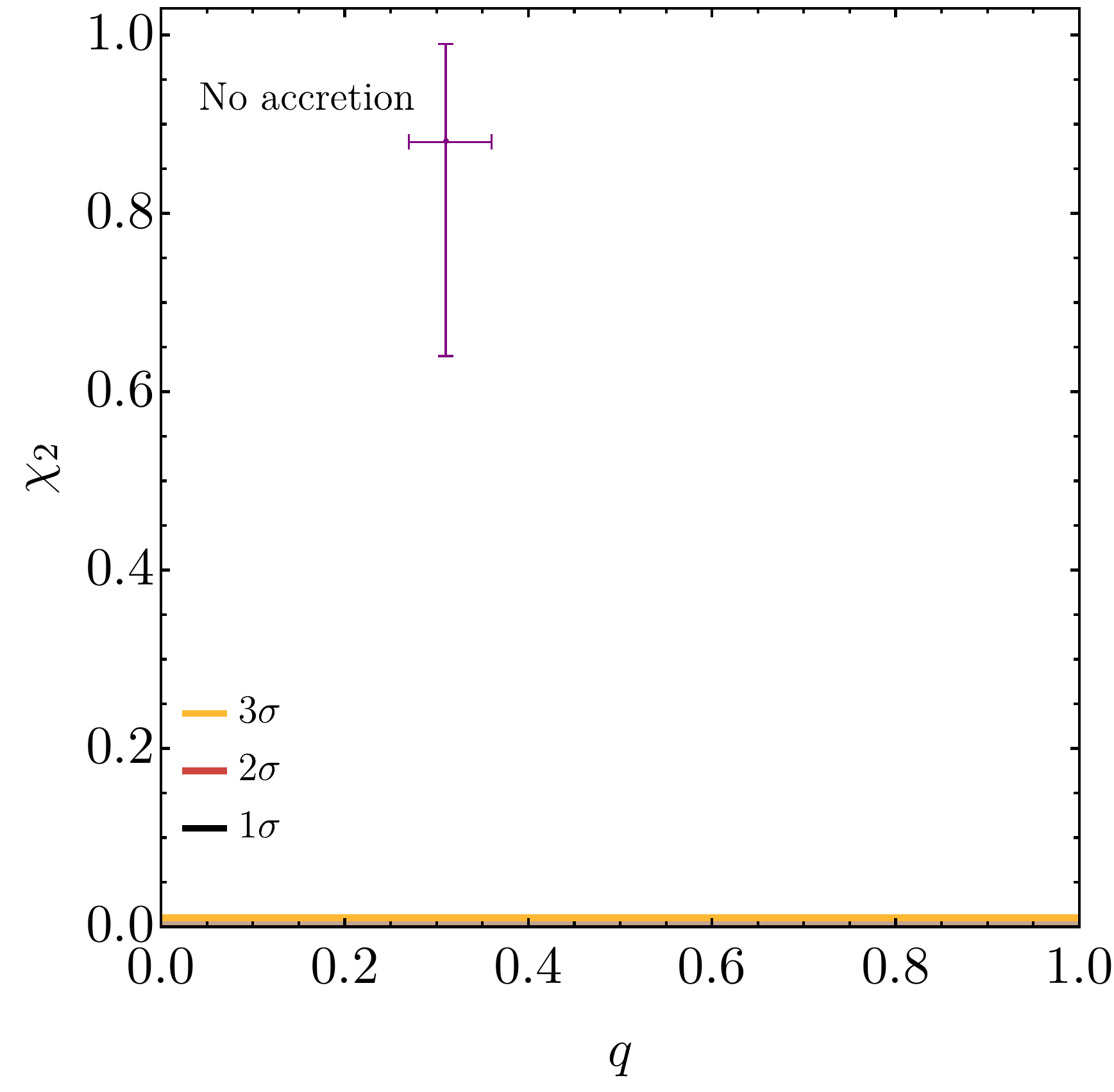}
	\includegraphics[width=0.24 \linewidth]{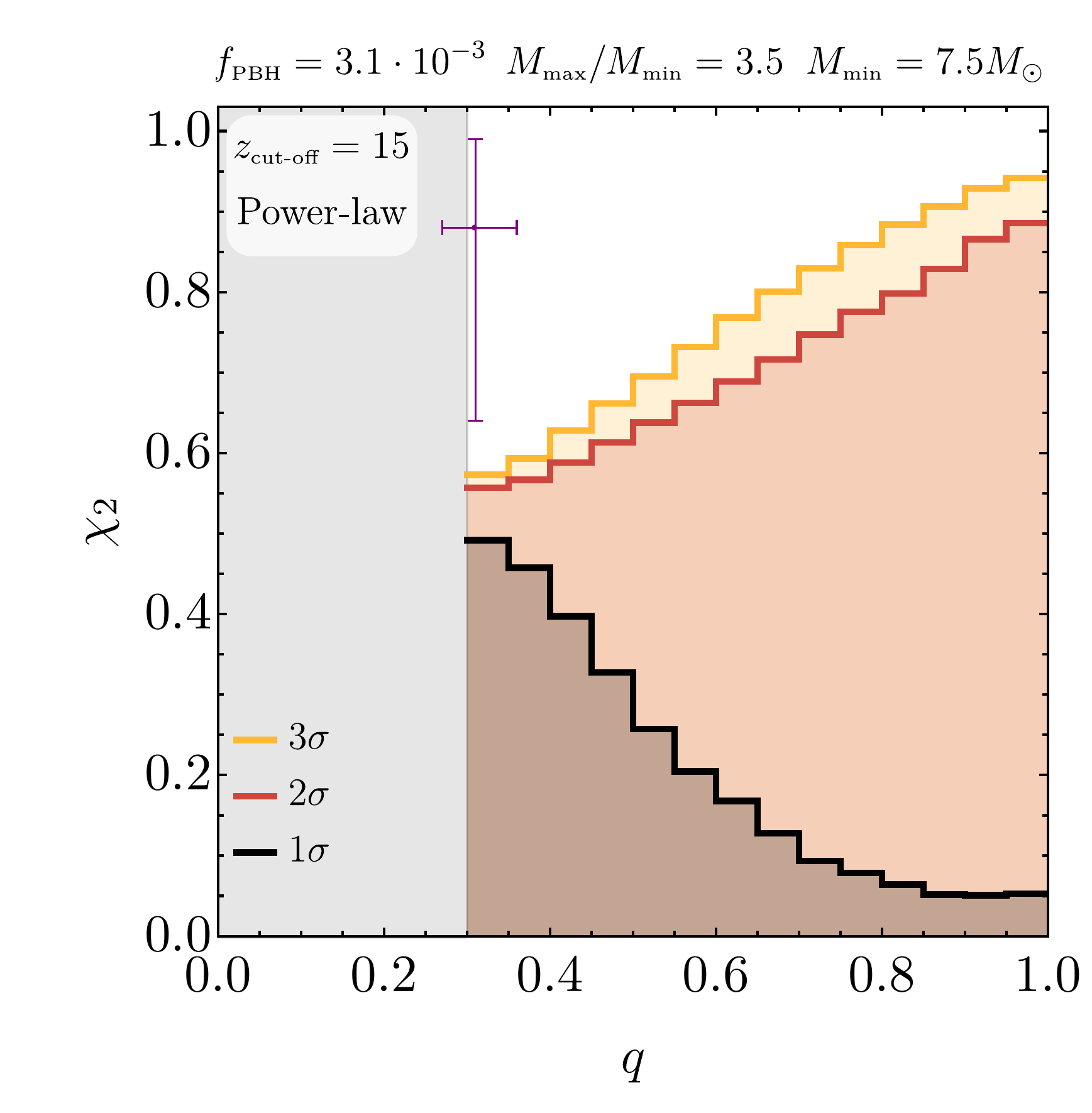}
	\includegraphics[width=0.24 \linewidth]{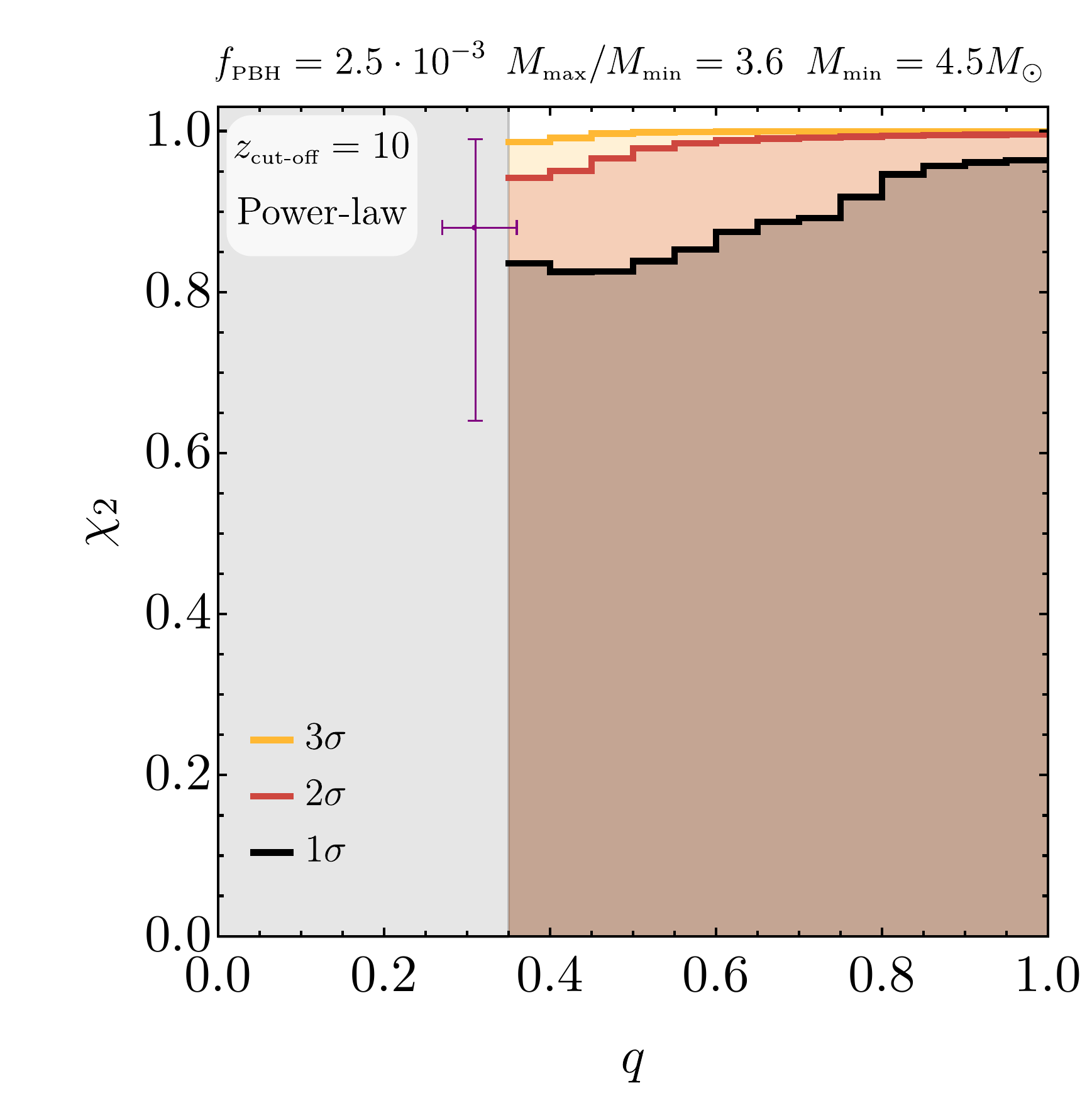}
	\includegraphics[width=0.24 \linewidth]{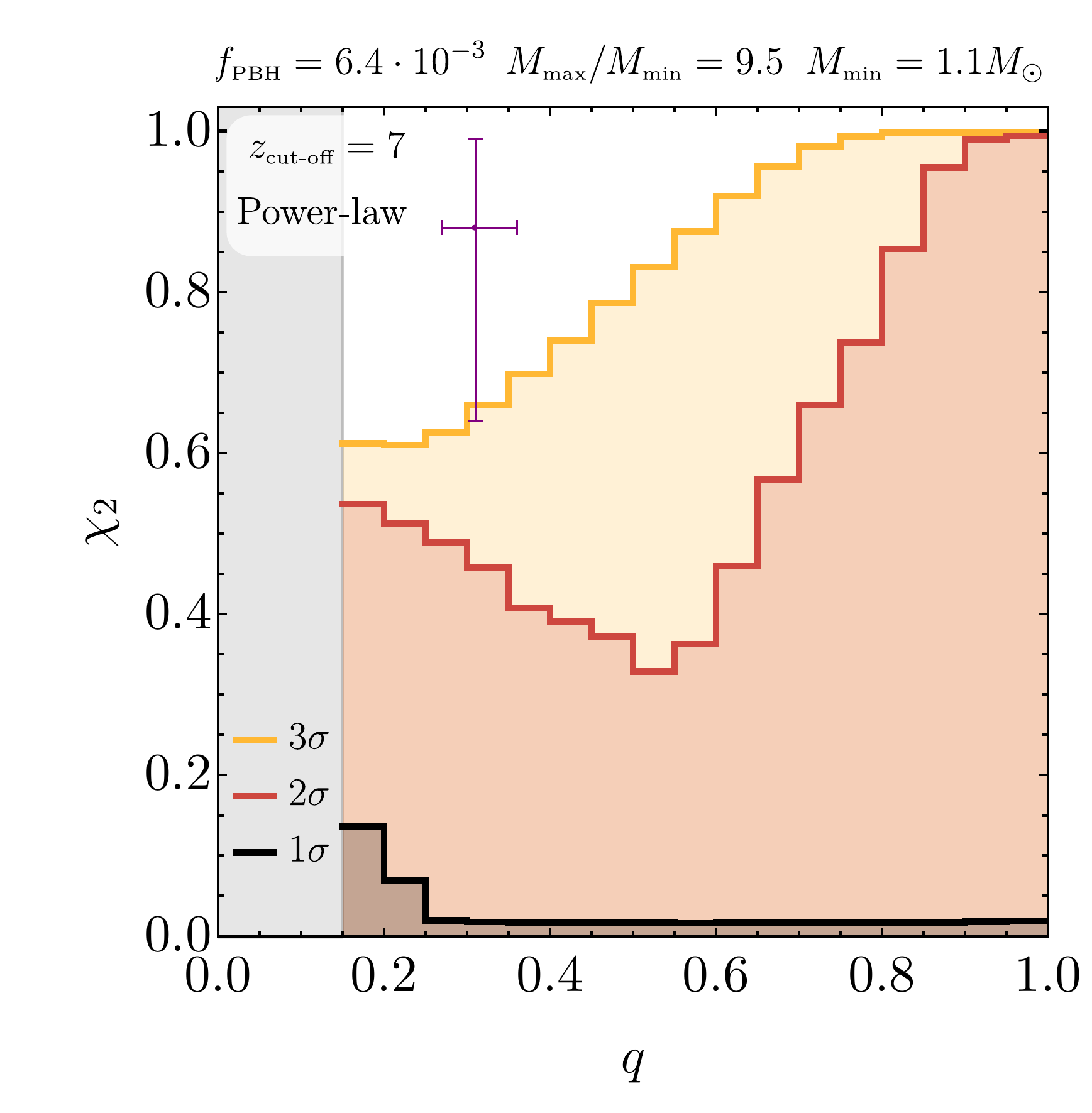}
	
	\includegraphics[width=0.23 \linewidth]{Plots/qchi1.pdf}
	\includegraphics[width=0.24 \linewidth]{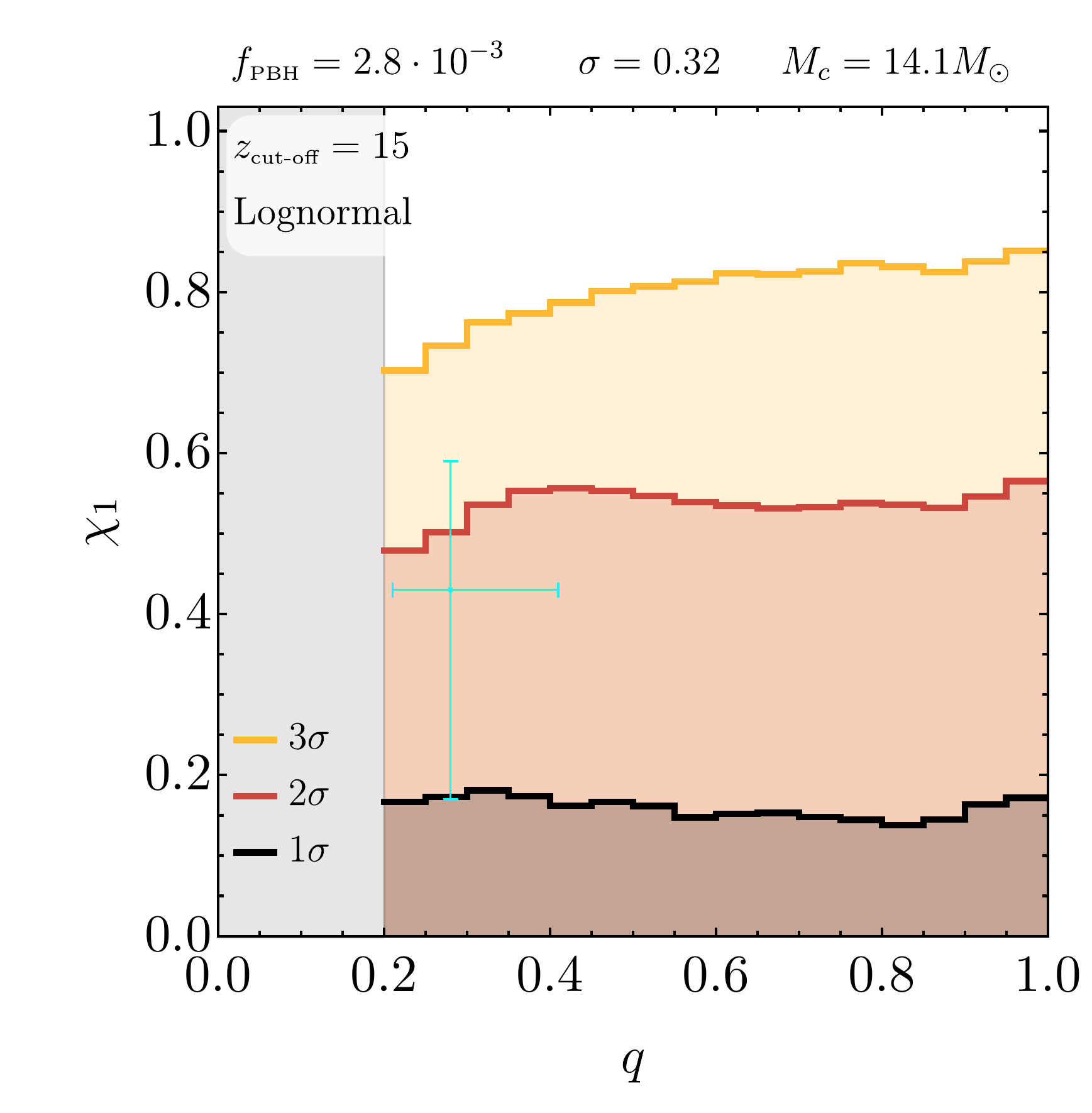}
	\includegraphics[width=0.24 \linewidth]{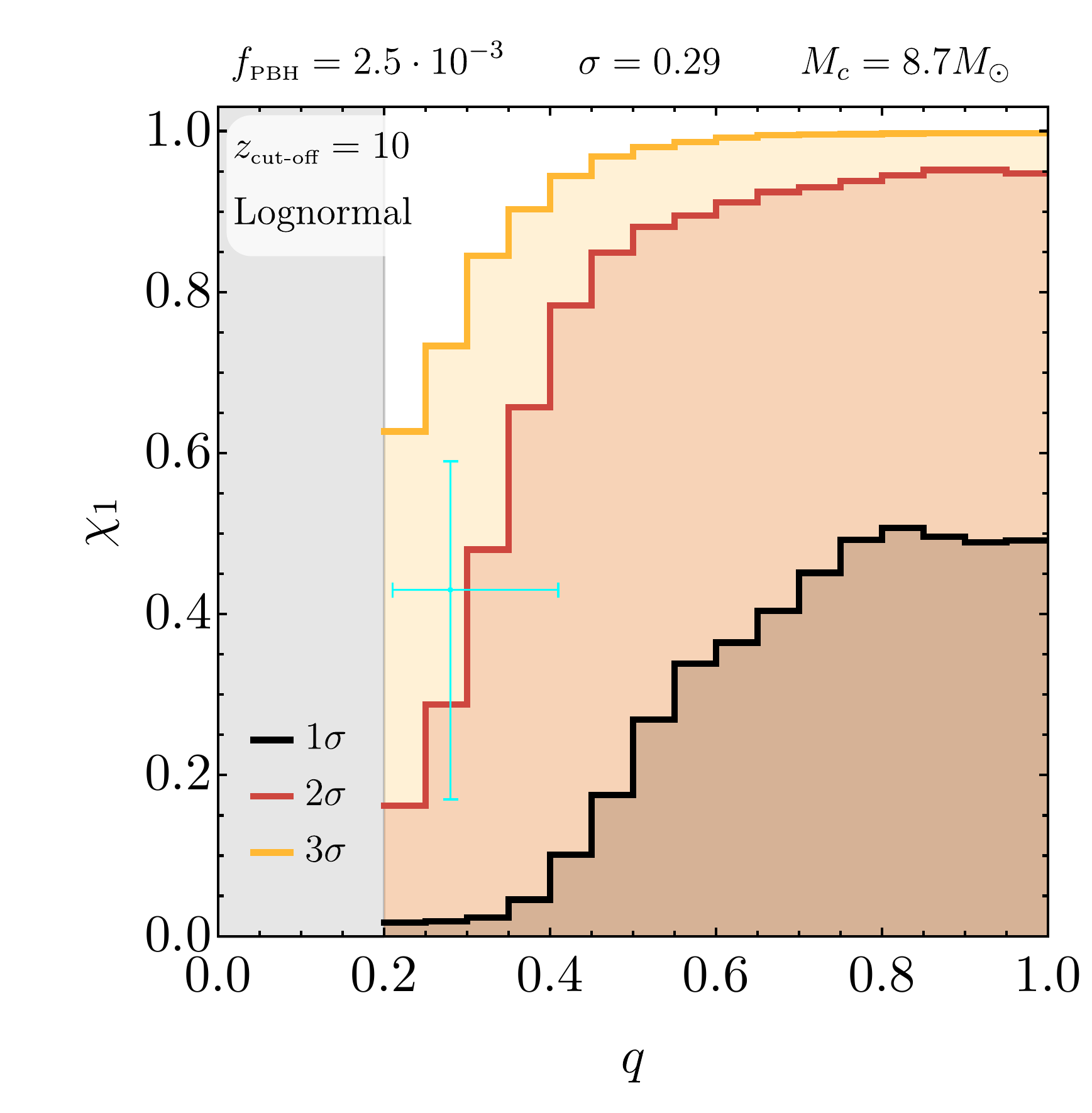}
	\includegraphics[width=0.24 \linewidth]{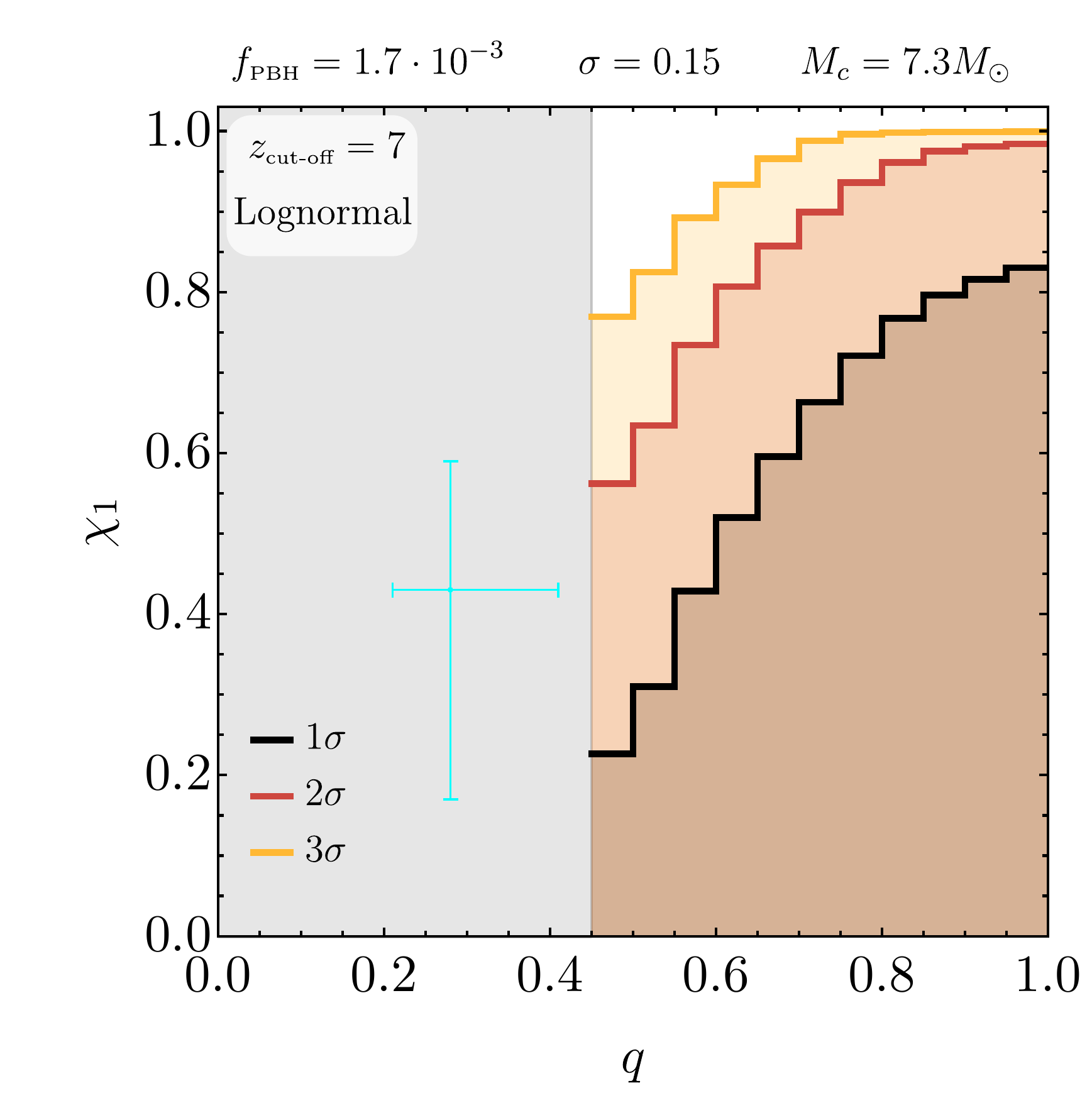}
	
	\includegraphics[width=0.23 \linewidth]{Plots/qchi2.pdf}
	\includegraphics[width=0.24 \linewidth]{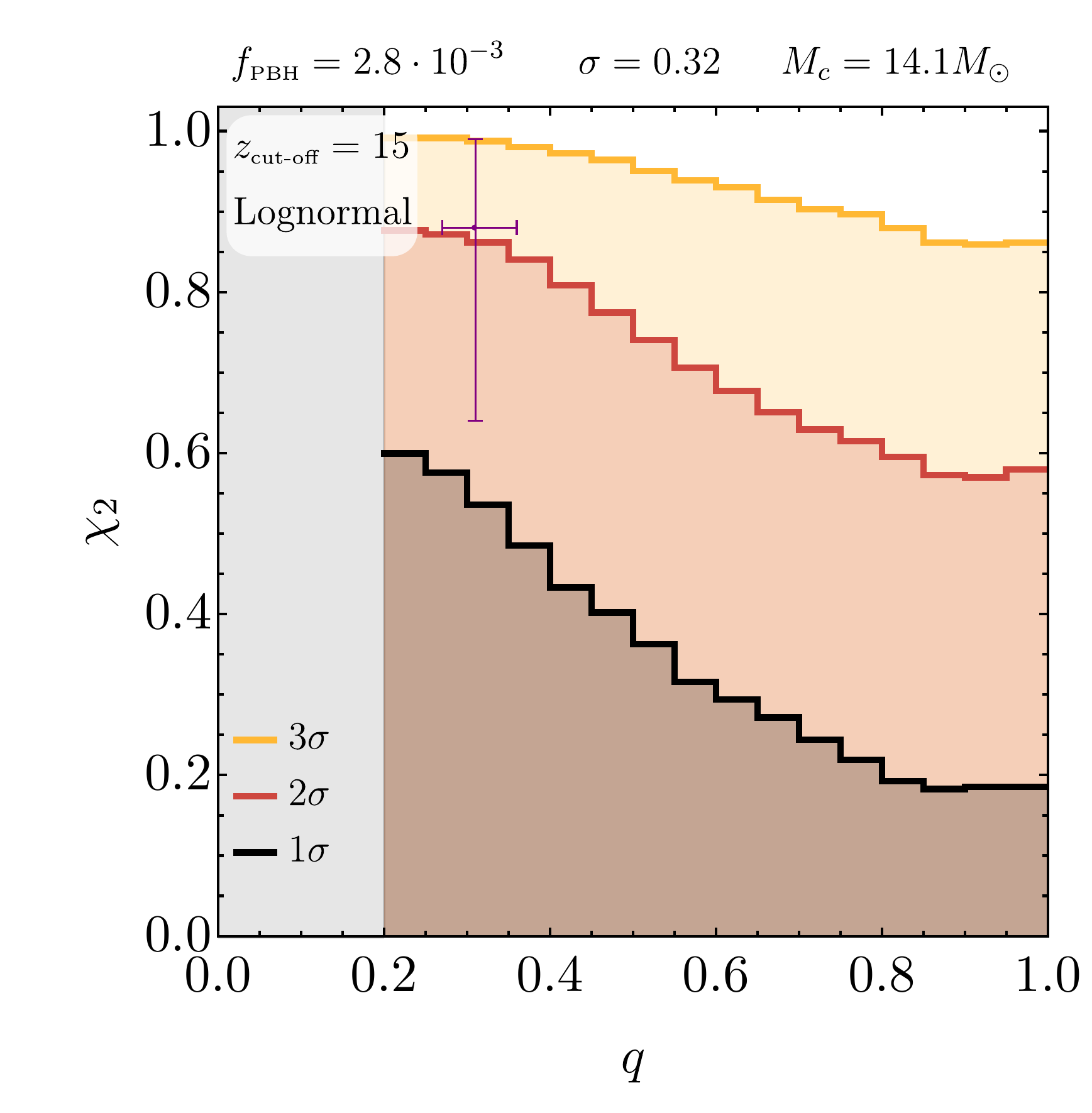}
	\includegraphics[width=0.24 \linewidth]{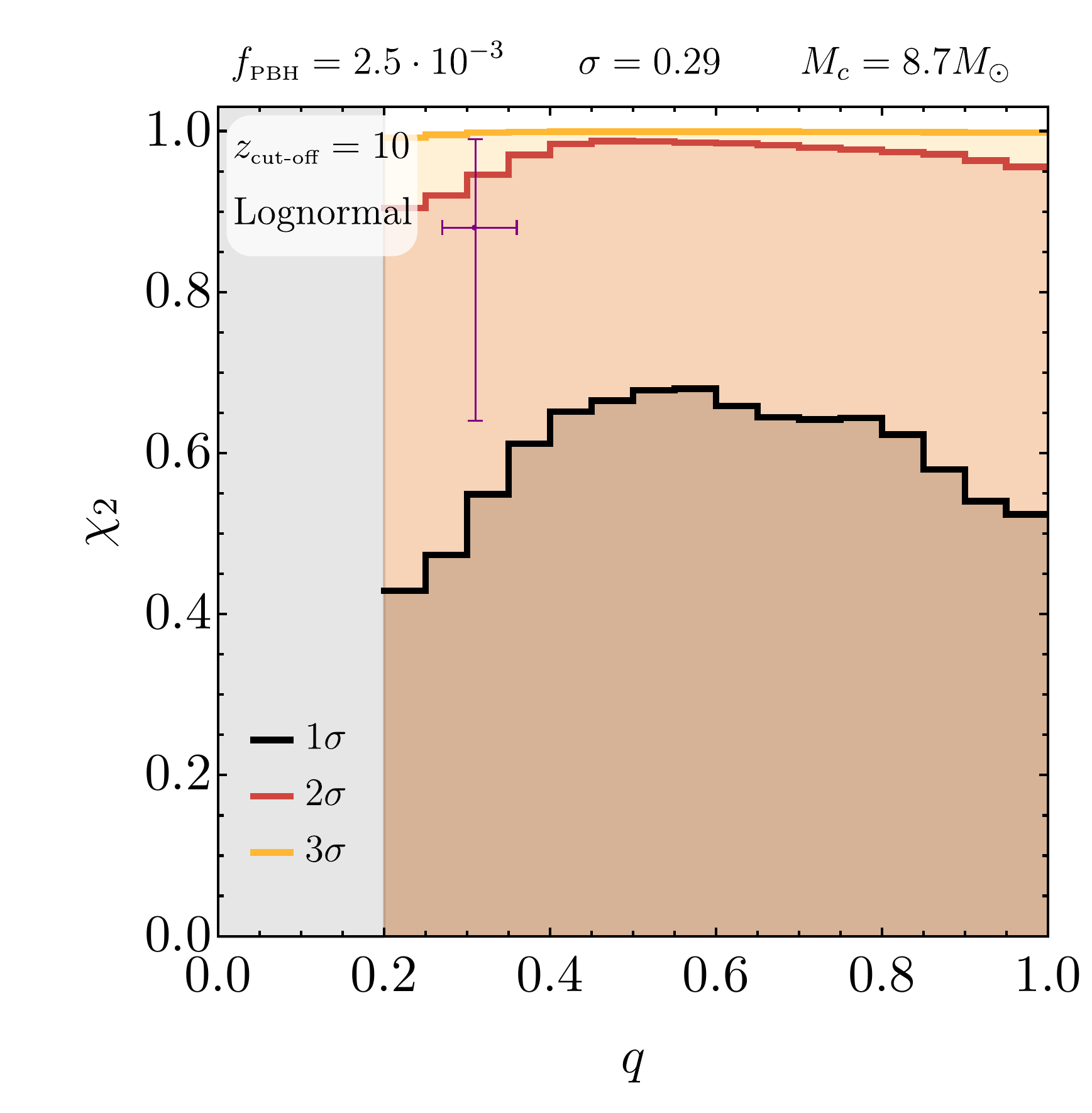}
	\includegraphics[width=0.24 \linewidth]{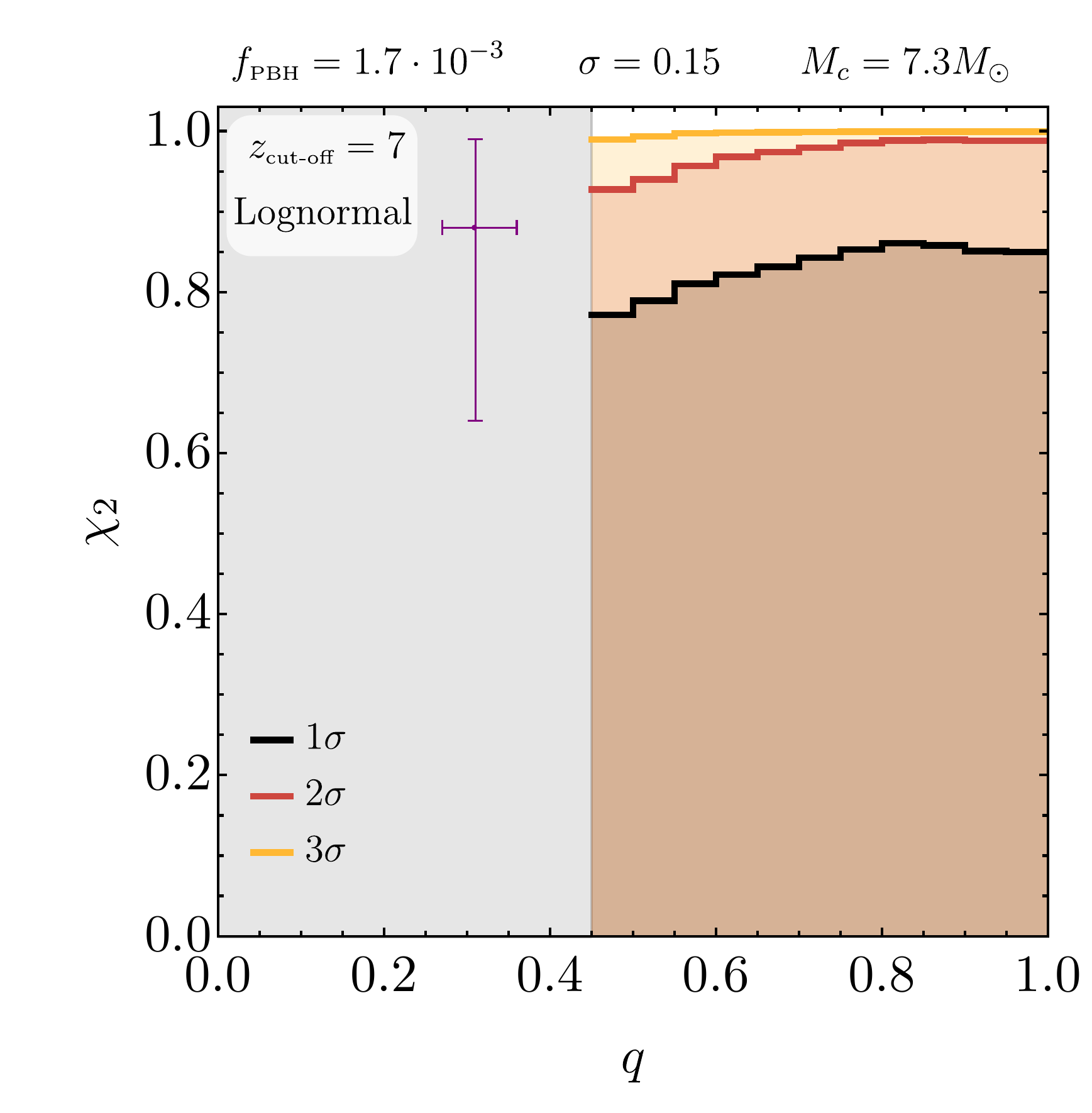}

\caption{\it Observable distribution of the individual spins as a function of $q$ for the power-law (first two blocks) 
and the lognormal (last two blocks) mass functions. The cyan data point refers to the measurement of $\chi_1$ for 
GW190412 provided in Ref.~\cite{LIGOScientific:2020stg}, whereas the purple data point to the measurement of $\chi_2$ 
for the same system as provided in Ref.~\cite{Mandel:2020lhv} (where a prior $\chi_1=0$ was imposed). The grey bands indicate those regions where the PBH model does not provide a number of observable events  with a sufficiently high statistical significance.
	}
\label{chis}
\end{figure}

Finally, let us now turn our attention to the distributions of the binary spins. In Fig.~\ref{chis} we plot the 
observable distributions of the individual spins as a function of $q$ with the same color code chosen to reflect the 1-, 
2- and 3-$\sigma$ regions of Fig.~\ref{chieffdistsf}. For each bin in $q$ we have simulated a number of binaries and 
built a distribution weighting the individual contributions by the relative detection rates. Notice that for the less 
massive PBH component of the binary, the corresponding spin $\chi_2$ is always larger than the spin $\chi_1$ of the more 
massive component. This is predicted by the spin evolution described in Sec.~\ref{subsec:spinevol} and is due to the 
fact that $\dot{m}_2$ is increased by a factor $q^{-1/2}$ with respect to $\dot{m}_1$. Also, for large values of $q$, 
the magnitude of the spin $\chi_1$ increases when accretion becomes stronger (smaller $z_\co$), showing a strong 
correlation between large values of $q$ and the spins.

It is interesting to note that at least one of the components of GW190412 is moderately spinning\footnote{The spins 
of the binary components in the O1-O2 events are less constrained. Among those events, only the effective spin of 
GW151226 is confidently different from zero~\cite{Miller:2020zox}.}. Indeed, assuming
priors that are uniform in the spin magnitudes and isotropic in the directions, the LIGO/Virgo collaboration estimated 
$\chi_1=0.43^{+0.16}_{-0.26}$, while $\chi_2$ is essentially unconstrained. On the other hand, using 
astrophysically-motivated priors, Ref.~\cite{Mandel:2020lhv} has imposed a non-spinning primary component\footnote{However, we expect
that the posterior on ${\vec \chi_2} \cdot {\hat L}$ might be affected
by a slightly different choice of the prior on $\chi_1$, e.g. a
uniform prior $\chi_1\in[0,0.1]$.} and a uniform prior
on ${\vec \chi_2} \cdot {\hat L}$, inferring ${\vec \chi_2} \cdot {\hat L}=0.88^{+0.11}_{-0.24}$. The latter can also 
be used as an estimate on $\chi_2$, assuming the orbit is not significantly tilted by the natal kick during the 
supernovae that produced the secondary.
As shown in Fig.~\ref{chis}, both cases inferred by the analyses of Refs.~\cite{LIGOScientific:2020stg,Mandel:2020lhv} 
are incompatible with a primordial origin for GW190412 unless: (i) accretion is 
significant during the cosmic history of PBHs; (ii) PBHs can be formed with non-negligible natal 
spin (as in some scenarios~\cite{Harada:2017fjm,Cotner:2017tir}), 
in contrast with the most likely formation scenarios~\cite{DeLuca:2019buf}, in which the spin is at the percent level; 
(iii) GW190412 is actually a higher-generation merger~\cite{Gerosa:2017kvu}, in which the spinning binary underwent a 
previous merger in the past. This possibility has been recently explored in Refs.~\cite{1795113,1795101,Olejak:2020oel}, 
under the hypothesis that GW190412 originates from first-generation or hierarchical mergers of astrophysical origin in 
the context of both the field and cluster formation scenarios.
%
However, in the case of PBHs the possibility of hierarchical merger is less likely, since the merger rates for 
higher-generation mergers are much smaller than those of first-generation mergers~\cite{Liu:2019rnx,Wu:2020drm,paper1}. 
The possibility that GW190412 was a higher-generation merger (of either astrophysical or primordial origin) requires a 
better assessment of its spin and a comparison with multiple-generation scenarios. A robust assessment will probably 
require to wait for more events like GW190412 in O3 or future observational runs\footnote{Recently, using a 
phenomenological model for BH mergers in dense stellar environments, Ref.~\cite{Kimball:2020opk} 
estimated that only a small fraction of the BH binaries in O1 and O2 can have negligible spin. If this result extends to 
the PBH scenario, it would provide further indication that non-accreting PBHs born with negligible spin can at most 
comprise a small fraction of the LIGO/Virgo merger events. We thank Christopher Berry for useful correspondence on 
this point.}.

\begin{figure}[t!]
	\centering
	 \includegraphics[width=0.237 \linewidth]{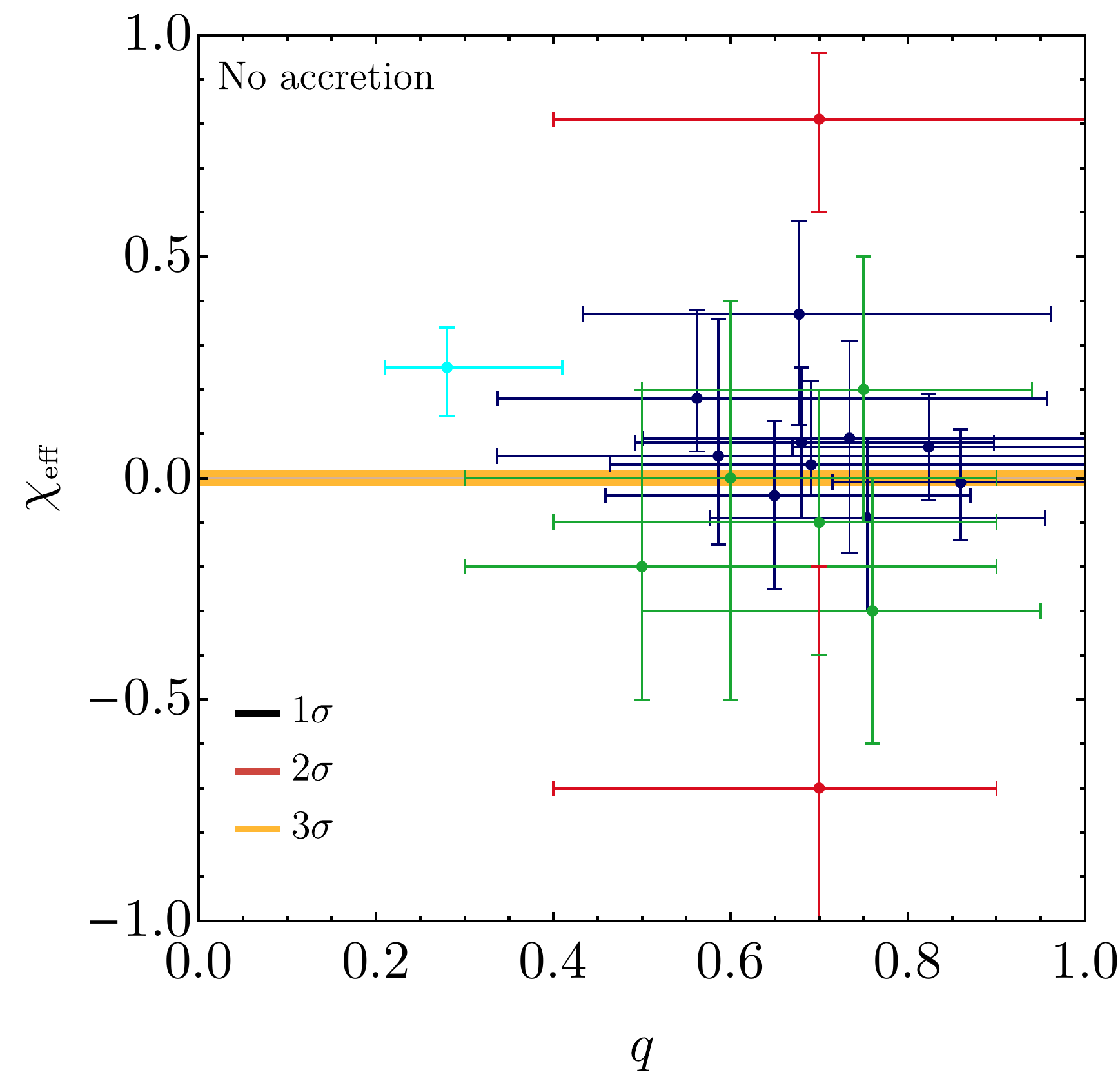}
	 \includegraphics[width=0.24 \linewidth]{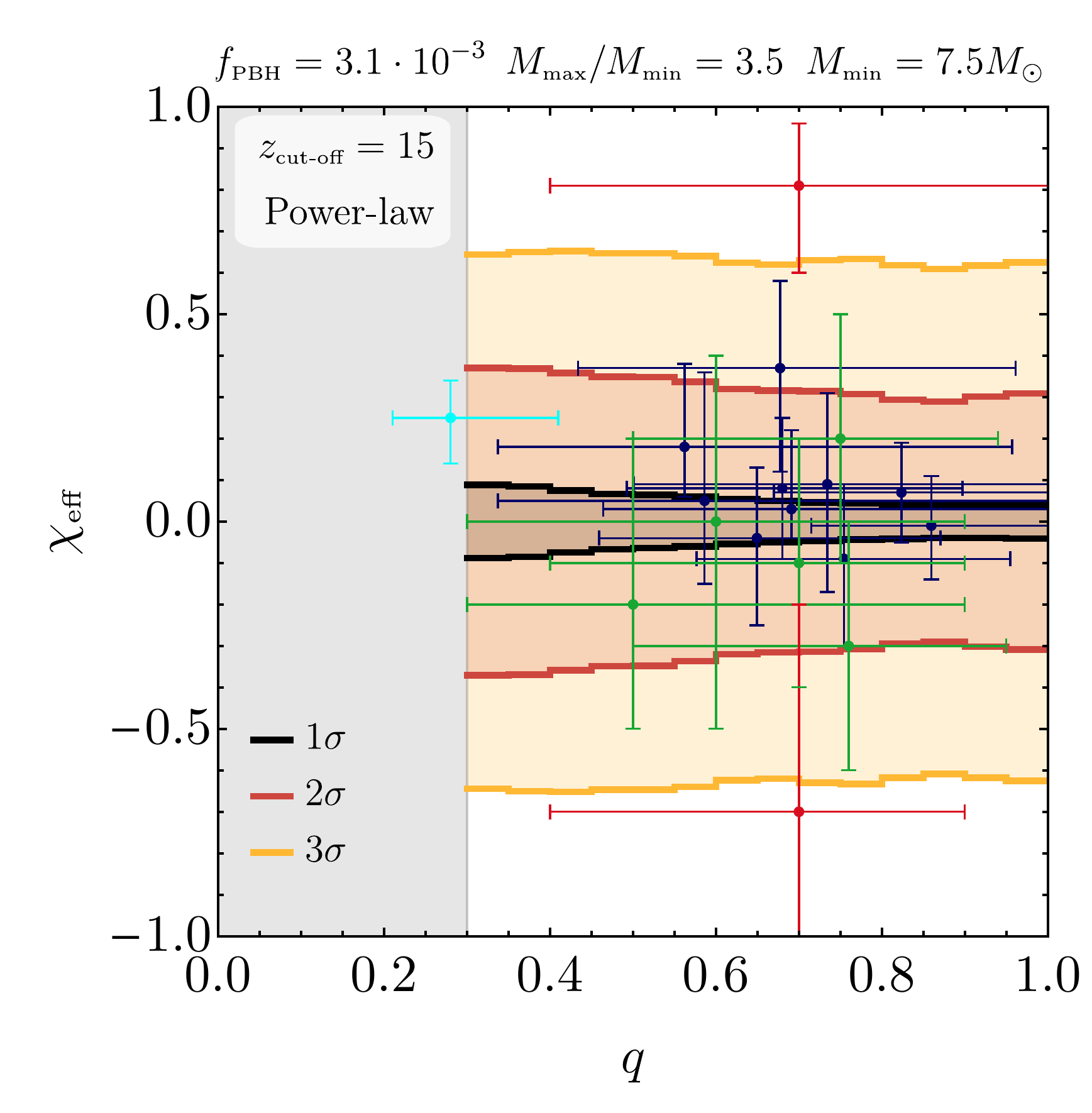}
	\includegraphics[width=0.24 \linewidth]{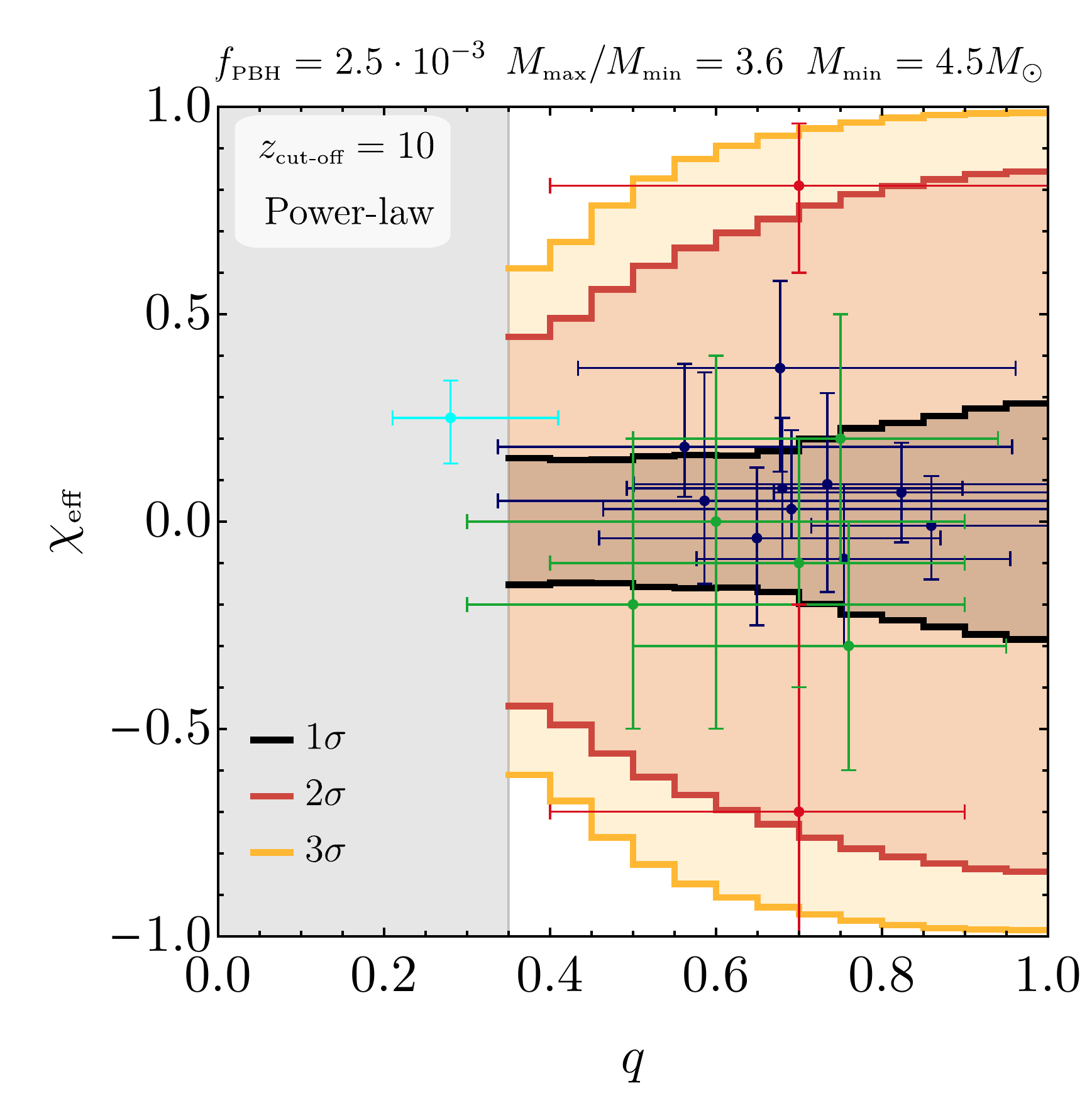}
	\includegraphics[width=0.24 \linewidth]{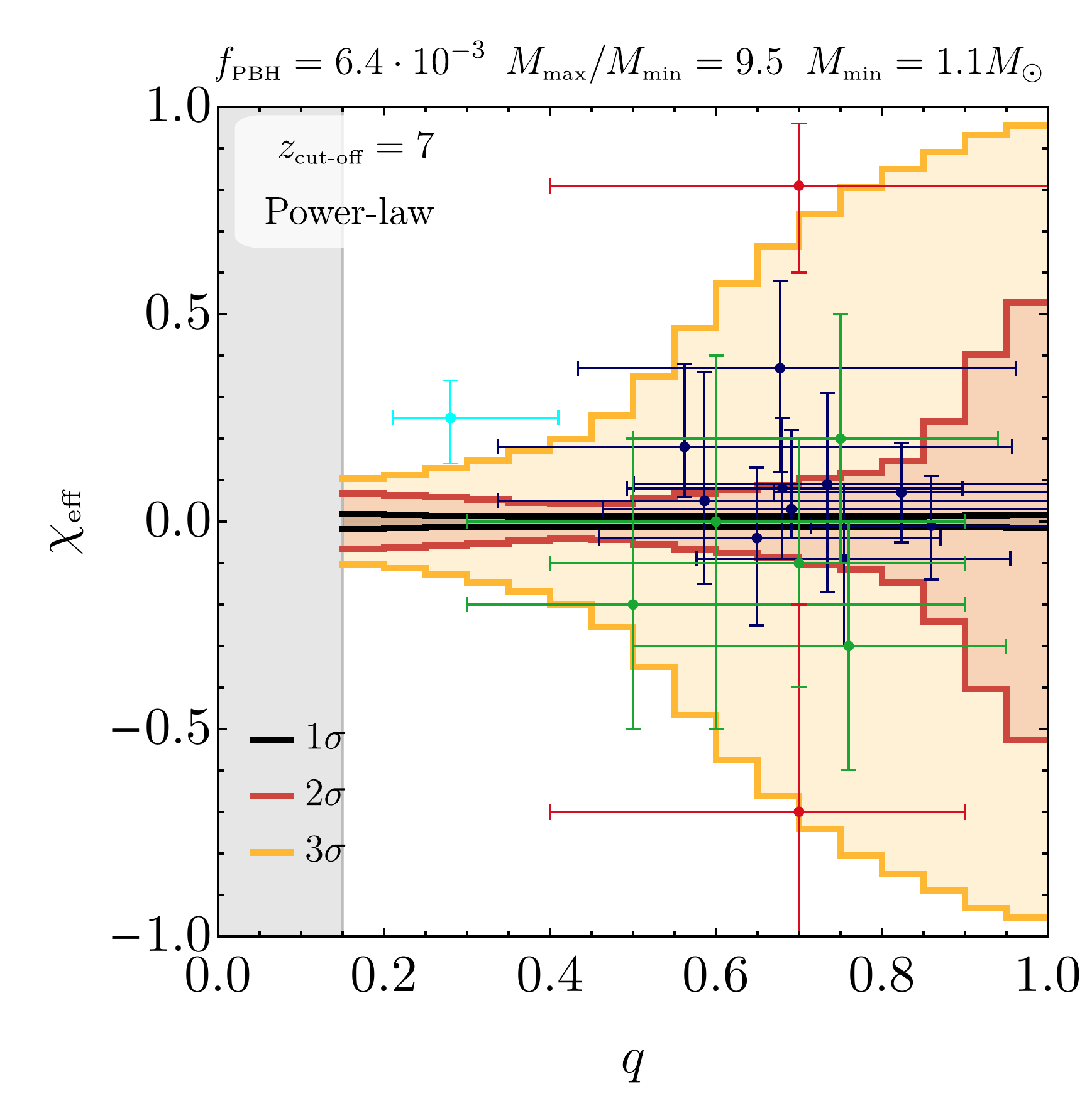}
	
	\includegraphics[width=0.237 \linewidth]{Plots/qchieff.pdf}
	\includegraphics[width=0.24 \linewidth]{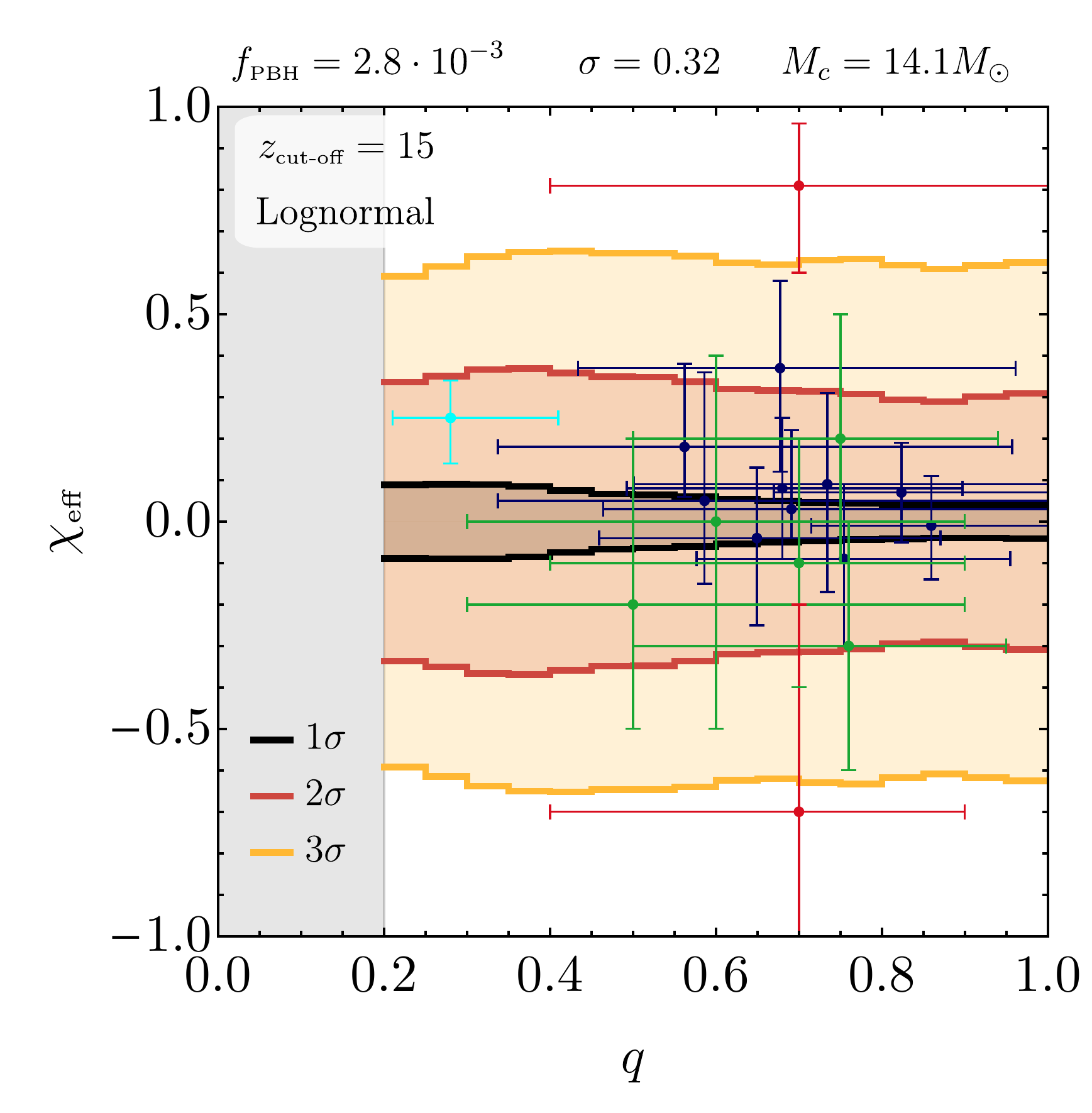}
	\includegraphics[width=0.24 \linewidth]{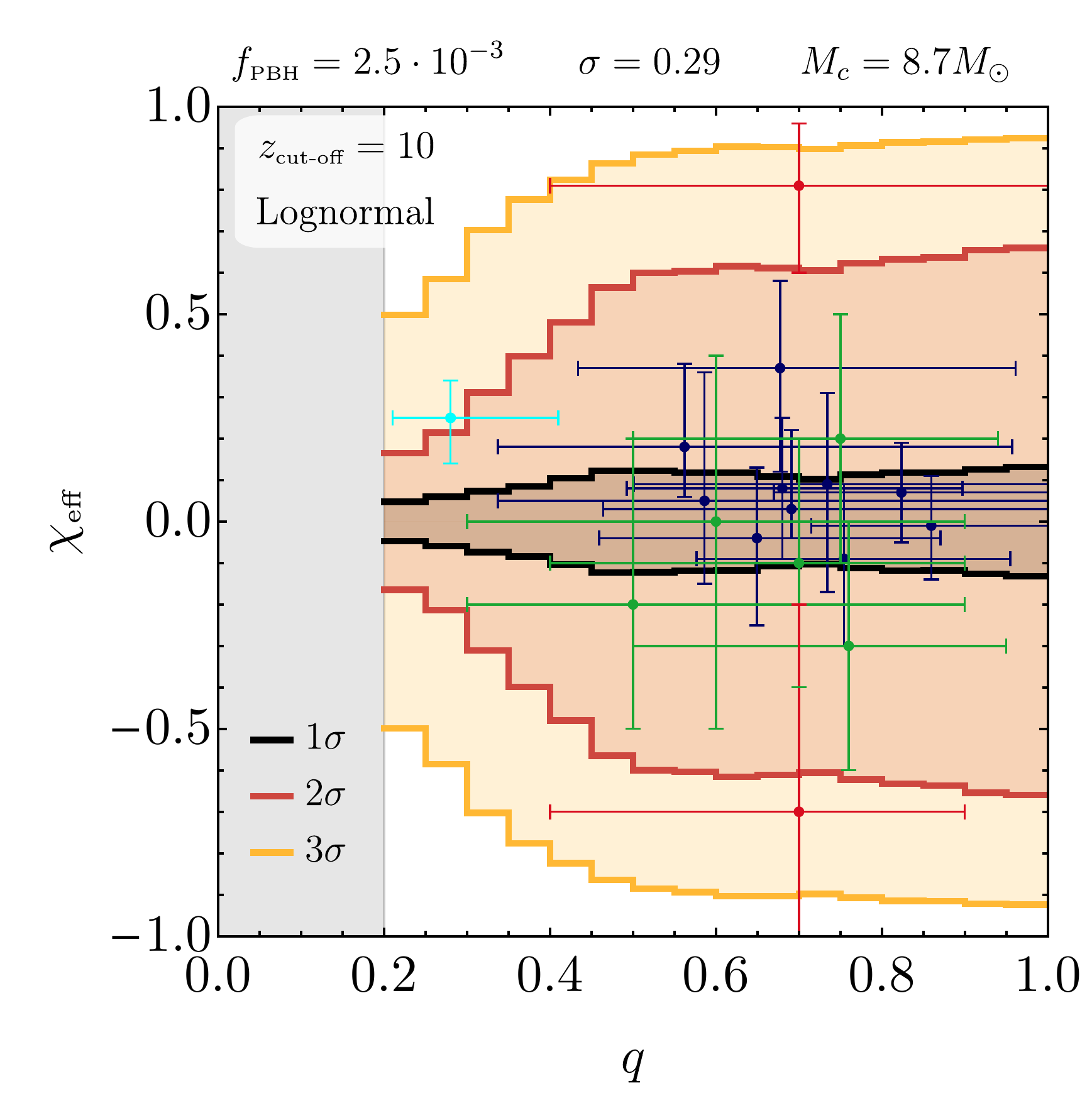}
	\includegraphics[width=0.24 \linewidth]{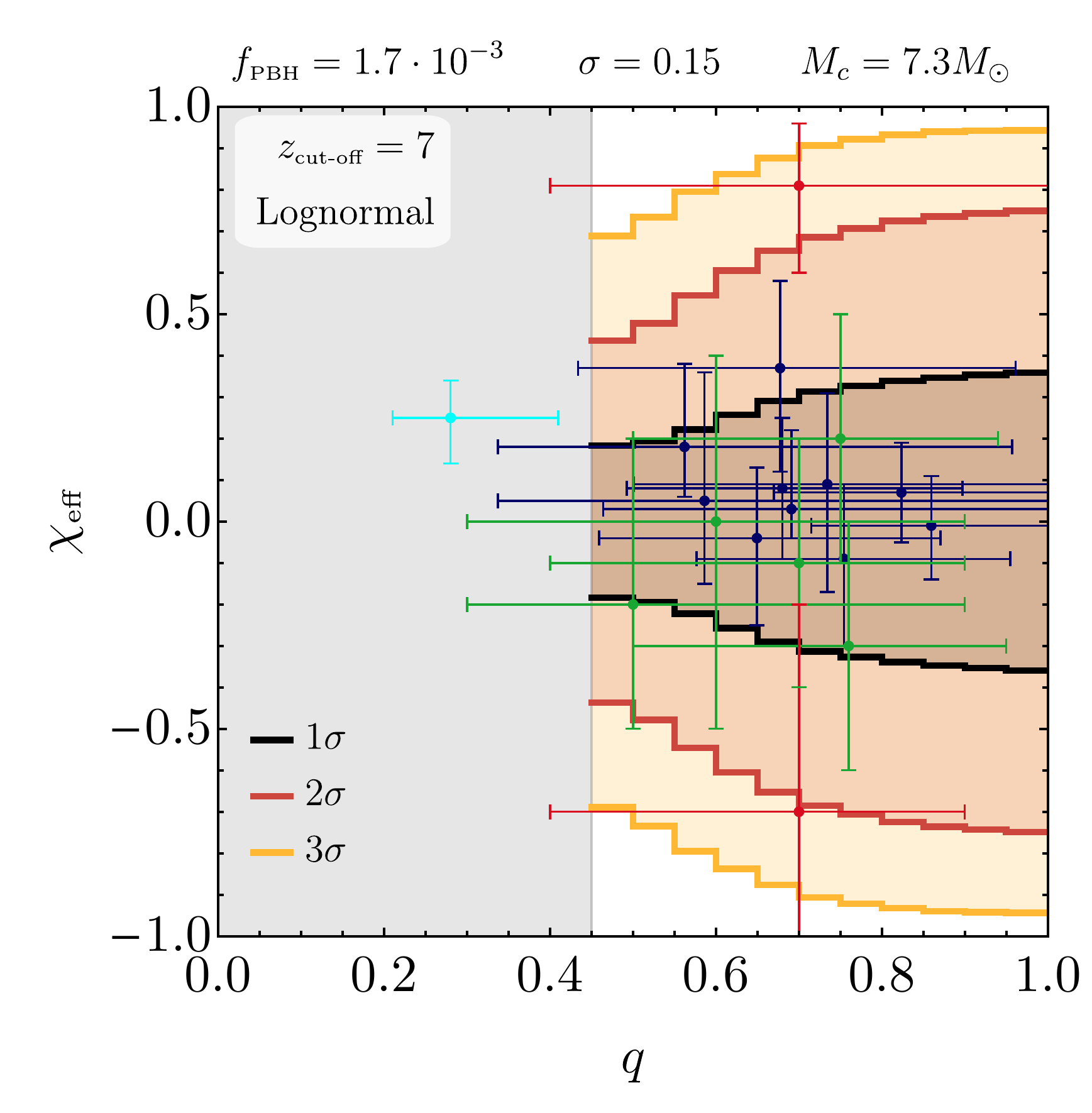}
	
	\caption{\it Observable distribution of the effective spin as a function of $q$ compared to the data 
available for the power-law (first block) and the lognormal (last block) mass functions. The structure of the panels 
and the grey bands have the same meaning as in  Fig.~\ref{chis}.
}
\label{chieff}
\end{figure}

Finally, in Fig.~\ref{chieff} we plot the observable distribution of the effective spin as a function of $q$ compared 
to the data available. These plots show, when accretion is important, a strong 
correlation between $\chi_{\text{\tiny eff}}$ and $q$. Obviously, in the case of no accretion, the effective spin 
parameter is vanishing, owing to the initial conditions of the individual spins.

\section{Conclusions: Key predictions of the PBH scenario for GW astronomy} \label{sec:keypredictions}
With the current and upcoming wealth of data on BH binaries from the LIGO/Virgo observatories, it becomes possible 
to perform model selection and rule out or corroborate specific formation scenarios for BHs in the LIGO/Virgo band.
We have investigated whether current data --~including O1, O2 and the recently discovered GW190412~-- are compatible 
with the hypothesis that LIGO/Virgo BHs are of primordial origin.

We conclude by summarising our main findings and listing the main predictions of the PBH scenario, which can be 
directly tested with current and future GW observations:

\begin{enumerate}
 \item For a given PBH mass distribution at formation, current merger rates set an upper bound on the PBH abundance 
at the level of $f_{\PBH}\lsim (10^{-2}\div 10^{-3})$, depending on the mass function. The best-fit values selected by the 
likelihood provide mass distributions which are optimally compatible with current events. For the case of a lognormal 
distribution and no accretion, our best-fit parameters are in agreement with the recent analysis in 
Ref.~\cite{Dolgov:2020xzo} within $1\sigma$. In addition, we found that accretion can alter 
quantitatively the distributions,  but not the qualitative aspects of this analysis. 
 \item Although not directly relevant for the LIGO/Virgo frequency range, it is intriguing that 
accretion can be very efficient for PBHs with initial mass $M^\text{\tiny i}\gtrsim 10 M_\odot$, increasing the 
latter by some orders of magnitude during the cosmic history. Thus, it might occur that PBHs formed with masses $\sim 
(10\div100)M_\odot$ can have much larger masses when detected at smaller redshift. These objects would be natural 
candidates for intermediate-mass BHs, which are sources for ET and LISA. If these objects were born in the stellar-mass 
range in the primordial universe, they should also have nearly-extremal spin and mass ratio close to unity.
 \item The fact that at least one of the components of GW190412 is moderately spinning is incompatible with a 
primordial origin for this event, unless: (i)~PBHs are born with non-negligible spin, in contradiction with the most 
likely formation scenarios; (ii)~at least the spinning component of GW190412 is a second-generation BH (however, this 
would require a better assessment of its spin and a comparison with multiple-generation merger scenarios, which are 
unlikely for PBHs~\cite{paper1}); (iii) accretion is 
significant during the cosmic history of PBHs.
 \item Despite the uncertainties in the accretion modelling, the role of accretion onto PBHs is manifold. Due to the 
effect of super-Eddington accretion at $z\approx 30$, and to the fact that accretion onto the smaller 
binary component is stronger than on the primary, the mass ratio of PBH binaries with total mass above a 
transition value, $M_{\text{\tiny tot}}\gsim 10M_\odot$ tends to be close 
to unity. The precise value of the transition depends on the cut-off redshift at which accretion ceases to be efficient. 
This produces a peculiar distribution of the binary chirp mass (Fig.~\ref{chirp}) and mass ratio
(Fig.~\ref{q}), which is absent in the stellar-origin scenario. However, for the best-fit values of the lognormal 
distribution, the effect of accretion on the expected distribution of $q$ for the detected events is small.

 \item Overall, in the case of efficient accretion the spins of the binary components at detection are non-negligible and 
can be close to extremality for individual PBH with final masses $M^\ii\gsim(10\div 30) M_\odot$. The transition value 
depends on 
the initial mass ratio and on the cut-off redshift.

 \item For the same reason as above, assuming PBHs are formed with negligible angular momentum, the spin of the smaller 
binary component at detection is always higher than the one of the primary, unless hierarchical mergers occur (which 
are, however, unlikely~\cite{paper1}).

 \item In all cases, the final spin of the secondary is close to extremality if the mass of the primary $M_1\gsim 40 
M_\odot$.
 \item It is worth noting that --~if accretion is inefficient~-- current non-GW constraints (in 
particular from the absence of extra CMB distortions~\cite{Ali-Haimoud:2016mbv,serpico}) already exclude that LIGO/Virgo 
events are all of primordial origin, whereas in the presence of accretion the GW bounds on the PBH abundance are the 
most stringent ones in the relevant mass range.
 \item Overall, a strong phase of accretion during the cosmic history would favour mass ratios close 
to unity, and a redshift-dependent correlation between high PBH masses, high spins, and $q\approx1$. These correlations 
can be used to distinguish the accreting PBH scenario from that of astrophysical-origin BH binaries. 
\item Extreme accretion scenarios (in our study represented by  
$z_\co\lesssim7$) predict that most of the events should  be clustered around 
$q\sim1$. This is in tension with the recent GW190412 data.  
\item The individual spin evolution results in a broad distribution of the effective spin parameter of the binary, 
which is compatible with the observed distribution of the GW events detected so 
far~\cite{LIGOScientific:2018jsj,LIGOScientific:2018mvr,Zackay:2019tzo,Venumadhav:2019lyq,Huang:2020ysn}, including 
GW190412~\cite{LIGOScientific:2020stg}. In particular, in the accreting PBH scenario the dispersion of 
$\chi_\text{\tiny eff}$ around zero grows with the mass (as GW data suggest~\cite{Safarzadeh:2020mlb}), with a dispersion $ 
 \approx0$ for low binary masses and ${\cal O}(1)$ at 
larger masses.
 \item The above properties produce a peculiar distribution of $\chi_i$ and $\chi_\text{\tiny eff}$ as a function of the mass 
ratio (Figs.~\ref{chis} and \ref{chieff}), which is absent in the stellar-origin scenario. 

\item According to our theoretical predictions, low mass PBHs, $M\lsim {\cal O}(10) M_\odot$, should have tiny spins. 
This property might help to distinguish PBH binaries from those composed by neutron stars~\cite{Fasano:2020eum}, in the 
case the spin of the latter is non-negligible. 
\end{enumerate}

It is worth mention that --~in confronting our theoretical predictions with GW observations~-- we have relied on the 
measurements obtained by the LIGO/Virgo collaboration. The latter are based on agnostic priors on the masses and spins. 
A different choice of the priors --~possibly motivated by a sound PBH scenario~-- might affect the posterior 
distributions of the observed parameters, especially for those which are poorly measured (see 
Ref.~\cite{Mandel:2020lhv} for an example in the context of astrophysical priors for the spin of GW190412). We plan to 
investigate this interesting problem in a future work \cite{inprep}.

Furthermore, given the important role that accretion on PBHs can play in the phenomenology of GW coalescence events and 
in relaxing current constraints on the PBH abundance, we advocate the urgent need of a better modelling of the 
accretion rate at redshift $z\lesssim30$. 

Finally, the question whether the PBHs may be responsible or not of the LIGO/Virgo data has the following  answer: while in the absence of accretion the LIGO/Virgo 
events are incompatible with the primordial nature of BHs,  the situation -- in the presence of  accretion -- is at the 
moment not conclusive, even though the theoretical predictions of the PBH scenario are rather  sharp. The next 
forthcoming data from the O3 campaign and from the Advanced LIGO might provide a more definite answer.


\acknowledgments
\noindent 
We are indebted to Christopher Berry and Luigi Stella for interesting correspondence,
and to Christian Byrnes for spotting a typo in the
value of $\sigma_M$ given in Ref.~\cite{raidal} which mildly affected the suppression factor used in the previous version of this manuscript.
We also thank Andreas Finke,  Francesco Lucarelli and Michele Maggiore for useful discussions.
Some computations were performed at University of Geneva on the Baobab cluster.
V.DL., G.F. and A.R. are supported by the Swiss National Science Foundation 
(SNSF), project {\sl The Non-Gaussian Universe and Cosmological Symmetries}, project number: 200020-178787.
G.F. would like to thank the Instituto de Fisica Teorica (IFT UAM-CSIC) in Madrid for its support via the Centro de Excelencia Severo Ochoa Program under Grant SEV-2012-0249.
P.P. acknowledges financial support provided under the European Union's H2020 ERC, Starting 
Grant agreement no.~DarkGRA--757480, under the MIUR PRIN and FARE programmes (GW-NEXT, CUP:~B84I20000100001), and 
support from the Amaldi Research Center funded by the MIUR program `Dipartimento di 
Eccellenza" (CUP:~B81I18001170001).



\end{document}